# Quantitative Methods in Research Evaluation: Citation Indicators, Altmetrics, and Artificial Intelligence

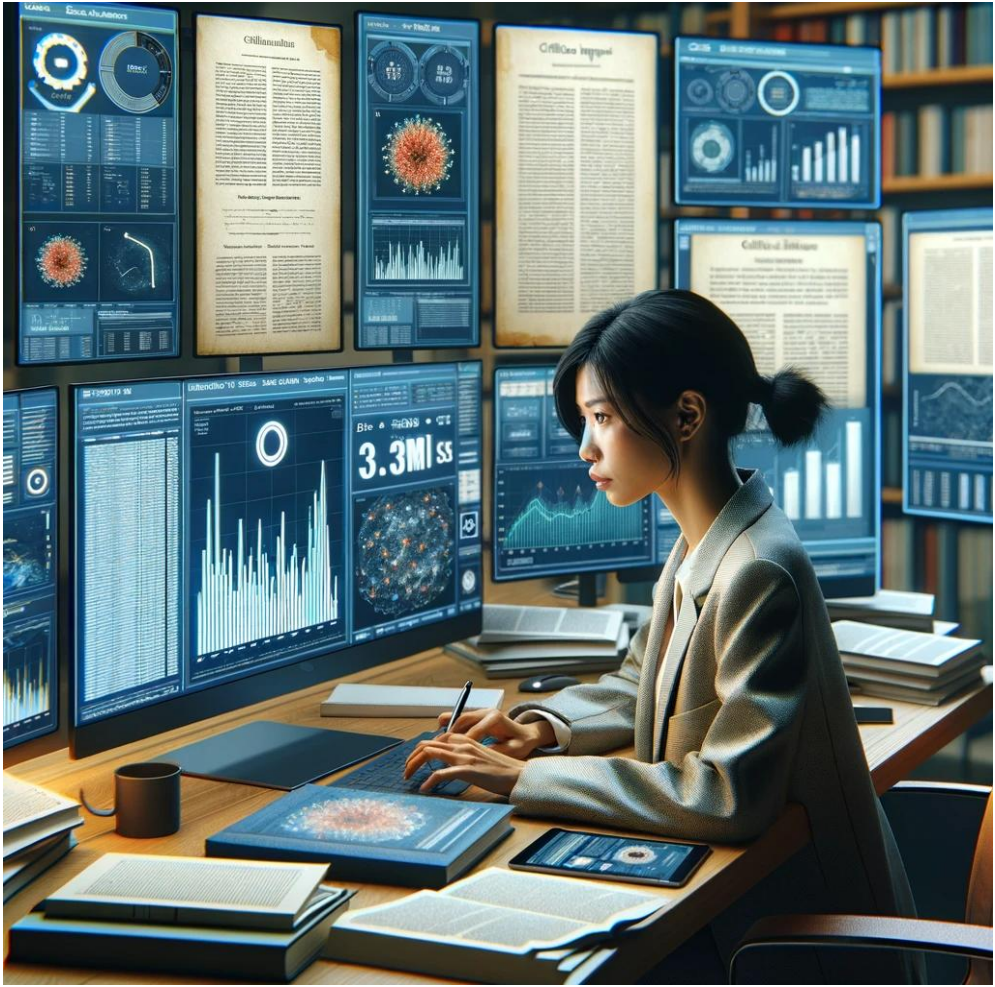

Mike Thelwall, University of Sheffield, UK.

10 April 2025 update

**Table of Contents**

# Quantitative Methods in Research Evaluation: Citation Indicators, Altmetrics, and Artificial Intelligence

**Abstract/Scope/Blurb**

This book critically analyses the value of citation data, altmetrics, and artificial intelligence to support the research evaluation of articles, scholars, departments, universities, countries, and funders. It introduces and discusses indicators that can support research evaluation and analyses their strengths and weaknesses as well as the generic strengths and weaknesses of the use of indicators for research assessment. The book includes evidence of the comparative value of citations and altmetrics in all broad academic fields primarily through comparisons against article level human expert judgements from the UK Research Excellence Framework 2021. It also discusses the potential applications of traditional artificial intelligence and large language models for research evaluation, with large scale evidence for the former. The book concludes that citation data can be informative and helpful in some research fields for some research evaluation purposes but that indicators are never accurate enough to be described as research quality measures. It also argues that AI may be helpful in limited circumstances for some types of research evaluation.

**Acknowledgements**

This book contains elements of research that was funded by Research England, Scottish Funding Council, Higher Education Funding Council for Wales, and Department for the Economy, Northern Ireland as part of the Future Research Assessment Programme (https://www.jisc.ac.uk/future-research-assessment-programme). It also contains research funded by the UK Economic and Social Research Council (ESRC) Metascience call. The content is solely the responsibility of the author and does not necessarily represent the official views of the funders.

# 1 Introduction to quantitative methods for research evaluation

Citation counts, citation-based indicators, and altmetrics are widely consulted by busy, worried, or ambitious scholars, research managers, journalists, and policymakers in the belief that they reveal something useful about academic achievements. They might do this informally and casually, such as by noticing a Journal Impact Factor (JIF) when deciding whether to submit to a journal. At the other extreme, they may also do it formally, such as by embedding citation-based information in university league tables or in national laws for promotion or remuneration. Despite all these applications, research indicator use is controversial, with strong advocates and detractors. Moreover, there are periodic calls to increase, reduce or eliminate their use in certain contexts. For example, many people have asked for the expensive and time-consuming Research Excellence Framework (REF) in the UK to be replaced by a metrics-based exercise (Wilsdon et al., 2015). In addition, there have been initiatives to assess whether artificial intelligence methods could replace human expert review for estimating research quality. In the opposite direction, several international initiatives have attempted to gain a consensus about reducing the importance of bibliometrics in academic evaluations.

This book focuses mainly on citation counts, citation-based indicators, and altmetrics, so these are defined here. Artificial intelligence terminology is introduced, when needed, later in the book.

**Citation index**: A bibliometric database that contains information about academic publications and the citations between them. Well known examples include the Web of Science, Scopus, Google Scholar, Dimensions, and OpenAlex, all of which index different, but overlapping, collections of journals, conferences, and other types of academic document. Some people consider citation indexes to be optional components of bibliometric databases, but this book uses the term to mean both combined.

**Citation count**: The number of times a document has been referenced in other documents. In other words, a citation count is the number of citing documents that cite a given cited document. Citation counts are normally specific to the citation index that they are derived from. These databases may all give different numbers and there may be citing documents that none find, so a document's "total" (and unknown) citation count is the overall number of citing documents. If a document is cited more than once in a citing document, only one citation is normally counted.

**Citation-based indicator**: Any formula that includes citation counts and is designed to give information about research quality or impact. If it does give some information, then it may be described as a *valid* citation-based indicator, otherwise it is an invalid or meaningless indicator.

**Altmetric**: A count of the number of mentions of, or citations to, academic research outputs derived from web sources, excluding traditional scholarly citation counts. Well known altmetrics include Twitter/X mentions and Mendeley reader counts.

**Artificial intelligence (AI)**: This term encompasses computer software designed to perform tasks that seem to require human-like expertise. One type of AI is machine learning, which involves programs that learn to perform tasks (e.g., peer review) after being fed with many examples (e.g., peer review reports).

## 1.1 Think about your goals and resources first!

Before using citation-based indicators or AI, it is important to consider the goals of the research evaluation. As this book argues, citation-based indicators and AI (so far) align with Global North style research quality goals. They are therefore less valuable for research evaluations with different types of goals, such as harnessing research methods to give value to the local community or to build a research capacity for education or future expansion. Each research institution should therefore consider its goals when deciding whether to harness quantitative methods. Of course, some institutions have the accrual of citations as an institutional goal, for example to climb local or global league tables that incorporate them.

A second consideration is the availability of the main alternative to quantitative approaches: human expert evaluators. Since for most purposes (except explicitly targeting citations) human expert judges are often thought to be the ultimate arbiters of research quality, the availability of such judges is important. The less they are practically available, the more valuable quantitative approaches are in comparison. For example, if a university has a ready supply of retired professors that are expert evaluators and willing to assess research for an affordable fee, then citation analysis might not be needed at all. At the other extreme of the spectrum, perhaps a new university starting on a shoestring budget in a Global South region previously without higher education has capable managers and administrators but no research evaluation experience, and no financial ability to pay others to perform this role. In this case, citation-based evaluations would tend to be more attractive.

Figure 1.1 summarises the above two considerations. The horizontal category is an oversimplification given that there are many other possible research goals. The remainder of this book is for people who do not fall into the top-left quadrant. It discusses methods that can be useful as well as the fields that they are most useful for.

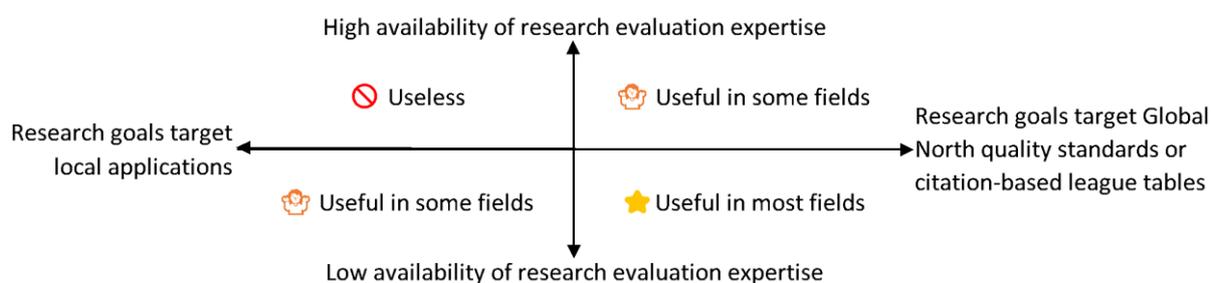

*Figure 1.1. The influence of research evaluation goals and the availability of expert research evaluators on the value of citation analysis or AI for research evaluation (source: author).*

## 1.2 Use more indicators!

Whilst the use of quantitative information in research evaluation is controversial, there are also calls for the use of AI for this. Those calling for greater use of indicators and/or AI in research evaluation can deploy strong common-sense arguments.

**Time saving**: Reading and evaluating academic research is a slow and expert task, so indicators or AI should be used to save the time of these experts.

**Facilitation**: Quantitative indicators and/or AI can make evaluations possible in contexts where the assessors, such as research managers or policy makers, lack the expertise to judge research or have too many outputs to read individually.

**Impartiality**: Automatically calculated indicators and AI may be less biased than human judges, who could be influenced by friendship, gender, or other partialities.

**Objectivity**: Citation counts are objective numbers and largely transparent, whereas all human judgements are necessarily subjective and can never be fully transparent.

## 1.3   Use fewer indicators!

Those opposing the use of indicators and/or AI for research evaluation can make a mix of common sense and analytical arguments.

**Oversimplification**: Research is complex, and all indicators oversimplify its value. Even at the most basic level, there is usually no fair way to decide how much credit to assign to each author in a collaborative project.

**Partial evidence**: Citations can only reflect influence within the scholarly publishing system, so citation counts can only provide partial evidence of the value of academic research. Similarly, altmetrics only provide partial evidence and AI may not be able to be fed with comprehensive enough information.

**Field or disciplinary differences**: Rates of citation vary between fields for organic reasons, such as the speed of research and the importance of monographs, so comparisons between fields are invalid. Also, citations may be irrelevant in some fields because they are not central to academic contributions. The same is true for altmetrics to an even greater extent and there is insufficient evidence to know whether it is also true for all major types of AI.

**Perverse incentives**: Using indicators in evaluation will create a perverse incentive to focus on indicator-generating scholarship at the expense of other tasks, narrowing the focus of academic activity.

**Gaming**: Those evaluated through indicators may try to game the system by self-citations and encouraging others to cite them, such as through citation cartels/clubs, or journals encouraging submitting authors to cite other articles from the same publication.

## 1.4   Use indicators responsibly!

Over the past decade, influential initiatives and reports have emphasised the need for careful consideration and responsibility when quantitative indicators are used in research. These have included the Leiden Manifesto (Hicks et al., 2015), the UK Metric Tide report (Wilsdon et al., 2015; Curry et al., 2022), and a European Union agreement on research assessment reform (CoARA, 2022). There is an influential parallel agreement to restrict the use of journal-based impact indicators in research assessment (DORA, 2020), and the More Than Our Rank (inorms.net/more-than-our-rank) initiative to emphasise that university rankings, some of which include citation-based indicators, do not reflect all important aspects of universities. These have mostly emphasised the limitations of quantitative indicators in research assessment and argued for their use to support rather than replace expert judgement as the primary evidence of research quality or value. This book tends to support these initiatives by providing evidence of the limitations of citation-based indicators for research quality evaluation. Nevertheless, this book is written primarily from a Global North perspective, and it is not clear that anything in it (or bibliometrics as a field) has any relevance at all to the sometimes-different priorities of Global South research evaluation.

### 1.4.1   Consider systemic effects!

Systemic effects are a controversial but important responsible consideration for research evaluations (Rushforth & Hammarfelt, 2023). As with any reward system, even if the only reward is prestige, the mechanism to achieve the overall goal can be substituted for the goal. Thus, academics might start to believe that generating citations is the same as producing high quality or impactful work. Thus, if academics are evaluated through citations, then some will chase citations (or highly cited journals), partly at the expense of conducting high quality research. This is part of the rationale for initiatives against the use of Journal Impact Factors in the reward system.

## 1.5   Use open data for indicators!

Research evaluations in recent years seem to have primarily used citation data from one of the two large specialist bibliometric databases: Scopus and the Web of Science, with Dimensions.ai having recently become a new competitor. These have similar (but incomplete) coverage of scientific documents (primarily journal articles). They also all offer features that are necessary for citation analysis, including citation counts and metadata for articles, document classifications (so that editorials are not mixed in with research articles, for example), and subject classifications. The choice of database might be based on coverage of the field or country to be analysed, cost, and the utility of the additional functionalities that they offer, some of which are available for an additional charge above the base price (e.g., InCites for the Web of Science, SciVal for Scopus, and various Apps for Dimensions). Google Scholar might also be used for informal citation analyses but lacks the features for most larger scale projects.

Against this background of commercial citation data providers, the open science movement (Bartling & Friesike, 2014) has led to calls for bibliometric indicators to be based on freely available open and transparent data, including the development of open citation indexes (Peroni & Shotton, 2020), with OpenAlex being a prominent current example. This supports the transparency goal of the research assessment reform initiatives discussed above as well as increasing global fairness by reducing the financial cost of citation analysis. At the time of writing, OpenAlex seems set to challenge traditional citation indexes (e.g., Culbert et al., 2024), and whilst there was some evidence that its document type classification was sub-optimal, it seems to be suitable for citation analysis (e.g., Thelwall, 2025).

## 1.6   What is an indicator and why might citation counts be one?

A discussion of the term "indicator" is useful to interpret citation-based and altmetric evidence. The term "metric" is more often used, but has connotations of greater accuracy and validity, so it is best avoided when possible. These two terms are defined here for clarity of discussion, although they also have other common language meanings and different definitions.

**Metric**: A numerical quantity that provides a reasonably accurate estimate of the value of something. For example, a word count might be a metric for the length of a journal article.

In research evaluation, metrics tend to have two meanings, but they are rarely disambiguated in practice, so people may talk at cross-purposes when discussing them. For example, a Scopus citation count might be described as a metric. It could be thought of as a reasonably accurate measurement of the number of citations to an academic work, at least as reflected within the academic database used. Thus, citation counts are citation metrics in this sense. Alternatively, Scopus citation counts might be described as research metrics, implying that they are somehow reasonably accurate "measures" of the quality or impact of the cited documents. Citation counts are *not* research metrics in the latter sense, as explained in this book, because they give too inaccurate estimates of quality or overall impact.

**(Valid) indicator**: A numerical quantity that provides information about, but not necessarily an estimate of, the value of something. For example, the citation count might be an indicator of the scholarly impacts of a set of journal articles.

The term indicator is much weaker than the term metric because every metric is an indicator, but most indicators are not metrics. Like metrics, however, an indicator could be valid for one thing (e.g., citation impact) but not for another (e.g., research quality). The phrase "information about" for indicators has the technical meaning of uncertainty reduction in the sense that knowing the indicator score increases the probability of knowing a value of the indicated document. For proposed research indicators the meaning tends to be more specific, however: a higher score on the indicator tends to associate with a higher value of the quantity indicated (or vice versa sometimes). For example, if JIFs are valid indicators of research quality, then knowing that an article was in a journal with a high JIF would increase the chance of the article being high quality (although it still could be weak), in the absence of other evidence. In this sense, in some fields JIFs might be indicators but not measures of research quality. An AI prediction of an article's research quality can also be a research quality indicator.

**Indicator strength**: The extent to which an indicator associates (e.g., correlates) with the quantity indicated. Whereas a weak indicator only has a small association, a strong indicator is more reliable, and a very strong indicator may be a metric. A proposed indicator with no association is not an indicator but for clarity will be described as an *invalid* indicator (of something).

As this book will argue, even if an indicator or metric is valid, it does not follow that all uses of it are helpful. For example, if an indicator introduces perverse incentives for researchers (such as JIF chasing), then its use may be undesirable whatever its strength (see Chapter 15). Nevertheless, much of this book is about assessing the strength of different types of citation-based indicators, altmetrics (which are also indicators), and AI in various contexts.

## 1.7 Positioning of this book

The goal of this book is to evaluate when citation-based formulae, altmetrics, and AI are valid and useful indicators of research impact or research quality, the strength of these indicators, and the factors that mitigate against their use, when valid, in various contexts (e.g., for evaluating people, departments, universities, countries). This book is in the same spirit as three other books about citation analysis in research evaluation (Gingras, 2016; Moed, 2006; Sugimoto & Larivière, 2018) but updates them with substantial new direct evidence about the relationship between citations and research quality, as well as covering a wide range of altmetrics and adding AI.

This book analyses quantitative methods in research evaluation rather than introducing them for newcomers seeking to conduct a quick analysis, but I hope that it is readable and informative for anyone with an interest in research evaluation. It is more advanced than the many useful introductions to bibliometrics that include sections on sources of bibliometrics and how to apply various formulae (e.g., Ball, 2017) but more specialist than books focusing on the wider field of informetrics (e.g., Qiu et al., 2017). Other books give excellent overviews of the history and core issues around citation analysis (De Bellis, 2009) and various uses of web indicators (Holmberg, 2015; Stuart, 2023). One of my previous books also covers indicator formulae for altmetrics in more detail (Thelwall, 2016). Finally, this is not a "how to" book in the sense of a step-by-step guide to identifying ready-made indicators from the platforms that support them or extracting citation count data and using it to form indicators, but there is information about this online and in introductory books.

## 1.8   Overview of the contents of this book

Research indicators elicit strong emotions from their advocates and opponents because they can be influential for academic careers and concern a topic about which the subjects of the indicators are experts: their own scholarship and that of others in their field. Moreover, the formulae used for indicators are usually very simple, so it seems reasonable to expect a broad international consensus on their value, rather than the strongly-held differing opinions that currently proliferate. The reasons for these disagreements may be the hidden complexity behind the simple formulae, loss of focus on the purposes of academic research, and a lack of empirical evidence about the relationship between citations and research quality. This book addresses the first two through explanations and the third through analyses of a large-scale science-wide dataset. The same approach is taken for AI in research evaluation. The empirical data presented is largely from peer reviewed articles.

The primary focus of this book is the research evaluation of journal articles, and most chapters exclusively discuss this issue. Journal articles are the primary scholarly outputs of most fields and active researchers, although conference papers, monographs or art are primary outputs in some fields, and book chapters are also important in a few fields. Research evaluation for anything other than journal articles is discussed in Chapter 0.

This book first discusses what is meant by research quality to set the context for the remaining chapters. It then introduces theoretical considerations for citation-based indicators and then the main types of formulae used to construct them. After this it evaluates the evidence available that citations have value for research quality indicators in all broad academic fields. The book then discusses factors other than the quality of articles that can influence the value of citations as research quality indicators. The next chapter focuses on journal citation rates and their value as indicators of the impact of their individual articles. After all this context, the value of aggregate citation indicators is assessed for academics, departments, universities, countries, and funders. This is the level at which they are primarily used for research evaluation. The types of biases that can exist in aggregate citation indicators are then discussed, partly with reference to the previously introduced factors that are known to influence citation rates. Next, alternative indicators for research impact are introduced and evaluated and the potential for AI to support research assessment is discussed. Finally, the book concludes with the perverse incentives that indicators can introduce into research evaluations.

The overarching themes of this book are disciplinary differences in everything related to indicators, the difference between citation impact and research quality, and the difficulties in measuring research quality in any meaningful sense.

## 2   Research quality and impact

Before discussing AI or indicators, the notions of research quality and research impact will be deconstructed since the goal is often to estimate an aspect of research quality or impact.

### 2.1   What is research quality?

The concept of research quality is widely used and thought of by researchers, usually without a precise definition but with an understanding of a range of factors that it might encompass. An important formal context is national research evaluation exercises where judges are explicitly asked to give quality scores to research outputs (e.g., UK, Italy, New Zealand, Australia). This entails formulating precise definitions of quality together with scale(s) (three sets of 1-10 in the Italian Valutazione della Qualita della Ricerca) or set(s) of categories (four main categories in the UK REF). Reviewers of academic outputs for journals or publishers may also be asked, implicitly or explicitly, to judge the quality of submissions against a set of guidelines or a checklist. These reviewers then usually register only action-oriented categories (e.g., accept, minor revisions, major revisions, reject) rather than explicit quality scores (an exception is SciPost: Thelwall & Hołyst, 2023). Also related is the task of reviewing funding bids, where explicit numerical scores are usually given against a set of criteria, which presumably encapsulate value for money, project plausibility, and expected quality of the proposed research (Langfeldt, 2001).

Although there is no universal definition of research quality and the practical operationalisations (guidelines/checklists) discussed above are highly varied, three underlying broad dimensions are common to most: rigour, originality, and impact/significance. The third dimension can be split into academic and non-academic impacts (Langfeldt et al., 2020). Thus, the highest quality piece of work might be rigorous, original, and influential on society and academia. An example of this might be a well-conducted study of an innovative treatment for a major disease that becomes adopted by society and analysed in academia largely based on the results. All three dimensions are subjective, even the first, and vary between fields.

**Rigour**: the extent to which the scholar has ensured that the output is plausible. Its nature varies between research types. For empirical research, it is the extent to which the methods reported support the conclusions. For analytical research (humanities style studies, literature reviews, mathematical arguments, or commentaries) it is the care with which an argument has been presented, perhaps encompassing the range and appropriateness of the sources cited and the depth of analysis presented. For artistic works (e.g., performances, paintings, musical scores), it may be the technical skill with which the object has been constructed and the appropriateness and range of influences displayed.

Methodological rigour for empirical research is probably thought of as binary in society: the study is valid or not. Apparently conflicting studies cause confusion (e.g., 99 studies say smoking associates with lung cancer, one does not) unless the reasons are understood. In practice, all empirical studies necessarily carry a range of assumptions, so none are 100% valid (Strevens, 2020). Moreover, much of science involves groups of researchers conducting studies that challenge the assumptions of prior work (Strevens, 2020). Thus, a more rigorous empirical study would tend to be one where all reasonable steps have been made to reduce the number and strengths of assumptions needed. For example, a survey of a random sample of children from all UK schools would be more rigorous than a survey of a random sample of children in one UK school because it would not need the (unlikely) assumption that the chosen school was representative of the country. Presumably, it would be impossible to randomly sample all UK schools, so any study that had put substantial efforts into selecting and sampling children from a representative large sample of schools would score high for this aspect of rigour. Its assumption would then be that its approach was reasonably effective at getting a representative sample, perhaps citing evidence in support of this claim. Of course, sampling is only one aspect of methodological rigour, and the study would have many other assumptions, including the extent to which the survey questions effectively elicited unbiased information about the knowledge or opinions of the children.

**Originality**: the extent to which some or all key aspects of the work differ from prior work. This seems likely to encompass methods and goals rather than presentation, especially in empirical research. This is clearly a subjective dimension, depending on the aspects of the output recognised by the judge and the judge's background knowledge. To give an extreme example, someone might believe an output to be original because they were unaware of a paper with identical methods and goals published the previous year. Conversely, a paper with novel methods and academic goals might be thought unoriginal by someone focusing on originality in value to society.

**Academic impact**: the influence the work has on the research of other scholars. This might be by triggering follow-up studies, supporting the methods or theory of future work or by closing off potential lines of enquiry or discrediting methods or theory. Counting citations seems to be a reasonable way of gathering evidence of academic impact in the former positive cases, but not in the latter negative cases. To give an extreme example, a paper that thoroughly debunked a theory (e.g., cold fusion, MMR-autism links, a hypothesised cause of Alzheimer's) might remain uncited because the issue disappears from research. It would be unfair to judge such a highly useful and impactful paper by its citation count.

**Societal impact**: the influence the work has had on society outside academia. This might be by supporting, enabling, or improving commercial, social, cultural, health, political, or other activities either directly (e.g., showing that a vaccine works) or indirectly (e.g., showing how firefighters access information, subsequently leading to improved firefighter information awareness training). The route between research and non-academic benefits is often long, tortuous and lacking documented evidence because there may be multiple direct and indirect influences, many of which are not recorded (e.g., Kuruvilla et al., 2006). Thus, identifying the societal impacts of an individual article is very difficult and so judgements about the likely societal impacts of typical research studies are necessarily usually speculative and therefore subjective.

## 2.2 Different types of research quality

Despite the above description of the three core dimensions of research quality, it has substantially different conceptualisations for different purposes, between fields and over time. Thus, an article agreed to be high quality in one context might be low quality in another.

### 2.2.1 Journal reviewing

For journal article reviewing, research quality might typically entail all three core dimensions, but some journals have rigour-only review (e.g., PLOS, 2022), albeit presumably requiring a degree of originality, and so would have a narrower conceptualisation of quality (Spezi et al., 2017). Journals may also largely ignore the originality dimension for replication studies.

### 2.2.2 Disciplinary differences

Within fields that rely on journal articles, and perhaps other reviewed outputs, the concept of quality might become tied to progress within the field, ignoring the societal dimension of research almost completely. In extreme cases, it has been argued that some fields have become tied to "ivory towers" research that is meaningless and unhelpful outside academia (Tourish, 2020). This seems particularly likely in basic research fields, such as pure mathematics or theoretical physics, where the goal is understanding rather than applications. In the opposite direction, applied journals and probably entire fields might orient on research with clear societal benefits. For example, in professional fields, benefits to the profession might be a widely agreed goal and therefore a recognised dimension of research quality (a societal impact) (Whitley, 2000). Thus, individual fields might largely ignore either the societal impact or scholarly impact aspects of research impact in their quality judgements.

At a finer grained level, there are probably also field differences in quality standards and the types of contribution that score highly for rigour, originality, and impact. For example, effective ethics might be an aspect of rigour in health fields but not in mathematics.

### 2.2.3 National research evaluation exercises

Quality for post-publication expert review seems to be conceived differently from quality for journal reviewing, perhaps for practical rather than for theoretical reasons. For example, whilst journal reviewers might be expected to be specialists on the topic evaluated, and hence capable of, and expected to, assess rigour, originality and significance, judges for national research evaluation exercises like the UK REF might only have general field knowledge for most outputs assessed.

In the UK case, about 1120 experts were selected to evaluate all 185,594 UK REF2021 research outputs (REF2021, 2022a). Even aside from time considerations, many or most of these outputs would have been scored by people with less specific expertise than the original reviewers. Thus, they may be frequently unable to fully assess rigour and originality because of a lack of specialist knowledge. For example, there were no bibliometric experts amongst the REF2021 assessors so those scoring the bibliometric submissions would have had to guess at these aspects. In this context and given that citation-based evidence was banned from most fields in the REF and probably used in a minor capacity in the remainder (Wilsdon et al., 2015), the aspect of research most easily assessed by non-experts might be possible societal impacts.

Thus, for national research evaluation exercises with expert reviewers, quality may be skewed towards ostensible societal benefits, with methodological and theoretical research being judged lower quality. This seems especially likely in fragmented fields with a wide variety of topics and methods (e.g., Whitley, 2000).

### 2.2.4 Global South perspectives and other international differences

Each country might agree to emphasise or include different aspects of research quality, so there can be international differences that would greatly affect national research evaluations and local interpretations of journal and field norms. Since research quality is not a universally agreed concept and all countries have different social and economic contexts, it would be reasonable for one to choose or emphasise different aspects of research quality than another.

From a Global South perspective, evidence of influence within science, such as given by citation counts, may not be very relevant to conceptualisations of quality (Barrere, 2020). Instead, local societal benefits and capacity building may be regarded as more central components (Lebel & McLean, 2018; Kraemer-Mbula et al., 2020). This perspective may even displace publications as central to research systems. In parallel, the research of Global South scholars is less likely to appear in the main citation indexes, which are based in the Global North, and so citation analysis results are misleading (Mills et al., 2023). Both issues also mean that the empirical results in this book, which use a Global North (UK) perspective, may have no relevance to Global South research. It also seems likely that the use of citation-based research indicators would be damaging to Global South scholarship by suggesting that their academics devote time to chasing activities that will help them benchmark themselves against the different priorities of the Global North. Bibliometrics may also be demoralising to some Global South scholars by devaluing their work by comparison: the Global North tends to score well since it has set the rules and controls the main sources of citation databases.

## 2.3 What is citation/research/scholarly/academic impact?

Although research impact has been introduced as an aspect of research quality, it is worth discussing again separately because it is frequently the focus of discussions of indicators. The idea that counting the citations to a paper might reflect the scholarly/scientific/academic impact of that paper in the modern era probably originates from the USA in the 1950s with Eugene Garfield's idea for a citation index (Garfield, 1955) and his subsequent computerisation of scientific information to create the world's first automated citation index (Garfield, 1979). This triggered the subsequent sociological theory of science from prominent sociologist Robert Merton (1973). In fact, the original aim of computerising citations was to help literature searchers to find citing papers for important works to ensure that their literature search was up to date (i.e., citation chaining) (Berry & Choi, 2012). Nevertheless, from this Garfield argued that more cited journals tended to publish better work (Garfield, 1972) and Merton developed a sociological theory to explain this.

Perhaps Garfield was the first person in the modern era to think about systematically counting citations. Previously, some scholars may have done this in an ad-hoc way for personal interest. Perhaps some had a notebook where they recorded the titles of papers that cited their own works or papers that they were interested in and kept a running total out of curiosity to see which had been referenced more often. With the computerisation of citations, this capability became instant rather than the results of a lifetime's curiosity.

Note that the term "citation impact" seems to be often used to refer to what citation-based indicators reflect. So, a person with many citations may be said to have high "citation impact". If "impact" here is intended to be a synonym of "count" then this is uncontroversial. Nevertheless, the phrase probably gets its power from people assuming that it is a measure of overall impact, so, by implication, a person with a high citation impact would then have had a high scholarly or overall impact. This is not necessarily true because of the non-impact reasons why citations can accumulate, as discussed below.

## 2.4 Summary

To summarise the above, the concept of research quality usually encompasses rigour, originality, and (societal and academic) significance. Nevertheless, there are many different definitions, reflecting the diverse set of factors that at least some scholars think are relevant. It should also be clear that research quality is not an absolute concept or truth but that it is socially constructed. This follows from the discussion of the way that different tasks can naturally lead to different notions of research quality. For simplicity of exposition, however, most of this book treats research quality as an absolute truth and dissects it only when necessary to make a point.

In contrast, whilst impact is a dimension of research quality and comprises both scholarly and societal impacts, in practice, scholarly impact on its own seems to be equated often with research quality. It is sometimes also equated with citation impact, and the problems with this are analysed below. The remainder of this book discusses the extent to which citation-based indicators, altmetrics and AI can inform judgements of research quality or any type of impact.

# 3  Citation-based indicators: Theoretical considerations

Most indicators used to evaluate research are currently derived from citations. This chapter discusses core issues around the use of citation-based indicators in research evaluation, mostly from a purely theoretical perspective. The next chapter mentions practical considerations and the one after introduces and discusses the strength of evidence in relation to these issues.

## 3.1  Merton's normative theory of citations

Merton used Newton's "standing on the shoulders of giants" idea to develop a theory to explain why more cited works might be more useful to science. On the basis that science is hierarchical, new research starts with the current state of knowledge and then builds on it to extend the boundaries of the known. The formal mechanism for building on existing knowledge is to reference that knowledge. Thus, work that has been referenced (i.e., cited) more has been built on more and has been more useful to science. For example, if a new published finding is interesting but no future researcher ever needs to cite it then it has been of limited value, whereas a heavily cited paper (e.g., Einstein's theory of relativity) has been extremely useful to science, with generations of scientists building on it for new discoveries. Thus, the hierarchical nature of science and the normative practice of referencing relevant prior work combine to suggest that citation counts reflect the scholarly impacts (i.e., contributions towards science building) of outputs.

Unfortunately, Merton's simple and intuitive idea has many flaws and exceptions that must be considered when trying to use citation counts or other citation-based indicators. In particular, not all science is hierarchical, not all citations reflect building on prior work, and reference lists are not comprehensive.

## 3.2  Flaws in Merton's normative theory of citations

First, a theoretical problem with Merton's normative theory is that not all science is hierarchical. Considering the wider meaning of "science" as all academic research, the arts and humanities are much less hierarchical and have a much weaker normative culture of using "building upon" type references (Lin, 2018). The social sciences and engineering probably vary between the traditional sciences and the humanities in the degree to which they are hierarchical. What does it mean to be non-hierarchical here? It means relying less on prior work when creating new knowledge or insights. In the humanities, citations for background, contrast and criticism may be used as well as different document types (Hellqvist, 2010). For example, a humanities paper might provide a detailed analysis of a minor medieval building (e.g., Impey, 2009). The primary purpose of the paper might be to provide richer detail into monastery life in Mediaeval Europe. To do this, it might cite books about similar topics and background information, such monographs about monasteries, as well as historical sources for facts, but it has little need to cite core prior work that it builds on because its primary topic (a single building) is very niche.

Even within traditional sciences, such as Newton's physics, applied research is important and this may aim to produce findings of societal value rather than basic knowledge. Such applied research may therefore be judged to have impact irrespective of whether it is cited. In the current era where scientists (especially in the UK) need to demonstrate the direct or indirect societal value of their work, it seems likely that there are few areas of scholarship where contributions to science alone should be used to measure the value of research.

Second, reference lists are not comprehensive for multiple reasons. Scholars may accidentally omit work they have built on, may prune reference lists to fit journal length requirements, may prefer to cite few papers, or may omit citations to seminal work that they consider common knowledge (obliteration by incorporation: McCain, 2011). The latter point is unavoidable in hierarchical subjects. There are also citation network issues in the sense if a paper A builds on work B that builds on C, then A may cite only B or both B and C. In the former case, C's impact would be underestimated by citations. In hierarchical areas of science this would be the norm rather than an exception. In an extreme case, perhaps paper C introduces innovative and groundbreaking research that B makes a minor improvement on, but everyone cites B as the newest and most comprehensive reference.

Related to the second problem, researchers have biased awareness of, or willingness to cite, relevant references. Previous studies have shown that academics are influenced by social factors (Lyu et al., 2021), such as being more likely to cite people that they know or that are based in the same country (Thelwall & Maflahi, 2015). Because of this, for example, work produced by academics in countries with few scholars starts with a citation disadvantage.

Third, academics cite work that they have not built upon. Work may be referenced because it demonstrates the importance of the problem, to argue that it is wrong, or for richer context (particularly in the humanities?) This can create systematic biases rather than just random noise in the system. For example, some academic papers are highly cited for stating that something is a problem when it first appears and then becoming a standard citation in the introductions of all papers about their topic. Literature review papers are another example of this in the sense that they do not introduce primary research but help others to find and make sense of the primary research; they may "steal" citations from that research by being cited instead.

Fourth, recall that research quality is generally thought to encompass three dimensions: rigour, significance, and originality (Langfeldt et al., 2020). Of these, citations probably reflect significance most and it is not clear that they are good indicators of rigour and originality (Aksnes et al., 2019). Moreover, citations do not reflect societal impacts (van Driel et al., 2007). Thus, even from a theoretical perspective, it seems unlikely that citation counts closely correlate with research quality within any field, unless its three dimensions usually coincide for some reason or if societal impact, rigour, and originality all frequently influence citing behaviours.

In summary, the extent to which citations are necessary to validate the usefulness of research varies, reference lists are incomplete and biased lists of the evidence built on, and are polluted by cited research that has not been built on.

## 3.3   Citation biases

Citation-based indicators seem to favour men, Global North researchers and perhaps also other demographic groups. The first two seem common sense given that citation-based rankings of individuals are dominated by men in many fields and often include few Global South researchers, except those that have emigrated to the Global North. Such biases are clear causes of concern for any use of bibliometric indicators.

Gender bias in citations has been hypothesised to occur because men disproportionately cite men, perhaps because they undervalue the work of women (Larivière et al., 2013). This is difficult to verify in practice with a high degree of confidence because there are gender differences in research topics and methods, which generates a natural tendency for a degree of gender self-citation. A discussion of some citation biases is in Chapter 10 and a range of factors that may influence citation counts is introduced in Chapter 7.

## 3.4   Citation indicators vs. peer review

The above sections have given many reasons why citation counts may give imperfect or misleading evidence about the value of academic research. Since the main alternative to citation counts is expert or peer review, it is important to also acknowledge the strengths and weaknesses of expert judgements.

Expert or peer review involves experts (senior scholars, usually from the same field) or peers (scholars from the same field) judging the quality of academic publications. Some reviews are careful, such as with appropriate narrow field experts judging against clear quality guidelines. In contrast, others are casual, such as with non-specialists quickly judging articles, perhaps from lists on job applicants' CVs. As mentioned in the quality section above, the type of quality assessed can vary. This section discusses careful expert peer review and assumes that a type of quality has been agreed upon.

The main advantage of peer review is that human judges can consider factors that citation counts cannot, including all the limitations mentioned above in this chapter. Whilst citation counts primarily (and imperfectly) reflect the scholarly impacts of each output, human judges can also evaluate its originality, rigour, and likely societal benefits. Thus, in theory, human judges are required to ensure that all the agreed core dimensions of research quality are considered in the quality judgements made or scores given.

For those that value citation counts above peer review, the above argument may seem tautological: humans are needed to judge the human-agreed dimensions of quality, so why not bypass all this human subjectivity and use citation counts? The answers are that (a) all judgements are subjective and even an agreement that citation counts somehow constitute quality would be a human agreement; thus we should follow the consensus or make strong arguments about why the consensus is wrong, and (b) quality judgements usually have a purpose, such as rewarding successful researchers or departments, and this purpose (such as progressing the field or benefitting society) has led to the three dimensions of quality being frequently considered to be important to assess. Even for publicly funded research, many and perhaps most countries seem to consider societal benefits to be a major goal of research and perhaps its ultimate purpose.

### 3.4.1   Biases and limitations of peer review

Despite the theoretical advantages of peer review, it has clear disadvantages. Perhaps most importantly, it consumes a large amount of expert time to do well. One of the reasons why citation-based indicators are used is that peer review is impractically slow and too expensive for some tasks. For example, a committee evaluating the publications on the CVs of a set of shortlisted candidates should really judge these publications individually but doing this for, say 20 publications on each of 5 candidates' CVs gives a task of evaluating 100 publications, which may take weeks to do well, whereas the interview panel may only be willing to allocate half a day to the shortlisting task.

Another prominent limitation of peer review is that even experts are likely to disagree and can be biased. Bias in peer review can be thought of as any judgment that systematically deviates from the true quality of the article assessed for an identifiable cause (Lee et al., 2013). Non-systematic judgement differences are also common (Jackson et al., 2011; Kravitz et al., 2010).

Sources of systematic bias that have been suggested for non-blinded peer review include malicious bias or favouritism towards individuals (Medoff, 2003), gender (Morgan et al., 2018), nationality (Thelwall et al., 2021), individual or institutional prestige (Bol et al.., 2018), and methods prejudice (Thelwall et al., 2023f). Systematic peer review bias may also be based on language (Herrera, 1999; Ross et al., 2006), and study topic or approach (Lee et al., 2013). There can also be systematic bias against challenging findings (Wessely, 1998), complex methods (Kitayama, 2017), or negative results (Gershoni et al., 2018). Studies that find review outcomes differing between groups may be unable to demonstrate bias rather than other factors (e.g., Fox & Paine, 2019). For example, worse peer review outcomes for some groups might be due to lower quality publications because of limited access to resources, or unpopular topic choices. A study finding some evidence of same country reviewer systematic bias that accounted for this difference could not rule out the possibility that it was a second order effect due to differing country specialisms and same-specialism systematic reviewer bias rather than national bias (Thelwall et al., 2021). Weak disciplinary norms can also create biases because there are different schools of thought about which theories, methods or paradigms are most suitable (Whitley, 2000). In extreme cases experts from one school of thought or tradition may regard all work from a competing tradition as fundamentally flawed and worthless (e.g., postmodernism: Callinicos, 1989).

Non-systematic differences between reviewers also occur. They may be due to unskilled reviewers, differing levels of leniency or experience (Haffar et al., 2019; Jukola, 2017), weak disciplinary norms (Hemlin, 2009), and perhaps also due to teams of reviewers focusing on different aspects of a paper (e.g., methods, contribution, originality). Another underlying reason for disagreements is that many aspects of peer review are not well understood, including the criteria that articles should be assessed against (Tennant & Ross-Hellauer, 2020). Thus, reviewers may disagree because they understand the task differently. Of course, human reviewers can make simple mistakes too. A judgement may be wrong because the assessor overlooked important information, was unaware of it, or did not have the appropriate expertise to assess some aspects of an output. Mistakes seem particularly likely when the assessors do not have expertise in the individual topics evaluated, have too little time for a thorough assessment, or lack evaluation and norm referencing experience or guidelines.

Because of the above issues, it is not obvious that peer review is always better than citation-based indicators. For example, in contexts where nepotism is the major threat to an academic system, citation-based indicators may be preferred on principle. Similarly, sufficient local peer review expertise may not be available or affordable in some situations, so indicators might be the only practical option.

### 3.4.2 Example: Peer review in the Research Excellence Framework 2021

This section is taken from (Thelwall et al. 2023c). The UK REF2021 can claim to be the largest scale, most expensive and most financially important science wide academic peer review exercise ever conducted in the world, and so is an interesting case study for expert academic review. The 34 disciplinary subpanels of the REF assessed 185,594 outputs (mainly journal articles) from 76,132 academic staff organised into 1876 submissions (each roughly a university department) as well as 6,781 impact case studies and information about the scholarly environments of 157 UK higher education institutions (REF2021, 2022a). The administrative cost of the prior REF2014 was estimated to be £240 million (Technopolis, 2015). REF2021 scores were expected to direct 16 billion pounds of research funding block grants over the eight years before REF2029 (UKRI, 2023).

From initial planning to eventual publication of results, each REF takes at least eight years. The care with which the REF is designed can be seen from the 13 public background documents that informed the transition from REF2014 (REF2021, 2018) to REF2021. The UK higher education sector is extensively consulted on any proposals for REF changes, with 388 responses to the one report alone (REF2021, 2018).

At the heart of REF2021 is the scoring of the 185,594 outputs by 1120 experts organised into 34 Units of Assessment (UoAs) from UoA 1 Clinical Medicine to UoA 34 Communication, Cultural and Media Studies, Library and Information Management. These experts are nominated by institutions following a public call for specific expertise areas (REF2021, 2021a; REF2021, 2021b). The experts are trained in systems, ethics, and assessment procedures and their working methods are outlined in a 106-page public document. This includes overall and panel-specific definitions of quality and their applicability to the four level scoring criteria used (REF2021, 2020). Each output is initially allocated by subpanel (UoA) chairs to two experts who independently score it, then consult and agree on a score on a nine-point scale, optionally consulting bibliometrics in cases of disagreement in 11 subpanels. These scores are then discussed collectively in each subpanel and there are also main panel calibration discussions combining multiple UoA subpanels, and REF-wide statistical checks on score distributions to norm reference the scores. At some stage the nine-point scale is narrowed down to the 4-point scale (plus 0 for out of scope) that is eventually published. The extensive norm referencing is essential to the credibility of the system and is useful for bibliometric uses of the data since it allows interdisciplinary analyses.

With the above process, each of the 185,594 REF outputs were allocated a quality score for "originality, significance and rigour" of 1* "recognised nationally", 2* "recognised internationally", 3* "internationally excellent", or 4* "world-leading". Outputs judged ineligible or below national quality were scored 0 instead (REF2021, 2020). In addition to this overall REF definition/interpretation of quality, there are more specific criteria for each of the four Main Panels, each of which contains multiple UoAs (REF2021, 2020). For example, the criteria below apply to the mainly health and life sciences UoAs 1 to 6 in Main Panel A:

> The sub-panels will look for evidence of some of the following types of characteristics of quality, as appropriate to each of the starred quality levels: • scientific rigour and excellence, with regard to design, method, execution and analysis • significant addition to knowledge and to the conceptual framework of the field • actual significance of the research • the scale, challenge and logistical difficulty posed by the research • the logical coherence of argument • contribution to theory-building • significance of work to advance knowledge, skills, understanding and scholarship in theory, practice, education, management and/or policy • applicability and significance to the relevant service users and research users • potential applicability for policy in, for example, health, healthcare, public health, food security, animal health or welfare. (REF2021, 2020)

The remaining three main panels have specific criteria for each of the starred levels. For example, the highest quality (i.e., 4*) Main Panel C (UoAs 13 to 24, mainly social sciences) guidance is:

> In assessing work as being four star (quality that is world-leading in terms of originality, significance and rigour), sub-panels will expect to see some of the following characteristics: • outstandingly novel in developing concepts, paradigms, techniques or outcomes • a primary or essential point of reference • a formative influence on the intellectual agenda • application of exceptionally rigorous research design and techniques of investigation and analysis • generation of an exceptionally significant data set or research resource. (REF2021, 2020)

In contrast, the lowest (i.e., 1*) grade for Main Panel C equates to the following:

In assessing work as being one star (quality that is recognised nationally in terms of originality, significance and rigour), sub-panels will expect to see some of the following characteristics: • providing useful knowledge, but unlikely to have more than a minor influence • an identifiable contribution to understanding, but largely framed by existing paradigms or traditions of enquiry • competent application of appropriate research design and techniques of investigation and analysis. (REF2021, 2020)

When UoA panels assess the same article, there are some systematic trends in terms of which UoA tends to give a higher score (Thelwall et al., 2023i). This could occur because of different levels of strictness or because of the different criteria used.

Despite the assessor expertise, the detailed guidelines and repeated norm-referencing, the REF2021 output scores are imperfect. The main reason is that the 1120 experts will have substantial topic knowledge gaps, with none having the expertise to assess some of the 185,594 outputs. For example, none of the assessors for UoA 34, which incorporates library and information science, was a bibliometrician. In addition, there may be institutional, gender or other biases in scores, or simple prejudices against competing research paradigms, topics, or methods. Another problem is that each output had two assessors, giving a workload of about 370 outputs to score per assessor, over about a year. This is a substantial task for busy academics, and this seems to preclude a detailed assessment of each output. On the other hand, the articles have already passed journal peer review, so the REF assessors can expect to be primarily reading polished, high-quality research.

## 3.5   What else do academics achieve?

A final theoretical consideration, which often seems to be overlooked, is that academics do not just write journal articles but also participate in other activities that may be more important but that in any case are useful and may completely or partly replace journal articles. Thus, citation-based indicators are irrelevant to the evaluation of some academics. For example, any list of the most cited academics based on citation data would not be a list of the best, most influential or most important researchers because, in addition to the limitation of citations as research quality evidence discussed throughout this book, it would not reflect non-publishing activities and would probably also primarily reflect journal articles rather than other types of research output.

The non-publishing important activities of academics include: teaching; giving expert advice to government, industry and non-commercial organisations; research management; research administration (e.g., grant reviewing, journal reviewing); writing software; collecting data and maintaining databases; mentoring junior academics; ethical oversight; and public engagement. For example, suppose that Professor Ada Yonath struggled to find as much time to do research after winning her chemistry Nobel Prize because of repeatedly accepting invitations to talk about the importance of women in science (Ramakrishnan, 2019). In this case, she would still be conducting as much valuable scientific work, but of a kind that would be hard to measure and that would not translate into citations.

## 3.6 Summary of theoretical considerations

This chapter has discussed a range of theoretical considerations that must be understood and considered when evaluating, applying, and explaining citation-based indicators in research evaluation. Perhaps the most important issue is that citations only directly reflect one aspect, scholarly impact, which is one of the three dimensions that are usually considered to be core to research quality. Moreover, even for this aspect they are many theoretical reasons why they are largely irrelevant in some fields, such as the arts and humanities, and imperfect in the remainder. The next chapter reviews a range of practical considerations that undermine the value of citation-based indicators in many contexts.

# 4   Citation-based indicators: Practical considerations

Following on from the previous chapter, this one introduces a range of factors that can influence the value of some or all citation-based indicators. Some of these factors, such as document length, influence the likelihood of individual articles being cited and therefore undermine attempts to compare the value of articles with their citation counts.

## 4.1   Field or disciplinary factors influencing the value of citations

Academic disciplines are sets of researchers that have shared research interests, supported by journals, conferences, degrees and departments (Sugimoto & Weingart, 2015). The organisation of disciplines varies substantially, but their core goals are influencing the knowledge, funding, and education within their remit. They never exert full control since external actors like funding agencies and governments are also influential. An important role for a discipline is reputational control: overseeing the credit given to members for their work (Whitley, 2000). For example, a discipline may reward people thought to be successful through funding, prizes, invited talks, and editorships. The factors that are considered important for reputation vary between fields, bibliometric indicators amongst them.

Although the term field is sometimes used interchangeably with the term discipline, the two are sometimes distinguished. In this case an academic field is a body of related research with a focus on a topic or methods. If a field is supported by conferences, degrees, and departments then it can also be called a discipline. In this book, the main interest is in fields rather than disciplines.

There are multiple reasons why the value of bibliometric indicators like citation counts as evidence of research impact may vary between fields. Many of the issues that limit the value of citation counts already mentioned vary in importance between fields, for example. Here is a summary.

**Differences in the basic/applied nature of a field**. Since research applications do not contribute to citation counts, citation data is inherently weaker for more applied fields. Thus, citation-based indicators would presumably be weaker for chemical engineering than for chemistry.

**Differences in the geographic fragmentation of a field**. Some fields are inherently generic, such as pure mathematics, whereas others are inherently regional, such as Law. In between, some fields have regional elements mixed with generic elements. An example of this might be public health, where national health systems are important alongside universal knowledge about generic procedures to support public health.

**Differences in the importance of journal articles**. Monographs and edited volumes are dominant methods of disseminating research in the humanities and some areas of the arts and social sciences, whereas performances and artworks are important for the performing and fine arts. This creates two problems. First, since citation databases typically have weaker coverage of books than journal articles (Sivertsen, 2014), more of the citation impact evidence is lost in book-based fields than in article-based fields. Thus, an article may have had a big impact on book-based scholarship but have a low citation count in all citation databases (Kousha & Thelwall, 2015). Second, a citation analysis of journal articles would give weak evidence of the impact of scholars in a department, university, or country for fields where journal articles are not dominant. This would not be a problem if non-article outputs were evaluated separately.

**Differences in field homogeneity**. There are variations in the extent to which fields are organised uniformly, in the sense of following broadly agreed goals and with accepted methods (Trowler et al., 2012; Whitley, 2000). Fields with high funding, such as clinical medicine and high energy physics, seem to be relatively homogeneous out of necessity to allocate and justify funding. In contrast, library and information science seems relatively diverse in terms of topics, with many niche specialisms. For example, whilst bibliometrics is highly quantitative, computational, and statistical, human information behaviour is typically qualitative, and information management is more analytical. A citation analysis of the field of library and information science would typically treat these and all other specialisms as close to equivalent, so the results would be unfair to some specialists.

**Differences in the extent to which field knowledge is built hierarchically.** In more hierarchical fields (Whitley, 2000), such as medicine and the physical sciences, the ratio of "influence" citations to other types can be expected to be higher. In contrast, arts and humanities articles (and books) contribute insights rather than building hierarchical knowledge. In social science, engineering, and other professional and applied specialties (and all fields, to some extent), research can often be useful by producing societal benefits without necessarily contributing to future academic knowledge.

**Differences in speed and obsolescence**. In relatively slow-moving fields, such as classics, it may take typical articles decades to exert their influence on the scholarly record through citations. In contrast, articles in fast moving fields may have their main value within a few years of publication. Since citation analysis usually needs to be conducted within a few years of research being published to have practical value, this disadvantages slow moving fields because they will have a lower proportion of citation evidence to assess. Field speed is sometimes measured through citation half-lives (Yang et al., 2023).

For the above reasons, it is usual to either only compare citation-based indicators within fields or to normalise them (see the next chapter) so that they are comparable between fields. These approaches implicitly assume that the value of all fields is equal, however, which is not necessarily true. For example, the top 1% of cancer research articles might be substantially better on any realistic criteria than the top 1% of bibliometrics articles but the normalisation or benchmarking approach of comparing like-for-like would hide this difference, which may be evident in the raw citation counts (Thelwall, 2025). Nevertheless, assuming that all fields produce the same quality research, on average, may be the only practical approach for many research evaluation exercises.

## 4.2   Document type differences

Scholarly outputs take many forms, from books, websites, and artworks to journal articles. Even with the focus on journal articles, there are many different types. There is not really a standard journal article, but perhaps a paper of 4,000 to 8,000 words introducing new findings or analysis, with a reference list and with or without figures and tables is a reasonable template. The following are common variations on this.

**Length**: Some journals accept short form articles, variously called letters (not to the editor), brief communications, or short articles. There may also be specialist short form articles, such as medical case reports. Others accept longer papers and may expect multiple results to be reported within individual articles (Kitayama, 2017). In parallel, some prestigious journals enforce strict short maximum lengths on articles but implicitly or explicitly require considerable supporting material to be available in online supplementary files. These are long form articles masquerading as short papers. Other factors being equal, longer articles (including short articles with extensive supplementary materials) have more content to cite and therefore have an advantage in fields publishing mixed length articles.

**Purpose**: Whilst academic journals typically publish some documents that are clearly not journal articles, such as editorials, letters to the editor (if not short form articles), commentaries, and annotated images (e.g., of unusual medical issues), these are typically excluded from citation analyses and do not cause problems, except with Journal Impact Factors (Seglen, 1997). A bigger issue is the literature review type of paper. Literature reviews tend to have higher citation rates than primary research, in part because they can hijack citations from the sources referenced. Their original value is in organising and synthesising the existing literature around a theme, but they are also cited as a type of concept marker, especially when their topic is not of primary relevance to the citing article. For these reasons, citation counts probably overvalue literature review papers. Some research evaluations, such as the REF, explicitly exclude them.

## 4.3   Time differences and citation windows

Citations take time to accrue because citing research may be conducted many years after the original article was published and there are also editorial, reviewing and publishing delays for the citing article. Thus, an article's first citation may take a year to appear, with early citations perhaps originating from the authorship team of the original article or others that were given pre-publication access to it. Citations may then continue to appear until the original article has been superseded by follow-up research, its topic is no longer actively investigated, its contribution is regarded as dated due to natural changes or it is just overlooked in literature searches because of its age. In practice it is impossible to be sure that an article will never be cited again and so the final citation count of an article is a theoretical rather than actual number.

Every citation analysis must get its citation counts at a specific point in time in the understanding that subsequent citations will be missed. In the unlikely event that all the articles analysed were published at the same instant then comparing citations between them would be fair. More usually, however, the articles originate from a range of years. This causes the problem that older articles have an unfair advantage over younger articles in that they have had more time to accrue their citations and are likely to have accrued a higher proportion of their theoretical eventual citation count. There are two approaches to address this issue, the citation window and benchmarking or field normalisation.

A **citation window** is a specified number of years over which citations will be counted (Wang, 2013). For example, if a set of articles from a ten-year period 2011-2020 is to be analysed with a citation window of three years, then this means that the citation count for each article in the set would only include citations from articles published within three years. The citations to an article published in 2011 would therefore count only articles published up to 2014 but the citations to an article published in 2020 would include articles published up to 2023. This is fairer on the newer articles than counting citations to date for all of them, but at the price of data loss.

Longer citation windows give more reliable citation indicators because of the extra citations that they include. For example, in Figure 4.1, the correlation between expert review scores and citation counts for REF data is higher for older articles because they have a longer citation window. In more detail, health or life science articles published in 2016 would have an average citation window of 4.5 years for citation data collected in January 2021. In contrast, health or life science articles published in 2020 would have an average citation window of 0.5 years for citation data collected in January 2021 (Figure 4.1). The shape of the graph suggests that 4.5 years is sufficiently mature for the power of citation counts as an indicator of research quality to be close to the maximum in the health and life sciences and perhaps also the social science. In contrast, longer may be needed in the physical sciences and engineering, and the arts and humanities. Shorter citation windows can be used in practice, but the indicators will be weaker.

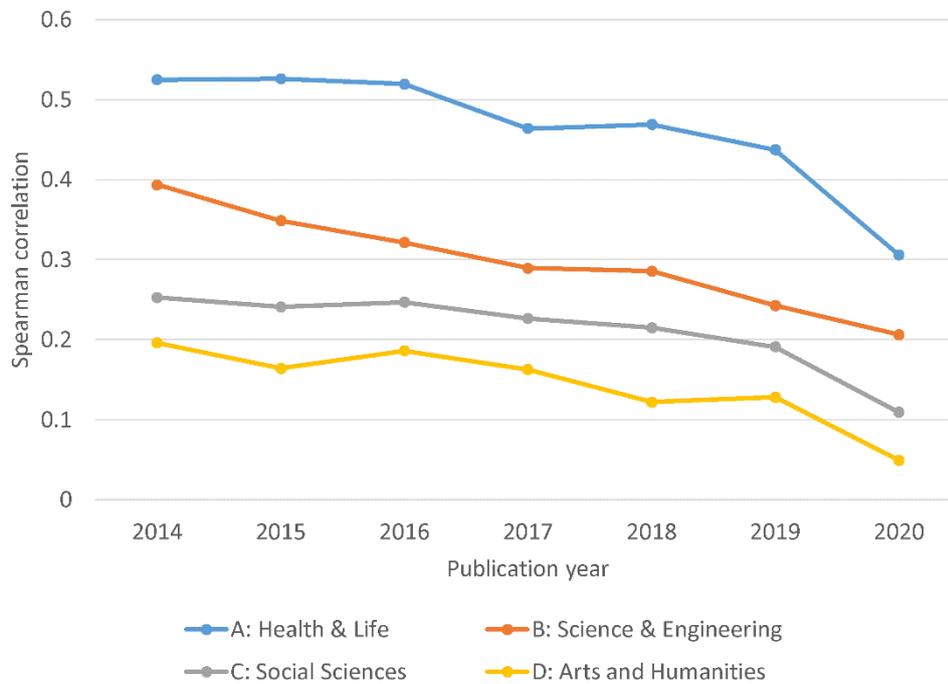

*Figure 4.1 Spearman correlations between Scopus citation counts from January 2021 and REF research quality scores for articles published between 2014 and 2020 (source: author).*

The concept of a citation window is also relevant to journal citation rate or impact calculations and Scopus and Clarivate may report impact factor calculations with different citation durations (e.g., [two year] impact factors and five year impact factors). Figure 4.2 suggests that long citation windows are not needed in the health and life sciences and perhaps also not in science and engineering. Overall, time is less important for journal impact calculations because they average citations across each journal and therefore accumulate more data.

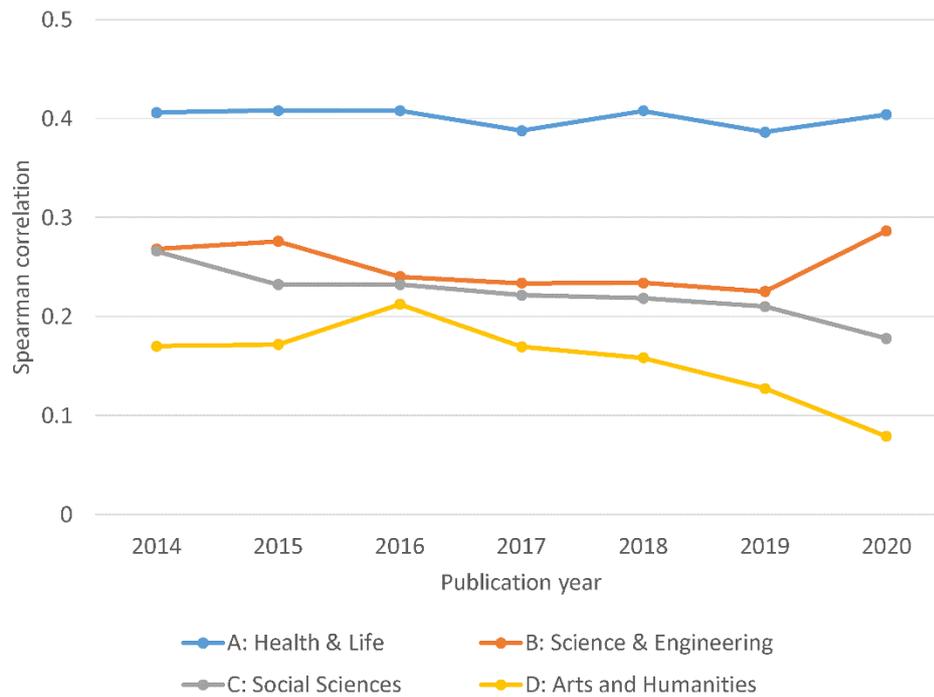

*Figure 4.2 Spearman correlations between average journal citation rates, as calculated from January 2021 Scopus data, and REF research quality scores for articles published between 2014 and 2020. The journal impact formula used is JMNLCS, calculated separately for each year based on that year's citation counts.*

The **benchmarking** approach avoids the citation window data loss by counting citations to date but only comparing articles published in the same year or by using an indicator formula that factors out the publication year. Thus, in the former case, citation counts for 2011 articles would be compared to citation counts for other 2011 articles and the citation counts for 2021 articles would be only compared against the citation counts for other 2021 articles. For the latter case, field and year normalisation formulae or percentiles (see the chapter on indicator formulae) would allow fair comparisons between 2021 articles and 2011 articles.

A currently only partly resolved issue is that of **within-year differences** in citation rates. In particular, an article published in January has almost a year more to accrue citations than an article published in December (de Araújo et al., 2012). This is a tricky issue because there is too little data to accurately benchmark on a daily or probably even monthly basis, so this difference is normally ignored. Ignoring it is fairly safe if it seems unlikely that there is a systematic difference between the sets of articles analysed. For example, it must be surely rare that a department primarily publishes research in the first or second half of a year. It would be possible to use a mathematical formula to compensate articles for late-year publication if publication dates are systematically available, but this does not seem to be common. Instead, the usual approach is to ensure that each article assessed is at least three to five years old (depending on field citation half-lives) so that the difference between early and late year publication is relatively small as a proportion of longer-term citations (Wang, 2013).

## 4.4 Categorisation differences

Citation-based indicators (should) all rely on implicit or explicit field classification schemes for the value of the information conveyed (see Chapter 5). At the simplest level, if citation counts are compared between articles produced by an economics department to find the apparently most impactful produced, then the implicit classification is economics. It does not make sense to find the most cited articles produced by a university because average citation rates differ so substantially between subjects that there would be little chance of recognition for articles from the arts, humanities, and most social sciences. For more complex indicators, article citation rates might be ranked by field (e.g., top 5% most cited psychology articles from 2023) or averaged by field (e.g., citation count divided by the average citation count of all other biochemistry articles from the same year).

Although there are well known names for parts of science (e.g., maths, medicine, fine arts), there is no definitive list of academic fields and the common fields of research, however defined, change over time. For instance, nursing, computer science, and psychology emerged in the last century as academic fields, botany has arguably become dwarfed almost out of existence by other life sciences and the horrible specialism of eugenics has nearly disappeared altogether (Sear, 2021). Equally fundamentally, much research combines inputs from, and makes contributions to, multiple different fields. Because of these issues, citation-based indicators must use heuristics to identify fields to examine and for delimiting articles to make fair comparisons. There are many different approaches, both journal-based and article-based. All seem to start from the idea that an academic field can be equated with a body of publications that have a common theme or set of related themes. As mentioned above, this is different from a discipline, which also has supporting journals, departments, and scholarly organisations.

The manual journal-based method to assign articles to fields is for a group of experts to decide on a classification scheme and then assign each journal to one or more of the pre-defined classes. The initial classification scheme might be produced by clustering journals together based on their citation relations and identifying clusters of journals with fields. The assignment may be primarily achieved from the article title, aims and scope, although bibliometric information about the journal may also be used, such as the extent to which it cites, and is cited by, articles in all fields. For example, the Journal of Physics might be assigned to both the theoretical and applied physics categories on the basis that its aims statement mentions both and it cites articles from both. The advantages of this manual journal-based approach include simplicity and relative transparency, but it is labour-intensive and is ineffective for journals that publish articles from multiple fields, such as Science and Nature, since all the articles in these get assigned to all the journal's fields, irrespective of whether they are good matches. Another problem is that journals sometimes publish out-of-field articles (e.g., Desai et al., 2018) and these will be misclassified by the journal approach. Nevertheless, it is used by Scopus and the Web of Science, albeit with different numbers and names of fields.

Articles can also be individually and automatically assigned to fields based on their references, citations, or metadata. For the text components, this can be done essentially by finding words commonly used in each field and assigning new articles to fields where their words match best. It can also be done through citations or references, assigning articles to fields with the highest share of references from or citations to the article. All three approaches can also be combined. Dimensions.ai uses this for its default article classification (Hook et al., 2018) and the Web of Science and Scopus offer article-level classifications as options for use in some of their citation-based indicators. Although article-level classification has the theoretical advantage of being finer grained and capable of higher accuracy than journal-based classification (since journals often contain articles from multiple fields), it can also generate systematic errors and anomalies due to shared references between different fields (e.g., for methods, background), polysemy, multidisciplinary papers, and articles close to the classification margins.

The categorisation process affects all citation-based indicators because of implicit or explicit norm referencing. For example, if a music physics article is incorrectly classified as music (arts) then it would gain a substantial citation norm referencing advantage by being primarily compared to articles from a very low citation specialism. Perhaps it has ten citations, which is average for physics from its publication year, but it is in the top 1% most cited for music research.

## 4.5   Database differences and limitations

The database used to count citations can influence citation-based indicators primarily because of the extent of its coverage of articles but also because of its classification scheme, if used. The coverage of academic databases varies substantially and non-systematically, so the results of any citation analysis are partly dependant on the database chosen for the citation counts used. There are many current options, including OpenAlex (Priem et al., 2022), Dimensions, Scopus, and the Web of Science (see also: Velez-Estevez et al., 2023).  Google Scholar usually not a practical choice, at least for large scale evaluations, because of a lack of subject categories and automated data access. There are also subject-specific databases with citation data or citation indexes. The following are some important dimensions of variation.

**Quality selectivity**: Scopus and the Web of Science have quality assurance procedures to decide which academic journals to index (Baas et al., 2020; Birkle et al., 2020). Thus, journals with predatory characteristics can't expect to be indexed. Irregularly publishing journals and journals with articles that are rarely cited may also not be indexed in some fields. The Web of Science partly compensates for its partial coverage with its Emerging Sources Citation Index supplement of additional journals that have their citations extracted (Somoza-Fernández et al., 2018) but are not part of its main indicators. At the other extreme Google Scholar and OpenAlex may have the least quality control on contents.

**Language selectivity**: Scopus and the Web of Science only index articles with English-language title and abstracts, or when translations are available of these. This creates a bias against non-English speaking nations that do not have widespread experience of working in English (Van Leeuwen et al., 2001). OpenAlex has the goal to be particularly universal in terms of language and countries of origin. This seems to give OpenAlex an advantage for citation analyses that focus on countries not well covered by other databases (Céspedes et al., 2025).

**Discovery selectivity**: There is not a single universal list of indexable academic journal websites and so every citation index needs a method of discovering journals to index. This might be through arrangements with publishers (Dimensions, Google Scholar, Web of Science, Scopus), perhaps with additional web crawling (Dimensions, Google Scholar, OpenAlex) and this creates the potential for substantial differences between citation indexes (Martín-Martín et al., 2021).

**Indexing accuracy**: Indexing errors are impossible to fully eradicate due to human error in the publishing chain. Nevertheless, data processing errors seem to be a relatively minor problem, except for journals when they are journal-based (e.g., losing many citations to a journal which does not publish DOIs). Similarly, if a specialist journal's references are in a format that citation indexes find hard to extract (e.g., footnotes with abbreviations and without DOIs) then articles in the same specialist field may tend to lose out on citations compared to others in the parent broad field. Data processing errors might be more frequent for newer and non-commercial databases because they take human labour to identify and correct (e.g., Orduña Malea et al., 2017).

**Document type classification accuracy and granularity**: When normalisation is used for citation analysis, it is important to compare like with like. For example, if editorials are rarely cited in a field, then it would be better to classify them separately from standard journal articles or the journal articles will have an advantage in the sense of higher normalised scores. At the time of writing, document type classification type classification was a problem for Google Scholar, which does not classify by type, and OpenAlex, which classifies all contents of journals as "journal-article", conflating editorials, commentaries and full articles (Haupka et al., 2024).

Database coverage issues can impact on citation evaluations in three ways. First, if the database has limited coverage of the documents to be analysed, then an evaluation might not be possible or, if conducted on the limited sample indexed in the database, might be biased by the extent of that sample. For example, if a sociology department in Cuba used a database that did not index its Spanish-language publications then its evaluation results would only depend on the quality of the research of its English-publishing scholars.

Second, citation counts will tend to be lower when counted in smaller databases or a database with many indexing errors, so a research evaluation will tend to be weaker through less data. Moreover, if articles are systematically missing (e.g., few Spanish language publications are indexed), the citation counts will tend to be biased against researchers working in the underrepresented areas (e.g., Spanish language topics) even if the researchers' work is included.

Third, and partly conflicting with the previous two points, citation counts from databases without quality assurance will include citations from low quality work and perhaps even fake or deliberately manipulative papers aiming to distort citation-based indicators. For example, the Web of Science periodically punishes and potentially deselects journals suspected of citation manipulation (RetractionWatch, 2024), showing that it does occur and that databases without the resources to tackle the issue are likely to retain the manipulated citations. This issue of indexing low quality sources can be partially alleviated using indicators that give lower weights to citations from less cited journals (Walters, 2024), such as the SCImago Journal Rank (González-Pereira et al., 2010). Indexing low quality work can also distort normalised citation indicators by creating unfair comparison sets. To give a simple example, suppose that a database adds a large collection of pharmacology magazines with popular articles that are never cited to its pharmacology category, doubling its size. This will mean that any genuine pharmacology article with 1 citation would be above average for any citation indicator in its category because half the articles are uncited. This would be unfair if, for example, without the magazines 10 citations were required to be cited. This is a tricky issue because citation indexes seem to frequently expand by adding less cited work, perhaps from formerly underrepresented countries. These have the effect of boosting the citation rank (whatever indicator is used) of most articles already in the system.

As this discussion shows, the issue of database coverage is not straightforward and bigger is not always better. A reasonable choice seems to be to use a database with adequate coverage of the publications to be analysed, and with as much quality assurance as possible. Moreover, if a database without quality control is used for an important evaluation, then the possibility that those evaluated will game the database should be considered. For example, they might try to increase their citation counts by publishing heavily self-citing trivial papers in places that the database would index.

## 4.6 Differences in individual contributions to articles

Modern research is usually collaborative, with most journal articles having more than one author, although solo authorship may be as common as collaboration in the social sciences (both claims are extrapolations from: Thelwall & Maflahi, 2020), and solo authorship may be the norm in the humanities. This complicates the situation when it is useful to differentiate between the author contributions, including when assessing individual scholars, departments, universities, or countries using aggregate citation data.

The fairest solution to the problem of attributing appropriate credit to each co-author in a team might be to allocate percentage contribution shares to each one when counting output or citation indicators. This is sometimes done for promotions, for example (I occasionally sign letters for my co-authors applying for promotion to testify to the percentage contribution that I made to our joint publications). This seems to be impractical for bibliometric exercises since this percentage contribution information is not recorded. Whilst some articles have an associated Contributor Roles Taxonomy (CRediT) statement (credit.niso.org; Larivière et al., 2021), it specifies the type and not the extent of each contribution and does not seem to be common enough to be used in large scale bibliometric exercises.

Despite the above considerations, bibliometric exercises seem to usually allocate full credit to each author of a document, as if they had written it alone. When aggregated indicators are calculated (e.g., departmental citation rates) then full credit might instead be allocated to each contributing unit (e.g., department, university, country), perhaps because this is the simplest approach and more complicated strategies might greatly increase the complexity (and cost of increased citation data access if the strategy is not possible through the standard price interface of the bibliometric database used) of the task. This must be acknowledged as a limitation, when used. The main strategies proposed include the following (Waltman, 2016).

- **Full counting**: each author (or contributing unit) is allocated full credit for each co-authored publication.
- **Fractional counting**: each author is allocated 1/n of the credit, where n is the number of co-authors.
- **Geometric/arithmetic/harmonic counting**: As for fractional counting, but a formula is used to allocate most credit to the first author and a decreasing share to subsequent authors. This is reasonable because the first author is the main contributor in most fields – alphabetical ordering and joint-first authorship are rare across science. This is complicated by corresponding authors being sometimes more important than first authors. A formula where the total credit allocated is greater than 1 might also be used to reward collaboration.
- **U-shaped counting**: As for fractional counting but a U-shaped formula is used to ensure that the first author gets most credit and last author also gets a substantial share of the credit. This is motivated by last authors being senior scholars that might have won the funding for a study, designed it, recruited the team and overseen its general operations despite having little direct involvement in the data collection, analysis and writing of a paper. There is empirical evidence to support the value of last authors in some fields (Larivière et al., 2016b; Sundling, 2023).

The above strategies are all heuristics to circumvent the problem that there is insufficient data on authorship contributions to perform a robust analysis of large sets of publications.

A deeper theoretical question that seems unlikely to be addressed concerns the best aspect to assess of the contribution of each author. Possible answers include time commitment (advantaging those conducting labour intensive but routine tasks), creative contribution (advantaging those designing research), or opportunity cost (how much weaker would the paper have been without a given author or how difficult would it have been to replace them).

Finally, the above discussion assumes that everyone contributing to a paper is allocated a co-authorship and those that don't are not. This is untrue in general due to gift authorship, ghost authors and field and probably national differences in decisions about the types of contributions that merit a co-authorship attribution.

## 4.7 Summary of practical considerations

This chapter has discussed a range of practical reasons why citation-based indicators can have less value in some contexts or can be misleading. The influence of some of these factors has been shown to be substantial (Thelwall, 2025). As subsequent empirical chapters show, this does not mean that citation-based evidence is irrelevant or can never be used in research quality evaluations. Instead, it means that they should always be used cautiously. The issues discussed in this chapter and the previous one all underline the importance of empirical evidence, such as through correlation tests, to validate indicators before they are used.

The next chapter provides more concrete information about citation-based indicators in the form of specific formulae. Most of the factors influencing citation counts discussed in the current chapter are not able to be considered within the formulae, with the main exceptions being field and time in the sense of publishing year. Subsequent chapters also give evidence for the value of citation-based indicators in different fields, primarily with the aid of UK-based evidence.

# 5  Citation-based indicator formulae

As the chapter about theoretical considerations for citation-based indicators shows, there are many factors that influence raw citation counts, so asking which article, or set of articles, has the highest citation count is usually unhelpful. For this reason, a wide range of formulae have been developed to remove some biases to allow fairer comparisons. The main considerations usually factored out are field and publication year, although document type is often considered indirectly by only allowing comparisons between documents of the same type. All of these approaches have the implicit assumption that all fields produce equal value research, however, which may not be true (Thelwall, 2025).

The two main types of indicators are probably percentiles and field normalised citation rates. Percentiles are intuitive: for example, a database might report that an article is in the top 5% cited for its field and year. Field normalised citation rates are slightly less easy to interpret; they are comparative citation rates that reflect the number of citations to an article compared to all other articles from the same field and year.

This chapter includes article-level formulae, as illustrated above, as well as aggregate formulae that could be applied to sets of articles, such as from departments, universities, countries, or journals. If necessary, the formulae be converted into weighted versions to allocate partial credit to authors within collaborations.

## 5.1  Percentiles

For any percentage X%, an article is in the Xth most cited percentile if it is in the top cited X% of articles from its field and year in the database used (often also of the same document type). For example, if there were 100 articles cited 0,1, 2, 3, … 99 times then the 99-cited article would be in all the percentiles, including the top 1% most cited, but the 1-cited article would only be in the top 100% (i.e., all articles) and top 99% most cited. Whilst whole number percentiles are usually calculated, the top 0.1%, top 0.01% or even top 0.001% cited may also be reported for large sets of articles.

Percentiles are directly comparable between fields and years, by design. For example, if a physics article from 1973 was in the top 50% cited for 1973 physics (i.e. the 50th percentile) but a history article from 2021 was in the top 10% cited from 2021 history articles then the history article would have the better citation record even if the physics article had been cited ten times more.

When sets of articles are compared with percentiles, the average percentile of the set could be calculated but it is more common to calculate the percentage of articles in each set that fall within one or more pre-specified percentiles. For example, it might be reported that a sociology department had 3% of its articles in the top 1% cited. This would be a good achievement given that the percentage is above the world average (1%) since the world average for the top X% cited is X% (assuming large enough numbers and few citation count ties, which would otherwise complicate this statement). If more detail is needed, then it might also be reported that the department had 70% of its articles in the 50% most cited (50% would be the world average) and 15% of its articles in the 5% most cited (5% would be the world average). This department would be performing well, in citation terms, by producing above average percentages of moderately cited, highly cited and very highly cited articles.

### 5.1.1 Limitations of percentiles

Percentiles have two main limitations. First, the thresholds used (e.g., the top 50%, 10%, 5% and 1% most cited) are relatively arbitrary and different thresholds may sometimes produce substantially different results, by the law of averages. Second, which is related to the first, percentiles lose a lot of the information about citation counts, resulting in a lossy simplification. For example, if the threshold to be in the most cited 1% is 100 citations, then this percentile indicator would not differentiate between an article cited 100 or 1000 times.

Other limitations include dependence on the field classification scheme used, problems with research that is not monodisciplinary, and the need for a substantial time window (several years) for the citation data to be mature enough to be used. Articles published early in a year would have a substantial advantage for selective percentiles (e.g., 1%) without a wide citation window.

## 5.2 Field and year normalised citation counts: MNCS

Field normalised citation counts address the main limitations of percentile indicators by not needing arbitrary thresholds and incorporating the full range of citation counts for all articles. Probably the most widely used is the Normalised Citation Score (NCS), which is the number of citations to an article, divided by the average number of citations to all articles published in the same field and year (Waltman et al., 2011). For example, if a history article from 2022 had been cited 10 times and all history articles from 2022 had been cited an average of 5 times each, then the NCS for the article would be 10/5=2. If an article had been cited an average number of times for its field and year, then its NCS would be 1. Values above one indicate an above average citation rate and values below 1 indicate a below average citation rate. Since this is true for all articles, irrespective of field and year, it is reasonable to compare the NCS for articles from different fields and years, if necessary.

If citation rates need to be compared between sets of articles, then this can be achieved in a straightforward and natural manner by calculating the NCS for each individual article and then averaging these values for each group. This gives the Mean Normalised Citation Score (MNCS). After this averaging it is still true that a value of 1 indicates world average citation rate for the field and year, with values above 1 being above the world average and values below 1 being below the world average.

### 5.2.1 Mathematical formulae

The Normalised Citation Score for an individual article with $c_0$ citations can be calculated as follows, where $\bar{c}$ is the average of the citation counts $c$ for all articles in the same field And year.

$$NCS_0 = c_0/\bar{c}$$

If the article has been assigned to multiple fields, then the denominator above is replaced by the average (arithmetic mean or harmonic mean: Waltman et al., 2011) of the $\bar{c}$ averages for each field.

For a set of $n$ articles with NCS values $NCS_1, NCS_2, \ldots NCS_n$, the Mean Normalised Citation Score (MNCS) is the arithmetic mean of the NCS values, as follows.

$$MNCS = (NCS_1 + NCS_2 + \cdots + NCS_n)/n$$

### 5.2.2   Limitations of the NCS and MNCS

The NCS and MNCS do not have the two main limitations of percentiles. First, there is no arbitrary threshold to decide on for the calculations. Second, no data is lost in the calculations: every citation makes an equal contribution.

NCS and MNCS calculations share two weaknesses with percentiles. They are dependent on the field classification scheme used, so a poor scheme can result in misleading values. This is a substantial problem in practice because field classification schemes are often crude (e.g., based on journals rather than individual articles) and in any case the field classification of articles is a difficult task that is complicated by cross-field research. A second weakness in some cases is that the articles need to be several years old to be robustly compared. A citation window of three years is usually recommended (e.g., Wang, 2013), but this means that recent research cannot be evaluated with the support of citation data.

The NCS has a limitation that does not apply to percentiles: a substantial majority of articles have a below average NCS. This is because the highly cited articles overload the denominator of the calculation. The MNCS also has an additional limitation that does not apply to percentiles: the likelihood of being dominated by a few highly cited articles and therefore giving values that are not representative of typical research. This occurs because sets of citation counts are typically highly skewed, with many 0s and a few high values (de Solla Price, 1976). The average in this context is dominated by the large numbers so a few highly cited articles can be highly influential.

### 5.3   Skew-corrected field and year normalised citation counts: MNLCS

The skewing limitation of the MNCS can be avoided by log transforming all citation counts with the formula $ln(1 + x)$ to calculate a Normalised Log-transformed Citation Score (NLCS) to replace the NCS. In the formula, 1 is included because otherwise the transformation would be undefined for uncited articles ($ln(0)$ is undefined). This solves one of the problems of the NCS for individual articles: whilst a clear majority of articles have an NCS below the world average because of the high values from highly cited articles in the denominator, about half of NLCS should be below average.

After this log transformation is applied to the raw data, averaging NLCS across a set of papers gives the Mean Normalised Log-transformed Citation Score (MNLCS) (Thelwall, 2017). The log transformation here largely eliminates the skewing problem of MNCS.

Although applying a log transformation to citation data is a statistically reasonable solution, it might be objected to on the grounds that each citation should be valued equally. Nevertheless, there are two good reasons not to do this. First, the log approach gives a more accurate measure of the central tendency of the data by largely eliminating in the influence of highly cited papers. Secondly, there is evidence to suggest that highly cited articles attract a proportion of their citations through imitation or they get recognised as the "concept marker" paper to cite for a particular function (Case & Higgins, 2000). On this basis, the individual citations of highly cited papers are much less important than the individual citations of less cited papers and the log transformation reflects this.

### 5.3.1   Mathematical formulae

The Normalised Log-transformed Citation Score for an individual article with $c_0$ citations is as follows, where $\overline{ln(1 + c)}$ is the average of the logged citation counts $ln(1 + c)$ for all articles in the same field and year as the article Assessed.

$$NLCS_0 = ln(1 + c_0)/\overline{ln(1 + c)}$$

If the article has been assigned to multiple fields, then the denominator above is replaced by the average (arithmetic mean) of the $\overline{ln(1 + c)}$ averages for each field.

For a set of articles with NLCS values $NLCS_1, NLCS_2, \ldots NLCS_n$, the Mean Normalised Log-transformed Citation Score (MNLCS) is the mean of the NLCS values, as follows.

$$MNLCS = (NLCS_1 + NLCS_2 + \cdots + NLCS_n)/n$$

If the articles averaged are from a single journal, then for convenience, the calculation is sometimes called the JMNLCS or Journal MNLCS.

### 5.3.2 Limitations of MNLCS

The NLCS and MNLCS have the same dependence on citation windows and field classification schemes as MNCS. They do not have the skewing problem of MNCS but have swapped this for a formula that is substantially less transparent to end users. Whilst the MNCS can be easily understood as the citation rate above or below the world average, the MNLCS is the more difficult concept of the logged citation rate above or below the world average. Thus, policy makers, research managers and academics may feel less confident to use and interpret it. Of course, lower confidence in citation indicators may be helpful to reduce the chance that they are given too much weight in research evaluations.

## 5.4  Network-based indicators

All the indicators analysed so far do not judge the value of individual citations but just count them, with or without a subsequent log transformation. In practice, not all citations are equal, and some are more influential than others. For example, a citation from a highly cited paper may be more important than a citation from an uncited paper, other factors being equal. Network-based citation indicators attempt to weight citations individually based on the role that they play in the network of citations between articles, typically through a heuristic to assess influence.

Approaches to calculate the network influence of citations are often motivated by Google's PageRank algorithm, which was designed to weight hyperlinks according to the importance of the page that they originated in (Page et al., 1998). In this algorithm, the importance of a page is estimated by counting links to it. It is an iterative heuristic with the result that each page (or article) gets a score that is determined by the number of links (or citations) to it and the importance of the originating webpages (or papers) (Ma et al., 2008). This is a heuristic because it depends on an arbitrary parameter but, from the success of Google, seems to be effective. Nevertheless, there are important limitations, such as high scores accruing to papers from trivial background citations in important papers. The SCImago Journal Rank, as used by Elsevier, is an indicator based on the idea behind PageRank (Mañana-Rodríguez, 2015), and the Relative Citation Ratio (Hutchins et al., 2016) is another prominent network-based indicator, but it exploits co-citation networks.

### 5.4.1  Limitations of network-based indicators

Network-based indicators have three main problems. Like the MNCS, they are highly influenced by a few highly cited articles. They are opaque and without an intuitive understanding, which makes them even harder to interpret than MNLCS. Finally, they can overweight trivial citations from influential papers, since not all citations are equally important. Because of this last point, it is not clear when (i.e., which fields and years) network-based indicators are more accurate (i.e., as indicators of either research quality or scholarly impact) than MNCS or MNLCS. Although network-based indicators seem promising in theory, their limitations mean that it does not seem reasonable to use them instead of simpler indicators until there is strong evidence of their greater value. Nevertheless, they may have extra value when a citation database is known to cover a range of low quality sources, since network indicators would work to discount citations from these.

## 5.5  JIF and other journal-based indicators

In principle, all the citation-based indicators can be applied to any individual articles or groups of articles but in practice the citation-based indicators used for journals tend to include specific journal-based aspects. A range of these are discussed here to connect them to the types of indicators mentioned above and to illustrate journal-specific features.

### 5.5.1  Journal Impact Factors

The oldest and best-known journal impact indicator is the Journal Impact Factor (JIF) from the Web of Science, as invented by citation analysis pioneer Eugene Garfield (Garfield, 2006). It is an attempt to assess the average citation rate of recently published articles in a journal on the basis that journals with higher averages may tend to publish better work. If this assumption is true, then it would help readers to prioritise journals to scan and authors to prioritise journals to submit to. When it was invented, it was an innovative way to move beyond tacit knowledge about the relative merits of journals in a field and towards quantification, taking advantage of the new availability of computing power for documentary analysis purposes in the 1960s.

The formula for the year X JIF, as used in Clarivate's Web of Science in 2023, is the number of citations to items published in year X-1 and X-2 in the journal from all articles indexed in the Web of Science in year X, divided by the number of citeable items (mainly journal articles) published in the journal in year X-1 and X-2 (Clarivate, 2023). It therefore gives the average citation rate of recently published articles from more recently published articles.

A technical criticism of the JIF is that the citations counted in the numerator include citations to document types that are excluded from the denominator, such as editorials and letters. Journals that publish lots of citable but not countable documents therefore have an advantage in this respect and editors might take advantage of this to game the system. The JIF also ignores the skewed nature of citations by using the arithmetic mean in its numerator, which is undesirable. Thus, JIFs change dramatically sometimes due to individual highly cited articles (Thelwall & Fairclough, 2015).

The JIF does not correspond to any of the citation indicator types discussed above because it is not year or field normalised. It is partially year normalised in the sense that it is calculated from a fixed citation window, so older JIFs do not have the advantage of a longer time to attract citations. Nevertheless, JIFs from different years are not comparable because science is expanding (hence more potential citing documents for later years). Moreover, increases in the speed of publishing have tended to increase early citations, so JIFs have tended to increase over time, probably also due to lengthening reference lists (Bergstrom et al., 2009). Although the JIF is not field normalised, in the Journal Citation Reports, the tables of journals and JIFs are organised by field and, when journals boast about their impact factors, they usually mention their JIF rank within one or more fields (e.g., 4th in the Applied Physics category). Thus, the lack of field normalisation is partially compensated for by comparisons tending to occur within fields.

### 5.5.2   Longer citation window journal impact formulae

The short citation window for the JIF has the advantage of currency: it reflects the citation rates of recently published articles. Nevertheless, it gives an advantage to journals that publish articles that tend to attract early citations and penalises journals publishing articles that tend to take longer to be cited (Liu & Fang, 2020). For example, if assessing journals using a five-year citation window, journals that disproportionately attract quick citations will tend to rank worse than for the traditional two-year citation window (Campanario, 2011). This time difference may be inherent to the specialities covered by the journal, or the types of articles published. In one small study (28 ophthalmology journals), a three- or four-year citation window gave results that correlated best with expert judgements of journal quality (Liu et al., 2015), but a shorter time window may now work as well, given the incorporation of early view citations into citation indexes.

Some commercial JIF variants use longer citation windows to be fairer to all journals and perhaps also for increased stability due to more data, but at the expense of reduced currency. Examples include Clarivate's 5-year Journal Impact Factor (incites.help.clarivate.com/Content/Indicators-Handbook/ih-5-year-jif.htm), and CiteScore from Scopus, which uses a four-year citation window (service.elsevier.com/app/answers/detail/a_id/14880/supporthub/scopus/).

### 5.5.3   Field and year normalised journal impact formulae

A few journal impact formulae consider field and/or year differences in citation practices. These include Scopus's Source Normalised Impact per Paper (SNIP), invented by the late great Henk Moed (Beatty, 2015) and Clarivate's Journal Citation Indicator (Szomszor, 2021). These are essential for cross-field and cross-year comparisons because they greatly reduce or eliminate field and year differences as sources of bias.

Whilst the Journal Citation Indicator uses an MNCS type formula, the SNIP method is unusual and different to all other formulae described in this book. It essentially compares the number of citations received by a journal with the average number of references available in the citing publications (Waltman et al., 2013). This is a type of field normalisation because some fields have long reference lists and articles in these fields would tend to have higher citation rates, other factors being equal. Its advantage over MNCS or MNLCS type formulae is that it does not need a journal categorisation scheme.

### 5.5.4   Network based journal impact formulae

Network-based journal impact formulae are currently rare, but Scopus's SCImago journal rank based on Google's PageRank is an exception (Mañana-Rodríguez, 2015). This gives higher weight to citations from more cited papers so a journal can get a high score from its papers receiving many citations or from its papers receiving a few citations from high citation impact papers. This adds complexity to the formula, and it is opaque in practice because of the recursive nature of the calculations involved. Nevertheless, it produces results that agree better with expert journal ranks than most other bibliometric formulae, at least in Business and Management (Walters, 2024).

## 5.6   The h-index and other indicator formulae

The sections above have covered the most used types of citation impact formulae used to give data on the citation impact of a body of research outputs. Hundreds of other formulae have been proposed, and a review can be consulted for many of these (Waltman, 2016).

One conspicuous omission from this chapter is the h-index. For a set of papers, the h index is the largest whole number h such that at least h papers all have at least h citations. This has become popular as a single simple number that combines the number and impact of the outputs of an academic, although the formula has also been used for other entities, including journals and universities. This formula is not recommended because it combines two separate quantities (output volume and citation impact) that are not directly related and are therefore better reported separately rather than in a single hybrid indicator, however convenient (Gingras, 2016).

## 5.7   Correlation tests to validate indicators

Citation-based indicators are usually validated mainly through correlation analysis. For example, the citation indicators could be compared against expert or Large Language Model assessments of article quality to assess how far they agree and if the degree of agreement is above random chance (e.g., Thelwall, 2025). It is important to understand the value and limitations of correlation tests. There are two main types of correlation formulae, Pearson, which tests for a linear relationship between two variables, and Spearman, which tests the extent to which the two datasets are in the same rank order. For most scientometrics datasets the two seem to give similar results but the Spearman test is a safer one to use because scientometric data can be skewed, so a linear relationship is not a default expectation.

Calculations for the Spearman's rank correlation coefficient start by independently ranking each of two datasets from the smallest to the largest. The test then puts these ranks into a mathematical formula to assess whether higher items from the first dataset tend to have higher ranks when items from the second dataset also do. For example, if the two datasets are citation counts and tweet counts then the formula would be checking whether the most cited papers also tend to be the most tweeted. The result of the formula is a number between -1 and 1. If the result is 1 then the two datasets are in the same rank order. For example, the 26th most cited paper would be the same as the 26th most tweeted paper. Conversely, if the correlation is -1 then the opposite would be true: the 26th most cited paper would be the 26th least tweeted paper. In the middle, a correlation of 0 indicates that there is no relationship between the two rankings. For example, the 26th most cited paper would have an equal chance of having any ranking for its tweet counts (in the absence of other information).

Correlation tests typically use a probability approach to report whether a correlation value is far enough away from 0 that it is unlikely to have been achieved due to the normal operation of chance. The test is usually made for $p<0.05$, at least to start with. So it is testing if the probability that there is no underlying (rank order) relationship between the two datasets (e.g., citation counts and tweet counts) is less than 0.05. Here, "underlying" can be interpreted as meaning either (a) the dataset examined is a sample from a larger population or (b) the dataset is an imagined sample from the theoretical population of reasonably possible articles under similar conditions. The latter is more common in citation analysis, although this is rarely explicit. For example, if a correlation is calculated based on all Web of Science articles from 2022 in chemistry, then this is a complete set of WoS rather than a sample. Nevertheless, it can be thought of as a sample in the sense of a subset of all Web of Science chemistry articles (including those before and after 2022) or as a sample in the sense of a collection of articles that was produced in 2022 under those specific conditions but accepting that if the conditions stayed the same then a different set of articles would naturally be produced each year. In any case, if the p<0.05 test is passed, it is reasonable to say that the correlation is statistically significantly different from 0 or "statistically significant" for short. Since statistically significant negative correlations are rare, the main interpretation task for scientometricians is evaluating the importance of statistically significant positive correlations.

The magnitude of a correlation depends partly on the strength of the underlying relationship between the two datasets (e.g., citation counts and tweet counts). In scientometrics, statistically significant correlations usually vary between 0.2 and 0.5, which is far from the theoretical maximum of 1, so how do we interpret such values? The following influencing factors need to be considered.

- For each variable individually, how much of its values can be expected to be influenced by things that cannot be measured, which might be thought of as chance factors. For example, the number of citations attracted by an article is influenced by the number of people that decide to conduct related research and how long they spend on literature searching, both of which are hard to systematically measure.
- Related to the above, how fine grained are the values? If a dataset consists primarily of 0s then its rank order will have lots of ties and it would be unreasonable to expect a high correlation with another dataset. This is possibly the reason why "rare" altmetrics like Wikipedia citations don't have high correlations with citation counts.
- How varied is the dataset? Any correlation for a dataset of paper with mixed quality can be expected to be higher than a correlation for a dataset of papers that are exclusively high quality, other factors being equal, because of the way in which the correlation formula works (Carretta & Rees, 2022). This is because small inaccuracies cause greater problems for the correlation formula if the difference between the values in the dataset is also small.

Unfortunately, it is not possible to give strong guidelines about how to interpret correlation coefficients for scientometric variables. This is an issue for all social science fields that use correlations for human-influenced data because of the natural variation between people. In psychology, one highly cited guideline for appropriate terms for correlation strength/effect size is: weak/small (0.2-0.3), moderate/medium (0.3-0.4) and strong/large (0.5+) (Cohen, 1988). This recommendation was heavily caveated and was based on the typical values found in psychology research. This same guideline seems reasonable for scientometrics research because it is broadly in line with the range of correlations observed and scientometric data is all ultimately human-derived.

It is important to remember the common saying, "Correlation is not causation". A non-zero correlation could occur because one variable influenced the other (causation) or because a third variable influenced both (not causation). In scientometrics, there are sometimes elements of influence, but third variable influences seem more likely, so it is important to be clear that correlation is not necessarily causation. For example, people might tweet about an article after finding it through a citation (citing influencing tweeting) or might cite an article after seeing it on Twitter/X (tweeting influencing citing) or a person might cite and tweet an article after finding it in Google Scholar and considering it useful (third variable, usefulness). Probably, most people tweeting an article don't cite it and most people citing an article don't tweet it though, so third variables like usefulness, quality or importance seem to be the most likely reason for any positive correlation between tweets and citations.

Finally, perhaps because of the correlation is not causation issue some scientometricians are sceptical of the value of all correlation tests. This may also be because most studies have correlated one indicator against another (e.g., citations against tweets) rather than an indicator against a direct measure of something (e.g., peer judgements of quality), so the information is tricky to interpret and does not give a clear answer to many research questions. Nevertheless, they are useful for showing whether two variables are related which gives information about them in the form of empirical evidence to support or refute conjectures about the value of new indicators. In particular, a statistically significant non-zero correlation with anything is evidence of non-randomness and that the two entities correlated are directly or indirectly related in some way . For a more detailed discussion of correlation issues, see (Thelwall, 2016a).

## 5.8   Indicator availability

The indicators discussed in this chapter all need a computer program to calculate. Whilst a practicing scientometrician probably will write or have access to relevant programs or use a spreadsheet for small scale analysis, others may rely on precalculated indicators from major databases. At the time of writing, OpenAlex, the basic versions of the Web of Science and Scopus provided metadata, citation counts, and broad subject classifications for articles but accessing versions of all the indicators discussed in this chapter requires additional purchases (e.g., InCites, Journal Citation Reports). In contrast, Dimensions offered two indicators in its basic version: Field Citation Ratio (similar to NCS) and Relative Citation Ratio (Hutchins et al., 2016). It also offers altmetric percentile indicators for its articles (e.g., "In the top 25% of all outputs scored by Altmetric").

The analytics offerings of the commercial databases include a range of metrics. For example, InCites reports top 0.1%, 1% or 10% papers and NCS type indicators (Category Normalised Citation Impact: CNCI) as well as a range of other bibliometric indicators that are not based on citation counts (e.g., percentage open access). These can be applied within the systems to collections of articles, such as all those from an individual university or department.

## 5.9   Summary

This chapter has introduced and rationalised three types of citation impact indicator: percentile-based, field and year normalised formulae (MNCS and the skew-corrected MNLCS), and network-based formulae. It has also discussed indicators for journals since these are calculated in the publishing industry and are the best-known citation-impact indicators. Each indicator has its own limitations, and most have additional weaknesses from the shortcomings of the classification schemes used to calculate them. Despite this, they are useful in some contexts where alternatives are impractical or if the limitations are not too severe. Empirical evidence of the value of some of these indicators in some contexts is discussed in many of the following chapters.

# 6   Citation-based indicators for individual articles: Empirical evidence

As the theoretical discussion in Chapter 3 shows, there are many factors that can influence the relationship between citation counts or citation-based indicators and research quality judgements or scores[1]. Moreover, the influence of these factors could vary between fields, topics, and countries and over time. Because it is impossible to reliably guess the relative strengths of these variables in any context, the value of citations as research quality indicators is rarely known with any precision. It is therefore important to get empirical evidence of the value of citation counts, or their relationship with research quality, in a wide variety of contexts. Unfortunately, there is limited evidence of this because expert quality evaluations of individual articles are rarely published. This chapter reviews the evidence from the rare exceptions. The approach taken in statistical: correlation or regression between citation-based indicators and research quality scores. This statistical approach is appropriate on the basis that each article is unique and never repeated so the most important knowledge is the overall strength of relationship between citations and research quality in different contexts.

As suggested above, a core theoretical basis for using citation counts as an indicator of research quality, or at least its scholarly impact dimension, is that citations serve to acknowledge some or all relevant or foundational prior work from other scholars. Nevertheless, there are many other reasons for citing and influencing factors that complicate the issue. A statistical response to this problem is to accept that there are reasons for citing that do not reflect impact or that reflect little impact, but that when citations are aggregated on a sufficiently large scale then these "imperfections" may tend to average out. This would allow citation-based indicators based on aggregating multiple journal articles to have some value, even if they do not work well for individual articles (van Raan, 2004). Since the amount of citation bias and the amount of "signal to noise" (or impact-reflecting citations to other citations) is unknown, the task of identifying the contexts, such as fields and years, in which it is appropriate to use citation-based indicators is statistical: assessing whether citation-based indicators predict or correlate to a sufficient degree with article quality scores.

A few studies have directly investigated the extent to which citation counts correlate with research quality for journal articles. The largest scale study before REF2021 (discussed in the next section) was non-academic (not peer reviewed, written by two professional statisticians) for a UK audit of the REF 2014 national research evaluation exercise. It investigated REF 2014 peer review scores for about 25,000 journal articles published in 2008 with citation-based indicators in 36 UK Research Excellence Framework (REF) Units of Assessment (UoAs) and reported weaker results for articles from 2013 (HEFCE, 2015). Overall, REF peer review scores for individual articles significantly and positively correlated (0.3) with Elsevier's field normalised citation impact metric Source-Normalised Impact per Paper (SNIP), Field-Weighted Citation Impact (0.3), and citation counts (0.2). There were large disciplinary differences within this overall figure, with the strongest correlations between citation counts and REF scores in Clinical Medicine (rho=0.7), Chemistry (0.6), Physics (0.6) and Biological Sciences (0.6). Correlations in most social sciences, arts and humanities were typically below 0.3 and some were negative but unreliable due to small sample sizes (e.g., 15 articles for one correlation) (HEFCE, 2015). Because of its goals, this study included duplicate articles (the same article submitted by authors in different institutions), which undermines the general (i.e., non-REF) value of the correlations because multiply-submitted articles can expect to be both higher quality and more cited because they have more authors (including in the UK: Thelwall & Maflahi, 2020). The reliance on REF self-classifications for articles is also imperfect because multidisciplinary authors and department members with out-of-field specialisms (e.g., medical statisticians) could result in articles submitted to inappropriate UoAs. The arts and humanities data also included a minority of articles due to a majority of missing DOIs. These problems were mostly fixed in the follow up larger-scale academic study of REF2021 reported in the next section.

---

[1] This chapter is based on a published analysis of the relationship between REF research quality scores and citations in the UK (Thelwall et al. 2023).

## 6.1 UK REF2021 case study

This section reports an analysis of the relationship between an appropriate citation-based indicator and research quality for journal articles, as judged by the UK REF2021 process. The indicator used is the field and year normalised, skew corrected NLCS, which is fair to compare between fields. This is important because the UoAs in the REF contain articles from multiple fields even though some are based on broad fields. To expand, improve and refine the HEFCE study mentioned above, multiple classification schemes were also used (Scopus and Dimensions field classifications as well as the original UoAs), more articles analysed, and duplicate articles eliminated. Methods details and limitations are discussed in the originating article (Thelwall et al., 2023).

### 6.1.1 Overall relationship between citations and research quality

From field and year normalised citation count indicators, articles with more citations tend to be higher quality in all fields of science. This is true whether classifying articles by the REF UoAs (Figure 6.1), Dimensions' article-level Fields Of Research (FOR) codes (Figure 6.2) or Scopus' journal-level All Science Journal Classification (ASJC) (Figure 6.3) broad fields. Surprisingly, given the prior HEFCE (2015) findings with smaller sample sizes and other issues, there are statistically significant positive correlations (i.e., 95% confidence intervals that do not include 0) even in most arts and humanities fields, including UoA 33: Music, Drama, Dance and Performing Arts (Figure 6.1), Studies in Creative Arts and Writing (Figure 6.2) and Arts and Humanities (Figure 6.3). Moreover, none of the correlations are negative, unlike in the HEFCE (2015) study. This gives the strongest evidence yet that citations are rarely *completely* irrelevant to research quality.

Comparing field three classification schemes (UoAs, Scopus, Dimensions), there are some patterns. First, the spread of field-based correlation magnitudes (ignoring the Scopus Multidisciplinary class) is substantially higher for the REF UoAs (0.55) than for Dimensions (0.36) and Scopus (0.39). Correlations for REF UoAs vary between 0.02 and 0.57 (Figure 6.1), for Dimensions they vary between 0.11 and 0.47 (Figure 6.2), for Scopus they vary between 0.06 and 0.45 (Figure 6.3). The relatively wide UoA spread of correlations suggests that the REF field classification scheme could be more effectively clustering articles by field, so that topics for which citations are better indicators of quality are less mixed with topics for which citations are worse indicators of quality. Alternative explanations are also possible, however, and the high correlation for the Scopus Multidisciplinary category is presumably due to predominantly high scores and citations for prestigious generalist journals like *Nature* and *Science* in comparison to lower scores and fewer citations to less well-known generalist journals.

From the perspective of individual subjects, the classification scheme has little influence on the strength of correlation for some subjects but a larger influence on others, when they are comparable. For example, for Mathematical Sciences, the correlations have a spread of 0.04: 0.35 (UoA), 0.32 (Dimensions) and 0.31 (Scopus Mathematics). In contrast, the Chemistry correlations have a four times larger spread of 0.17: 0.57 (UoA), 0.40 (Dimensions Chemical Sciences) and 0.42 (Scopus). Again, there are alternative plausible explanations, but it is possible that purer categories allow higher correlations by avoiding work for which citations have little relevance to quality (e.g., perhaps chemical engineering for chemistry).

In terms of the overall disciplinary patterns shown, physics, chemistry, biology, and medicine are the areas with the consistently highest correlations under all three schemes, followed by other natural and health sciences. In contrast, arts and humanities topics have the weakest correlations under all three schemes, with social sciences and engineering tending to be between these two groupings.

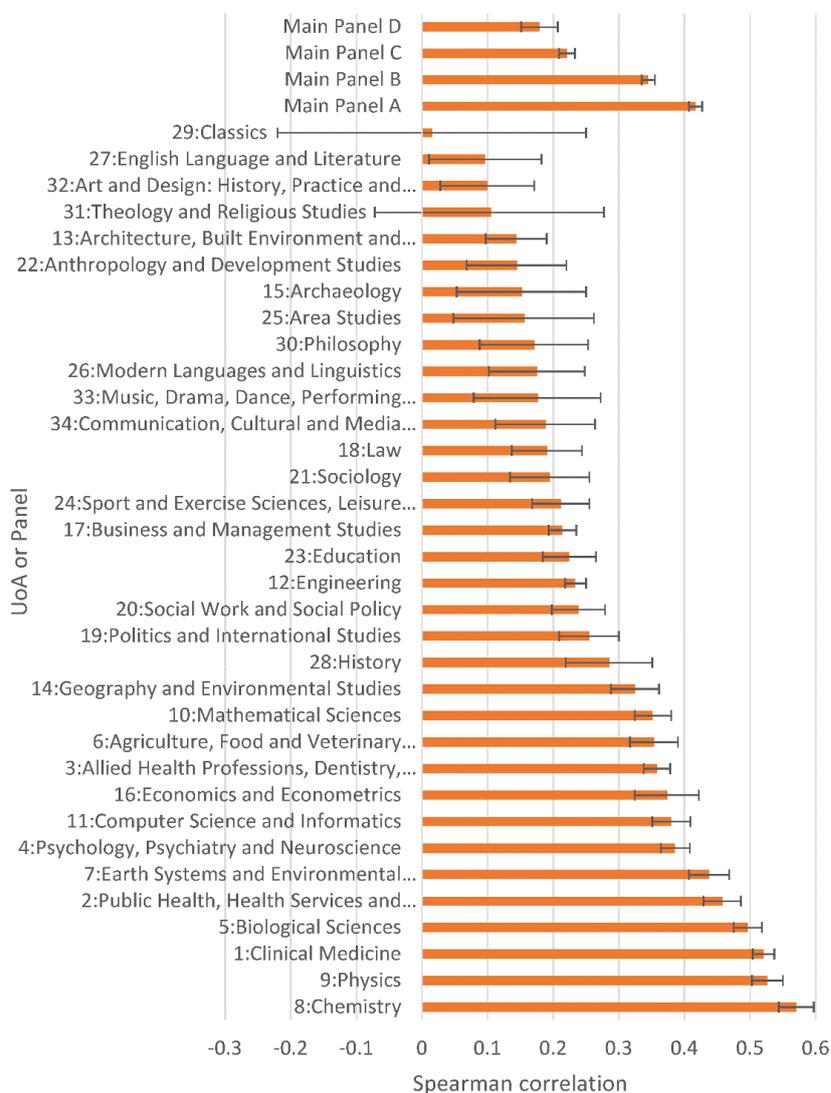

*Figure 6.1 Spearman correlations between field and year normalised citation counts (NLCS) and REF scores for 2014-18 journal articles submitted to UK REF2021 by submitting Unit of Assessment or Main Panel. Error bars indicate 95% confidence intervals (Thelwall et al., 2023).*

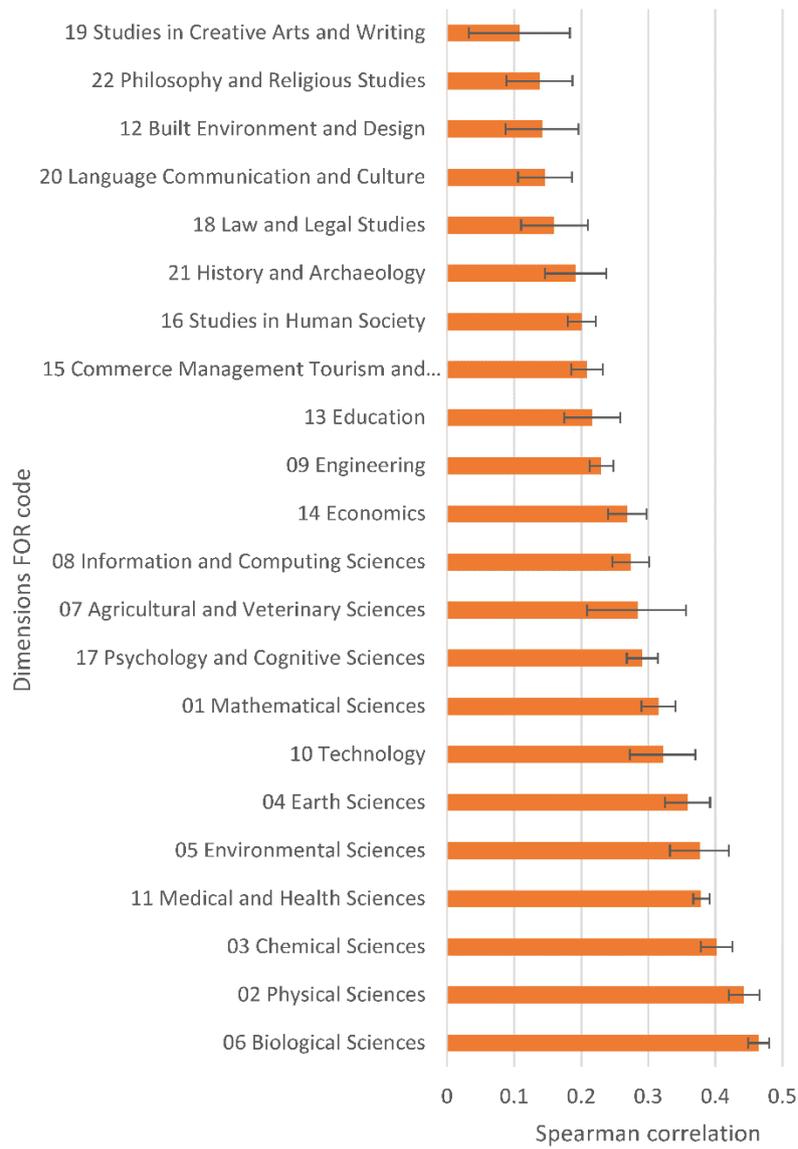

*Figure 6.2 Spearman correlations between field and year normalised citation counts (NLCS) and REF scores for 2014-18 journal articles submitted to UK REF2021 by Dimensions FOR code (n=22). Error bars indicate 95% confidence intervals (Thelwall et al., 2023).*

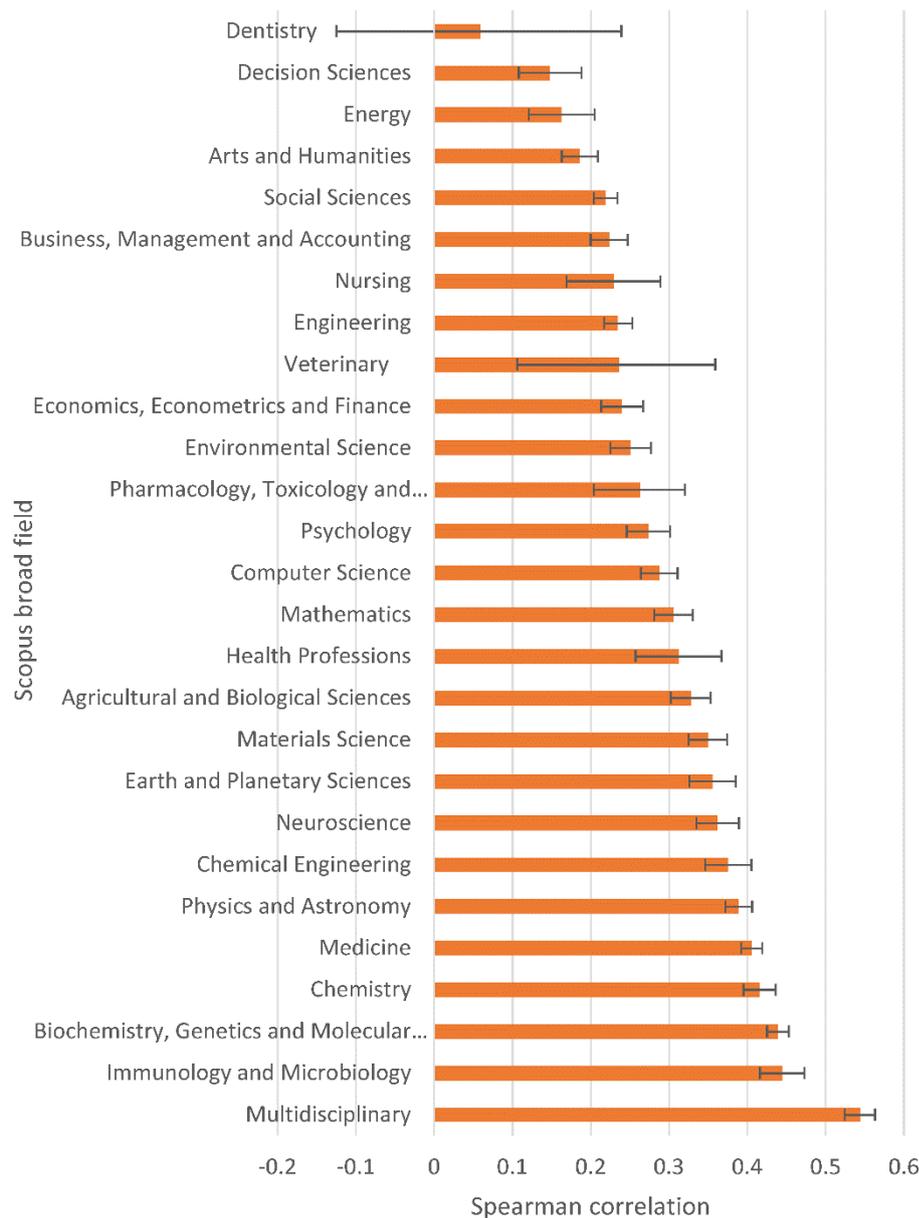

*Figure 6.3 Spearman correlations between field and year normalised citation counts (NLCS) and REF scores for 2014-18 journal articles submitted to UK REF2021 by Scopus broad field (n=27). Error bars indicate 95% confidence intervals (Thelwall et al., 2023).*

Whilst it was already known that the relationship between research quality and citation counts varied between fields at the institutional level (e.g., Franceschet & Costantini, 2011; HEFCE, 2015, Mahdi et al., 2008) and suspected for articles (HEFCE, 2015), the universally positive nature had not been suggested by previous datasets. Although not all correlation confidence intervals excluded zero, the correlations were positive for all 34 UoAs, all 22 FOR codes and all 27 Scopus broad fields. Out of these, only three (very wide) confidence intervals contained 0 and these covered few articles (UoA 29 Classics [n=70] and UoA 31 Theology [n=124] in Figure 6.1 and Dentistry [n=115] in Figure 6.3). Thus, whilst not fully proven, the results are consistent with a positive relationship occurring across all broad academic fields and give strong evidence that the relationship is near universal. The statistical power of the large numbers of articles in many fields supports this conclusion even for fields where the correlation is weak.

The unexpected positive (albeit weak) associations between citations and quality across the arts and humanities has multiple plausible explanations. It is possible that all arts and humanities categories in all three schemes had a degree of pollution by social science or science articles, which was enough to create a detectable association. Alternatively, citations to articles may reflect influence (i.e., an aspect of quality) often enough in the arts and humanities to be detectable amongst the noise of other types of citation. Since the association is weak, a lot of empirical evidence, such as from citation motivation surveys, would be needed to distinguish between the two.

### 6.1.2   Do high citations guarantee high quality in any field?

Although the above section shows the overall relationship between citations and REF research quality, it is sometimes argued that the most cited articles in some fields are always high quality. This section investigates this issue by checking whether the most cited 25 articles in any UoA were always given the highest REF score, 4*.

The results show that there were no UoAs in which the 25 most cited (i.e., highest NLCS) articles all had the highest REF2021 quality score (Figure 6.4). Thus, there is no reasonable citation threshold in any UoA that guarantees the highest quality score, since a bucket size of much smaller that 25 would be statistically hard to check.  Seven UoAs are close to 100% 4*, however. Thus, in some fields a very high citation rate makes an article very likely to be high quality, but there are always exceptions.

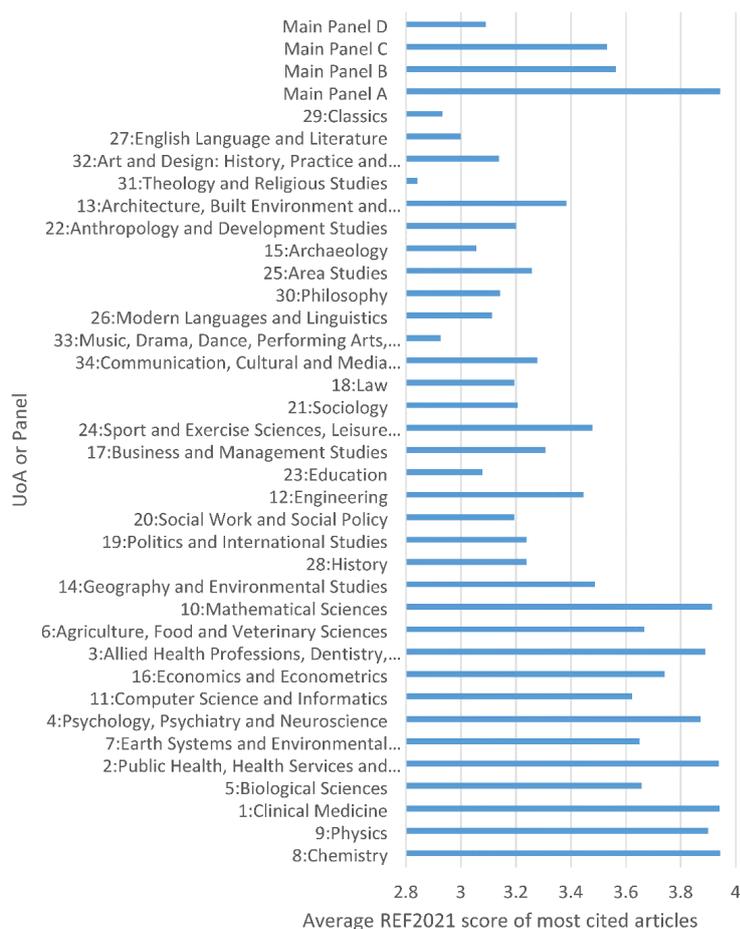

*Figure 6.4 Mean REF scores for 2014-18 journal articles submitted to UK REF2021 for the 25 articles with the highest field and year normalised citation counts (NLCS) by submitting UoA or Main Panel. Sort order as for Figure 6.1 (Thelwall et al., 2023).*

Whilst it is well known that articles occasionally become highly cited for negative reasons (e.g., the MMR/autism study: Godlee et al., 2011; the cold fusion article: Berlinguette et al., 2019), the results suggest that is it in fact *normal* for occasional articles in all fields to become extremely highly cited without having world leading quality. Moreover, in many fields (most UoAs) an extremely highly cited article is likely to be *not* world leading (e.g., averages below 3.5 in Figure 6.4). This does not undermine the use of percentiles in research evaluation, such as reporting the percentage of articles in the most cited 1% for a country (e.g., Rodríguez-Navarro, & Brito, 2024) but it cautions against fully equating highly cited with world leading research in any fields at the individual article level.

### 6.1.3   Overall shape of the relationship between citations and research quality

This section gives more context to the correlations reported in the first section of this chapter. A positive Spearman correlation can reflect many different underlying shapes, so it is informative to examine the underlying relationship between citations and research quality. The clearest way to do this is to plot average REF2021 scores against NLCS values, bucketing articles into groups with similar NLCS and taking the mean REF2021 score. This hides the variation between articles with similar NLCSs but shows the underlying trend. This is a problematic approach because the scores 1* to 4* are ordered but not scalar. Nevertheless, it is at least plausible to interpret 1* to 4* as a scale 1 to 4 and this assumption is routinely made for departmental Grade Point Averages (GPAs) constructed from REF scores. Given that this aspect of the calculation of GPAs does not seem to be challenged in the UK, it seems reasonable to make the same assumption here.

In all cases where there is a positive correlation above 0.1, the underlying shapes are close to straight lines, but some are more consistent with approximate logarithmic curves: relatively rapid increases in average REF scores for NLCS increases at lower NLCS values and smaller increases in average REF scores for NLCS increases at higher NLCS values. The steepness of the increase and the range of average REF scores differs substantially between UoAs, however.

In fields with higher correlations (e.g., Figure 6.5), the increase in average REF score for higher NLCS values is relatively steep, ending close to 4. Although there are variations within each NLCS range, in these fields, citation scores seem to be good indicators of quality and it would be possible but surprising to find an excellent little cited article or a non-excellent highly cited article.

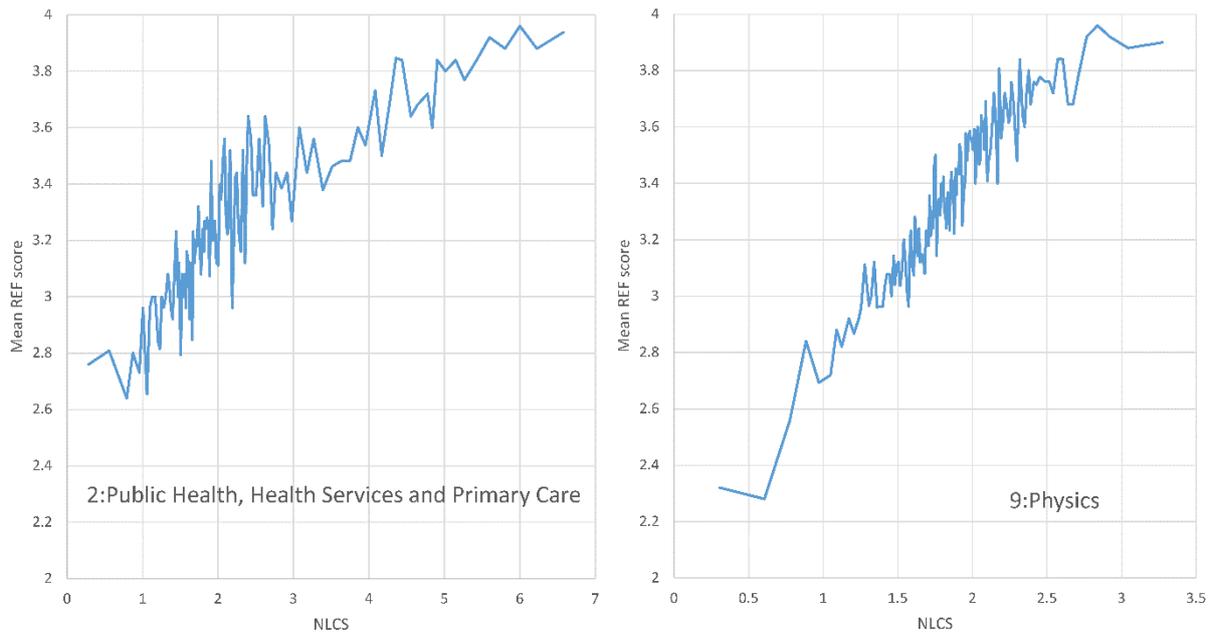

*Figure 6.5 Mean REF scores for 2014-18 journal articles submitted to UK REF2021 in UoAs 2 and 9 against field and year normalised citation counts (NLCS). Articles are bucketed into groups of at least 25 with similar NLCS (Thelwall et al., 2023).*

In UoAs where the correlation between NLCS and REF scores is more moderate, the slope of the broadly linear trend between NLCS and average REF2021 scores is less steep but still clear and does not get as close to the maximum (Figure 6.6). In these fields, whilst there is a tendency for more cited articles to be higher quality, many articles break this trend.

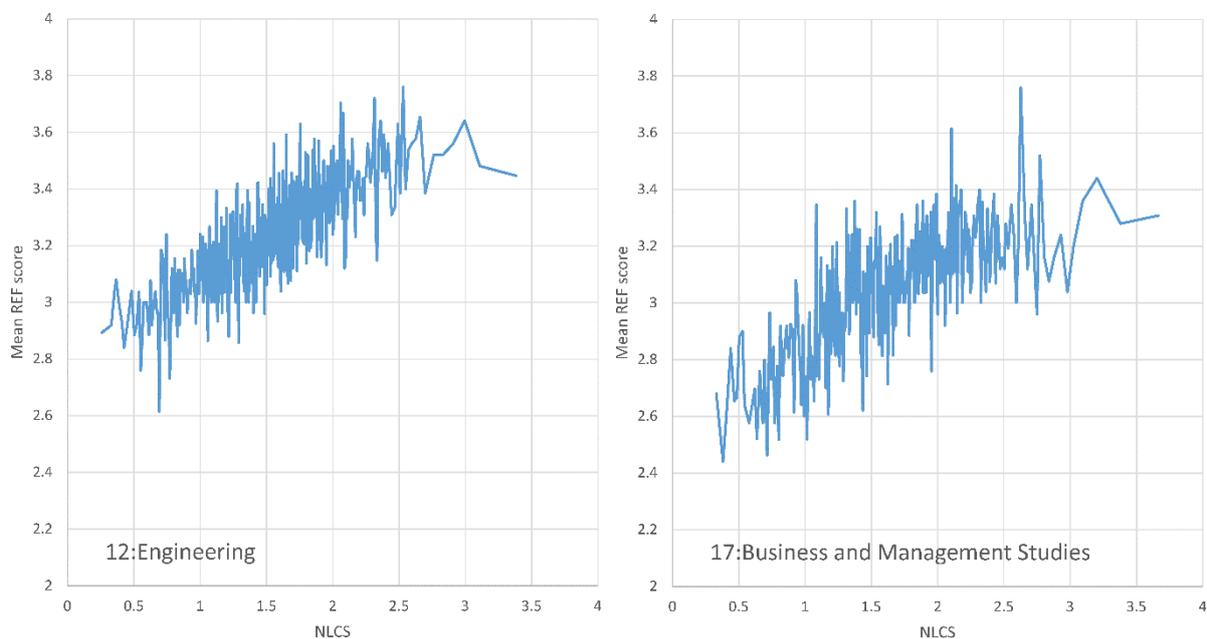

*Figure 6.6 Mean REF scores for 2014-18 journal articles submitted to UK REF2021 in UoAs 12 and 17 against field and year normalised citation counts (NLCS). Articles are bucketed into groups of at least 25 with similar NLCS (Thelwall et al., 2023).*

In UoAs where the correlation between REF2021 scores and NLCS is close to 0, this probably reflects a very shallow increasing tendency rather than a more complex relationship (e.g., not a U-shaped curve) (Figure 6.7). A shallow general slope like that for UoA 26 may reflect combinations of fields, some of which have no relationship between citations and quality (e.g., modern languages) and others that have some relationship (e.g., computational linguistics). The shapes for the remaining 28 UoAs are broadly similar to one of the three pairs of figures in this section.

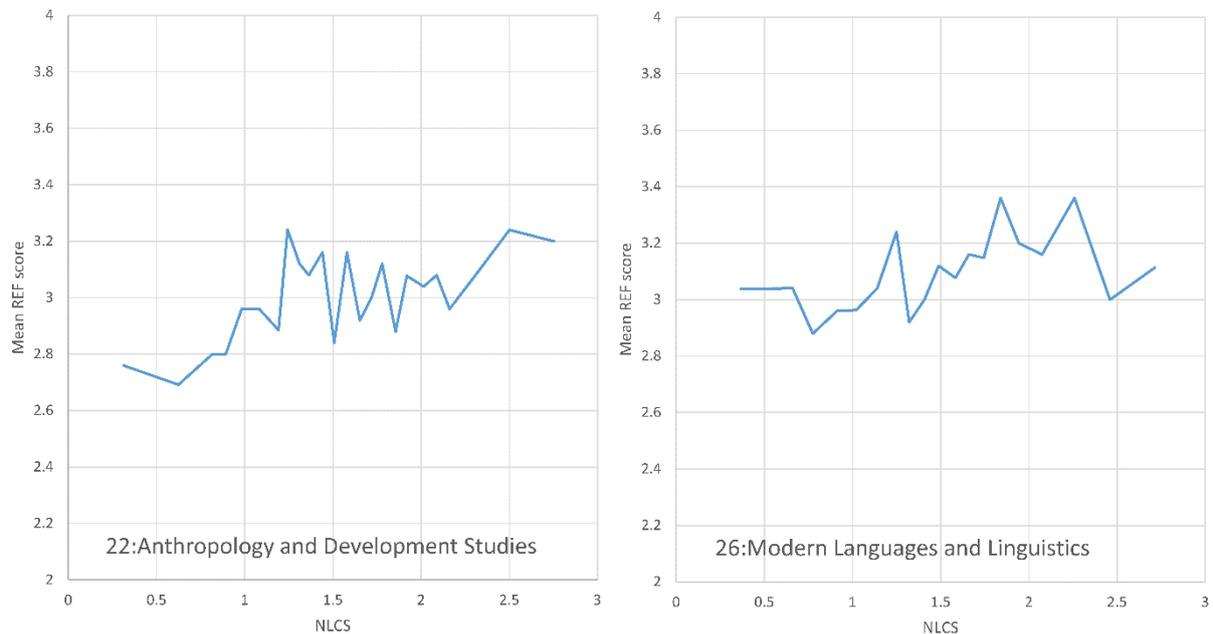

*Figure 6.7 Mean REF scores for 2014-18 journal articles submitted to UK REF2021 in UoAs 22 and 26 against field and year normalised citation counts (NLCS). Articles are bucketed into groups of at least 25 with similar NLCS (Thelwall et al., 2023).*

The primary value of these graphs is their monotonically increasing nature, showing that the positive association mentioned above occurs at all or most citation levels. The exact shapes should not be interpreted at face value for two reasons. First, REF scores are ordinal rather than forming a scale: it is not clear that the gap between, say, 1* and 2* is the same as the gap between 3* and 4*, or even that the concept of gap width in this context is meaningful. In the absence of evidence to the contrary, it is reasonable to at least hypothesise that the scores form a numerical scale. Nevertheless, the citation counts are log transformed as part of the NLCS calculation, so the x axes of Figure 6.4 to Figure 6.7 are effectively log-transformed. If the x-axes were reverse log transformed, expanding the difference between the higher numbers, then the graph shapes would be close to logarithmic. Thus, it is reasonable to hypothesise that the underlying relationship between research quality and citation counts is logarithmic, with citation counts providing diminishing returns in terms of increased probability of higher quality at higher values. This would fit with the rich-get-richer phenomenon by which highly cited articles are believed to attract new citations partly because they are highly cited rather than for their intrinsic value (Merton, 1968).

## 6.2 Limitations

The results shown above should not be taken to be universal or fully convincing for several reasons. First, all journal articles analysed are from the UK and the relationship between citations and quality (and its different operationalisations) might be different in other countries, such as those that value research applications more highly than scientific contributions.

Second, the articles are self-selected and represent the outputs considered by the authors to be their best work. The relationship might be different for lower quality research. Third, the field normalisation is limited by the primarily journal-based categorisation scheme of Scopus, which might generate anomalies through interdisciplinary journals. For example, the large *Journal of the Acoustical Society of America* is classified as both Acoustics and Ultrasonics (in Physics and Astronomy) and Arts and Humanities (misc.). Its relatively highly cited articles in the latter category (e.g., "Grid-free compressive beamforming" with 99 Scopus citations) will greatly add to the denominator of the field normalisation calculations and reduce the field normalised scores of genuinely arts and humanities research in its category. Finally, there may well be narrow fields (and other output types) for which the relationship between citations and research quality is inverted or null.

## 6.3 Summary

The universal positive association between citation scores and research quality support the appropriate use of citation-based indicators to aid research quality evaluations. They also suggest that there are no broad fields of scholarship for which citations are *completely* irrelevant. Nevertheless, the wide variation between fields in the strength of the relationship confirms that citation-based indicators need greater levels of aggregation to yield useful information in some fields than others. For example, in fields with correlations above 0.5 at the article level, very strong aggregate correlations between average citations and average quality might be expected for small departments or small journals whereas the same aggregate correlations might only appear for very large departments or very large journals in other fields. Thus, the argument against inappropriate use of citations should not be that they are completely irrelevant in a field but that it is not reasonable to use them at a too low level of aggregation. Of course, if there are systematic biases in the citation data that field normalisation cannot eliminate, such as against qualitative research in a mixed methods field, then citation-based indictors would need to be used very cautiously in any context.

The fact that extremely high citation counts do not guarantee the highest research quality in any field and are not a high probability indicator of it in most (at least at the level of REF2021 UoAs) is another important point. This should be remembered when journal articles are ranked by citations to identify the most influential articles in a field (Shadgan et al., 2010). Moreover, some research evaluations count the proportion of articles in the top 1% cited as an indicator of capacity to produce excellent research ("the vanguards of science": Wagner et al., 2022). In these contexts, it should always be acknowledged that articles can become highly cited for reasons other than research excellence.

# 7   Factors affecting citation impact indicators for individual articles

Many empirical studies have assessed whether various metadata and other properties of articles that are not aspects of research quality can affect, or associate with, their quality or citation impact [2]. Sometimes the direction of causality (i.e., affect vs. reflect research quality/impact) has been unclear. For example, if article titles containing colons tend to be more cited, is that caused by colon titles being more understandable or informative, therefore attracting extra readers or citers to the article? Or perhaps article titles with colons are more cited because they belong to high citation specialities or journals that expect long, complex article titles? If the former is true then knowledge of the relationship is helpful for authors but in the latter case, it is only helpful for predicting future citations. When cause-and-effect is unclear, authors might use their own judgements about whether to consider a factor.

   Most investigations into this topic have implicitly or explicitly used citation counts as proxies for article quality, with the assumption that more cited articles tend to be higher quality. This chapter discusses separately the minority of studies that have used expert review quality scores instead of citation counts. These are particularly important in fields where citation counts are poor proxies for research quality, such as the arts, humanities, and some social sciences.

   With a few exceptions, studies of documentary factors associating with the quality or citation impact of research have focused on evidence that can be automatically extracted from publications, such as title length or the presence of a question mark in a title. They have typically used correlational approaches to find associations, regression to identify associations and make predictions, or machine learning to make predictions. The chapter ignores discipline-specific factors, such as the influence of hierarchies of evidence in evidence-based medicine, to focus on science-wide issues. The chapter is split into two connected parts: evidence of associations between document features and citation counts or quality scores; and predicting citation counts or quality scores from document features. The main value of this overview is to identify when factors associating with document quality or impact are universal or show clear patterns, and when these factors are context dependent and without clear patterns. Unfortunately, the latter is dominant.

## 7.1   Article content properties associating with citation counts

Many studies have analysed document features, such as title length or the number of references, by correlating them with citation counts. For example, one study examined article text features (title length, number of figures, tables, equations, and characters with no spaces), metadata (number of authors and number of views) and citation counts from high and low cited papers (each 100 articles, n=200) published by MDPI in 2017, finding significant positive associations between citation counts and the number of views, tables, and authors and a negative significant correlation with title length (Elgendi, 2019). A meta-analysis of 262 studies found that there were associations between various document properties and citation counts (Pearson's r < ±0.2) (Mammola et al., 2022). Multiple properties have sometimes been investigated together through a regression model, which has the advantage of assessing their relative contributions. For example, a regression approach would be needed to distinguish between the citation associations of article lengths and reference list lengths and to assess whether one was a by-product of the other.

---

[2] This chapter is based on Kousha and Thelwall (2024).

No studies have proved that the property investigated *caused* citation count increases. In all cases, it the root cause of any association could be a third factor, such as research specialism or audience engagement, that also affects citation counts. For example, a more understandable abstract might attract more readers and hence more citations. A more understandable abstract might also reflect a higher quality article with simple important findings (all the + options in Figure 7.1), or be a requirement of the top journals in a field. Conversely, in some fields a higher quality article may tend to have a more complex theoretical component with lengthy jargon terms, leading to a less understandable abstract. This less understandable abstract may attract also more citations by flagging the theoretical terms for academic literature searchers (all the − options in Figure 7.1). More generally, all combinations of positive negative and neutral relationships in Figure 7.1 seem possible. There could also be indirect connections between the three factors. For example, better or more cited authors might tend to write more/less readable articles in some fields.

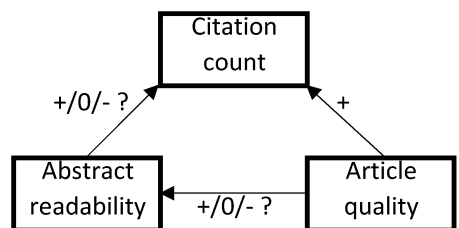

*Figure 7.1 Possible relationships between abstract readability, article quality and citation counts (source: author).*

### 7.1.1 Article title lengths and citation rates

Interesting, informative, keyword rich, or easy to understand titles may attract the attention of other researchers, making the articles more likely to be found and read and then cited. For example, articles with more conclusive titles are more likely to be cited in six biomedical areas (Urlings et al., 2021).

Investigations into the relationship between citation counts and article title length, measured in words or characters, have generated mixed results for unknown reasons so there is not a simple and universal relationship between the two. The most general result for contemporary research is that in highly cited journals, shorter titles tend to be more cited, whereas for less cited journals, longer titles tend to be more cited (Sienkiewicz & Altmann, 2016). Nevertheless, there are exceptions and possibly disciplinary differences and changes over time (Jiang & Hyland, 2022). Also, the precise reason for the association can realistically only be speculated about and may be a second order effect of article type so cause-and-effect is unknown even when a relationship exists.

### 7.1.2 Longer articles tend to be more cited

Longer papers may tend to attract more citations because they contain more citeable content, but some prestigious journals require short articles and so the relationship is may be inverted in some cases. All studies reviewed here measured article length using the total number of pages, even though these depend on page layouts and printing formats and are not relevant to online-only articles. Word counts could be a better indicator of article length but ignore figures, and article full text is needed for such analyses because no major bibliometric database reports word or character counts.

Despite the above caveat, **the evidence is almost unanimous that longer articles tend to be more cited** (e.g., Haustein, Costas, & Larivière, 2015; Xie et al., 2019). Nevertheless, at least two articles have pointed out that expected proportional increase in citations for longer articles is less than their proportional increase in length (Abt, 1984; Haslam & Koval, 2010b).

### 7.1.3 Articles with longer abstracts tend to be more cited

Abstracts have become nearly universal for journal articles over the past half century (Thelwall & Sud, 2022). They help potential readers to understand the topic and results of an article efficiently before they read the full article. Informative abstracts can presumably help relevant research to be quickly identified, and this may be influenced by length, structure, or readability. In theory, longer abstracts might associate with more citations because they are more informative, or fewer citations because they are harder to digest.

The evidence so far suggests that articles with longer abstracts tend to be more cited. For a million abstracts from eight subject areas, longer abstracts and more sentences in abstracts associated with more citations in all fields (Weinberger, Evans, & Allesina, 2015). This aligns with similar findings for Biology and Biochemistry, Social Sciences, and Chemistry (Didegah & Thelwall, 2013b). At the journal level this relationship mostly persists: a very large study of 4.3 million papers from over 1500 journals also showed that abstract length positively correlated with citation counts in nearly all journals (Sienkiewicz & Altmann, 2016). In contrast, another large-scale investigation of 300,000 highly cited articles between 1999 and 2008 (30,000 papers per year) found that articles with longer abstracts received fewer citations at the journal level (Letchford, Preis, & Moat, 2016), so the overall relationship may be different for highly cited articles.

Positive associations between citation counts and abstract length might be a statistical side-effect of a minority of small articles having very short abstracts. These could be errors (e.g., corrections published as articles), or short articles or comments with a few summary sentences instead of a detailed abstract.

### 7.1.4 Articles with less readable abstracts are more cited

More readable abstracts might be expected to associate with higher citation rates, but the evidence mostly finds the opposite. For example, for 264,156 articles from five American universities 2000–2009, articles with more readable abstracts were less cited (Gazni, 2011). A large-scale analysis of 4.3 million papers from over 1,500 journals also found that articles with more readable abstracts were less cited (Sienkiewicz & Altmann, 2016). The one major exception to these findings is that Economics Letters articles 2003–2012 with more readable abstracts were more cited (Dowling, Hammami, & Zreik, 2018).

### 7.1.5 Articles with more, more recent, and higher impact references are more cited

Articles with more references have been found to be more cited in many studies (e.g., Urlings et al., 2021). This seems logical because a longer reference list suggests connections to a wider literature (and therefore potentially more widely relevant), higher quality research (because more justified through references) and a longer article (hence more content to cite). Articles with more recent references also tend to be more cited (Ahlgren, Colliander & Sjögårde, 2018; Onodera & Yoshikane, 2015). Newer references presumably indicate a more current topic that is more likely to be cited by subsequent articles. Articles with more high impact references also tend to be more cited (Boyack & Klavans, 2005; Lancho-Barrantes et al., 2010; Peng & Zhu, 2012). Citing highly cited references may indicate tackling important topics, leveraging ground-breaking prior research, or working within a high citation topic.

### 7.1.6  Other article features

This section reviews article features that have occasionally been investigated for associations with citation counts.

**Images**: Higher-impact PubMed articles have more diagrams per page and a higher proportion of diagrams but a lower proportion of photos (Lee et al., 2017).

**Review articles are more cited**: Review articles tend to be more cited than other research articles, although there are some disciplinary differences (Blümel & Schniedermann, 2020). For instance, a very large-scale study of 14.2 million records from Science Citation Index Expanded database during 2000–2015 across 35 science subject areas found that reviews received 1.3 to 6.7 times more citations than standard research articles, depending on the subject area (Miranda & Garcia-Carpintero, 2018). Citing a review article can be a useful shortcut to reference a body of literature when a detailed analysis is not needed.

**Article methods**: Individual methods may be more cited than average, including questionnaires (Fairclough & Thelwall, 2022), structural equation modelling (Thelwall & Wilson, 2016) and interviews, focus groups and ethnographies, although the degree has changed over time (Thelwall & Nevill, 2021). For biomedical research, methods-focused papers are heavily overrepresented (90%) in the top 100 cited papers (Small, 2018).

**Language**: Articles in English or in English-language journals may tend to be more cited (for a review, see: Tahamtan et al., 2016), perhaps because English is currently the main international language of scholarly communication, so more scholars can read it. They may also be more cited because a higher proportion of non-English articles address local issues, or because citation indexes mainly index English-language journals (Mongeon & Paul-Hus, 2016), so a greater proportion of other language citations may be lost.

**Open Access**: Open Access (OA) articles seem to have a citation advantage from being more widely accessible. This is difficult to check because there are many types of open access, and there are journal-level factors because high- and low-quality journals may be fully OA or fully non-OA. Moreover, it is impossible to account for author decisions, such as if scholars are more likely to ensure that their best work is (or isn't) OA. Perhaps because of these factors, together with possible disciplinary differences, current evidence is inconclusive about whether OA advantages exist (Langham-Putrow et al., 2021).

**Topic growth**: Articles in a rapidly expanding area, such as a new hot topic, are likely to be more cited than average for the field because the expanding pool of publications has a smaller pool from which to cite (Sjögårde & Didegah, 2022).

## 7.2  Authorship team associations with citation counts

This section reviews evidence of associations between authorship team properties and citation counts. Disciplinary differences occur here because of differences in average team sizes and the extent to which equipment and collaboration is essential or beneficial for research.

### 7.2.1  Articles with more authors tend to be more cited

Many studies of various types have found that articles with more authors tend to be more cited, so this seems to be an almost universal and reasonably strong phenomenon. For example, a large-scale investigation across 27 broad subjects from the 10 countries with most journal articles during 2008-2012 found that increased collaboration associated with more citations for all countries and most subjects, but China and a few fields (e.g., computer science; business, management and accounting) had much lower associations between author numbers and citation counts. There was a significant increase in the average citation impact of research from single to two authored articles with a subsequent linear rise with additional authorship, giving overall logarithm-like shape (Thelwall & Maflahi, 2020). There may be a stronger association between citations and research collaboration for developing countries (r= 0.180) than for developed countries (r= 0.112), however (Shen et al., 2021) and co-authorship may not associate with more cited research in some fields (e.g., Slyder et al., 2011).

In theory, larger numbers of authors may generate more interest for an article not because collaborative articles are better, but they may attract more readers through the authors' friends and acquaintances, an audience effect (Wagner et al., 2019; Rousseau, 1992). The one study that has investigated the three-way relationship between author numbers, research quality, and citation impact found that more authors associated with higher quality in most fields, and there may also be an audience effect that partly explains higher citation impact in some, but not all fields (Thelwall et al., 2023g).

## 7.2.2  Articles with more country and institutional affiliations tend to be more cited

International collaboration has increased from rare (5% of articles in 1980) to common (26% of articles in 2021), perhaps reflecting an appreciation of its value or the ease of electronic communication (Aksnes & Sivertsen, 2023). Internationally co-authored papers tend to attract more citations than domestic articles in most contexts tested so far (e.g., Leydesdorff, Bornmann, & Wagner, 2019). This may be due to wider audiences for the research (more people knowing the authors: Wagner, Whetsell, & Mukherjee, 2019), more varied expertise, or more funding (assuming that international collaboration is often triggered by grants). Most investigations of this phenomenon have factored out team size so that internationalism is counted separately from the number of authors.

Although there does not seem to be empirical evidence for this, funding may sometimes be the critical factor with internationalism being a side-effect, particularly in expensive fields. For example, an international health consortium may have passed multiple peer review stages for its components to be funded, giving much stricter quality control before the project starts than a similar scale national collaboration, even if also funded.

The citation advantage of international collaboration may depend on which countries are involved. For example, biochemistry articles in 2011 (n=13,578), research collaboration with the U.S. associated with increased scholarly impact for published research, whereas co-authorship with some other countries including India and China associated with reduced impact (Sud & Thelwall, 2016).

Articles with more institutional affiliations also tend to be more cited (e.g., Bordons et al., 2013; Didegah & Thelwall, 2013b), perhaps because they are more likely to be funded, or the researchers are more likely to be higher profile to attract extra-institutional collaborators.

## 7.2.3  Author publication and citation records

Authors with a good track record of publishing or attracting citations seem to be more likely to write future highly cited papers. It is hard to fully assess this with career-level analyses, but there is some evidence in favour of the hypothesis from many fields (e.g., Didegah, 2014; Qian et al., 2017). There are disciplinary differences in the strength of association, however: a unit increase in the h-index associates with a higher increase in citations in mathematics (6.6%) and economics & business (5.1%) than in immunology and materials science (both 0.8%) (Didegah, 2014).

From a related perspective, a science-wide analysis of the association between the journal impact (as a proxy for article quality) and authorship properties found that authorship teams publishing more research and higher impact research were more likely to publish in higher impact journals. For this, publishing more cited research was more important than publishing more articles. A first author publishing highly cited research is a science-wide advantage in this regard, whilst a productive first author is sometimes a disadvantage. A possible explanation is that in some fields, junior first authors might be PhD students conducting particularly careful studies (Thelwall, 2023).

### 7.2.4 Author nationality, institution, and gender

The average citation impact of academic research varies substantially between nations (Confraria et al., 2017). Although there are field differences in this, with countries having high citation specialisms (Elsevier, 2017), essentially richer countries tend to publish more cited work, presumably because of greater infrastructure and resources for research (Confraria et al., 2017). Thus, the national affiliations of the authors of a paper associate with its citation impact.

The average citation impact of academic research also varies substantially between institutions within a nation, as evidenced by international citation-based league tables of universities (Waltman et al., 2012). This is likely to be due to some institutions having better researchers and/or more resources and prestige than others. Differences are likely to be greater in countries like the UK that encourage a hierarchy of universities than in countries like Germany where they are intended to be more equal. The relative citation impacts of universities also vary between specialisms. Thus, the institutions of the authors of a paper associate with its citation count.

Many researchers have found author gender (male vs. female) differences in average citation counts for journal articles, with some studies finding that male first authored articles tend to be more cited (Larivière et al., 2013) and others the reverse (see below). For instance, for over 13,000 research articles and reviews published 2015-2019 in 14 high-impact (greater than 5) general medical journals, the median number of citations per year was 5 for female first authors compared with 6.8 for male first authors (Sebo & Clair, 2022). This issue is complicated by averaging citation counts by the arithmetic mean favouring males whereas averaging citation counts after first taking the natural log, which is statistically better due to the highly skewed nature of citation counts, favouring females (Thelwall, 2018). Using the most precise approach, the most comprehensive study found a small tendency for female first authored articles to be more cited within the seven English-speaking countries examined 1996-2018 (Thelwall, 2020a), but a follow up analysis of disciplinary differences within six English-speaking countries 1996-2014 found some country/field/year combinations reversing the trend, such as a male citation advantage for Canadian medicine for most years (Thelwall, 2020b).

## 7.3 Articles in higher impact journals tend to be more cited

Since the journal impact factor is calculated from the citation rates of the articles in a journal, it is logical to expect articles to be more cited when they are in a journal with a higher journal impact factor. This relationship is not certain, however, since individual highly cited articles may be the cause of a high journal impact factor and the impact factor calculation exclusively includes short term citations. Nevertheless, there is strong evidence from many studies of different fields that articles in higher impact factor journals tend to be more cited (e.g., Boyack

& Klavans, 2005) and the same is probably true for all other formulae that estimate the average impact of the articles published in a journal.

## 7.4 Factors associating with higher journal article quality

Higher quality articles tend to be more cited than others from the same field and year in all fields, at least for UK research, with the highest (and strong) correlations being in health, life sciences and physical sciences and the lowest (and weak) being in the arts and humanities (Thelwall et al., 2023c). The overall correlations may hide the fact that citations primarily reflect the impact component of research quality, rather than the soundness and originality dimensions (Aksnes et al., 2019). Nevertheless, bibliometric indicators of quality slightly advantage female first authored research, at least in the UK, and especially in the social sciences, physical sciences, and engineering (Thelwall et al., 2023b). Overall, however, citation counts are indicators of research quality, with substantial disciplinary differences.

The journal citation rate (surprisingly) also associates with article quality in all fields of science, at least for the UK. A correlation analysis of REF2014 peer review scores and Elsevier's SNIP (Source Normalised Impact per Paper) journal citation impact indicator (HEFCE, 2015; see confidence intervals in Figure 1 of: Thelwall et al., 2023d) for 2008 articles found three out of 27 fields to have negative correlations, but all fields either had statistically significantly positive correlations or had correlation confidence intervals containing positive values. More conclusive evidence was found with a subsequent larger scale study with a finer-grained journal impact calculation. This found weak (0.11) to moderate (0.43) positive correlations between peer review REF2021 quality scores and average journal citation rates (not the JIF, but a similar type of calculation) for all 27 Scopus broad fields and all except one Scopus narrow fields. The correlations were strongest in the medical and physical sciences (and economics) and weakest in the arts and humanities (Thelwall et al., 2023d).

Articles with more authors tend to be higher quality in some but not all fields, at least in the UK. There are moderately strong Spearman correlations between author numbers and REF2021 quality scores (0.2-0.4) in medicine and the health, life, and physical sciences, but little or no positive association in engineering and the social sciences. In contrast, there was no evidence of association in the arts and humanities, and the decision sciences seemed to benefit from fewer authors (Thelwall et al., 2023g). For the UK, after controlling for the effect of collaboration, having international (rather than national) co-authors associates with higher quality research in 27 out of the 34 Units of Assessment, with collaboration with other advanced economies being particularly advantageous and collaboration with weaker economies tending to be a disadvantage from a (possibly flawed) Global North quality perspective (Thelwall et al., 2024).Finally, UK articles declaring a funding source tend to be higher quality in all fields, irrespective of team size, and this seems particularly advantageous for health fields (Thelwall et al., 2023h).

## 7.5 Summary

The studies reviewed here show that a wide range of factors derived from article text (e.g., length of articles, titles or abstracts, number or impact of cited references and article readability) might be related to the scientific impact of journal articles or conference papers as reflected by citation counts. However, there are disciplinary differences in almost all the results, often without a general pattern, and some findings could be biased by journal style norms that associate with higher or lower impact factors. Some of the associations also varied over time or between journals. Thus, whilst there are general trends for some properties, there are no universal laws for most, or too little evidence to speculate about such patterns. An additional risk with text mining to predict citation counts is that it is likely to work best by identifying highly cited topics, predicting higher citation counts for all articles on these topics. A successful prediction model for one year might be invalid for the next one due to topic changes, so text mining may need rebuilding each year to identify the new hot topics.

In terms of general trends, it seems that more cited research is likely to have more authors, be published in higher cited journals, be longer, and list more and higher impact references. Other potential factors are more variable between disciplines and/or countries, including international collaboration and inter-institutional collaboration. These tend to associate with higher citation counts but there are many exceptions. Moreover, there does not seem to be a general pattern in the association between title and abstract properties and citation counts. In parallel, higher quality research tends to be more cited and in more cited journals, especially in medicine, health, and physical sciences. Other potential factors are more variable between disciplines and/or countries, including author numbers and international collaboration.

The associations found rarely have a clear cause-and-effect relationship. For example, it is not clear whether team size associates with more cited research because larger team research is intrinsically better, funders often insist on large teams, or better researchers find it easier to attract collaborators. Thus, even the clearest findings are only suggestions about what researchers might consider when attempting to design or report the highest quality or impact research. As a practical recommendation, researchers might identify the factors found to associate with more citations or higher quality in their field and critically evaluate which, if any, are relevant to their research. For example, given that longer articles tend to be more cited, a scholar might consider whether this fact might nudge them towards describing their research in more detail, conducting more substantial studies, or reporting multiple studies in one paper.

Finally, several machine learning and regression analyses have shown that it is feasible to predict the long-term citation counts of papers from the factors discussed in this chapter to some extent, although with likely substantial disciplinary differences (e.g., Abrishami & Aliakbary, 2019; Haslam et al., 2008; Li et al., 2019a; Kousha & Thelwall, 2024). It is difficult to quantify the disciplinary differences due to the differing methods, scopes and accuracy measures of the experiments reviewed. The most important inputs are probably journal properties, authorship team properties and field/topic properties, with early citation information being especially useful, for predicting long term citation counts a few years after publication.

# 8 Journal citation impact indicators for articles: empirical evidence

Formulas to calculate the average citation rate of articles in a journal, such as the Journal Impact Factor (JIF) were originally designed to help academics to find important journals in their field (Garfield, 1972) [3]. As mentioned above this is reasonable on the basis that journals attracting more citations are more likely to contain articles that could be cited and may even tend to publish higher quality articles. JIFs and similar journal rankings have since been used to evaluate researchers (McKiernan et al., 2019) and have become targets for academics seeking to publish in the most prestigious outlets (e.g., Salandra et al., 2021; Śpiewanowski & Talavera, 2021; Walker et al., 2019) or get recognition for their work (Brooks et al., 2021). This is supported by evidence that publishing in higher ranked journals associates with career success in some fields (e.g., finance: Bajo et al., 2020).

The widespread misuse of JIFs and similar journal-level indicators as proxies for article quality has led to initiatives to restrict their use in evaluation, such as the San Francisco Declaration on Research Assessment (DORA, 2020), which is now accepted in the UK (UKRI, 2020). DORA emphasises that the value of an article should not be reduced to the value of its publication venue. Nevertheless, the continued importance of JIFs for academics is suggested by their prominent appearance on many journal websites, except where there are agreements to avoid them (e.g., Casadevall et al., 2016).

## 8.1 Advantages of journal impact indicators

In research evaluation contexts, JIFs have the advantage of being relatively transparent compared to informal ideas of journal prestige shared within a research community and compared to article-level expert review when the reviewer does not need to provide a rationale (e.g., REF2021). In addition, they are relatively objective and draw on many individual academic decisions by editors, reviewers, and citing authors (Waltman & Traag, 2020).

In fields where citation counts are reasonable indicators of research quality, journals with more citations per article would tend to publish better articles, so journal citation rate calculations would give (imperfect) indicators of research quality. They may also be better indicators of the quality of an article than article citations in some fields (Waltman & Traag, 2020). Moreover, in fields where JIFs are well regarded, competition to publish in higher-JIF journals would form a positive feedback loop (Drivas & Kremmydas, 2020) in which higher JIF journals increasingly monopolise research that the field regards as high quality. This may even change the field by encouraging authors to standardise on research conforming to the expectations of reviewers for the high impact journals (e.g., Kitayama, 2017).

## 8.2 Disadvantages of journal impact indicators

JIFs have most of the disadvantages of citation counts, as discussed above, and especially in Chapter 3. Most importantly, in fields where citations are not indicators of research quality, such as the arts and humanities, they are irrelevant (Fuchs, 2014; Thelwall & Delgado, 2015). Moreover, they are often inappropriately compared between fields, despite large natural variations in field citation rates. There are many technical problems too, such as failure to deal appropriately with the skewed nature of citation counts in most, calculation errors, and discrepancies between the numerator and denominator in calculations that allow journals to game the system by overpublishing citable non-article outputs, such as editorials (Jain et al., 2021; Lei et al., 2020; Seglen, 1997; Thelwall & Fairclough, 2015). Thus, despite the simplicity and intuitive appeal of JIF-like calculations, they should be interpreted cautiously.

---

[3] This chapter is based on Thelwall et al. (2023).

Journal quality control processes associated with more cited journals can also be detrimental for some types of research. In particular, highly original research may tend to be published in journals with lower impact factors (Wang et al., 2017).

A focus on journal impact can have systemic negative effects by pushing academics away from their preferred publishing styles, research topics (Brooks et al., 2021) and locally relevant research (Lee & Simon, 2018). It can also undervalue less cited specialties through their associated journals (Stockhammer et al., 2021). Moreover, journals can manipulate citations to inflate impact factors (Chorus & Waltman, 2016; Heneberg, 2016) and entire fields may be devalued by impact factor chasing (Tourish, 2020). Thus, whilst journal-level impact evidence may have value in some academic fields, it is problematic and may be useless in some contexts.

## 8.3   Journal citation rates and journal quality rankings

Intuitively, journal impact indicators seem to broadly reflect informal journal quality hierarchies in some fields where these hierarchies exist. For example, all journal citation rate formulae seem to give high values for prestigious journals like *Cell*, *Lancet* and *NEJM*, which perhaps gives the indicators some credibility. The value and limitations of journal citation rates are evident in the methodology for the well-known Association of Business Schools (ABS) journal ranking list, which is composed by subject experts but uses JIFs to support the judgements needed (Kelly et al., 2013; Vogel et al., 2021). This reflects a belief amongst some business researchers that JIFs have value but are imperfect. Similarly, national journal ranking lists are sometimes also informed by JIFs (e.g., Pölönen et al., 2021). Of course, journal rankings constructed by experts have their own flaws because academics tend to give higher ratings to journals in their own field (Serenko & Bontis, 2018) and the results vary by country (Taylor & Willett, 2017).

Empirical research assessing whether academics in a field find JIFs to be credible vary between those that find broad acceptance (implicit in: Currie & Pandher, 2020) or rejection (e.g., Hurtado & Pinzón-Fuchs, 2021; Meese et al., 2017). There are two issues: whether journals in a field can be credibly ranked and whether rankings produced by JIF-like calculations agree with expert rankings. Of course, academics are frequently sceptical about expert-based journal rankings too (Bryce et al., 2020) and different expert rankings may disagree substantially (Meese et al., 2017) so there is no "gold standard" against which citation-based journal rankings can be compared.

Using the expert-based Australian journal quality strata, one study has systematically compared expert rankings with journal impact rankings. Elsevier's Source Normalised Impact per Paper (SNIP) correlated better than the JIF with human judgement in 27 field-based categories. The SNIP advantage may be its normalisation for field differences that makes it more appropriate in large categories containing multiple fields. In the 26 monodisciplinary broad categories checked, the correlations were close to zero in the arts and humanities (0.2), and weak in the social sciences (0.2-0.4) but stronger elsewhere (0.4-0.8), ignoring the multidisciplinary category (Haddawy et al., 2016). This seems to be the most definitive evidence so far of the field-based credibility of journal impact indicators. The results may vary over time for a field, however (Walters, 2017).

## 8.4 Journal citation rates and article quality ratings

Although the previous section discussed whether journal quality ratings (when they exist) correlate with journal citation rates, for research evaluation the more important issue is whether and when journal citation rates associate with the quality of the articles that they publish. There is evidence that journal citation rates positively associate with article quality ratings in some fields from a few studies. An analysis of the correlation between peer review quality ratings and field/year normalised journal citation rates (SNIP) for 19,130 articles from REF2014 in 36 UoAs published in 2008 found Spearman correlation strengths being zero or negative in four UoAs: Classics (-0.8); Art and Design: History, Practice and Theory; Theology and Religious Studies (-0.1), Arts Area Studies (0). The strongest remaining correlations occurred for Clinical Medicine, Chemistry (all 0.5) and Biological Sciences (0.6) and Economics and Econometrics (0.7) (HEFCE, 2015, Table A18).

An analysis of Italian research has also compared journal citation rates and article quality scores. An investigation into the VQR (Valutazione della Qualita della Ricerca) research evaluation 2004-2010 has combined journal impact and citation count data, comparing it with peer review scores from two or three experts using a four-point scale. It analysed 590 economics, management and statistics (Area 13) journal articles for which the VQR process produced both bibliometric and peer review scores. The bibliometric method used a combination of article and journal citation rates and the peer review method used two independent reviewers, who may have been influenced by bibliometrics (especially since they were known to be important for the VQR). The peer review and bibliometric approaches agreed only moderately (weighted Cohen's kappa of 0.54), but at a higher rate than for the agreement between two independent reviewers (0.40). There was a suggestion of disciplinary differences in the results (Bertocchi et al., 2015).

In summary, the evidence so far suggests that field/year normalised journal impact is an imperfect indicator of article quality in most academic fields, its value varies greatly between fields, and it is close to useless in some arts and humanities fields. The next section confirms this with larger scale and more specific information.

## 8.5 UK case study

This section reports an analysis of the relationship between journal citation rates and research quality as judged by the UK REF2021 process. Methods details and limitations are discussed in the originating article (Thelwall et al. 2023). This is similar to the Italian case in the previous section but on a larger scale.

The Spearman correlations between average journal impact (MNLCS) and REF score (1* to 4*) for REF articles matching Scopus journal articles 2014-18 are positive for all UoAs, although the 95% confidence intervals contain 0 in four cases (Figure 8.1). The corelations tend to be very low for Main Panel D (mainly arts and humanities), with the unexpected exception of History. The correlations tend to be highest for Main Panel A (mainly health and life sciences). There are large variations within Main Panels B, C and D.

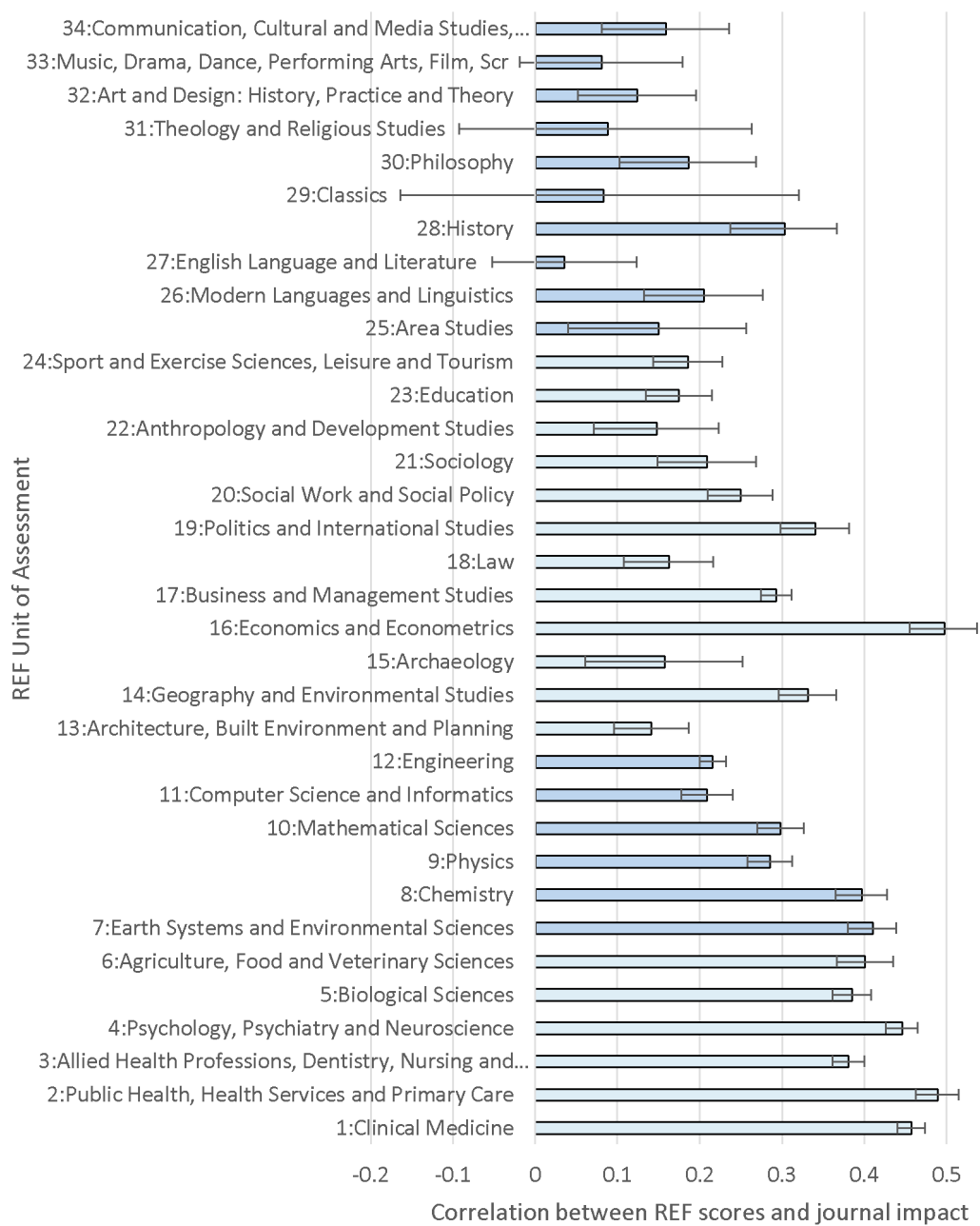

*Figure 8.1 Article-level Spearman correlations by UoA between average journal impact (MNLCS) and UK REF provisional score for UK REF2021 articles matched with a Scopus journal article published 2014-18 (n=96,031). Error bars illustrate 95% confidence intervals. Slight colour changes indicate main panels A, B, C, D (Thelwall et al., 2023).*

For the 27 Scopus broad fields, the correlations between REF scores and journal impact are above 0.1 in all fields and only the Veterinary confidence interval contains 0 (Figure 8.2). There are large variations within each of the four Scopus top-level categories (Health Sciences, Life Sciences, Physical Sciences, Social Sciences). Combined with the UoA results above, this confirms that there is not a simple disciplinary rule about the types of scholarship in which journal impact associates most strongly with article quality.

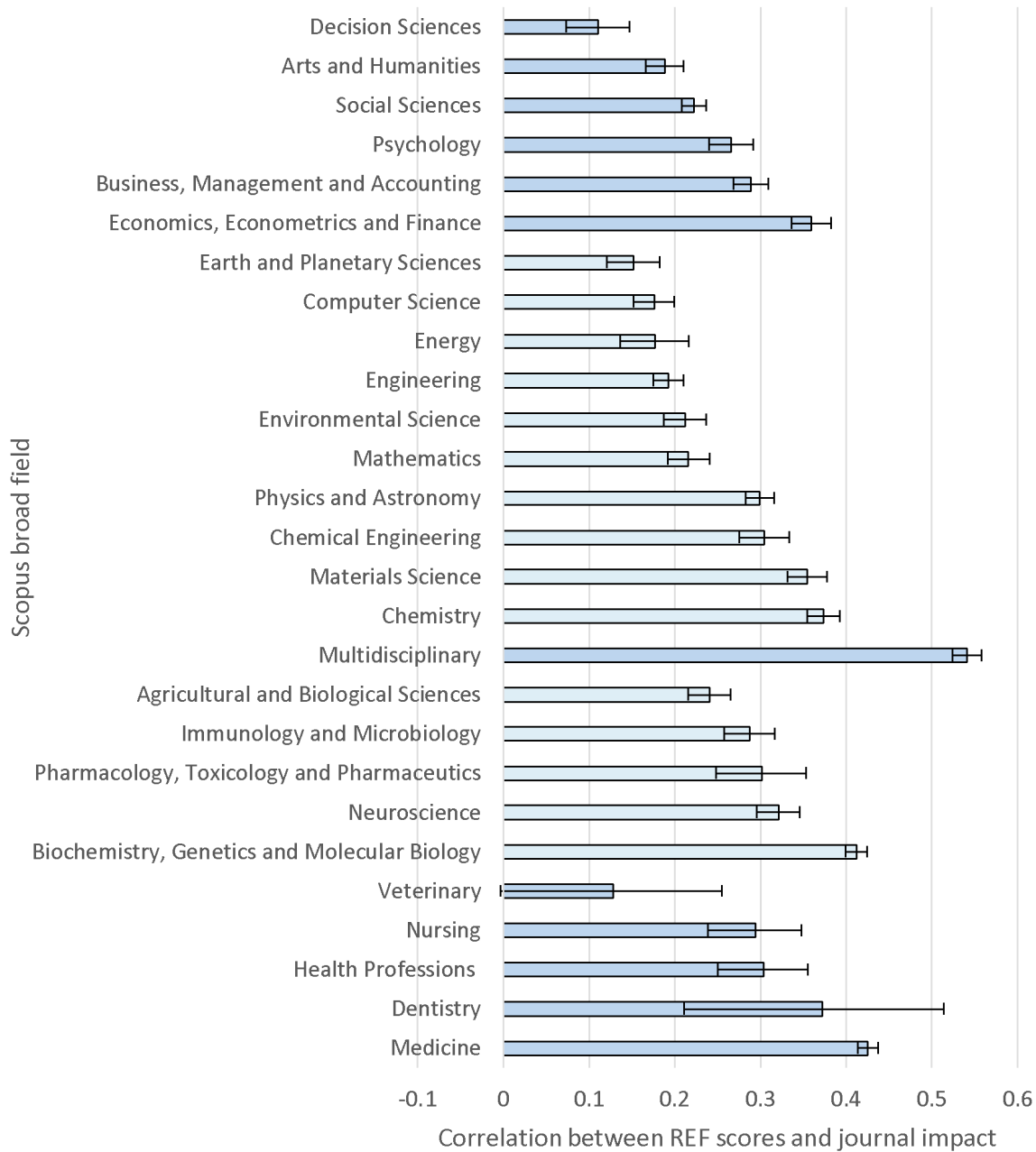

*Figure 8.2 Spearman correlations by Scopus **broad field** between average journal impact (MNLCS) for UK REF provisional scores for UK REF2021 articles matched with a Scopus journal article published 2014-18 (n=169,555, double counting articles in multiple broad fields). Broad fields are ordered by correlation within the four Scopus Top-level areas, plus Multidisciplinary (indicated by slight colour changes). Error bars illustrate 95% confidence intervals (Thelwall et al., 2023).*

Violin plots for the Scopus broad fields (Figure 8.3) show that there are large overlaps in journal impact between the four different quality ratings in all cases, despite mean journal impacts tending to be higher for articles with higher REF scores. These large overlaps occur even in the field with the highest correlations, medicine. In all fields, an article in a substantially above average citation impact journal has a reasonable chance of scoring 3* instead of 4* and in nearly all fields it might also score 2*. Perhaps more importantly, low or moderate citation impact journals host 4* ("world-leading") articles in all fields.

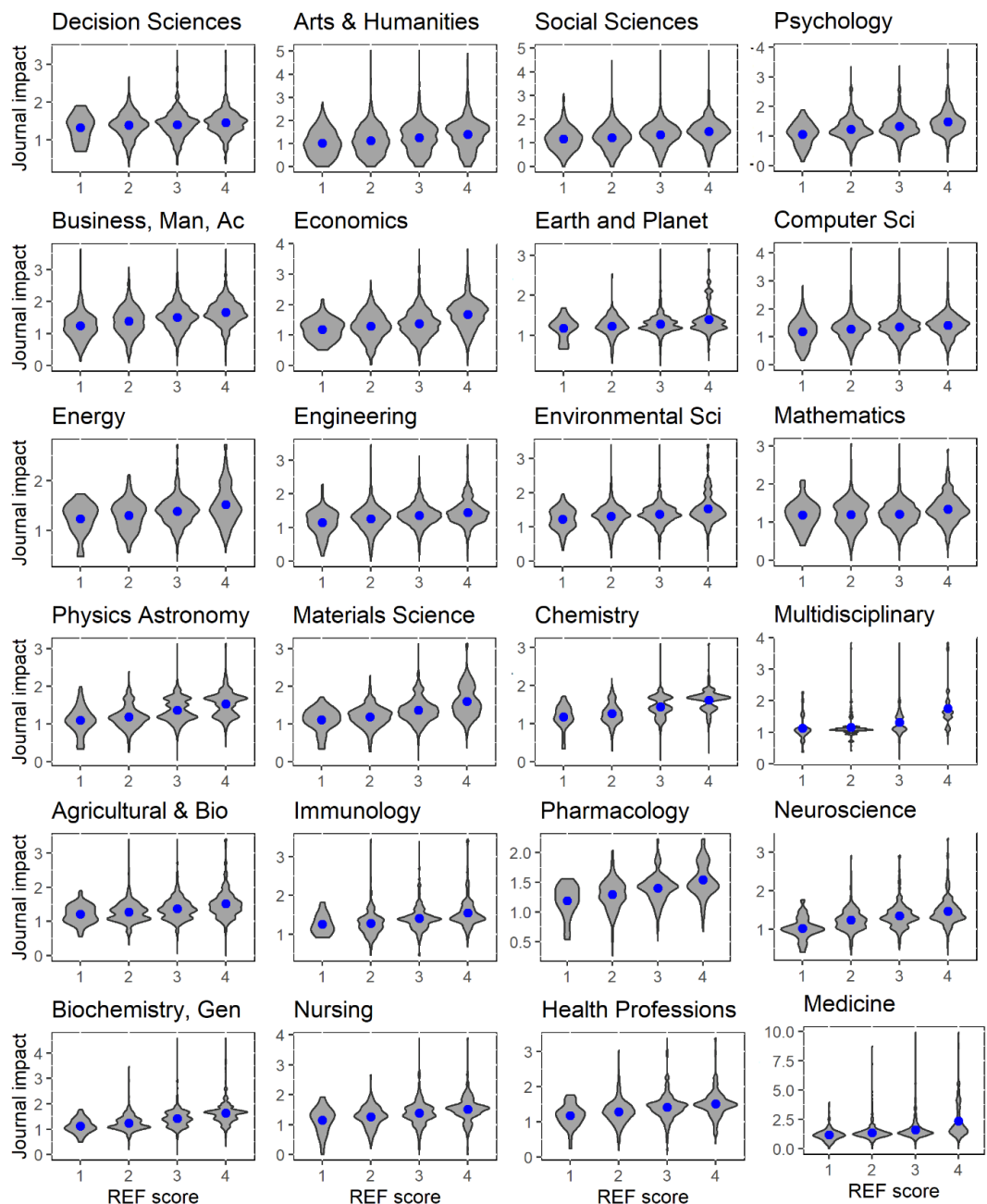

*Figure 8.3 Violin plots and means (blue dots) for journal impact (MNLCS) against REF score for 24 of the 27 Scopus broad fields. The sample sizes are small for the REF score 1\* in all cases, so the shape of the first violin is coarser than for the other scores. Two fields have been redacted for small sample sizes and one for data protection purposes. Journal MNLCS values tend to exceed 1 on average in all graphs as a second order effect of UK research having above average citation impact and REF articles being the self-selected best outputs of UK academics (Thelwall et al., 2023).*

Almost all Scopus narrow fields with at least 750 REF articles 2014-18 have a positive correlation between REF scores and journal impact (MNLCS) (Figure 8.4 to Figure 8.7). The only exception is Computer Science (all) (Figure 8.6). This is one of the two unusual types of narrow field in Scopus. The "all" and "miscellaneous" narrow fields that occur within Scopus broad fields are not academic fields but are instead categories for articles that do not fit neatly within narrow fields. Thus, the correlations are positive for all genuine narrow fields in Scopus.

The almost universally positive narrow field correlations add weight to the evidence that journal impact associates with article quality at least a small amount in all areas of scholarship. The arts and humanities can still be an exception, however, since the numbers were small in arts and humanities fields because most REF submissions in these areas were not journal articles.

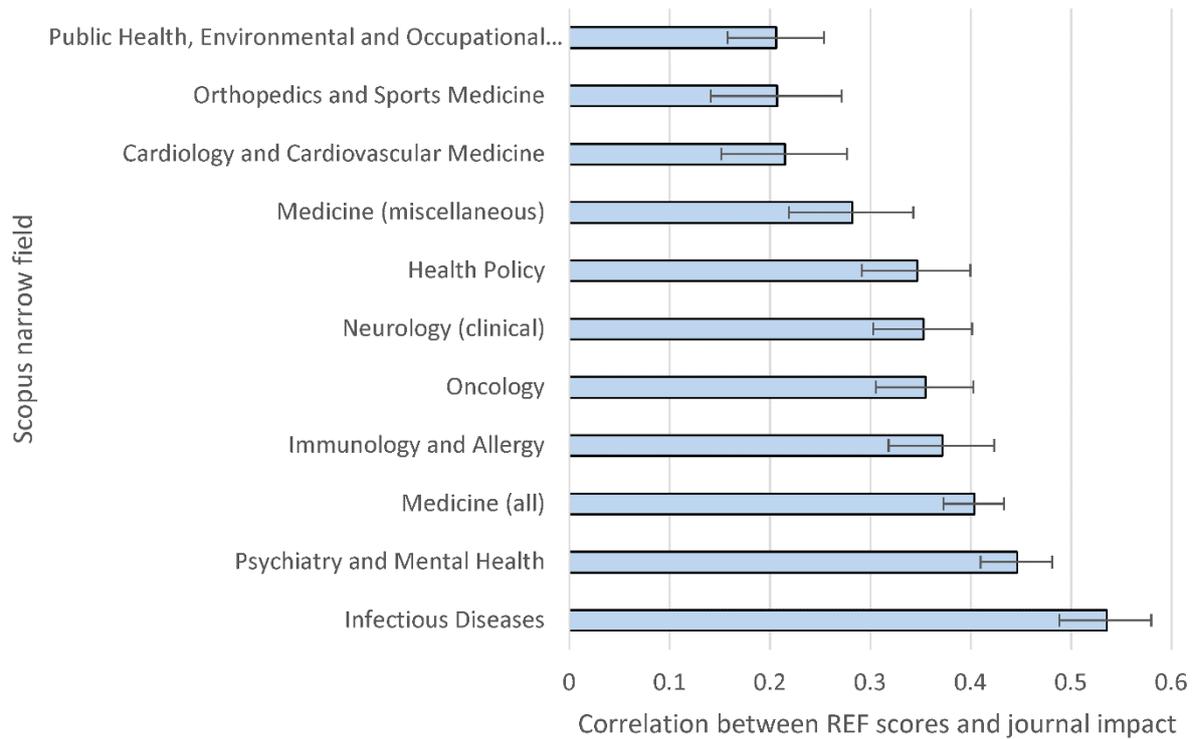

*Figure 8.4 Spearman correlations by **narrow field** between average journal impact (MNLCS) for UK REF provisional score for UK REF2021 articles matched with a Scopus journal article published 2014-18 within a **Health Sciences** broad field. Error bars illustrate 95% confidence intervals. Qualification: At least 750 articles with REF scores (Thelwall et al., 2023).*

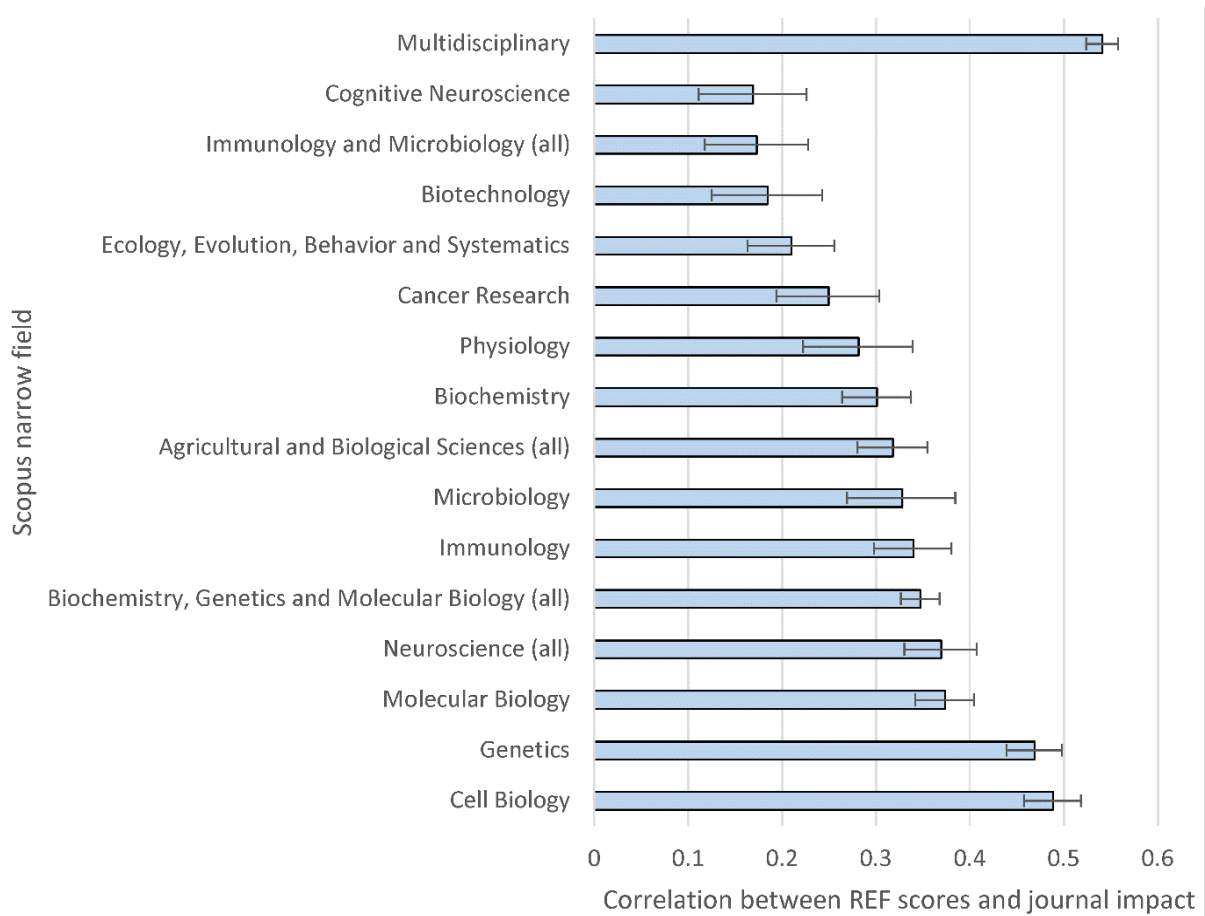

*Figure 8.5 . Spearman correlations by **narrow field** between average journal impact (MNLCS) for UK REF provisional score for UK REF2021 articles matched with a Scopus journal article published 2014-18 within Multidisciplinary or a **Life Sciences** broad field. Error bars illustrate 95% confidence intervals. Qualification: At least 750 articles with REF scores (Thelwall et al., 2023).*

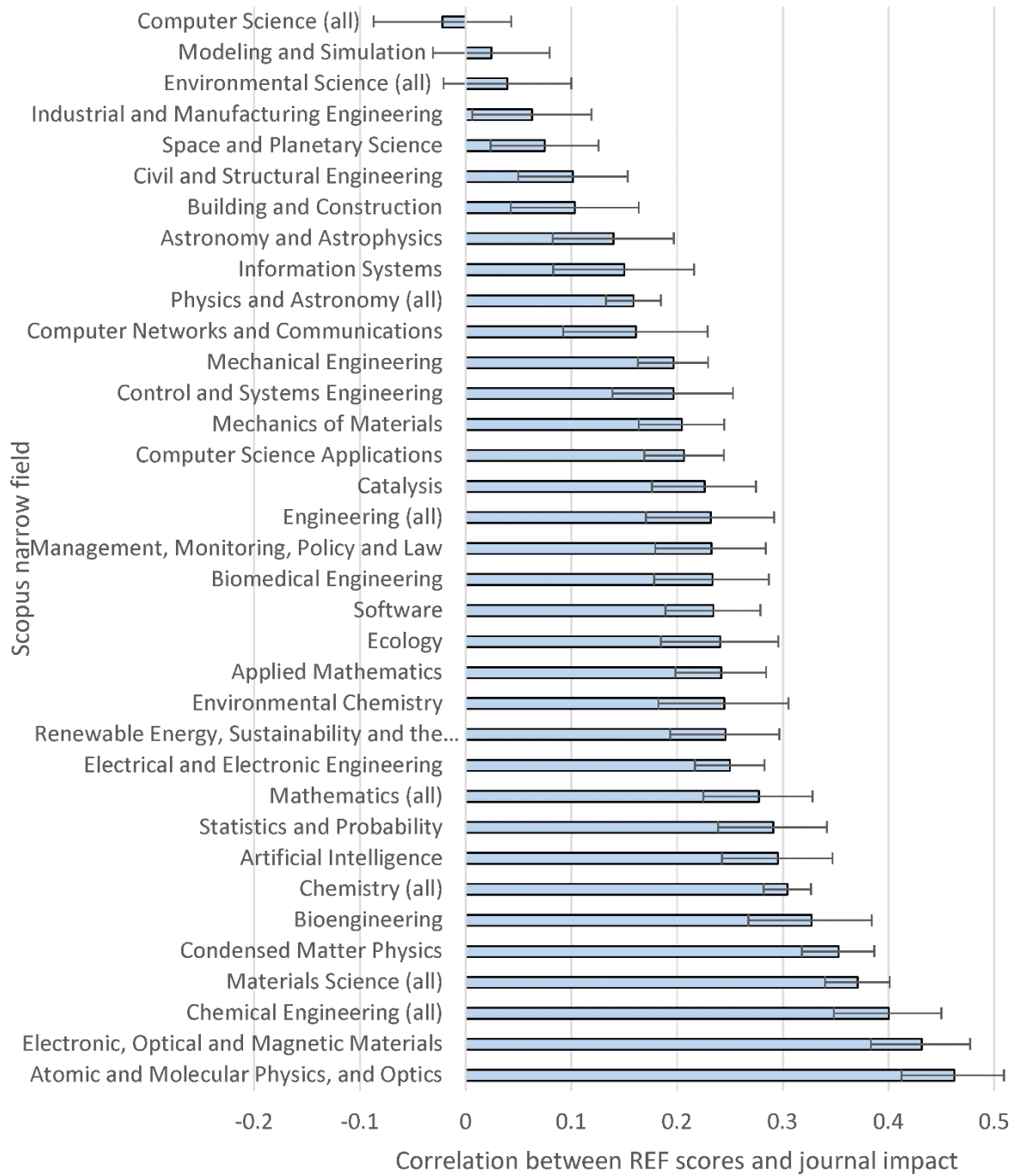

*Figure 8.6 Spearman correlations by **narrow field** between average journal impact (MNLCS) for UK REF provisional score for UK REF2021 articles matched with a Scopus journal article published 2014-18 within a **Physical Sciences** broad field. Error bars illustrate 95% confidence intervals. Qualification: At least 750 articles with REF scores (Thelwall et al., 2023).*

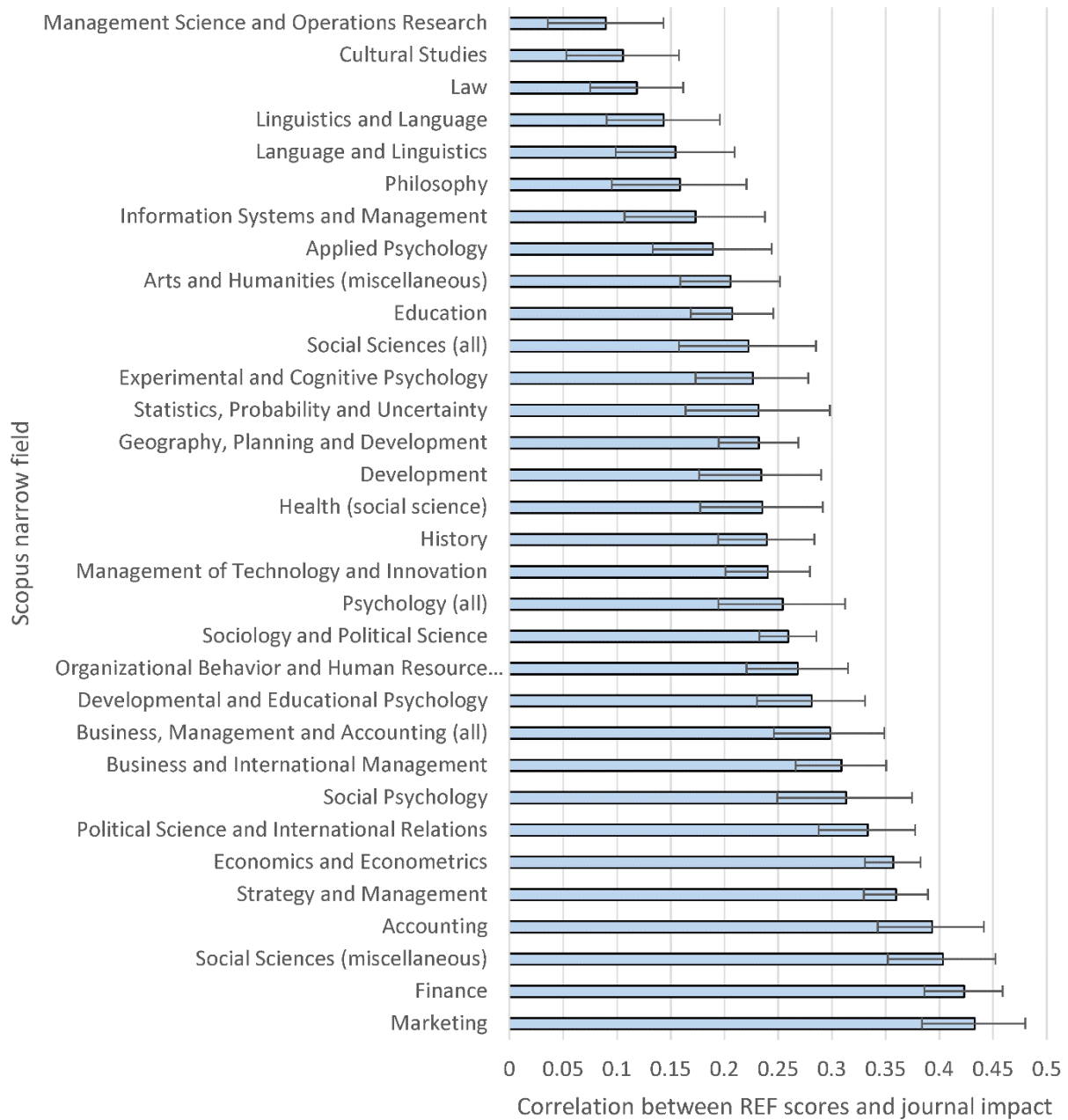

*Figure 8.7 Spearman correlations by **narrow field** between average journal impact (MNLCS) for UK REF provisional score for UK REF2021 articles matched with a Scopus journal article published 2014-18 within a **Social Sciences** broad field. Error bars illustrate 95% confidence intervals. Qualification: At least 750 articles with REF scores (Thelwall et al., 2023).*

### 8.5.1 Limitations

The results consider a single period (2014-18) and country and may change in the future as journals and fields evolve. They are also restricted to a single country, and other country evaluators may consider different criteria when judging the quality of an article (Taylor & Willett, 2017), such as its value for practical solutions or with a Global South perspective. Whilst the UK REF is almost an ideal case in the sense of large-scale expert judgements by people explicitly and repeatedly told ignore the reputation of the publishing journal, individual sub-panel members in some UoAs may have disregarded this advice or have been subconsciously influenced by journal reputations, based on their own perceptions of their fields.

Ihe journal impact calculation is also a limitation. Since the MNLCS method used here was designed to be optimal for fair assessments of average journal impact in multi-disciplinary contexts, correlations may well be lower for the journal impact indicators available from Scopus and the Web of Science, especially because they do not use log normalisation, so can give misleading averages. A weakness of the MNLCS field normalisation component is that it relies on the Scopus narrow field categorisation scheme, which may be imperfect for this purpose.

Another limitation is that some of the articles are in journals that are multidisciplinary, generalist, or from rapid publishing open access publishers where the impact factor may be less useful as a quality indicator than for typical journals because of their relatively wide scope.

### 8.5.2 Comparison with previous results

The likely theoretical reasons for the disciplinary differences in the results can be found in Chapter 3. The results mostly align with the REF2014 journal comparison from HEFCE (2015), except that all correlations were positive for 2014-18, perhaps due to larger data set (almost five times more articles), and a much higher correlation was found for Economics and Econometrics (0.63 for 2008 rather than 0.5 for 2014-18). The difference may be due to the impact factor used if published impact factors are consulted and considered important to economists, to changes in the Scopus-indexed journal set, or to the evolution of economics as a discipline (e.g., changing methodological orientations: Cherrier & Svorenčík, 2018).

The results also broadly align with previous research using expert rankings of journals rather than expert rankings of their articles. They align with Australian findings that expert rankings of journals positively correlate with journal impact for all 27 Scopus broad fields (Haddawy et al., 2016). They also tend to confirm numerous previous studies showing that the expert-judged value of a journal tends to correlate positively with its citation impact, including broadly the wide field differences found.

### 8.5.3 UK case study summary

The results for articles submitted to UK REF2021 show that higher quality articles tend to be published in higher impact journals in all REF UoAs (n=34), all Scopus broad fields (n=27), and nearly all Scopus narrow fields of science (n=94 shown), with the sole exception not being an academic field. The correlations are very weak (0.11) to moderate (0.43) for broad fields, and stronger (0.54) for Multidisciplinary, perhaps due to competitive generalist journals like *Science* and *Nature* being mixed with more accessible generalist journals. Weaker correlations may reflect non-hierarchical subjects, where journal specialty is more relevant than any journal prestige. As the violin plots (Figure 8.3) suggest, the weak correlations have no practical value in helping to assess the likely quality of individual articles. Moreover, at the level of aggregation needed to average out the noise (van Raan, 1998; see also Chapter 9) it is not clear that journal citation data for the weak correlation fields would be helpful in any role. Even the stronger correlations may reflect partial patterns in some cases, and especially for multidisciplinary journals, articles, and fields, since positive correlations can occur due to partial relationships (e.g., a subset of the journals in a field reliably publish high/low quality articles, but the rest do not).

## 8.6   Summary

Based on the results discussed in this chapter, it seems reasonable for scholars and evaluators to take journal citation rates into consideration when making relatively *minor* decisions in the fields where the correlations are not too weak, especially when there is a lack of expertise, time, or impartiality to fully evaluate individual articles or when only aggregate scores are needed for large sets of articles. For example, the results do not conflict with the minor role of journal citation rates in a prominent business ranking (Kelly et al., 2013) and the supporting role of impact factors in creating national lists of journals for evaluation purposes (Pölönen et al., 2021). Since there is not a simple rule about the fields in which journal impact is the least weak indicator of article quality, the graphs in this chapter may serve as a reference point to lookup the level of importance that may be attributed to journal impact in any given field. Of course, any use of journal impact data should first carefully consider unintended consequences, such as undervaluing work in low citation specialties within a field or encouraging researchers to migrate to high citation specialities or high citation impact generalist journals that are not the best fit for their work.

Finally, the lack of a strong correlation between article quality and average journal impact within any fields in the UK case study (e.g., never above 0.5 for any UoA, never above 0.42 for any broad field, never above 0.54 for any large narrow field) shows that journal impact is never an accurate indicator of the quality of individual articles. This result supports DORA's advice "Do not use journal-based metrics, such as Journal Impact Factors, as a surrogate measure of the quality of individual research articles, to assess an individual scientist's contributions, or in hiring, promotion, or funding decisions" (DORA, 2020).

# 9  Aggregate citation indicators for scholars, departments, institutions, funders, and journals

The value of citation-based indicators for the impacts of individual research outputs differs from the value of aggregate citation-based indicators for collections of outputs. This is true whether the collection is the works of a scholar, department, institution, country, or funder. The main reason is fundamentally probabilistic: under some assumptions average citation-based indicators, such as the citation rate of articles from a department, may be much more accurate than citation-based indicators for individual articles. This is possible in theory under the assumption that whilst indicators are misleadingly high for some articles and misleadingly low for other articles, these overestimates and underestimates tend to cancel out over large collections. This is discussed in more detail below with some empirical evidence about the extent to which it is true.

To address the above aggregation issue, this chapter introduces the fundamental principles of aggregate citation indicators and applies them to a range of contexts. Whilst aggregate indicators typically enable comparisons, these can be of different types, including the following.

- Time: Has a group improved since last assessed? Or what is the research performance trend for a group? Here the research unit (e.g., department, university, country) is compared against itself at a previous point or points in time.
- Competitors: Is the group better or worse than a set of comparable groups? For example, all the chemistry departments might be compared against each other as part of a periodic national assessment of chemistry research.
- Ranking: Constructing a league table to rank a set of comparable groups.
- World average: Is a group better or worse than the world average? Exceeding the world average might be a group's core goal, for example.

## 9.1  The probability model for aggregate citation-based indicators

There are many valid criticisms of citation-based indicators, as reviewed in Chapter 3, but some are less important when applied to sets of articles rather than individual articles, if certain assumptions are made or known to be true. For example, a strong article might be rarely cited because it is on a niche topic, but in fields where appropriately field and year normalised citation counts are positive indicators of research quality, it would be reasonable to expect departments producing high quality research to also have high average citation scores, even if a few articles were rarely cited. Such rarely cited articles might be thought of as statistical "noise" that, in the absence of bias, would tend to be cancelled out by occasional extremely highly cited articles that somehow have more citations than expected for their quality (van Raan, 1998).

To illustrate this key probabilistic issue, suppose that a citation-based indicator suggests that the quality of a journal article is 1.5 times the world average. Since citations are only crude indicators of research quality, then there is a reasonable chance, say 20%, that the article is in fact of below average quality (i.e., that the "real" score should be < 1). On the other hand, if the same citation-based indicator is averaged over 1000 articles and suggests that the average quality of these articles is 1.5 times the world average then, assuming there is no systematic bias, the true average quality of these articles is likely to be quite close to 1.5 and almost certainly above 1. This is because all citation-based formulae are error prone indicators of research quality. So, in the absence of bias, it is likely that the errors that overestimate the quality of individual articles approximately cancel out those that underestimate the quality of individual articles.

To make the above example more concrete, suppose that there are 10 articles that all have research quality exactly 1.5 times the world average and this quality is estimated with citation indicators, giving the following scores.

<div align="center">

0.5  0.5  1  1  1.5  1.5  2  2  2.5  2.5

</div>

In this case, if one article was picked at random then the citation indicator would have a 2 in 10 (i.e. 20%) chance of correctly estimating the research quality to be 1.5 times the world average. On the other hand, it would also have a 20% chance of incorrectly estimating the research quality to be below 1. In fact, 40% of the time the citation-based indicator overestimates the research quality at 2 or 2.5 and 40% of the time it underestimates the research quality at 0.5 or 1. If we average all these estimates then we get 1.5, which is exactly the correct answer for the average research quality of these articles because we assumed this to be 1.5.

In practice, we can't assume that the errors will exactly cancel out, even for a large sample of articles. Nevertheless, in the absence of bias, and especially for non-skewed indicators, the larger the sample size, the closer the average of the citation-based indicators should tend to be to the average quality of the articles. This is related to the Central Limit Theorem in statistics and is justified statistically in the next subsection.

## 9.1.1 Statistical formulation of the probability model

This section is an optional bonus subsection for readers who like statistics. You can safely skip it if this is not you. Suppose for statistical simplicity that there is a citation-based research indicator that estimates the quality of research articles on an infinite continuous positive and negative scale. Suppose that if an article has quality q then the citation-based indicator is normally distributed with mean µ=q and standard deviation σ=1. If sets of n such articles are sampled from an infinite set, then from the normal distribution, it is known that the population mean of these samples of n is still µ=q but the standard deviation has shrunk to $\sigma/\sqrt{n} = 1/\sqrt{n}$. Thus, the larger the sample of articles, the closer the mean indicator value of that sample tends to be to the true average quality of the sample.

The situation is tricker to formulate mathematically if considering a set of articles with varying individual quality but average quality q. Nevertheless, the conclusion should be the same: the larger the sample the closer the average of the indicators should be to the true average quality of the articles. Again, this is subject to the important assumption that there is no bias that would make it likely that most of the citation-based indicators underestimated (or overestimated) the quality of the articles.

## 9.1.2 Problems with the probability model

The probability model for citations suggests that aggregating citation-based indicators tends to give more precise and reliable results for larger numbers of articles. There is no simple rule to connect the sample size with aggregate indicator precision, however, so judgement must be used. Adding to this problem, the variability in quality for individual articles is certainly greater than the variability in average quality for groups of articles. Because of this, a more precise indicator is needed to be useful for comparisons between groups of articles than for comparisons between individual articles.

Despite the above consideration, the main practical issue that must be considered when analysing average citation scores in the light of the probability model is that systematic bias is possible and should be checked for when evaluating averages. Examples are given in each subsection below, but the main task for analyst is always the same: look for reasons why there might be systematic bias and then report them alongside the citation scores. The extent of bias can rarely be quantified, so its effect must be judged subjectively.

As a hypothetical example of bias, suppose that two groups of papers from a common field are compared based on their average citation scores but the second group of papers is from a lower citation specialty than the first group. It could be expected that articles in the second group tend to be less cited partly because they in a low citation specialty, even if the average quality was the same in both groups. In this context, increasing the sample size would not help to address the specialty bias in citations.

### 9.1.3   Testing the probability model and practical advice

The probability model is simple to test crudely in the sense that if we have expert human judgements for articles from several groups then we can compare the expert scores with scores obtained through bibliometrics both individually and in groups. Here, a stronger positive correlation between the aggregate scores for groups than for the individual articles would give evidence that the probability model works, at least to some extent, and in that context. This would show whether there is a higher chance that the average bibliometric scores are close to the aggregate human judgements than that the same is true for individual articles. Unfortunately, tests like this do not show that there is no bias or little bias.

Despite the above strategy, the probability model is impossible to validate in advance for any *new* dataset in the sense that it is impossible to know the relative strengths of systematic noise and biases in advance. For example, perhaps there is one country where a single psychology department specialises in a useful but low citation topic, generating systematic bias in its average citation indicators. It would be impossible to rule this out, even if previous tests of the probability model for psychology departments in other countries had all been successful.

Based on this discussion, tests of the probability model can perhaps best be used to generate an expectation about the extent to which it can be applied to any new context. This expectation should then be accompanied by subjective assessments about whether any of the new groups of publications to be assessed are unusual in any way. For example, the name of a psychology department might give away an unusual specialism that should trigger extra care being given to their bibliometrics.

## *9.2   Citation indicators for scholars*

For an individual scholar, whilst it is possible to calculate citation-based indicators for their work, there do not seem to be any reasonable choices. As mentioned above, the well-known h-index is not recommended for several reasons. First, it conflates output with impact, which should be assessed separately because they are different dimensions of work. Second, it is not fair to compare between academics with different service lengths or fields, so in almost all cases where academics are compared, it is biased. Finally, it is sexist and ableist because it disadvantages people with unavoidable career gaps, such as for carer responsibilities or illness.

In theory, a reasonable approach would be for a scholar to report a field normalised citation rate indicator, which would be fair to compare between academics, and the number of outputs, if the latter was contextualised with career information, such as unavoidable gaps. I am not aware of this ever having been done, however. This is presumably because few academics would be aware appropriate field normalisation approaches and those that understood the need and could access a source of field normalised data would probably not be confident that their evaluators (e.g., job interviewers) would understand the data. The standard approach is probably to list publications on a CV without citation-related information, hoping that the assessors will understand the value of the work from article titles and publishing journals. Alternatively, impact or quality claims might be made in an accompanying cover letter or as part of a narrative CV (Bordignon et al., 2023). As an example of the latter case, in 2023 UK research funders asked for narrative CVs that included lists of applicants top five research outputs, with explanations of their value (UKRI, 2023).

In practice, some academics report the JIFs of their publications on their web pages or CVs, either with the belief that JIFs are (or are believed by evaluators to be) accurate measures of journal quality or with the more reasonable belief that JIFs are useful approximate indicators of the quality of academic journals. In the latter case, JIFs might serve as evidence to assessors that the candidate had published in good quality journals, even if the assessors did not know the journal names. This latter perspective is statistically reasonable in health and physical sciences and to some extent in the social sciences and engineering, where journal citation rates have moderate or strong associations with the quality of the articles they publish (see Chapter 8). Nevertheless, JIFs are not field normalised so this is technically not a good strategy, and evaluators may not be aware of field normalised citation rate formulae and would probably struggle to interpret them even if they knew of them. This is because they would need to know the range of values of the journal formula for all candidates to know whether the values reported by one were exceptional. If JIFs are reported by a candidate then this also generates the risk of alienating assessors that dislike the JIF on principle, perhaps due to initiatives like DORA (2020). Thus, there does not seem to be an ideal solution here. Perhaps a candidate believing that evaluators will not recognise the journals that they have published in might report their JIFs on the basis that they are unlikely to get the job without providing some evidence (however weak) that they have published in strong journals.

An alternative to reporting JIFs might be to report the quartiles of the journals in which the author has published as part of their cover letter or the narrative part of their CV. This has the advantages that quartiles are easily understandable and are field normalised by definition. For example, a scholar might write, "All of my 15 journal articles are in Web of Science indexed journals and 5 are in the top quartile (Q1) of the category Chemistry (Applied)." If they wanted to be extra careful about anti-JIF sentiments, they might include a caveat like, "Whilst I acknowledge the limitations of JIFs and quartiles, this is presented as indicative evidence to support my claim that I have published my research in high quality competitive journals." A researcher might also reasonably make statements about the ranking position of publishing journals if they were low numbers, such as, "Two of my papers have been published in a journal ranked third in the Chemistry (Applied) category." This strategy might be helped by a caveat about journal rankings and might be considered appropriate in contexts where the researcher expected the evaluators to be unable to assess the quality of the work and to be unaware of the journals in question.

In summary, explaining the value of their work is probably better than reporting bibliometrics for most researchers but quartiles or ranks within categories (when low numbers) might be reasonable to report for those who believe that evaluators will not otherwise give them adequate credit for the journals that they have published in, especially if accompanied by caveats about the unreliability of bibliometrics.

## 9.3 Citation indicators for departments: Moderately large sets of articles focused on a field

Perhaps the most serious (in terms of reputational and financial impact) use of bibliometrics for groups of outputs is for sets of academic departments, or for sets of similar sized groupings of researchers. This could be to compare units within an institution or to compare departments in the same field across a country as part of a national research evaluation. Such evaluations may well be based on peer review, but with bibliometrics to support the judgements. For example, bibliometric-based scores or rankings might form a starting point for discussion, a tie-break for difficult cases, or a second opinion to look for individual departments that may have been scored unfairly by the assessors.

In the absence of bias, according to the probability model, bibliometric averages for departments that are not small should be based on enough data (journal articles) to make them reasonably accurate. It is not possible to quantify the expected level of accuracy, however, without information about the accuracy of article-level judgements, which is rarely available. The precision of a departmental citation-based indicator depends not only on the size of the department (number of articles) but on the field of the department. In health and physical sciences fields where citation-based indicators are the strongest article-level indicators of research quality, a smaller set of articles will generate the same level of precision as for a social sciences department. Here there is a "signal to noise" issue in the sense that a higher proportion of citations to social science articles are likely to be unrelated to research quality and so there is more noise in their citations. For arts and humanities departments, there is so much noise that bias is likely to drown out the "signal" even for large departments, so citation-based indictors would be unhelpful.

A more serious problem when comparing departments is that many have a degree of specialism, and this can cause systematic citation biases. For example, if a set of psychology departments is compared but one is in a low citation speciality, such as social psychology, and another is in a high citation speciality, such as neuropsychology, then these are likely to be ranked last and first, respectively, whatever the quality of their work. It would take the unlikely circumstance of all psychology departments having equal numbers of outputs on the same specialities for comparing citations to be unbiased. Whilst this problem is reduced using field normalised indicators, it is not eliminated because the categories used in the normalisation process seem to be large enough to allow high and low citation specialities to co-occur within them.

From the above argument, the probability model cannot be invoked to claim that any departmental citation indicators are precise, even for very large sets of articles. Nevertheless, the likely power of these indicators can be assessed using the results of previous studies that have compared citation-based indicators with expert ratings of departments in various contexts, and these are reviewed below. These studies usually correlate human scores with bibliometric averages, but the strength of these correlations varies between fields, countries, and indicators, as well as depending on citation windows and varying over time. This variety makes it difficult to draw general conclusions. They are summarised in the next subsection anyway.

### 9.3.1 Empirical evidence

Many studies have compared published aggregate evidence of research quality with average citations for collections of outputs, with mixed results. Rankings of UK departments based on average peer review scores for their outputs have been compared to average citation-based rankings, with correlations being very strong (rho=0.9) for psychology (Smith & Eysenck, 2002) library and information science (rho=0.8) (Norris & Oppenheim, 2003; Oppenheim, 1995), Archaeology (rho=0.7), Genetics (rho=0.7), and Anatomy (rho=0.5) (Seng & Willett, 1995), and Music (rho=0.8) (Oppenheim & Summers, 2008) and political science (partial correlation: 0.5) (Butler & McAllister, 2009). A larger scale study found strong associations between average citations and average REF scores for journal articles in life and health sciences (except nursing), business and economics, but weak associations in the social sciences (Mahdi et al., 2008). High correlations (0.7-0.8) between departmental REF2014 output rankings and median citations per paper have also been found for ten UoAs (Pride & Knoth, 2018).

One UK analysis also compared the difference between automatic rankings of departments based on citation data and other bibliometric information processed by machine learning. and human rankings for department. This went further than previous UK studies by producing by far the most accurate bibliometric assessments of average departmental research impact and by asking a sample of the assessors in focus groups whether the results were acceptable. Despite the high level of accuracy and the potential time saving with the approach, the assessors were mostly not persuaded of the value of the bibliometric approach because even small changes in average scores could produce large ranking changes for individual departments, which was perceived as unfair. For example, the predicted departmental score averages were amongst the most accurate in Unit of Assessment 1, Clinical Medicine (a Spearman correlation of 0.90 between predicted and actual average score, and a Spearman correlation of 1.00 between predicted and actual total score), but all except one department had a reasonable chance (at least 1 in 10) of having its ranking changed by this highly accurate and relatively conservative approach (Figure 9.1) (Thelwall et al., 2023a).

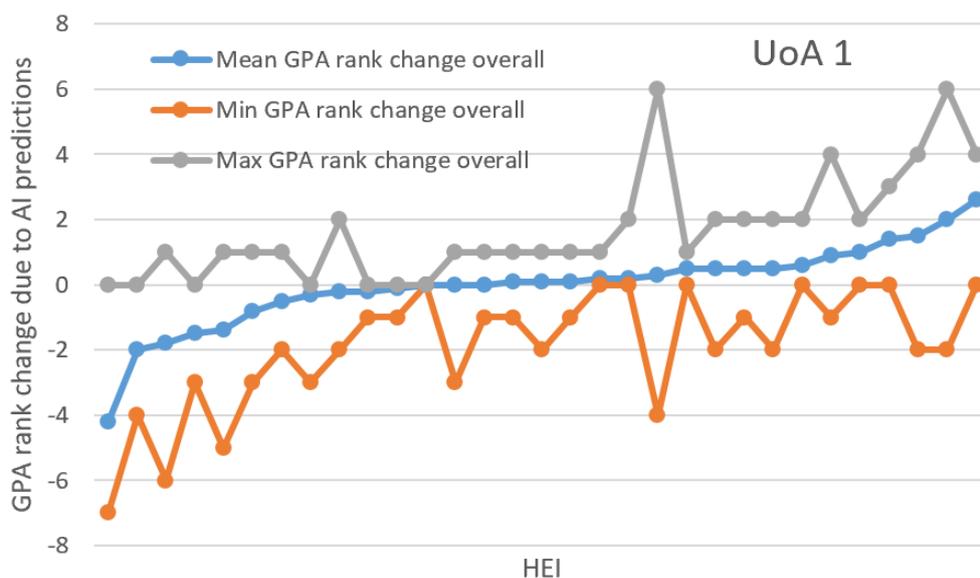

*Figure 9.1 Average REF institutional grade point average (GPA) gain by replacing human scores with machine learning predictions for half of the older journal articles, retaining the human scores for the remainder. Lower and upper lines are based on 10 iterations of the system (Figure 3 of: Thelwall et al., 2022).*

Outside the UK, an investigation into 12,000 Italian research articles correlated institutional average peer review scores with institutional average citations per paper in ten fields. There were strong correlations in most, including Physics (rho=0.8), Earth Sciences (0.8), Biology (0.7), and Chemistry (0.6) (Franceschet & Costantini, 2011), but weaker correlations have been found with a different method for Italy (except medicine, 0.5: Abramo et al, 2011; Baccini & De Nicolao, 2016). High correlations have also been obtained for the Netherlands (Rinia et al., 1998; van Raan, 2006). From a related perspective, panel ratings had weak correlations with citation-based indicators for research groups within an institution in Norway (Aksnes & Taxt, 2004). These studies give little information about the strength of article level correlations within fields, however, because correlations increase in magnitude when data are aggregated, with the degree of increase depending on the size of the aggregation units. Thus, it is not possible to draw conclusions about article-level correlations from institution-level correlations.

### 9.3.2   Summary

In summary, citation-based indicators can never be relied upon to accurately measure the average research quality of a department's journal articles. They are more useful in the sense of likely to be more accurate in the health and physical sciences but are useless in the arts and humanities. They are also more useful when the objective is to estimate the average quality of a set of departments rather than to rank the departments because even small errors can have a large influence on departmental rankings. For important decisions, human judges, if reliable enough, should always have the final say because of the high probability of citation biases against some departments, even with appropriate field normalisation of indicators and anti-skewing formulae.

Finally, the biases discussed in this section do not prevent a department from using bibliometric indicators to help analyse its changes over time, assuming that its specialities change little. This is because self-comparisons factor out (control for) biases in the absence of specialty changes.

## 9.4   Citation indicators for universities: Large multidisciplinary sets of articles

The caveats about the potential biases when comparing departments with citation-based indicators also apply to the situation of comparing universities, with some changes of emphasis. On the positive side, the overall biases should be weaker, because universities are large and contain many departments from different fields so biases against some of a university's departments are likely to be cancelled out partially by biases in favour of other departments. Nevertheless, the biases remain and there can also be systematic biases if a university has a high or low citation flavour to its research, such as tending to take a humanities orientation on most subjects or emphasising the technical aspects of research.

On the negative side, whereas analyses of departments can choose whether to use bibliometrics, depending on the subject, universities would be most naturally compared across all their departments. This would force the comparisons to include, for example, the arts and humanities departments where citation indicators are meaningless and social science departments where they are weak. Thus, the overall university average would be a combination of relatively meaningful figures from health and physical sciences departments with close to random numbers from arts and humanities, and with other departments contributing something in between. This would statistically introduce strong bias against universities that had arts and humanities strengths and weak bias against universities having social sciences strengths. In addition, it would tend to reduce the differences between universities by including random "noise" in their averages. Another negative is that some of a country's universities may have an applied focus, making it unreasonable to compare them against other institutions based on publications alone.

Because of the above factors, it does not seem reasonable to use bibliometrics to assess the performance of universities or to rank them. The most statistically justifiable approach might be to rank universities using only their bibliometric-friendly departments, such as all health sciences, all physical sciences, or the two combined. Whether this is a useful thing to do is a different question and my answer would be no because all universities have different field strengths so the results would be misleading. In any case, university rankings can't reflect the varied contributions of a university, whether it is based on bibliometrics, surveys or other data. This is the reason for the More Than Our Rank initiative, which was "developed in response to some of the problematic features and effects of the global university rankings. It provides an opportunity for academic institutions to highlight the many and various ways they serve the world that are not reflected in their ranking position" (inorms.net/more-than-our-rank).

As for departments, the biases discussed in this section do not prevent a university from using bibliometric indicators to help analyse its changes over time if its field composition does not greatly change. This is again because self-comparisons will effectively factor out biases.

## 9.5   Citation indicators for countries: Very large multidisciplinary sets of articles

At one aggregation level greater than universities, countries are sometimes compared with bibliometric indicators, or even groups of countries (e.g., EU vs. China and the USA: DGRI, 2022). This causes a slightly different set of problems to those for comparing universities. On the positive side, the much greater amount of aggregation when universities are combined within a country to give an overall national score gives substantial sample size advantages for average citation impact formulae. Nevertheless, whilst it might also be suspected that field specialisms within individual universities would tend to even out across entire countries, reducing this university-level problem, this is not necessarily true. In fact, countries tend to have research specialties in the sense of topics that they research to a much greater extent than the world average. Moreover, since professional social science literature and arts and humanities articles may be written in local languages, some countries may have little arts and humanities research and be underrepresented for the social sciences in international citation indexes. This can cause biases for comparisons between countries because of the varying relationship between research quality and citation indicators in different fields.

It would also be unfair to compare the journal articles of countries that are more focused on publishing with those of countries that tend to emphasise other goals, such as commercialisation or societal benefits. For example, Vietnam's health system coped much better with Covid-19 than that of the USA, despite the latter's highly cited biomedical research strength (e.g., DGRI, 2020) and Vietnam having only 0.3% as many journal articles in Scopus for 2022 classified as Medicine (14,399 vs. 5,047,653). One possible explanation amongst many is that Vietnamese health experts tend to be less focused on academic publishing.

There is an additional technical limitation for tiny countries: a substantial minority of a small country's outputs may be large scale international collaborations that deliberately have co-authors from all the world's countries or from all countries in a specific region. Such articles, if included in a tiny country's citation data might make the overall average misleading.

As for departments, the above reasons mean that it is not very meaningful to compare countries, although it would be reasonable to conduct self-comparisons, for example to assess whether a country's average citation impact has tended to increase or decrease over time or after an important national research policy change. International changes in coverage of a citation database or large changes in other countries' publishing performance can make such analyses difficult to interpret, however.

## 9.6  Citation indicators for funders: Small multidisciplinary sets of outputs

Research funders may want to use citation analysis to help investigate whether their funded projects were successful. They may also want to compare themselves with other similar funders, compare their own different funding streams, or assess whether a policy change has led to better outcomes. Of course, funders may well require awardees to report extensive information about their outcomes, including not just publications but also patents, inventions, and public-facing events. This may also include training events if the funding has a researcher development component and is likely to include, or be centred on, a narrative explanation of the impact generated by the project. In this context, the main self-evaluation of a research funder may be a primarily or purely qualitative report.

Since all research projects are different, it is difficult for a funder to draw strong conclusions about the value for money or success of an entire funding stream based on a large collection of qualitative reports. Thus, it is tempting to include quantitative data, such as bibliometrics, to provide suggestions about differences or trends. For example, a funder might produce graphs of the number of journal articles per project over time, or the field normalised citation rate of funded papers over time or for comparisons with other funders (e.g., Thelwall et al., 2016).

The factors that influence the extent to which it is reasonable to use citation rate indicators to analyse the success of research funders overlap with those for departments and universities. In fact, the department-related factors would be more applicable to narrowly focused funders, whereas the university factors are more relevant to larger funders with a broader remit. Many funders seem to have a wide remit without being science-wide, however, such as all medical research or all social science research. The size of a funder is also important in terms of the number of projects funded. Bibliometrics would be useless for small funders with wide field coverage because there is little to compare and so averages would probably fluctuate widely by year. In contrast, bibliometrics could be most useful for large narrowly-focused funders, other factors being equal.

An additional factor for funders (and perhaps also for funding-focused departments), is that funded research may vary dramatically in common topics between years due to changes in what is considered "hot" in the research community or by the funders themselves. Thus, for example, medical funders might have seen substantial increases in their bibliometric indicators during Covid-19 if they reacted quickly to the pandemic, but this increase would reflect a (hopefully) temporary phenomenon rather than a long-term increase in the quality of their funding procedures.

For the above reasons, the value of bibliometric evidence for funders varies from useless to moderate at best, depending on funder size and mission (i.e., the importance of non-publication goals). In the best case where bibliometrics has some value it should not be taken at face value but reported instead as suggested trends or comparisons, with the caveat that irrelevant factors might be the cause of any patterns found.

## 9.7   Citation indicators for journals: Small specialist sets of articles

Although journal-based indicators have already been discussed above as indicators for the value of their articles, this section discusses them from the perspective of using aggregate citation-based indicators to compare journals rather than articles. There are two types of journal impact indicators, field normalised and non-normalized. Field normalised journal impact indicators like JNMLCS and SNIP have similar issues to departments or universities, depending on their nature and size. Based on the probability model, in the absence of bias, citation-based indicators for larger journals should tend to be more precise. As for departments, the precision for a given journal size will tend to be highest in health and physical sciences fields where the article-level correlation between citations and research quality is biggest. Similarly, there may be close to zero precision in the arts and humanities where even small biases are likely to drown out the small article-level correlations.

Non-normalised journal impact indicators, like the Journal Impact Factor (JIF) do not consider the specialty of the articles that they publish. It is therefore not valid to compare journals with these unless they are from the same field and have the same specialties, if any. Journal rankings often circumvent this by ranking journals only against others in the same field. This is an imperfect solution  because most journals have different specialities or niches, some are more generalist than others, and some are general.

Empirical evidence of the extent to which various types of journal citation-based indicators agree with expert evaluations of journals is summarised in Section 8.3. This suggests that there can be a moderate degree of correlation between the two in some fields.

## 9.8   Summary

Many factors need to be taken in to account when using citation-based indicators to help evaluate the citation impact of scholars, departments, universities, countries, funders, and journals. Whilst the importance of field differences in the relationship between citation impact and research quality applies to all cases, as well as the generic limitations of citation indicators (e.g., problems with field classifications), there are also specific considerations in each case. These relate to scale (the number of outputs evaluated), the field/specialty mix of the articles (if any), and the goals of an evaluation (e.g., self-comparisons over time, comparisons between different units).

Each evaluation will also have its own unique problems that should be considered. These are likely to include the data not being comprehensive, such as only including journal articles when those evaluated also produce important monographs, refereed full conference papers, or artworks. They are also likely to include the consideration that the outputs evaluated are not the only important contributions made by those assessed. Thus, evaluators should always question and report local limitations as well as the more general issues discussed in this chapter.

Finally, as discussed above, every bibliometric analysis is likely to include collaborative articles with authors inside and outside of the unit assessed (e.g., co-authorship with scholars outside the department assessed by a departmental evaluation) and there is no fair way that can take this into account on a large scale unless the authors have provided evidence of their percentage contribution to each article assessed. Thus, all such assessments must assume that whichever counting method is used for publications and citations (e.g., fractional or whole counting) does not systematically bias the assessment to the extent that it renders the exercise meaningless. This issue would arise if some of the assessed publications were from huge scale particle physics and astronomy collaborations where authors can be assigned by prior agreement rather than by contribution. The same problem could occur for large scale international health collaborations where authorship might reward many small local contributions. In such cases, evaluators might need to agree to remove certain types of articles from the assessment or devise an alternative strategy to get a fairer result.

# 10 Gender, departmental, and interdisciplinary biases

This chapter[4] analyses whether there may be departmental quality, author gender, or article interdisciplinarity biases when using bibliometrics to support or replace peer review. It investigates each from two perspectives: journal-level and article-level bibliometric data. Since departmental scores are often the most important research assessment outcomes, it is important to know whether there are systematic gains or losses from journal-level or article-level bibliometrics. Gender differences are both an ethical (natural justice) issue and an efficiency concern if half of all researchers are devalued. Finally, interdisciplinary research is both difficult to evaluate and widely encouraged in the belief of its scientific and societal value and so needs special attention (e.g., as given the REF). For clarity, this chapter is not about the (admittedly more important) issues of systematic and individual bias against individuals and groups, such as through sexism, racism, homophobia, transphobia, islamophobia, antisemitism, elitism, and poverty. Instead, it is concerned with the narrower type of bias that might occur when citation indicators inform or replace peer review. Of course, peer review may also be biased.

The current trend for national research evaluations is to rely on peer review, with bibliometric data sometimes providing a supporting role (DORA, 2020; EU, 2022; Hicks et al., 2015; Wilsdon et al., 2015), but there are moves towards a greater role for artificial intelligence or other data driven approaches with bibliometrics (e.g., Jisc, 2022; ARC, 2022) and indicator-only exercises are still common (e.g., Belgium, Croatia, Denmark, Estonia, Finland, Norway, Poland, Slovakia, Sweden: Sivertsen, 2023). It is therefore important to assess whether bibliometric indicators would introduce biases in these roles. For example, if they have institutional, author gender, or study type biases, then they could push an assessment into mistaken and/or unethical outcomes.

Although bibliometrics are primarily used to inform peer review, this chapter does not directly investigate how assessors exploit bibliometric data to aid their judgements but instead identifies the directions of the changes likely if peer review decisions are informed or replaced by bibliometric data. For example, if women score more highly on bibliometrics than on peer review then it would be reasonable to assume that bibliometric-informed peer review would give higher scores to women than would peer review alone. From the institutional perspective, overall score shifts may be more important than biases for individual researchers or outputs, which they subsume, but the latter biases are still important because they may systematically disadvantage departments with atypical gender combinations or interdisciplinary research contributions.

As mentioned in Chapter 3, the core rationale behind using citation counts as an indicator of the value, quality or impact of an academic article is that scientists sometimes cite to acknowledge prior influences so that an article's citation count may partly reflect its influence on subsequent research (Merton, 1973). The Journal Impact Factor (JIF) and other average citation impact indicators for journals are also widely consulted in formal and informal research evaluations, so bias in these is also a consideration. At the informal level, academic appointment committees lacking the time to read the candidates' papers might use JIFs to help make quick decisions about the quality of research described in a CV (McKiernan et al., 2019). Individual researchers may also consult JIFs when deciding where to publish (Beshyah, 2019; Sønderstrup-Andersen & Sønderstrup-Andersen, 2008). More formally, some national evaluation systems reward scholars for publishing in journals meeting a JIF threshold or include JIFs in performance-based funding formulae (Sivertsen, 2017), although many countries construct bespoke stratified lists of journals to assess or reward research (Pölönen et al., 2021).

---

[4] This chapter is largely based on Thelwall et al. (2023).

## 10.1 UK case study background

Subsections within this chapter report evidence of bias derived from the UK REF2021 process. Methods details and limitations are discussed in the originating article (Thelwall et al. 2023b). The number of articles 2014-17 and HEIs varies considerably between UoAs (Table 10.1). Overall, the number of articles and HEIs per UoA varies greatly, interdisciplinary research is rare, and men dominate Main Panel B UoAs 7-12. This data is analysed below in this chapter.

*Table 10.1 The number of articles, Higher Education Institutions (HEIs), first author male/female genders, and interdisciplinary or monodisciplinary articles analysed. All were submitted to REF2021 and matching a Scopus journal article 2014-17.*

| # | UoA or main panel | HEIs | Female | Male | Interdisc | Monodisc | Articles |
|---|---|---|---|---|---|---|---|
| 1 | Clinical Medicine | 31 | 1948 | 2695 | 682 | 5289 | 5971 |
| 2 | Public Health, Health Services & Primary Care | 33 | 936 | 984 | 336 | 2043 | 2379 |
| 3 | Allied Health Professions, Dentistry, Nursing & Pharmacy | 89 | 2296 | 2124 | 850 | 5031 | 5881 |
| 4 | Psychology, Psychiatry & Neuroscience | 92 | 1997 | 2134 | 499 | 4496 | 4995 |
| 5 | Biological Sciences | 44 | 1252 | 1841 | 305 | 3535 | 3840 |
| 6 | Agriculture, Food & Veterinary Sciences | 25 | 610 | 686 | 203 | 1621 | 1824 |
| 7 | Earth Systems & Environmental Sciences | 40 | 519 | 1091 | 342 | 1936 | 2278 |
| 8 | Chemistry | 41 | 458 | 1088 | 294 | 1817 | 2111 |
| 9 | Physics | 44 | 282 | 1218 | 229 | 2913 | 3142 |
| 10 | Mathematical Sciences | 54 | 354 | 1909 | 314 | 2819 | 3133 |
| 11 | Computer Science & Informatics | 89 | 335 | 1615 | 423 | 2406 | 2829 |
| 12 | Engineering | 88 | 1128 | 4670 | 1330 | 9408 | 10738 |
| 13 | Architecture, Built Environment & Planning | 37 | 321 | 745 | 99 | 1345 | 1444 |
| 14 | Geography & Environmental Studies | 56 | 561 | 1005 | 105 | 1844 | 1949 |
| 15 | Archaeology | 24 | 105 | 139 | 22 | 276 | 298 |
| 16 | Economics & Econometrics | 25 | 140 | 718 | 25 | 997 | 1022 |
| 17 | Business & Management Studies | 107 | 1753 | 3771 | 361 | 6487 | 6848 |
| 18 | Law | 67 | 391 | 501 | 59 | 933 | 992 |
| 19 | Politics & International Studies | 56 | 422 | 856 | 84 | 1354 | 1438 |
| 20 | Social Work & Social Policy | 75 | 892 | 722 | 168 | 1674 | 1842 |
| 21 | Sociology | 37 | 376 | 380 | 142 | 708 | 850 |
| 22 | Anthropology & Development Studies | 22 | 193 | 231 | 29 | 489 | 518 |
| 23 | Education | 82 | 881 | 730 | 202 | 1670 | 1872 |
| 24 | Sport & Exercise Sciences, Leisure & Tourism | 60 | 404 | 862 | 154 | 1403 | 1557 |
| 25 | Area Studies | 20 | 107 | 120 | 49 | 214 | 263 |
| 26 | Modern Languages & Linguistics | 41 | 263 | 196 | 51 | 487 | 538 |
| 27 | English Language & Literature | 79 | 206 | 162 | 61 | 352 | 413 |
| 28 | History | 76 | 212 | 330 | 44 | 549 | 593 |
| 29 | Classics | 17 | 19 | 34 | 9 | 48 | 57 |
| 30 | Philosophy | 35 | 92 | 333 | 18 | 469 | 487 |
| 31 | Theology & Religious Studies | 22 | 23 | 57 | 16 | 74 | 90 |
| 32 | Art & Design: History, Practice & Theory | 69 | 225 | 230 | 58 | 530 | 588 |
| 33 | Music, Drama, Dance, Performing Arts, Film & Screen Studies | 68 | 120 | 148 | 46 | 258 | 304 |
| 34 | Communication, Cultural & Media Studies, Library & Information Man. | 54 | 220 | 244 | 57 | 471 | 528 |
| A | Main Panel A (UoAs 1-6) | 128 | 8168 | 9350 | 2519 | 19951 | 22470 |
| B | Main Panel B (UoAs 7-12) | 105 | 2973 | 11260 | 2778 | 20748 | 23526 |
| C | Main Panel C (UoAs 13-24) | 126 | 6294 | 10379 | 1395 | 18755 | 20150 |
| D | Main Panel D (UoAs 25-34) | 129 | 1477 | 1840 | 402 | 3430 | 3832 |

## 10.2 Departmental biases in bibliometrics

Researchers may suspect that there are peer review biases towards departments for prestige reasons, but this is difficult to prove. Studies that have correlated average citations with average quality scores aggregated at the departmental level have tended to find positive correlations varying in strength from 0.2 to 0.8 (Abramo et al, 2011; Baccini & De Nicolao, 2016; Franceschet & Costantini, 2011; Pride & Knoth, 2018; Rinia et al., 1998; van Raan, 2006), but these show a citation association with quality rather than the existence of bias. Most previous investigations of the relationship between departmental average numbers of citations and RAE/REF scores have also found statistically significant positive correlations, although with disciplinary differences (e.g., Mahdi, D'Este & Neely, 2008; Jump, 2015; Traag & Waltman, 2019), and for the same reason these do not show bias. The next subsection addresses this gap.

### 10.2.1 UK case study

This section investigates whether journal-level citation-based indicators favour high quality departments compared to peer review scores in any fields, using UK REF2021 data. With one exception, the results below show that departments in Higher Education Institutions (HEIs) with *lower* Grade Point Average (GPA) quality scores tend to gain from the bibliometric predictions. They always gain from journal citation rate (JMNLCS) predictions and usually gain from article citation rate (NLCS) predictions, except in UoA 9. This tendency is moderate or strong in all UoAs except two (1 and 9). The correlations are statistically significantly different from 0 in all cases except UoA 1 Clinical Medicine (NLCS and JMNLCS), UoA 9 Physics (NLCS), and UoA 25 Area Studies (NLCS). Thus, journal-level and article-level bibliometrics disadvantage higher scoring departments in almost all fields (Figure 10.1).

For the article-level NLCS, the most likely reason behind the result for lower numbered UoAs (health, biological and physical sciences) is that citations imperfectly reflect research quality so that weaker articles occasionally become highly cited, whereas stronger articles sometimes attract few citations. A department consistently producing high quality research could therefore expect to have some rarely cited articles and a department constantly producing lower quality research could expect to have some highly cited articles. Thus, whilst generally higher scoring departments tend to produce more cited research, they tend to be more consistent at producing high quality research than highly cited research. Chemistry is an exception. In this case there is a very weak tendency for the highest scoring departments to be more consistent at producing highly cited work than high quality work. It is possible that some departments selected articles partly on bibliometrics, in the knowledge that they may be consulted in REF evaluations. For higher numbered UoAs, where citations and impact factors have little correlation with research quality, the negative correlations are a statistical effect of replacing genuine scores with almost random noise and then averaging both. Thus, although the magnitudes of the correlations are similar across all UoAs and the practical implications are the same (bias against higher scoring departments), there are at least two distinct causes.

For the journal-level JMNLCS, the above argument largely applies, except for the chemistry exception. Again, for lower numbered UoAs, departments tending to produce high quality research tend to be more consistent in producing high quality articles than in getting them published in high impact journals. Clinical Medicine is a partial exception, in that good departments are almost equally able to produce consistently high-quality research and publish in consistently high impact journals.

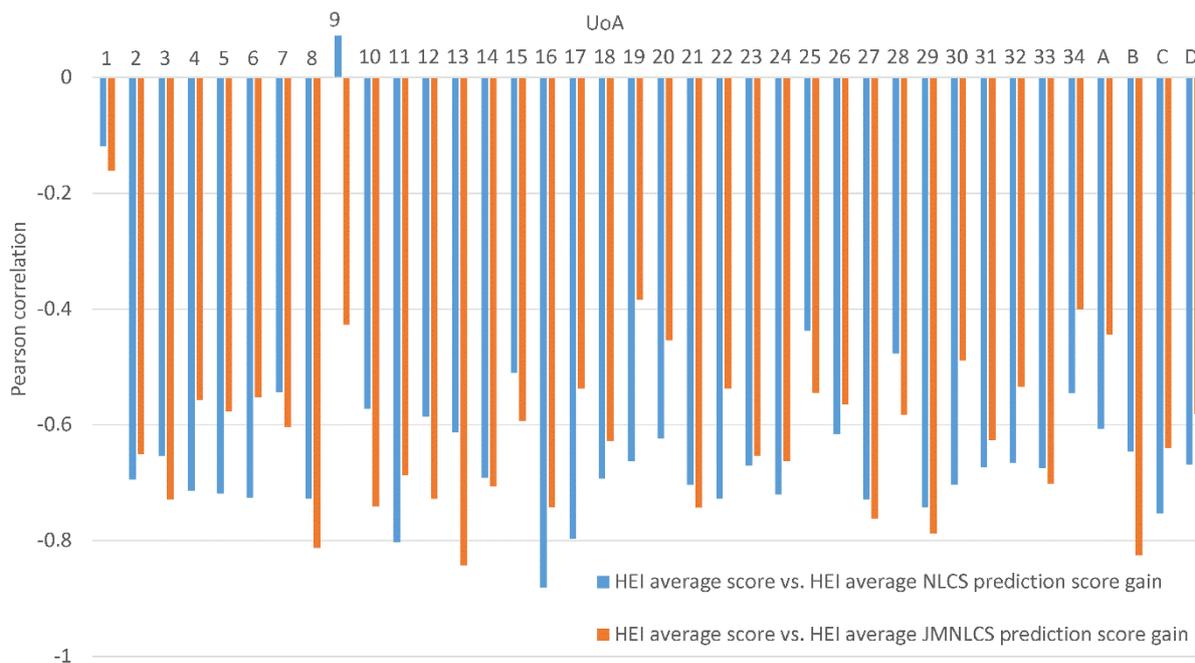

*Figure 10.1 Pearson correlations between HEI average scores and HEI average prediction gains from allocating scores only with NLCS or JMNLCS (Thelwall et al., 2023).*

These results are limited by the bibliometric ties being randomly allocated higher or lower scores. Whilst this simulates how the bibliometrics would have to be used if the exact number of articles in each star rating class is predetermined, such a use would be unrealistic in practice unless an assessment had fixed quotas for quality scores, such as to norm reference between fields in the assessment practice. Thus, this random assignment could be the reason why the bibliometrics have a damping effect in most UoAs. Nevertheless, the problem of ties would need to be resolved somehow if bibliometrics were to be used, and there does not seem to be a fairer solution.

## 10.3 Gender bias in academia, peer review and bibliometrics

There is wide suspicion that sexism affects evaluations of the work of female academics because sexism is not yet eradicated from society and women are underrepresented globally in senior roles (UNESCO, 2022) and for academic prizes (Meho, 2021). Many lists of highly cited scholars are also male dominated. In support of this, for the Italian VQR research evaluation 2004-2010, outputs (of all types) submitted by women were less likely to receive the top score from post-publication peer review or bibliometrics than research submitted by men, even after accounting for age, seniority, and compulsory maternity leave. This result was not affected by reviewer genders (Jappelli et al., 2017). Nevertheless, the extent of the impact of sexism on peer review scores and citation counts in academia is contested. Many studies show that females are and are not discriminated against in evaluations of their research, with no clear outcome (Begeny et al., 2020; Ceci & Williams, 2011). Moreover, overall career statistics and prizes favour men because of shorter female career lengths and a greater likelihood of carer leave (Huang et al., 2020). It is therefore possible that female-authored research is generally fairly judged in some fields but not others, such as those generating "chilly climates" for women (Biggs et al., 2018; Overholtzer & Jalbert, 2021). Intersectional factors may well also be relevant, with women that are also from other disadvantaged groups being particularly affected in some or all fields (Banda, 2020; Wilkins-Yel et al., 2019).

Many studies have investigated whether female-authored papers tend to be less cited than male-authored papers, with the suspicion of direct citation sexism through men preferentially citing men in some or all fields (e.g., Wang et al., 2021). Sexist citation practices may also be indirect, if the achievements of male authors are more celebrated, making their work more likely to be noticed and cited (Merton, 1968). Similarly, if men tend to cite their friends and these are men then this would generate a second order sexist citation bias against women. The empirical evidence for sexist citation is mixed, however, with the largest-scale evaluation with the most robust citation indicator suggesting a small female citation advantage in six out of seven large predominantly English-speaking countries (Thelwall, 2020a). These national averages may hide individual fields where females are slightly less cited, however (e.g., Andersen et al., 2019; Maliniak et al., 2013). Moreover, since female first authored research attracts disproportionately many readers than citers, citations might still systematically underestimate its wider value (Thelwall, 2018).

The strongest easily evidenced gender difference in academia is between fields rather than in citations. In many countries women numerically dominate some fields (e.g., nursing, allied health professions, veterinary science) and men numerically dominate others (e.g., mathematics, philosophy, physics, engineering) in terms of personnel (UNESCO, 2022) and publications (Thelwall et al., 2019; Thelwall et al., 2020). Africa may have the least gender variation between fields (at least in terms of students: UNESCO, 2022) and there is paradoxically greater gender inequality between fields in countries where there is less gender inequality overall (Stoet et al., 2018; Thelwall & Mas-Bleda, 2020). There is also a male/female gender differentiation within fields, with women more likely to engage in people-related topics, to use qualitative methods (Thelwall et al., 2019; Thelwall et al., 2020) and to have societal progress goals (Zhang et al., 2021). These factors all cause second order effects in bibliometric studies and perhaps also for peer review. Second order gender effects in bibliometrics are likely to occur because topics and fields have different citation rates. Thus, even for a set of researchers within a field, if one gender is more cited than the other then this could be because of differing research methods or specialties rather than sexism affecting choices of citations (e.g., Downes & Lancaster, 2019). It is impossible to fully differentiate between the two because all research is different and narrowing down to a specific enough topic to avoid the likelihood of topic or methods differences is likely to generate too few articles to statistically identify any gender difference, given that it is likely to be small (i.e., large samples are needed to detect small effect sizes).

## 10.3.1 Italian Case study

Bibliometric scores have been compared with post-publication peer review scores for 7500 outputs assessed in the 2010-2014 Italian VQR (Jappelli et al., 2017) to give insights into possible gender biases in either peer review or bibliometrics. For the peer review component, 14 field-based panels of about 30 experts sent each output to two external reviewers (three, when there was a discrepancy) and adjudicated on the results to give a four-point score. Reviewers were asked to check three quality dimensions: originality/innovation, relevance, and internationalisation. Bibliometrics were used for most journal articles in areas primarily producing English-language journal articles: natural and life sciences, engineering, mathematics and statistics, computer science, and economics. The bibliometric scoring system used a combination of citations and journal impact factors. Essentially, each output was given one of four scores by comparing its citation counts to three citation-based thresholds for its field and the same for its journal impact factor. A 4x4 matrix of outcomes was then used to judge what score to assign or whether to apply peer review instead. For example, if both methods (citations and JIF) got the same score, then this bibliometric score was used but if there was a large disagreement then peer review was called on instead (Ancaiani et al., 2015). A sample of 7500 journal articles assigned bibliometric scores was also double assessed by also subjecting these articles to peer review. By design this sample excluded articles where the two bibliometric indicators disagreed (e.g., low cited articles in high impact journals).

Separate regressions for the peer review and bibliometric scores given to the 7500 double assessed articles found that articles submitted by women tended to be scored lower by peer review than by bibliometrics, even after accounting for age, academic rank, and co-authorship (Jappelli et al., 2017). This suggests that it was easier for an Italian woman, compared to an Italian man, to get a highly cited article in a high impact factor journal than for her article to be judged to be excellent by two reviewers (of any gender). This study did not reveal field differences, or the effect of journal impact or citation counts independently, however.

## 10.3.2 UK case study

This section investigates whether grades based on (article-level or journal-level) bibliometrics favoured females compared to grades based on peer review in any or all fields, using UK REF2021 data. This is based on a study with similar goals to the Italian example mentioned in the previous subsection but is more recent, uses a larger dataset, covers more fields, and is not selective based on any bibliometric criteria.

There is some evidence of a weak tendency for female first-authored research to gain from bibliometric score allocation in some fields, relative to REF2021 peer review scores (Figure 10.2). The error bars include zero in almost all UoAs and the difference is marginal for the exceptions. It is not reasonable to draw strong conclusions for individual UoAs in this case because the marginal results are to be expected whenever many confidence intervals are drawn, even if there are no underlying differences (Rubin, 2017). Nevertheless, the female advantage is positive in 26 out of 34 UoAs for NLCS (p=0.001 for a post-hoc binomial test for gender difference α=0.5) and in 25 out of 34 UoAs for JMNLCS (p=0.001 for a post-hoc binomial test for gender difference α=0.5), giving statistical evidence of an overall female gain from bibliometrics. Moreover, the difference is statistically significant and positive for journal impact (JMLNLCS) in Main Panel B (mainly physical sciences and engineering). It is also statistically significant and positive for article citations (NLCS) in Main Panel C (mainly social sciences).

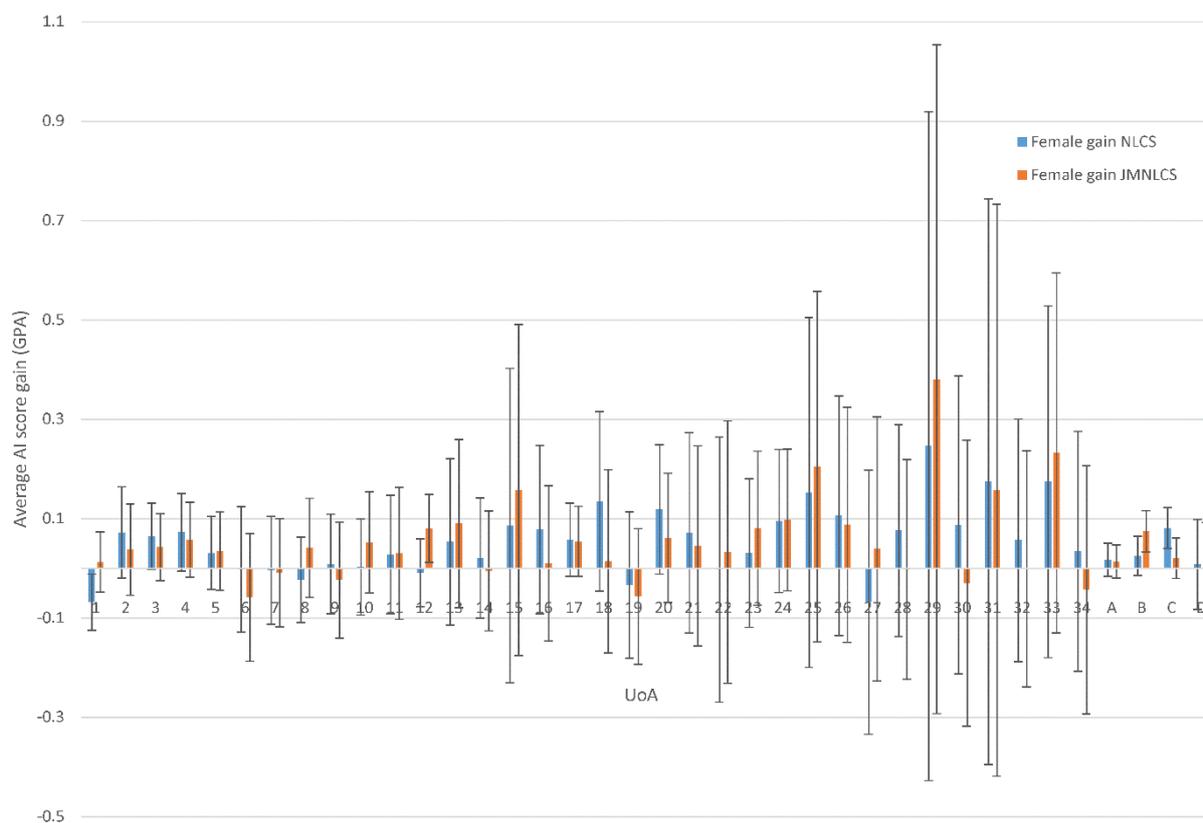

*Figure 10.2 Prediction gains for female researchers compared to male researchers from allocating scores with NLCS or JMNLCS instead of peer review (Thelwall et al., 2023).*

The minor gender bias in favour of women from both article-level and journal-level citations compared to peer review aligns with prior research of small gender citation advantage of women compared to men in the UK (Thelwall, 2020a). It extends this by suggesting that citations slightly overestimate the quality of female first-authored research, as judged by peer review. This partly conflicts with a previous suggestion that citations underestimate the significance of female first authored research because it tends to be more read than cited (Thelwall, 2018). Thus, either the peer review scores have a slight male bias, such as by insufficiently considering wider societal value, or the previous argument based on readership information was incorrect, perhaps because it did not consider non-educational impacts, such as commercial value. The results agree with a related finding for Italy 2004–2010 discussed in the previous subsection. It used a bibliometric heuristic combining journal impact factors and citation and also factored out age and seniority (Jappelli et al., 2017; see also: HEFCE, 2015). It is also possible that the results hide other bibliometric gender biases through gender differences in team contributions. For example, perhaps bibliometrics favour senior male last authors compared to peer review.

The study reported here has important limitations, however. For example, the gender detection may have introduced a second order bias related to ethnicity for names that were not detected with the algorithm. Moreover, the study has not investigated the cause of the gender bias in bibliometrics, relative to peer review, and knowledge of this might help to judge whether the underlying bias is in the peer review or the bibliometrics.

## 10.4 Difficulties evaluating interdisciplinary research

The term "discipline" is sometimes used to denote a research field (e.g., common topics and/or methods) and sometimes to denote a mature field backed by journals, conferences, and departments (Sugimoto & Weingart, 2015). The former, more general, sense is used in this section but the discussion also applies to the latter sense. Depending on how it is defined, interdisciplinary research combines theories, methods and/or personnel from multiple disciplines/fields to address a common goal (Aboelela et al., 2007; Arnold et al., 2021; Wagner et al., 2011). For the REF, interdisciplinary research is effectively defined as research that needs expertise from multiple UoAs to evaluate (REF2021, 2019ab). This does not directly match existing definitions of interdisciplinarity because it is evaluation-focused rather than input- or goal-focused, but it seems likely to have a large overlap in practice because both definition types involve multiple disciplines.

Interdisciplinarity is useful for applied research to address societal issues and for basic science that targets such issues as a longer-term goal (Gibbons et al., 1994; Stokes, 1997). Citation counts are likely to be less useful for evaluating interdisciplinary research than for single discipline research because its significance is more likely to be at least partly determined by non-academics judging its societal value (Gibbons et al., 1994; Whitley, 2000). Thus, factors unrelated to citations seem likely to be more important for interdisciplinary research quality judgements (Huutoniemi, 2010).

Citation analyses of interdisciplinary research have tended to evaluate the extent to which average citation counts for interdisciplinary research relate to the average citation counts of the constituent fields. It has been shown, for example, that interdisciplinary research citation counts can tend to be greater or less than the average of the constituent fields, depending on the fields in question (Levitt & Thelwall, 2008). Using three dimensions of diversity (Stirling, 2007), combining a greater number of fields associates with more citations but combining dissimilar fields associates with fewer citations (Yegros-Yegros et al., 2015). Nevertheless, this is only one study of a complex issue so, the effect of interdisciplinarity on citation counts is unclear. Moreover, there is little information about the overall relationship between interdisciplinarity and research quality, except as reported in the next section.

## 10.4.1 UK case study

This section investigates whether grades based on (article-level or journal-level) bibliometrics favour interdisciplinary research compared to grades based on peer review in any or all fields, using UK REF2021 data. There is some evidence of a moderate tendency for interdisciplinary research to gain from bibliometric score allocation compared to REF2021 peer review in some fields, but not overall (Figure 10.3). Whilst the error bars contain zero in most cases, interdisciplinary research in UoAs 16 (Economics & Econometrics) and 18 (Politics & International Studies) has statistically significant moderate advantages from both NLCS and JMNLCS. There is a weak but statistically significant bibliometric interdisciplinary gain for Main Panel B JMNLCS and Main Panel C NLCS. The overall UoA pattern is not statistically significantly different from equal, however, with 20 out of 34 UoAs having an advantage with NLCS (p=0.081 for a post-hoc binomial test for gender difference α=0.5) and 21 with JMNLCS (p=0.054 for a post-hoc binomial test for gender difference α=0.5). Nevertheless, it is possible that interdisciplinary differences found are second order effects from large strong or weak HEIs using it differently from average.

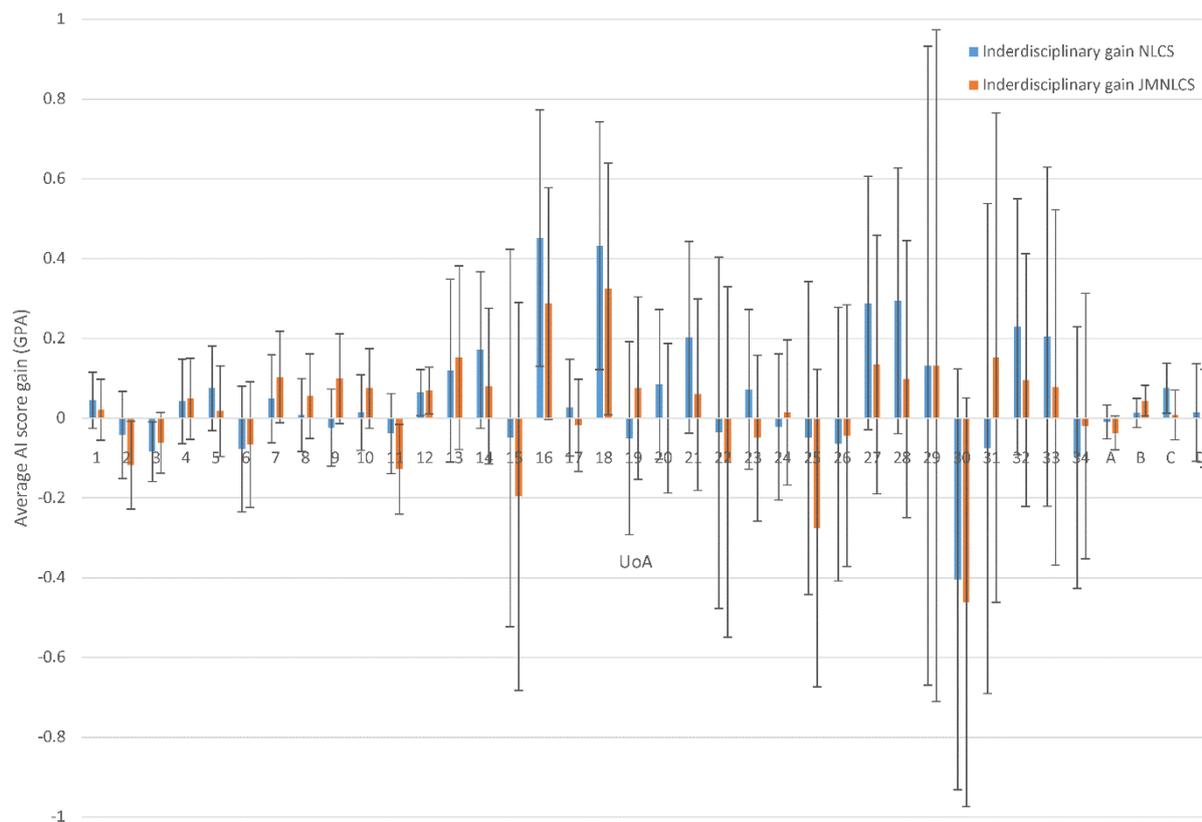

*Figure 10.3 Score gains for interdisciplinary research compared to monodisciplinary research from allocating scores with NLCS or JMNLCS instead of peer review (Thelwall et al., 2023).*

The lack of an overall trend in the relationship between interdisciplinarity and any citation advantage aligns with prior arguments that interdisciplinarity is complex, with no simple quality pattern (Huutoniemi, 2010; Yegros-Yegros et al., 2015). The existence of exceptions in smaller UoAs may be due to relatively stable and highly valued interdisciplinary topics, such as econophysics, with high levels of citation (Sharma & Khurana, 2021), or relatively stable low-citation interdisciplinary topics.  The interdisciplinary evidence used in this study was partial, however, since it was based on self-declarations by the institutions submitting the articles. Thus, it is possible that a relationship exists but has been hidden by the method of flagging interdisciplinary research. Longer citation windows are also sometimes needed to assess interdisciplinary research citations (Chen et al., 2022a), which may also have been a factor.

## 10.5 UK case study limitations

Whilst the UK case study results provide the largest scale evidence of the presence or otherwise of biases in the relationship between research quality and citation-based indicators, they should be generalised cautiously for several reasons. Most importantly, the UK situation may be different from other countries in terms of differing human or bibliometric biases. Moreover, the details of the citation impact formulae may also influence the existence or the size of biases. The next paragraph discusses these factors in more detail.

Bibliometrics may have a different value relative to peer review outside the UK and for different peer review goals. UK researchers submitted only their best work, and academics producing outputs thought to be lower quality may have been transferred to teaching contracts to avoid having their outputs assessed (affecting average scores in rankings tables). Other field/year normalised indicators may have produced different results, particularly if they did not take citation skewing into account. The 34 UoAs used in REF2021 are relatively broad and a different categorisation scheme may have produced slightly different outcomes. The results may also change over time, and particularly for the journal-level analysis with the continued rise of large broad scope open access megajournals like PLoS One (Spezi et al., 2017).

## 10.6 Summary

These results suggest that the use of citation-based indicators can introduce biases. Departments producing better research (as judged by peer review) tend to be disadvantaged when bibliometrics are used, even in fields where bibliometrics have high correlations with quality scores. This may be due to the damping effect of randomly assigning tied bibliometric scores to higher or lower classes. Thus, evaluation exercises relying on bibliometrics should be aware of this potential deficiency and either accept it or take steps to remedy it. This applies equally to exercises like the REF, where bibliometrics are used to support peer review rather than to replace it. For example, if the bibliometric information does not help make a quality decision in cases where REF peer reviewers disagree, it would be logical to favour a quality score that aligned with the departmental average. This would give a small nudge to partly offset the bibliometric bias.

The minor gender advantage for females compared to males for bibliometrics in the UK relative to expert review scores should be reassuring for those seeking to use bibliometrics to support research assessment in the sense that it is unlikely to introduce a bias against women, at least compared to peer review. Given the additional obstacles faced by women in society and academia, a small citation bias in their favour may help to reduce systemic biases against them.

The results also suggest that interdisciplinary research is not disadvantaged overall by bibliometrics compared to peer review. Nevertheless, evaluators should be watchful for individual high or low citation interdisciplinary fields in which bibliometrics may be misleading.

For individual-level research evaluations consulting bibliometrics, such as for appointments, promotions and tenure, the results imply that article-level and journal-level citation rate information will not disadvantage women or interdisciplinary researchers overall. This supports the continued use of article-level bibliometrics in these contexts, when appropriate.

# 11 Altmetrics and other alternative indicators

Many sources of evidence other than citation indicators and peer review are sometimes used to help evaluate published academic research, or to support evidence claims on CVs (Piwowar & Priem, 2013) or about non-academic research impacts (Kousha et al., 2021). This is important because the public funds much academic research and it is reasonable to try to assess the benefits it generates in terms of practical applications or influence. In particular, the inability of academic citations to reflect societal impacts has led to attempts to generate indicators to fill this gap, with patent citations (Hammarfelt, 2021) being an early example. This chapter discusses the most widely known type: altmetrics (Priem, 2015), derived from the former webometrics (Almind & Ingwersen, 1997; Aguillo et al., 2008). An altmetric is a web-derived quantitative indicator (or attempted indicator) of some type of research impact, including scholarly and societal. This is an umbrella term for a wide range of indicators of different types from different sources and the term has broadened from its initial restriction to social web indicators (Erdt et al., 2016). Since all relevant information now seems to be online, the term altmetric now encompasses all research indicators not based on scholarly citation counts.

This chapter introduces some altmetrics and the evidence for their value. Although altmetrics in theory could be used as inputs into the various citation indicator formulae introduced above, with one main exception they have been almost exclusively reported as raw counts, so this is the focus used here. The chapter first introduces sets of related indicators with initial evidence of their value and the next chapter summarises empirical evidence of their value as research quality indicators. As in all chapters, the focus is on journal articles rather than conference papers or books, although there are some book-specific altmetrics (e.g., Maleki et al., 2023).

## 11.1 Tracing research impacts

Although a key goal of much academic research is to generate societal benefits in the short or long term, it is extremely difficult to track the benefits of individual publications. Even tracing the scholarly influence of academic research is very difficult because each study builds on many previous studies. For example, whilst a paper cited by a seminal article might claim the citation as evidence of substantial scholarly influence, it would be difficult to assign a value to each reference (e.g., from essential to trivial background) or the opportunity value of the citation, in the sense of whether the citing seminal study would have been possible or similarly strong without it.

A fundamental problem for tracing the societal impacts of research is that even a unitary research invention with societal value may need multiple studies to establish it. For example, a new drug may need many clinical trial stages as well as replication studies and other follow-up investigations and challenges before it is accepted. Each of these academic studies may in tern rely on a wide body of prior academic research for methods, knowledge, and interpretation.

Probably the most common problem with tracing the societal benefits of research is that even work directly influencing an innovation seems to be rarely cited, especially outside of medical fields. For example, many innovations may be based indirectly on research that the innovators learned about during their education and these innovators might not even know about the original research publications.

Because of the above problems, whilst in theory, an altmetric might capture a mention of research in a context that demonstrates societal impact, such mentions are likely to be rare exceptions. For example, whilst a professional organisation might tweet about an article that had influenced a new policy, such direct connections are surely rare, and it seems likely that they would tweet a review in preference to the original research and would not mention foundational papers if there were more recent studies. Moreover, even in this case the organisation might also tweet about many other publications that they found interesting but did not lead to policy changes. Similarly, citations to academic research from patents seem to have been originally thought of as a strong source of evidence about which academic research had commercial value, but many areas of industry rarely patent and citations can be added by patent examiners for context rather than by inventors, so do not necessarily reflect influence.

For all these reasons, alternative indicators seem unlikely to capture more than a small fraction of the societal impacts of research. In addition, interpreting indicator values is likely to be tricky because few online mentions of academic research provide strong evidence of its societal value.

## 11.2 Public interest indicators? Tweets, Weibos, Facebook walls, Blogs

Academics, funders, and governments sometimes attempt to generate public interest in their research in the belief that people have a right to know about what they have supported through their taxes, they may benefit from the knowledge culturally or educationally, or that it may help specific decisions in their work. In parallel, universities routinely generate press releases to generate positive news stories about their scholars' work. This publicity may enhance the prestige of the university and perhaps attract more students and industry collaborations. Whilst public interest could be investigated indirectly through media coverage (e.g., Sun, 2014), this has the disadvantages that it does not directly count the number of people reading a story and excludes non-news methods of sharing research. Counting mentions of research in social web sites could therefore provide an alternative source of evidence about public interest in academic research.

General social web sites like Facebook, Twitter/X and Weibo are a potential source of evidence of public interest in research on the basis that people reading an article may post about it to declare their interest or recommend the article to their followers. Thus, counting general social web site posts mentioning an article could be used as a public interest indicator, at least in theory. In practice, few people reading an article will post about it and they may only post about it if there is an ulterior motive, such as that their friends might find it interesting, useful, or impressive. Moreover, students and academics can also use general social web sites and so a count of social web site mentions reflects unknown proportions of academic and public interest. In addition, academics, journals, and publishers may promote their own articles on social media, further undermining the value of counting social media mentions.

Blogs are different from social media sites because they tend to be extended discussions of articles, often to translate them for a non-academic audience (Shema et al., 2015). Thus, blog mention counts, like mainstream media mentions, could be taken as an indicator of potential public interest. This assumes that the bloggers, like mainstream journalists, have audiences that read their work and/or are good at identifying academic research that the public can be interested in.

## 11.2.1 Validity checks

Because of all the above limitations of blogs and general social web sites as sources of evidence of public interest, it is important to find ways to check the value of the altmetrics as indicators of public interest. Ideally, this would be done by taking a large sample of journal articles with altmetric scores and then asking a representative sample of the public if they had read the articles or found them interesting. This seems impossible, however, since news may be quickly forgotten, and it would not be reasonable to ask random members of the public to read large sets of academic papers to decide if they were interested in them. Because of this, there have only been very indirect empirical checks of the value of altmetrics as public interest indicators.

The first type of check made has been for face validity, and only for tweets. Some papers have analysed samples of tweets mentioning academic research to see who the authors were or whether the tweets suggested that the research had been valuable. A survey of tweeters of medical articles suggested that half were non-academic (Mohammadi et al., 2018), but the fraction is probably higher for non-medical research because it has less relevance to typical members of the public. In parallel, content analyses of tweets have found that they typically just mention a paper's title or give a summary without any evaluative information or use context (Holmberg & Thelwall, 2014; Thelwall et al., 2013a). If the same is true for other general social web sites, then it seems likely that academics dominate social web mentions of research and individual posts rarely give evidence of the value of the studies mentioned.

A more common but weaker indirect strategy to evaluate social media mentions has been to correlate altmetrics with citation counts to check for positive values. This does not give any evidence that the altmetrics reflect public interest, but statistically significant positive correlations would show that altmetric data is non-random, which is a step forwards. Moreover, since academic research varies in quality it might be expected that higher quality research generates more public interest and more academic interest (hence more citations), so a positive correlation between the two should be expected. Several studies have indeed found positive correlations between citation counts and altmetrics (e.g., Costas et al., 2015). Probably for statistical reasons, these correlations have tended to be highest for the social web sites with the most data (e.g., higher for Twitter/X than for Facebook), higher for fields with more public interest, and higher for fields with higher citation counts (Haustein et al., 2014a). Overall, they have tended to be positive but very low (e.g., 0.1) with some being moderate (0.5). A more direct test is reported in the next chapter.

## 11.3 Scholarly and educational impact: Reference sharing altmetrics

Social reference sharing sites like Mendeley and Zotero allow readers of research to publicly share information about what they are reading or citing. The emergence of these sites has allowed altmetrics to be created that count the reference shared in them. Assuming that these sites are primarily used by students and academics, this would then give indicators that reflect a combination of research and educational impact. The main limitation is that a small minority of academics and students use these sites, either not using a reference manager or using one that does not share data, such as EndNote. This small minority is also probably biased by age, nationality, and field. Because of these limitations, it is important to empirically assess the value of refence sharing site altmetrics as indicators of any type of impact.

### 11.3.1 Validity checks

The value of reference sharing altmetrics has been assessed in two ways: face validity and correlational. For face validity, a survey of Mendeley users asked them if they had read the articles in their Mendeley libraries, finding that most had either read them or intended to read them (Mohammadi et al., 2016). This shows that Mendeley counts give evidence of article readership, albeit only partial since it excludes non-users and articles not added by users to their Mendeley libraries.

As for social web mentions, the value of reference sharing altmetrics has also been assessed by correlating them against citation counts. This again gives evidence that they are non-random. The corelations also show how close they are to citation counts, although not how well they reflect educational impacts. The Mendeley correlations tend to be stronger than for social web sites (partly because of more data for Mendeley) (Thelwall et al., 2013b) and vary between fields (Thelwall, 2017), probably also partly due to statistical reasons. For 325 different fields and articles that were at least 5 years old, correlations between Mendeley readers and Scopus citations in one study were always positive and usually above 0.5, varying from 0.255 to 0.829, and averaging 0.672 (Thelwall et al., 2017). This shows that Mendeley readers are good indicators of citation impact in most fields. Moreover, Mendeley readers appear before Scopus citations, so they are useful as early indicators of longer-term citation counts (Thelwall, 2018). This allows them to be used to help evaluate the (primarily) scholarly impact articles that are too young for citation counts to be useful.

In contrast to the situatiIor Mendeley, the citation correlations for Zotero are low due to a lack of data (Thelwall et al., 2013b).

## 11.4 Grey literature citations and Overton

Research may aim to influence government policy or the policies and procedures of other organisations. Whilst this influence may often occur in an undocumented way, such as by personal recommendations by researchers or an unacknowledged drawing upon scholarly sources by policy makers, it is sometimes formally acknowledged through a citation from a grey literature source. Grey literature is a broad term for informally published documents that have not been subjected to scholarly peer review. This includes government policies and meeting transcripts, and business and consultancy reports and white papers, as well as a wide range of other document types. Depending on the document in question, the importance of a citation may vary from a suggestion of interest in the cited work (e.g., in a listing of recently published reports circulated in a regional business newsletter) to reasonably conclusive evidence of substantial impact on key public policy (e.g., a citation in government document underpinning a new law).

Since grey literature citations are rare and vary substantially in importance, their main use to support research evaluation is individually to support specific impact claims. For example, in the impact case studies from the UK REFs, they were used to support narratives that made specific claims of impact for bodies of research (Kousha et al., 2021). The grey literature indexing service Overton.com seems particularly suited to this type of application because it helps organisations to find grey literature that their work has been cited in (Szomszor & Adie, 2022). Nevertheless, in the realm of policy making, citations in Overton tend to originate in scientific advice documents rather than policy documents (Pinheiro et al., 2021), weakening, but not invalidating, the evidence for the influence of grey literature citations from Overton on policy.

There are several technical problems with finding grey literature citations. First, the definition of grey literature is loose, and so it is not simple to distinguish between grey literature and other documents. Second, it could be published anywhere on the web (or even offline), and the few services that attempt to index it, such as Overton and Altmetric.com, must restrict themselves to grey literature found in a limited number of websites because they cannot crawl the whole web. This is different from the situation for scholarly databases of journal articles like Scopus and the Web of Science because in these have large markets of regular users, allowing the indexing companies to devote substantial resources to creating adequately comprehensive databases. Journal article databases also have the advantage that publishers benefit from indexing and so are largely prepared to support the process by providing data in a standard format (articles were typed in or scanned in the past, however). Because of these limitations, counting grey literature citations may be less useful than identifying individual important citations and using them to justify a specific impact claim (for grey literature citation count correlations with research quality, see also: Thelwall et al., 2023).

## 11.4.1 Clinical practice guidelines

Clinical practice guidelines are a type of grey literature is worth singling out for its high value and systematic availability. Some countries produce evidence-based national guidelines for health professionals, and these provide strong evidence of the translation of research into practice, or the societal benefits of research. For example, if a paper describing a method of controlling diabetes was cited in the UK's guideline for diabetes treatment, then the authors of the paper could reasonably boast that their work had improved the health of people with diabetes in the UK. This would not be true if the paper had been cited as background information, however (Traylor & Herrmann-Lingen, 2023).

The UK systematically creates health guidelines for major health conditions. They are overseen by the National Institute for Health and Care Excellence (NICE), who put together a team of experts on a topic to review the evidence and to make a series of recommendations that are published as a guideline with associated citations to the evidence used. The guidelines can be found on the NICE website (www.nice.org.uk). Other countries following similar procedures include the Philippines (Clinical Practice Guidelines: e.g., Ona et al., 2021), The Netherlands, and Germany.

Although the value of clinical guideline citations as impact indicators seems clear, their disadvantage is partial coverage: health research may be valuable but not cited in any guidelines for various reasons including not being in scope for any. Moreover, there may be a national bias in the selection of papers to cite from a guideline and basic research papers tend not to be cited (Grant et al., 2000; Lewison & Sullivan, 2008) which undermines their general value. It is possible that guidelines from medical societies might cite a higher share of basic research, however, since this was found by an analysis of European Society for Medical Oncology (ESMO) guidelines (Pallari et al., 2018). Despite these biases, clinical guideline citations correlate positively with scholarly citations from journal articles (Thelwall & Maflahi, 2016), giving supporting evidence of their value as an impact indicator.

Overall, it seems that clinical practice guideline citations have potential to help assess the health impact of clinical research, especially since Overton's clinical guideline coverage and the emergence of at least one specialist database (ci.minso.se) for this task has lowered the technical threshold for data access (Eriksson et al., 2020).

## 11.5 The Altmetric.com donut

Perhaps the best known altmetric indicator is the colourful Altmetric.com donut that is displayed on the Altmetric.com website, in Dimensions.ai article pages, and in the article pages of subscribing publishers. This reports an overall "attention score" for each article as well as the main sources of altmetric citations. The score is a weighted sum of most of the different sources of citations collected by Altmetric.com (Liu & Adie, 2013). The weights in the sum are not theoretically informed or evidence based in any transparent way, so the attention score has little credibility as an impact indicator. Nevertheless, its job is to give a quick impression to the potential readers of an article about how much general interest the article has attracted, and it may perform this role well. Despite the limitations of the Altmetric.com attention score, it positively correlates weakly to moderately with the quality of scholarly publications (Thelwall et al., 2023).

## 11.6 Generic limitations of altmetrics

Altmetrics have all the field delimitation issues of citations and most also have substantial coverage and gaming limitations. The coverage limitation is that all altmetrics cover only a fraction, often a tiny fraction, of the phenomenon that they are used as an indicator of. For example, if using Tweet counts as a public interest indicator, then the coverage of public interest is extremely partial because most people do not use Twitter/X and almost certainly, most Twitter/X users post about a tiny proportion of the research that they have read (if any) or heard about. The partial coverage comes with biases due to the people left out. For example, people that a Twitter/X public interest altmetric would be biased against includes Chinese (it is banned in China), Russian (it is less popular in Russia) and older people (younger people currently use it more). A similar case for partial and biased coverage can be made for all altmetrics. This is one of the reasons why correlation tests are useful: they can give evidence that the biases do not render an altmetric meaningless.

The gaming limitation is that most altmetrics can be easily manipulated and this can be expected if altmetrics are used in formal evaluations. This can be deliberate, in the form of researchers paying bot farms to inflate their social media altmetrics, or accidental in the form of self-publicity or legitimate educational uses. As an example of the latter, a library instructor might set their students a Mendeley exercise involving bookmarking their latest article.

Despite these two limitations, altmetrics can be used for self-evaluation if they are interpreted cautiously. They can also be used in formal evaluations, at least in theory, if those evaluated are not told in advance of their use. Extreme caution of interpretation would be needed for this, however. This makes them typically less reliable than citations and rules them out from research evaluation exercises where those evaluated are told the mechanism in advance.

## 11.7 Summary

Although the main practical value of altmetrics is probably as early attention indicators for new articles in digital libraries, they may also have some value for research evaluation is as indicators of non-scholarly impact. In this context, their utility is difficult to assess because they are typically indicators of phenomena that are hard to define and have no practical way of measuring. In addition, all altmetrics have partial and biased coverage as indicators and most are easily gamed. The main validity problems include the following.

- Low coverage of the type of impact

- Face validity (e.g., tweets about research not reflecting public interest)
- Irrelevant mentions, spam, and gaming

Despite the above issues, this chapter has reviewed some evidence that altmetrics are non-random and even indicators of either scholarly impact or overall value. In contrast, the next chapter covers direct evidence of the value of altmetrics as research quality indicators.

# 12 Altmetrics as research quality indicators

Following the mainly theoretical discussion in the previous chapter, including evidence about how altmetrics correlate with citation counts, this chapter introduces data concerning the level of validity of various altmetrics as research quality indicators.[5] This support is needed for them to be interpreted effectively (Haustein et al., 2014c; Sud & Thelwall, 2014) due to the potential for many altmetrics to be gamed or infiltrated by irrelevant data.

As mentioned in the previous chapter, some altmetrics have been hypothesised to reflect different dimensions of attention or impact, and especially societal impact (Priem et al., 2011; Kousha, 2019). Most also have the advantage of appearing before citation counts, giving earlier evidence of interest or impact. There is substantial evidence that one altmetric, Mendeley reader counts, is a scholarly impact indicator and a partial educational impact indicator for journal articles primarily because Mendeley reader counts correlate moderately or strongly with citation counts for articles in most academic fields (Thelwall & Sud, 2016) and can be used as early scholarly impact indicators (Zahedi et al., 2017). Nevertheless, the value and best interpretations of all other altmetrics are uncertain. Tweeter/X counts, for example, although having moderate correlations with citation counts in some fields (Costas et al., 2015; Haustein et al., 2014a), seem to reflect academic interest and author/publisher dissemination activities in many fields rather than the initially hypothesised public interest (e.g., Lemke et al., 2022), despite most Twitter/X users being non-academics. Biomedical research might be an exception because this research is widely tweeted by the public (Mohammadi et al., 2018; see also: Haustein et al., 2014b).

The main reason for the ongoing uncertainty about how to interpret altmetrics is a lack of relevant data. Although there are many ways to partially evaluate altmetrics (Sud & Thelwall, 2014), there is no large-scale systematic evidence of the attention given to, or societal impact of, academic research. Thus, there is no direct way to check which altmetrics can reasonably be claimed to be indicators of these, or in which fields. Given this absence, the most common approach has been to correlate altmetric scores with citation counts, as an indicator of scholarly impact, as discussed in the previous chapter, on the basis that positive correlations would at least indicate that altmetric scores are non-random and scholarly-related to some extent. This is almost a paradox since the value of most altmetrics would be in being different from citation counts, but an overlap could nevertheless be expected for any scholarly-related indicator (Thelwall, 2016a). Other methods previously used have included content analyses of individual sources (e.g., tweets: Holmberg & Thelwall, 2014), and predicting future citation counts from early altmetric scores (Akella et al., 2021; Thelwall & Nevill, 2018).

Since citation counts are not direct measures of scholarly impact, a better way to evaluate altmetrics would be to correlate them against peer review quality scores for journal articles. This is more direct and may reveal altmetrics that reflect dimensions of quality not well captured by citations. This is plausible since significance (i.e., impact, whether scholarly, societal or other) is one of the three core components of quality, with the other two being rigour and originality (Langfeldt et al., 2020). Although there have been many departmental level comparisons (e.g., Bornmann & Haunschild, 2018; Bornmann et al, 2019) only one (non peer reviewed) publication has previously compared altmetrics with quality scores at the article level, in addition to the (peer reviewed) data presented in this chapter. The prior analysis correlated a range of altmetrics, including Mendeley reader counts and tweet counts from Altmetric.com, with REF expert peer review quality scores for 19,580 journal articles published in 2008 from 36 field-based UoAs. It found only relatively low correlations with quality scores, with the highest in Clinical Medicine (Spearman's rho=0.441) and Biological Sciences (rho=0.363) (HEFCE, 2015).

---

[5] This chapter is based on Thelwall et al. (2023e).

The current chapter updates the REF2014 technical report (HEFCE, 2015) with more, and newer REF2021 data on the basis that altmetrics have matured over time and Altmetric.com data may be more comprehensive after 2011. Data maturation is likely because the only year previously analysed, 2008, precedes Altmetric.com's foundation in 2011 and immediately follows Mendeley's creation in 2007 (Henning & Reichelt, 2008). Thus, remainder of this chapter reports an analysis of the relationship between altmetric scores, and research quality as judged by the UK REF2021 expert assessors. Methods details, sample sizes, and limitations are discussed in the originating article (Thelwall et al. 2023e).

## 12.1 Mendeley readers: Scholarly impact?

Both Scopus citations (not an altmetric but included for comparison purposes) and Mendeley readers have similar levels of correlation with REF2021 provisional quality scores in most UoAs, but the Scopus correlations are higher in all except two (17, 34) (Figure 12.1). Mendeley readers seem to be particularly weak in the humanities. This might be because Mendeley is a reference manager and humanities reference styles are often based on discussions in footnotes rather than standard format references. Thus, Mendeley is less useful to such scholars and its records may be sparser (Thelwall, 2019). The relatively low correlations for Mendeley in Mathematical Sciences and Computer Science and Informatics are presumably due to the LaTeX document formatting language commonly used in these areas (also in parts of Physics), for which Mendeley would be less use. The results confirm that citations and Mendeley readers have the most information value in medicine, health, physical sciences, in comparison to having moderate value in mathematics, engineering, and social sciences, and little in the arts and humanities.

As an aside, it is surprising that raw citation counts are slightly better indicators of research quality than NLCS in most UoAs (exceptions: UoAs 10, 16, 18, 21, 25, 28, 31, 32, 33). This is unexpected because field normalised indicators are designed to be fairer than raw citation counts by considering the publication field, so an article does not have an advantage for being published in a high citation area. In this case the correlations are calculated within field-based UoAs, so field normalisation should make little difference. Nevertheless, articles submitted to UoAs by their UK authors can be interdisciplinary or submitted to out-of-field UoA (e.g., because the author is a statistician in a medical department), which the NLCS normalisation process should help with. Mostly lower correlations for NLCS suggest that the field normalisation process is flawed, presumably because Scopus categorises articles by journal, but article-level classifications more closely align with underlying topics (Klavans & Boyack, 2017).

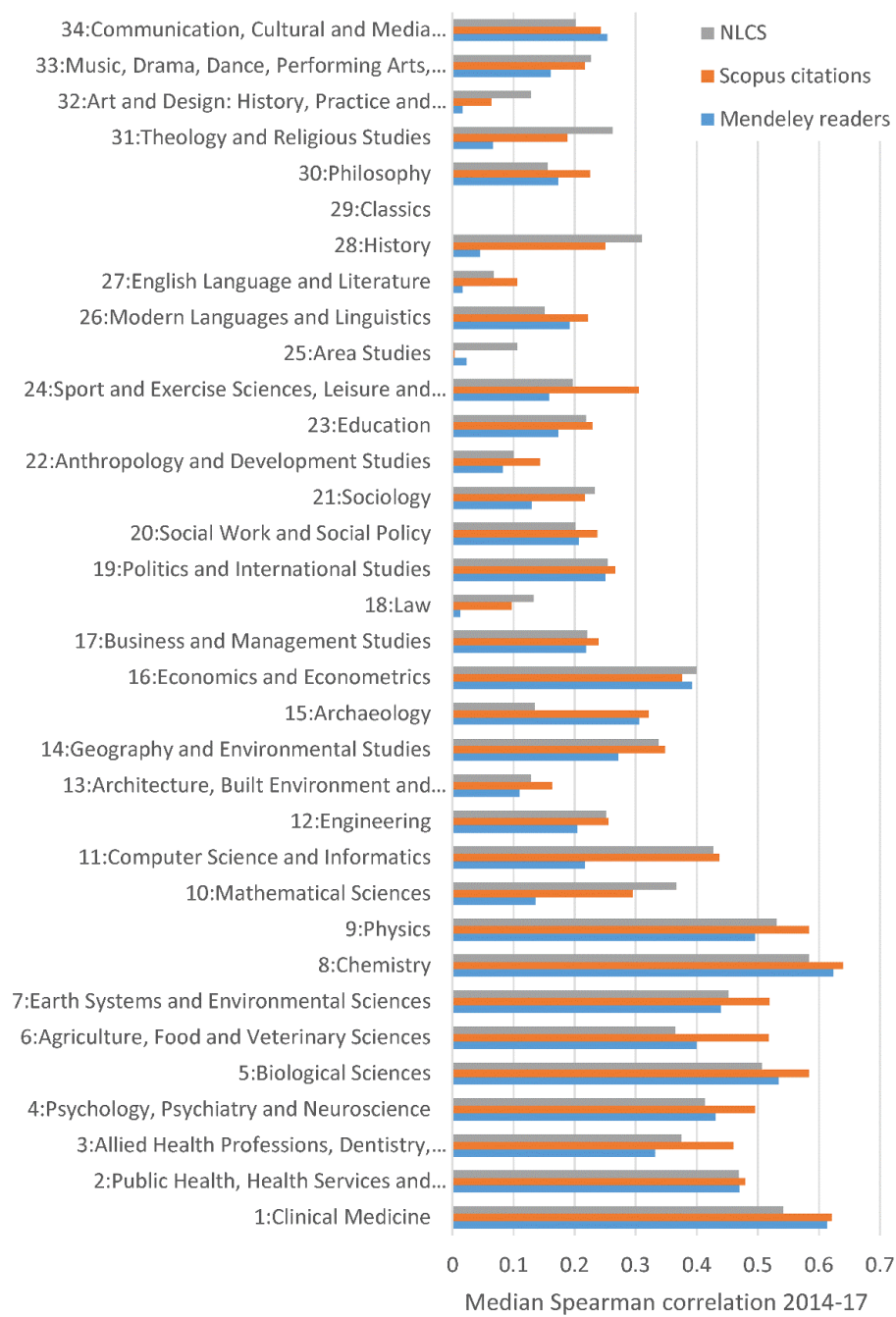

*Figure 12.1 Scopus citations (count and NLCS) and Mendeley readers (from Mendeley API) for 2014-17 articles: Spearman correlations with provisional REF2021 scores, calculated separately for each UoA and year, with the median across years reported. UoA 29 results have been removed for single figure sample sizes (Thelwall et al., 2023).*

The Altmetric.com data for Mendeley gives similar correlations to data from the Mendeley API (Figure 12.2), even though the latter should be comprehensive, and the former isn't. Altmetric.com claims to have counted readers from CiteULike until December 2014 (Altmetric, 2022b), but its CiteULike data became sparse for 2020 publications (data not used), so it may have ceased collecting new CiteULike data at the end of 2019. Nevertheless, this partial coverage and CiteULike's use by fewer people are the likely causes of lower correlations with REF2021 scores in all UoAs except Area Studies. Combining the CiteULike with the Mendeley counts to give Total Readers does not tend to improve on the Mendeley reader count correlation, so Mendeley readers do not need to be supplement by CiteULike data.

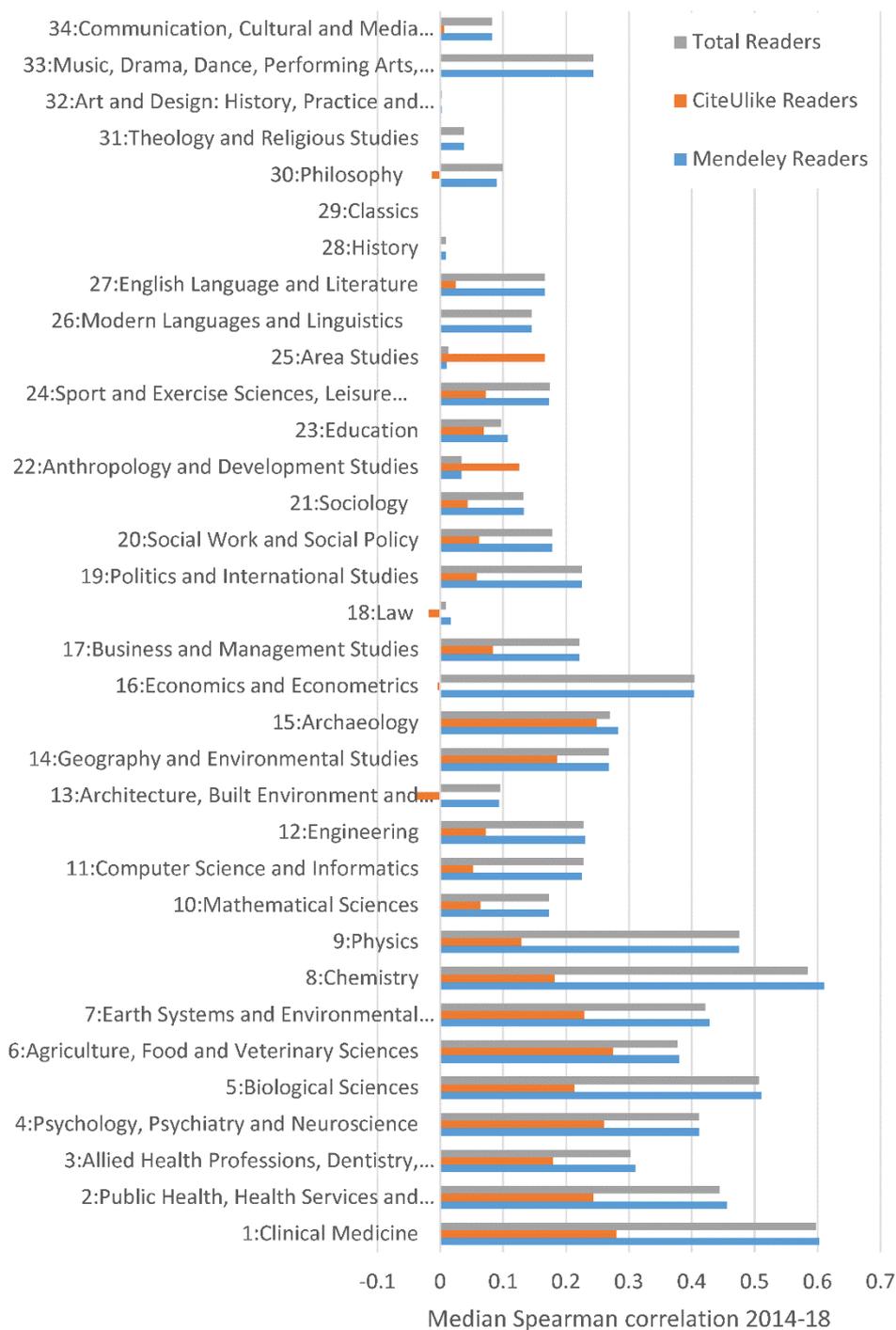

*Figure 12.2 Altmetric Scores and online reference manager readers from Altmetric.com 2014-18 articles: Spearman correlations with provisional REF2021 scores, calculated separately for each UoA and year, with the median across years reported. UoA 29 results have been removed for single figure sample sizes (Thelwall et al., 2023).*

A previous study claimed that Mendeley readers were as useful in the arts and humanities as elsewhere on the basis of their correlations with citation counts (Thelwall, 2019). The above results suggest that this is false because Mendeley is of little use in the arts and humanities as a quality indicator. It is even substantially less useful than citations, which are themselves very weak research quality indicators.

## 12.2 News, attention and information altmetrics?

Of the news/attention related sources, Tweeters (a count of Twitter/X users tweeting an article URL, although not a complete set: Altmetric, 2022a) seems to be the best indicator of research quality (Figure 12.3). Nevertheless, Blog and news citations (both from curated lists of sources: Altmetric, 2022a) also have moderate strength as research quality indicators in many UoAs. Facebook Wall links (from a curated list of walls: Altmetric, 2022a) are the weakest quality indicator, presumably due to smaller numbers of academically-relevant walls curated. Twitter/X is weaker than Altmetric's Mendeley readers as a research quality indicator in over three quarters of UoAs. The exceptions are mostly in the social sciences, arts, and humanities: UoAs 6, 13, 14, 18, 22, 25, 28, 30, 34.

The REF2021 Twitter/X correlations reported here are much stronger than the previously reported corresponding REF2014 correlations (Tables A54 of: HEFCE, 2015). For 2008, the highest Twitter/X correlation was 0.23 for Art and Design: History, Practice and Theory, the second highest was 0.17 for Public Health, Health Services and Primary Care, and the remaining correlations were below 0.15, with an average of 0.06. This is only a third of the average correlation above (0.18). This is most likely due to the maturation of the Altmetric.com data between the two studies.

The relatively high correlations for health-related fields may reflect widespread public interest in potentially impactful medical research (Mohammadi et al., 2018). This increases the amount of altmetric data but also suggests that the public tends to be interested in higher quality research to some extent. This is despite public interest in health research being very topic driven, for example with particular concern for cancer and especially breast cancer (Lewison et al., 2008).

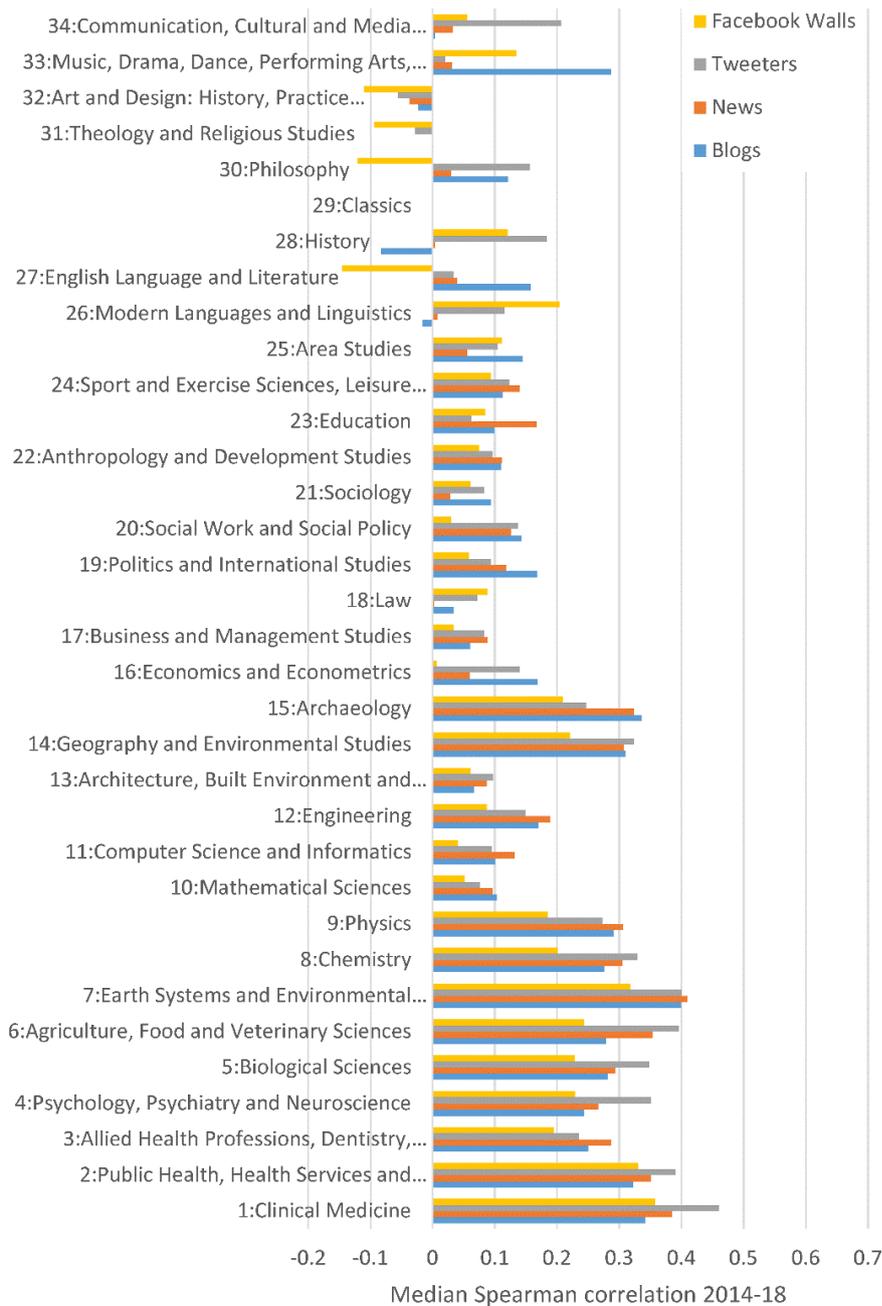

*Figure 12.3 Social network and news sites for 2014-18 articles: Spearman correlations with provisional REF2021 scores, calculated separately for each UoA and year, with the median across years reported. UoA 29 results have been removed for single figure sample sizes (Thelwall et al., 2023).*

Reddit mentions, Wikipedia citations and research highlight reviews ("Recommendations of individual research outputs from Faculty Opinions": Altmetric, 2022a) are all weak indicators of research quality in all fields, presumably for their scarcity. Nevertheless, Wikipedia citations have a moderate correlation with research quality in Archaeology and perform well compared to Mendeley Readers and Tweeters in some arts and humanities subjects (Figure 12.4).

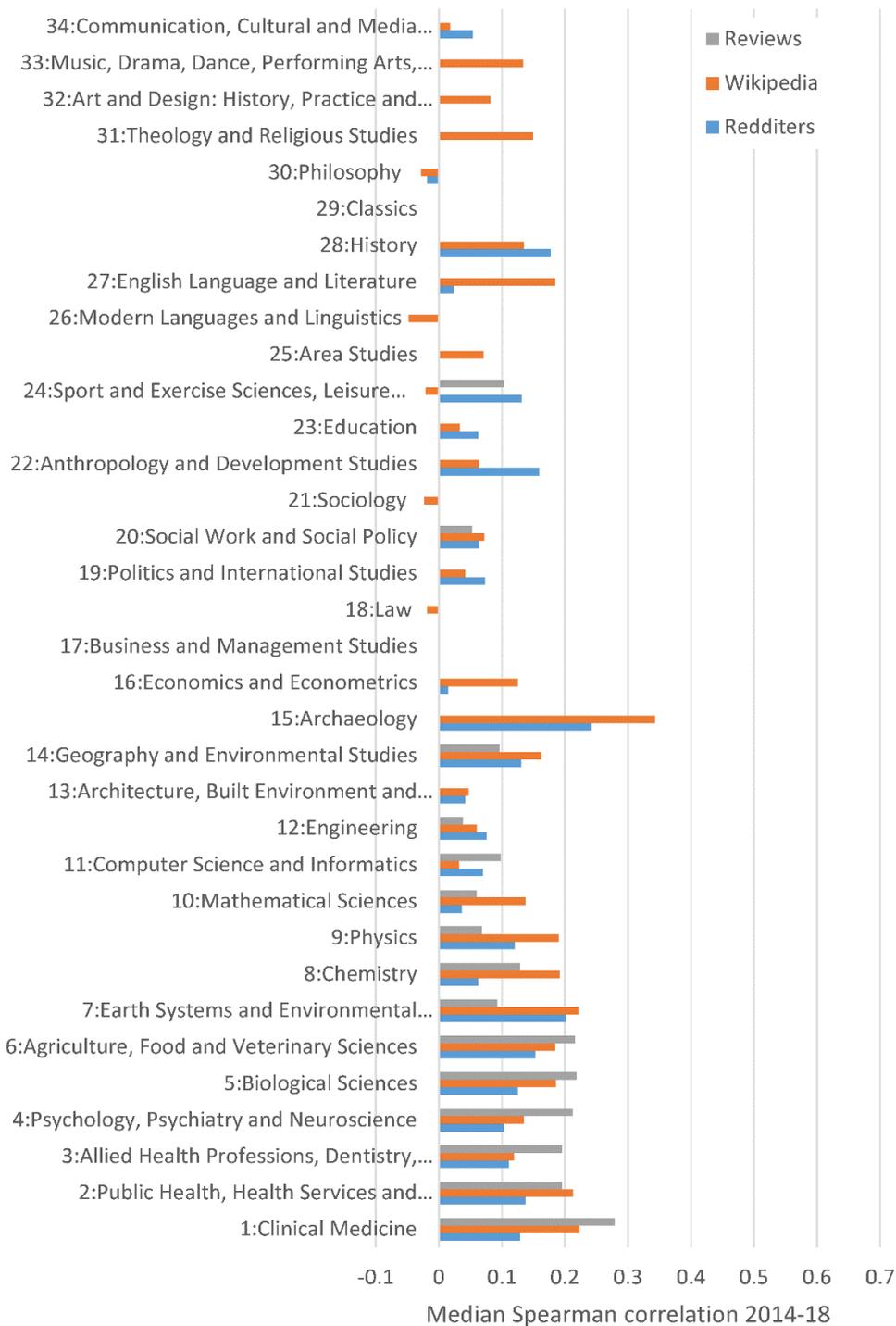

*Figure 12.4 Wikipedia, Reddit and facultyopinions.com research highlight reviews for 2014-18 articles: Spearman correlations with provisional REF2021 scores, calculated separately for each UoA and year, with the median across years reported. UoA 29 results have been removed for single figure sample sizes (Thelwall et al., 2023).*

## 12.3 Limitations

The graphs in this chapter are limited REF2021-related factors that affect many of the evidence chapters in this book. In particular, the restriction to the UK and the REF2021 articles analysed being self-selected by academics to be their best work from 2014-20 are samlg issues. The relatively low proportions of weaker research in the sets used for correlation probably reduces the strength of the correlations because correlations tend to be higher when a wider range of values are included, other factors being equal (Carretta & Ree, 2022).

In particular, there are few low quality 1* articles and the absence of a substantial proportion of low quality articles that may well score of 0 on all indicators would reduce all correlations. Another limitation is the use of the REF concept of research quality. Whilst it incorporates the main three quality dimensions, other dimensions or interpretations are also valid, the evaluations are likely to be imperfect because not all articles will have an assessor expert enough to reasonably assess them (Sayer, 2014).

There are also some limitations specific to altmetrics. Since the UK is a heavy user of social media, including Mendeley, Twitter/X and Facebook, it is likely that similar correlations would be lower for most other countries due to lower overall altmetric scores (and hence the evidence being less fine-grained). The value of altmetrics will also change over time as the demographics of their users shift. For example, the desktop version of Mendeley started to be phased out in September 2022 (Shlyuger, 2022), which may have alienated some users.

## 12.4 Summary

The above results suggest that the Altmetric.com altmetrics that are most useful as article-level indicators of research quality are, in descending order, Mendeley readers, Tweeters, Facebook Wall citations, News citations, Blog citations, Wikipedia citations, Redditers and Research Highlight Review citations. Of these, Mendeley reader counts are close to Scopus citation counts in power as research quality indicators (but worse in the arts and humanities and some social sciences), and Tweeter/X counts are clearly the best of the social web indicators. The last three (Wikipedia, Reddit and Research Highlights) only have minor value. This supports the continued use of altmetrics as attention indicators by publishers even though the evidence for some is weak. Doubt had been previously cast on Twitter/X due to its use for publicity and spam, but the UK REF evidence suggests that these uses have either declined, been filtered out by Altmetric.com, or naturally align with the quality of articles (e.g., if scholars publicise their best work more or journals publishing better work tend to use more online publicity).

In terms of field differences, altmetrics have the most value in health fields, and the physical sciences and the least value in the arts and humanities. It is particularly disappointing that altmetrics have little value in the arts and humanities, since it plausible, in theory, that higher quality arts and humanities research would be more newsworthy or attract more public attention.

Despite the above, none of the correlations are strong enough to claim that altmetrics "measure" research quality in any way. Instead, they are weak or moderate strength *indicators* of research quality, meaning that a high score on them weakly or moderately associates with higher quality, but is far from guaranteeing it, especially given the possibility of manipulation. In particular, because of the potential for gaming, altmetrics should not be used for important evaluations when stakeholders are aware of the methods in advance (Roemer & Borchardt, 2015; Wouters & Costas, 2012).

The data also suggest that none of the altmetrics are as effective as citation counts as research quality indicators, despite Mendeley being a close second. It is reasonable to continue to use Mendeley readers as a substitute for citations as an early impact or quality indicator when the citation window is too narrow for citations, however. Unfortunately, the data in this chapter does not directly address one of the main claims for altmetrics: that they might reflect a dimension of societal impact that is not well captured by citations (Priem et al, 2011). This hypothesis remains unproven.

# 13 Can AI help to estimate journal article research quality?

This chapter[6] reports the sole large-scale multidisciplinary attempt to use machine learning to estimate the quality of journal articles, although many previous studies (reviewed in: Thelwall et al., 2023a) have attempted to predict long term citation counts. Automated research quality estimation has been suggested for, but not yet implemented in, labour intensive national periodic research evaluations. These are simultaneous for all fields (e.g., Australia, New Zealand, UK: Buckle & Creedy, 2019; Hinze et al. 2019; Wilsdon et al., 2015), or rolling evaluations for departments, fields, or funding initiatives (e.g., The Netherlands' Standard Evaluation Protocol: Prins et al., 2016). Peer/expert review, although imperfect, seems to be the most desirable system because reliance on bibliometric indicators can disadvantage some research groups, such as those that focus on applications rather than theory or methods development, with bibliometrics only recommended for a supporting role (CoARA, 2022; Hicks et al., 2015; Wilsdon et al., 2015). Nevertheless, this requires a substantial time investment from experts with the skill to assess academic research quality and there is a risk of human bias. In response, some systems inform peer review with bibliometric indicators (UK: Wilsdon et al., 2015) or automatically score outputs that meet certain criteria, reserving human reviewing for the remainder (as Italy used to: Franceschini & Maisano, 2017). A third approach would be to use machine learning to estimate the score of some or all outputs in a periodic research assessment. The accuracy and feasibility of this third approach is evaluated here with post publication peer review quality scores for a large set of UK journal articles.

A previous study (Thelwall, 2022) reported experiments with machine learning to predict the citation rate of an article's publishing journal as a proxy article quality measurement. Accuracy varied substantially between the 326 Scopus narrow fields that it was applied to. Some previous studies have attempted to estimate the quality of computational linguistics conference submissions (e.g., Kang et al., 2018; Li et al., 2020). This is a very different task to post publication peer review for journal articles across fields because of the narrow topic and lack of post-publication citation counts.

This chapter assesses whether it is reasonable to use machine learning to estimate any UK REF output quality scores for journal articles. The purpose of the analysis is to test whether computers can use any available inputs to guess research quality scores accurately enough to be useful in any contexts. Two approaches are tested: (a) human scoring for a fraction of the outputs, then machine learning predictions for the remainder; (b) human scoring for a fraction of the outputs, then machine learning predictions for a subset of the rest where the predictions have a high probability of being correct; and human scoring for the remaining articles. The approaches are assessed with expert peer review quality scores for most of the journal articles submitted to REF2021.

The research questions are as follows, with the final research question introduced to test if the results change with a different standard classification schema. The research questions mention "accuracy", but the experiments instead measure agreement with expert scores from REF2021, which are therefore treated as a gold standard even though they are also estimates. Whilst the results focus on article-level accuracy to start with, the most important type of accuracy for departmental evaluations is institutional-level (Traag & Waltman, 2019), as reported towards the end.

---

[6] The chapter largely replicates Thelwall et al. (2023a), including most figures and tables. It includes more methods details than other chapters because these are at the heart of machine learning.

- How accurately can machine learning estimate article quality from article metadata and bibliometric information in each scientific field?
- Can higher accuracy be achieved on subsets of articles using machine learning prediction probabilities?
- How accurate are machine learning article quality estimates when aggregated over institutions?
- Is the machine learning accuracy similar for articles organised into Scopus broad fields?

## 13.1 Methods

The research design was to assess a range of machine learning algorithms in a traditional training/testing validation format: training each algorithm on a subset of the data and evaluating it on the remaining data.

### 13.1.1 Data: Articles and scores

Two data sources were used: Scopus and REF2021. First, records were downloaded for all Scopus-indexed journal articles published 2014-2020 in January-February 2021 using the Scopus Application Programming Interface (API). This matches the date when the human REF2021 assessments were originally scheduled to begin, so is from the time frame when a machine learning stage could be activated. Reviews and other non-article records in Scopus were excluded for consistency. The second source was a set of 148,977 provisional article quality scores assigned by the expert REF sub-panel members to the articles in 34 UoAs, excluding all data from the University of Wolverhampton. This was confidential data that could not be shared and had to be deleted before 10 May 2022. The distribution of the scores for these articles is online (Figure 3.2.2 of Thelwall et al., 2022). Many articles had been submitted by multiple authors from different institutions and sometimes to different UoAs. These duplicates were eliminated, and the median score retained, or a random median when there were two (for more details, see: Thelwall et al., 2023a).

The REF data included article DOIs (used for matching with Scopus, and validated by the REF team), evaluating UoA (one of 34), and provisional score (0, 1*,2*,3*, or 4*). The REF scores were merged into three groups for analysis: 1 (0, 1* and 2*); 2 (3*) and 3 (4*). The grouping was necessary because there were few articles with scores of 0 or 1*, which gives a class imbalance that can be problematic for machine learning. This is a reasonable adjustment because 0, 1* and 2* all have the same level of REF funding (zero), so they are financially equivalent.

The REF outputs were matched with journal articles in Scopus with a registered publication date from 2014 to 2020 (

Table 13.1). Matching was primarily achieved through DOIs. Articles without a matching DOI in Scopus were checked against Scopus by title, after removing non-alphabetic characters (including spaces) and converting to lowercase. Title matches were manually checked for publication year, journal name, and author affiliations. When there was a disagreement between the REF registered publication year and the Scopus publication year, the Scopus publication year was used. The few articles scoring 0 appeared to be mainly anomalies, seeming to have been judged unsuitable for review due to lack of evidence of substantial author contributions or being an inappropriate type of output. These were excluded because no scope-related information was available to predict score 0s from.

Finally, a sample of articles without an abstract containing at least 500 characters was also examined because these tended to be non-standard outputs that would be difficult to automatically score. The most accurate predictions were found for the years 2014-18, with at least two full years of citation data, so these are reported for the main analysis as the highest accuracy subset.

*Table 13.1 Descriptive statistics for creation of the experimental dataset (Thelwall et al., 2022).*

| Set of articles | Journal articles |
|---|---|
| REF2021 journal articles supplied. | 148,977 |
| With DOI. | 147,164 (98.8%) |
| With DOI and matching Scopus 2014-20 by DOI. | 133,218 (89.4%) |
| Not matching Scopus by DOI but matching with Scopus 2014-20 by title. | 997 (0.7%) |
| Not matched in Scopus and excluded from analysis. | 14,762 (9.9%) |
| All REF2021 journal articles matched in Scopus 2014-20. | 134,215 (90.1%) |
| All REF2021 journal articles matched in Scopus 2014-20 except score 0. | 134,031 (90.0%) |
| All non-duplicate REF2021 journal articles matched in Scopus 2014-20 except score 0. | 122,331 [90.0% effective] |
| All non-duplicate REF2021 journal articles matched in Scopus **2014-18** except score 0. These are the most accurate prediction years. | 87,739 |
| All non-duplicate REF2021 journal articles matched in Scopus **2014-18** except score 0 and except articles with less than 500 character cleaned abstracts. | 84,966 |

The 2014-18 articles were mainly from Main Panel A (33,256) overseeing UoAs 1-6, Main Panel B (30,354) overseeing UoAs 7-12, and Main Panel C (26,013) overseeing UoAs 13-24, with a much smaller number from Main Panel D (4,209) overseeing UoAs 25-34. The number per UoA 2014-18 varied by several orders of magnitude, from 56 (Classics) to 12,511 (Engineering), as shown below in a results table. The number of articles affects the accuracy of machine learning and there were too few in Classics to build machine learning models.

### 13.1.2 Machine learning inputs

Textual and bibliometric data were used as inputs for all the machine learning algorithms. This included all inputs shown in previous research to be useful for predicting citations counts, as far as possible, as well as some new inputs that seemed likely to be useful. Inputs used in previous research were also adapted to use bibliometric best practice, as described below. The starting point was the set of inputs used in a previous citation-based study (Thelwall, 2022) but this was extended. The non-text inputs were tested with ordinal regression before the machine learning to help select the final set.

The citation data for several inputs was based on the Normalised Log-transformed Citation Score (NLCS) or the Mean Normalised Log-transformed Citation Score (MNLCS) (for detailed explanations and justification, see: Thelwall, 2017). The NLCS of an article is field and year normalised by design, so a score of 1 for any article in any field and year always means that the article has had average log-transformed citation impact for its field and year. The following were calculated from the NLCS values and used them as machine learning inputs.

- Author MNLCS: The average NLCS for all articles 2014-20 in the Scopus dataset including the author (identified by Scopus ID).

- Journal MNLCS for a given year: The average NLCS for all articles in the Scopus dataset in the specified year from the journal. Averaging log-transformed citation counts instead of raw citation counts gives a better estimate of central tendency for a journal (e.g., Thelwall & Fairclough, 2015).

**Input set 1**: **bibliometrics**. The following nine indicators have been shown in previous studies to associate with citation counts, including readability (e.g., Didegah & Thelwall, 2013a), author affiliations (e.g., Fu & Aliferis, 2010; Li et al., 2019a; Qian et al., 2017; Zhu & Ban, 2018), and author career factors (e.g., Qian et al., 2017; Wen et al., 2020; Xu et al., 2019; Zhu & Ban, 2018). The first author was selected for some indicators because they are usually the most important (de Moya-Anegon et al., 2018; Mattsson et al., 2011), although corresponding and last authors are sometimes more important in some fields. Indicators based on the maximum author in a team were also used to catch important authors that might appear elsewhere in a list.

1. **Citation counts** (field and year normalised to allow parity between fields and years, log transformed to reduce skewing to support linear-based algorithms).
2. **Number of authors** (log transformed to reduce skewing). Articles with more authors tend to be more cited, so they are likely to also be more highly rated (Thelwall & Sud, 2016).
3. **Number of institutions** (log transformed to reduce skewing). Articles with more institutional affiliations tend to be more cited, so they are likely to also be more highly rated (Didegah & Thelwall, 2013a).
4. **Number of countries** (log transformed to reduce skewing). Articles with more country affiliations tend to be more cited, so they are likely to also be more highly rated (Wagner et al., 2019).
5. **Number of Scopus-indexed journal articles of the first author** during the REF period (log transformed to reduce skewing). More productive authors tend to be more cited (Abramo et al., 2014; Larivière & Costas, 2016), so this is a promising input.
6. **Average citation rate of Scopus-indexed journal articles by the first author** during the REF period (field and year normalised, log transformed: the MNLCS). Authors with a track record of highly cited articles seem likely to write higher quality articles. Note that the first author may not be the REF submitting author or from their institution because the goal is not to "reward" citations for the REF author but to predict the score of their article.
7. **Average citation rate of Scopus-indexed journal articles by any author** during the REF period (maximum) (field and year normalised, log transformed: the MNLCS). Again, authors with a track record of highly cited articles seem likely to write higher quality articles. The maximum bibliometric score in a team has been previously used in another context (Van den Besselaar & Leydesdorff, 2009).
8. **Number of pages of article, as reported by Scopus, or the UoA/Main Panel median if missing from Scopus.** Longer papers may have more content but short papers may be required by more prestigious journals.
9. **Abstract readability**. Abstract readability was calculated using the Flesch-Kincaid grade level score and has shown to have a weak association with citation rates (Didegah & Thelwall, 2013a).

**Input set 2**: **bibliometrics + journal impact.** Journal impact indicators are expected to be powerful in some fields, especially for newer articles (e.g., Levitt & Thelwall, 2011). The second input set adds a measure of journal impact to the first set. The journal MNLCS was used instead of Journal Impact Factors (JIFs) as an indicator of average journal impact because field normalised values align better with human journal rankings (Haddaway et al., 2016), probably due to comparability between disciplines. This is important because the 34 UoAs are relatively broad, all covering multiple Scopus narrow fields.

10. **Journal citation rate** (field normalised, log transformed [MNLCS], based on the current year for older years, based on 3 years for 1-2 years' old articles).

**Input set 3**: **bibliometrics + journal impact + text.** The final input set also includes text from article abstracts. Text mining may find words and phrases associated with good research (e.g., a simple formula has been identified for one psychology journal: Kitayama, 2017). Text mining for score prediction is likely to leverage hot topics in constituent fields (e.g., because popular topic keywords can associate with higher citation counts: Hu et al., 2020), as well as common methods (e.g., Fairclough & Thelwall, 2022; Thelwall & Nevill, 2021; Thelwall & Wilson, 2016), since these have been shown to associate with above average citation rates. Hot topics in some fields tend to be highly cited and probably have higher quality articles, as judged by peers. Whilst topics easily translate into obvious and common keywords, research quality has unknown and probably field dependent translation into research quality (e.g., "improved accuracy" [computing] vs. "surprising connection" [humanities]). Thus, text-based predictions of quality are likely to leverage topic-relevant keywords and perhaps methods as indirect indicators of quality rather than more subtle textual expressions of quality. Article abstracts were pre-processed with a large set of rules to remove publisher copyright messages, structured abstract headings, and other boilerplate texts (available: https://doi.org/10.6084/m9.figshare.22183441).

Journal names were also included on the basis that journals are key scientific gatekeepers and that a high average citation impact does not necessarily equate to publishing high quality articles. Testing with and without journal names suggested that their inclusion tended to slightly improve accuracy.

11-1000. **Title and abstract word unigrams, bigrams, and trigrams** within sentences (i.e., words and phrases of 2 or 3 words). Feature selection was used (chi squared) to identify the best features, always keeping all Input Set 2 features. **Journal names** are also included, for a total of 990 text inputs, selected from the full set as described below.

## 13.1.3 Machine learning methods

Machine learning stages were used that mirror those of a prior study predicting journal impact classes (Thelwall, 2022) with mostly the same settings. These represents a range of types of established regression and classification algorithms, including the generally best performing for tabular input data. As previously argued, predictions may leverage bibliometric data and text, the latter on the basis that the formula for good research may be identifiable from a text analysis of abstracts. In total, 32 machine learning methods were used, including classification, regression, and ordinal algorithms (

Table 13.2). Regression predictions are continuous and were converted to three class outputs by rounding to integers and rounding down (up) to the maximum (minimum) when out of scale. These include the methods of the prior study (Thelwall, 2022) and the Extreme Gradient Boosting Classifier (Klemiński et al., 2021). Deep learning was not used because there was too little data to exploit its power (Kousha & Thelwall, 2024). Accuracy was calculated after training on 10%, 25% or 50% of the data and evaluated on the remaining articles. These percentages represent a range of realistic options for the REF. Training and testing was repeated 10 times, reporting the average accuracy.

*Table 13.2 Machine learning methods chosen for regression and classification. Those marked with /o have an ordinal version. Ordinal versions of classifiers conduct two binary classifications (1\*-3\* vs. 4\* and 1\*-2\* vs 3\*-4\*) and then choose the trinary class by combining the probabilities from them (Thelwall et al., 2023).*

| Code | Method | Type |
|------|--------|------|
| bnb/o | Bernoulli Naive Bayes | Classifier |
| cnb/o | Complement Naive Bayes | Classifier |
| gbc/o | Gradient Boosting Classifier | Classifier |
| xgb/o | Extreme Gradient Boosting Classifier | Classifier |
| knn/o | k Nearest Neighbours | Classifier |
| lsvc/o | Linear Support Vector Classification | Classifier |
| log/o | Logistic Regression | Classifier |
| mnb/o | Multinomial Naive Bayes | Classifier |
| pac/o | Passive Aggressive Classifier | Classifier |
| per/o | Perceptron | Classifier |
| rfc/o | Random Forest Classifier | Classifier |
| rid/o | Ridge classifier | Classifier |
| sgd/o | Stochastic Gradient Descent | Classifier |
| elnr | Elastic-net regression | Regression |
| krr | Kernel Ridge Regression | Regression |
| lasr | Lasso Regression | Regression |
| lr | Linear Regression | Regression |
| ridr | Ridge Regression | Regression |
| sgdr | Stochastic Gradient Descent Regressor | Regression |

The most accurate classifiers were based on the Gradient Boosting Classifier, the Extreme Gradient Boosting Classifier, and the Random Forest Classifier. All are based on large numbers of simple decision trees, which make classification suggestions based on a series of decisions about the inputs. For example, Traag and Waltman (2019) proposed citation thresholds for identifying likely 4\* articles (top 10\* cited in a field). A decision tree could mimic this by finding a threshold for the NLCS input, above which articles would be classified as 4\*. It might then find a second, lower, threshold, below which articles would be classified as 1\*/2\*. A decision tree could also incorporate information from multiple inputs. For example, a previous VQR used dual thresholds for citations and journal impact factors and a decision tree could imitate this by classing an article as 4\* if it exceeded an NLCS citation threshold (decision 1) and a MNLCS journal impact threshold (decision 2). Extra rules for falling below a lower citation threshold (decision 3) and lower journal impact threshold (decision 4) might then classify an article as 1\*/2\*. The remaining articles might be classified by further decisions involving other inputs or combinations of inputs. The three algorithms (gbc, rfc, xgb) all make at least 100 of these simple decision trees and then combine them using different algorithms to produce a powerful inference engine.

## 13.2 Results

### 13.2.1 Primary machine learning prediction accuracy tests

The accuracy of each machine learning method was calculated for different year ranges, and separately by UoA and Main panel. The results are reported as accuracy above the baseline (accuracy-baseline)/(1-baseline), where the baseline is the proportion of articles with the most common score (usually 4* or 3*). Thus, the baseline is the accuracy of always predicting that articles fall within the most common class. For example, if 50% of articles are 4* then 50% would be the baseline and a 60% accurate system would have an accuracy above the baseline of (0.6-0.5)(1-.50)=0.2 or 20%. The results are reported only for the years 2014-18 combined, with the graphs for the other years available online, as are graphs with 10% or 25% training data, and graphs for Input Set 1 alone and for Input Sets 1 and 2 combined (Thelwall et al., 2023e). The overall level of accuracy for each individual year from 2014 to 2018 tended to be similar, with lower accuracy for 2019 and 2020 due to the weaker citation data. Combining 2014 to 2018 gave a similar level of accuracy to that of the individual years, and so it is informative to focus on this set. With the main exception of UoA 8 Chemistry, the accuracy of the machine learning methods was higher with 1000 inputs (input set 3) than with 9 or 10 (input sets 1 or 2), so only the results for the largest set are reported.

Six algorithms tended to have similar and the high levels of accuracy (rfc, gbc, xgb, or ordinal variants) so the results would be similar but slightly lower overall if only one of them had been used. Thus, the results slightly overestimate the practically achievable accuracy by cherry-picking the best algorithm.

Articles 2014-18 in most UoAs could be classified with above baseline accuracy with at least one of the tested machine learning methods, but there are substantial variations between UoAs (Figure 13.1). There is not a simple pattern in terms of the types of UoA that are easiest to classify. This is partly due to differences in sample sizes and probably also affected the variety of the fields within each UoA (e.g., Engineering is a relatively broad UoA compared to Archaeology). Seven UoAs had accuracy at least 0.3 above the baseline, and these are from the health and physical sciences as well as UoA 16: Economics and Econometrics. Despite this variety, the level of machine learning accuracy is very low for all Main Panel D (mainly arts and humanities) and for most of Main Panel C (mainly social sciences).

Although larger sample sizes help the training phase of machine learning (e.g., there is a Pearson correlation of 0.52 between training set size and accuracy above the baseline), the UoA with the most articles (12: Engineering) had only moderate accuracy, so the differences between UoAs are also partly due to differing underlying machine learning prediction difficulties between fields.

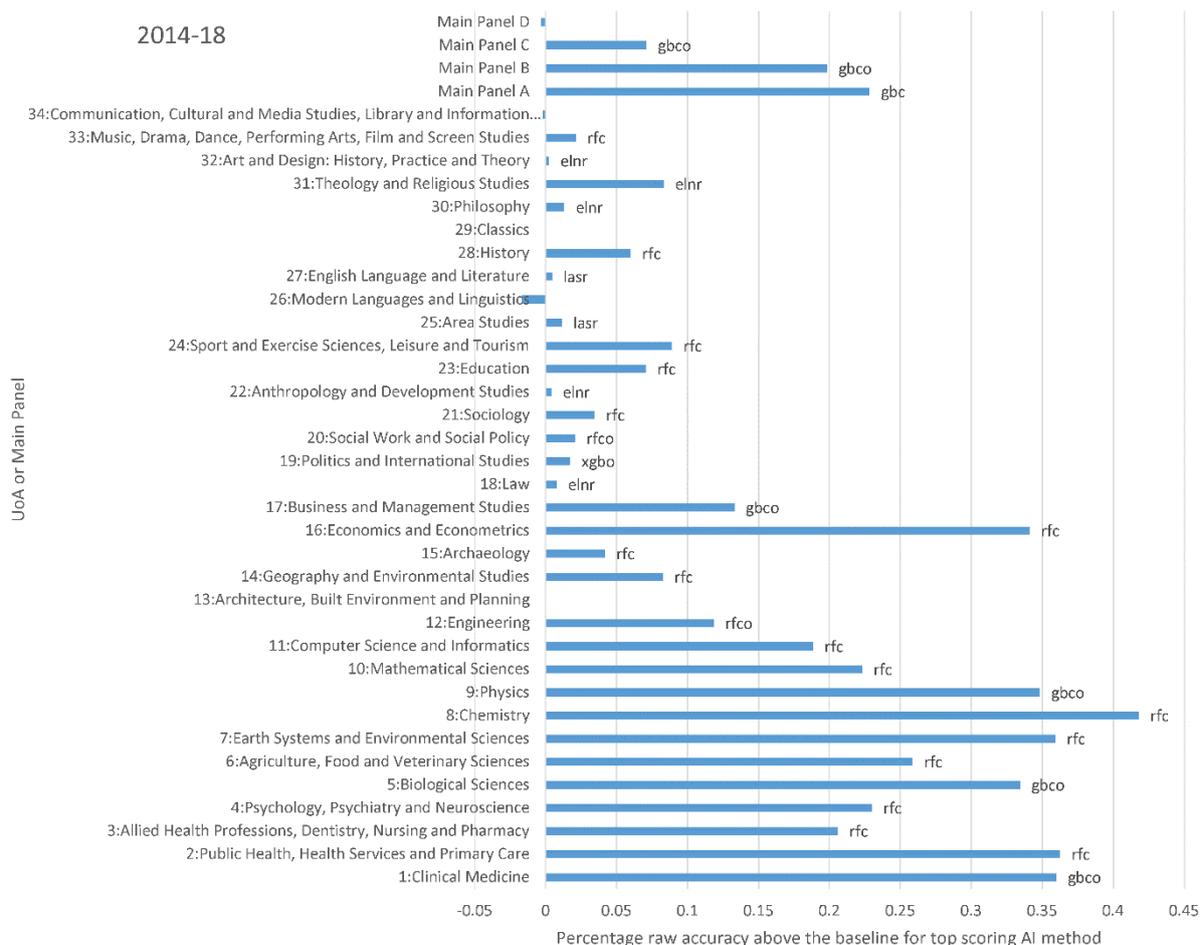

*Figure 13.1 The percentage accuracy above the baseline for the most accurate machine learning method, trained on **50%** of the 2014-18 Input Set 3: Bibliometrics, journal impact and text, after excluding articles with shorter than 500-character abstracts **and excluding duplicate articles within each UoA**. The accuracy evaluation was performed on the articles excluded from the training set. No models were built for Classics due to too few articles. Average across 10 iterations (Thelwall et al., 2023).*

The individual inputs were not tested for predictive power but the three input sets were combined and compared. In terms of accuracy on the three sets, the general rule for predictive power was: bibliometrics < bibliometrics + journal impact < bibliometrics + journal impact + text. The differences were mostly relatively small. The main exceptions were Chemistry (bibliometrics alone is best) and Physics (bibliometrics and bibliometrics + journal impact + text both give the best results) (for details see: Thelwall et al., 2023e).

The most accurate UoAs are not all the same as those with highest accuracy above the baseline because there were substantial differences in the baselines between UoAs (Figure 13.2). The predictions were up to 72% accurate (UoAs 8,16), with 12 UoAs having accuracy above 60%. The lowest raw accuracy was 46% (UoA 23). If accuracy is assessed in terms of article-level correlations, then the machine learning predictions always positively correlate with the human scores at rates varying between 0.0 (negligible) to 0.6 (strong) (Table 13.3). These correlations roughly match the prediction accuracies. The correlations are much higher when aggregated by institution, reaching 0.998 for total institutional scores in UoA 1 (Table 13.4).

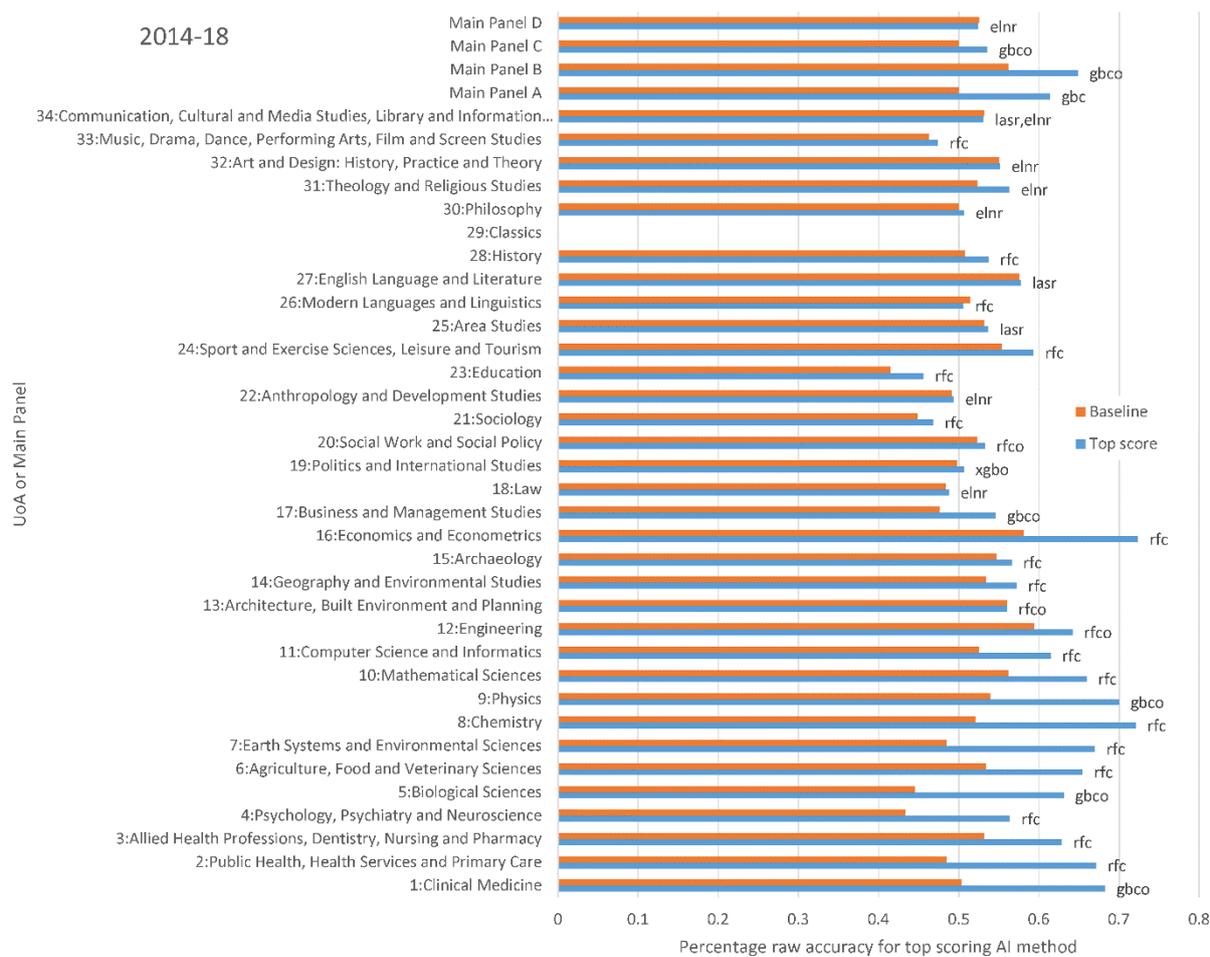

*Figure 13.2 As for the previous figure but showing raw accuracy (Thelwall et al., 2023).*

*Table 13.3 Article-level Pearson correlations between machine learning predictions with 50% used for training and actual scores for articles 2014-18, following Strategy 1 (averaged across 10 iterations). L95 and U95 are lower and upper bounds for a 95% confidence interval (Thelwall et al., 2023).*

| UoA | Articles 2014-18 | Predicted at 50% | Pearson correlation | L95 | U95 |
|---|---|---|---|---|---|
| 1:Clinical Medicine | 7274 | 3637 | 0.562 | 0.539 | 0.584 |
| 2:Public Health, Health Services & Primary Care | 2855 | 1427 | 0.507 | 0.467 | 0.545 |
| 3:Allied Health Professions, Dentistry, Nursing & Pharmacy | 6962 | 3481 | 0.406 | 0.378 | 0.433 |
| 4:Psychology, Psychiatry & Neuroscience | 5845 | 2922 | 0.474 | 0.445 | 0.502 |
| 5:Biological Sciences | 4728 | 2364 | 0.507 | 0.476 | 0.536 |
| 6:Agriculture, Food & Veterinary Sciences | 2212 | 1106 | 0.452 | 0.404 | 0.498 |
| 7:Earth Systems & Environmental Sciences | 2768 | 1384 | 0.491 | 0.450 | 0.530 |
| 8:Chemistry | 2314 | 1157 | 0.505 | 0.461 | 0.547 |
| 9:Physics | 3617 | 1808 | 0.472 | 0.435 | 0.507 |
| 10:Mathematical Sciences | 3159 | 1579 | 0.328 | 0.283 | 0.371 |
| 11:Computer Science & Informatics | 3292 | 1646 | 0.382 | 0.340 | 0.423 |
| 12:Engineering | 12511 | 6255 | 0.271 | 0.248 | 0.294 |
| 13:Architecture, Built Environment & Planning | 1697 | 848 | 0.125 | 0.058 | 0.191 |
| 14:Geography & Environmental Studies | 2316 | 1158 | 0.277 | 0.223 | 0.329 |
| 15:Archaeology | 371 | 185 | 0.283 | 0.145 | 0.411 |
| 16:Economics and Econometrics | 1083 | 541 | 0.511 | 0.446 | 0.571 |
| 17:Business & Management Studies | 7535 | 3767 | 0.353 | 0.325 | 0.381 |
| 18:Law | 1166 | 583 | 0.101 | 0.020 | 0.181 |
| 19:Politics & International Studies | 1595 | 797 | 0.181 | 0.113 | 0.247 |
| 20:Social Work & Social Policy | 2045 | 1022 | 0.259 | 0.201 | 0.315 |
| 21:Sociology | 949 | 474 | 0.180 | 0.091 | 0.266 |
| 22:Anthropology & Development Studies | 618 | 309 | 0.040 | -0.072 | 0.151 |
| 23:Education | 2081 | 1040 | 0.261 | 0.203 | 0.317 |
| 24:Sport & Exercise Sciences, Leisure & Tourism | 1846 | 923 | 0.265 | 0.204 | 0.324 |
| 25:Area Studies | 303 | 151 | 0.142 | -0.018 | 0.295 |
| 26:Modern Languages and Linguistics | 630 | 315 | 0.066 | -0.045 | 0.175 |
| 27:English Language and Literature | 424 | 212 | 0.064 | -0.071 | 0.197 |
| 28:History | 583 | 291 | 0.141 | 0.026 | 0.252 |
| 29:Classics | 56 | 0 | - | - | - |
| 30:Philosophy | 426 | 213 | 0.070 | -0.065 | 0.203 |
| 31:Theology & Religious Studies | 107 | 53 | 0.074 | -0.200 | 0.338 |
| 32:Art and Design: History, Practice and Theory | 665 | 332 | 0.028 | -0.080 | 0.135 |
| 33:Music, Drama, Dance, Performing Arts, Film & Screen Studies | 350 | 175 | 0.164 | 0.016 | 0.305 |
| 34:Communication, Cultural & Media Studies, Library & Information Management | 583 | 291 | 0.084 | -0.031 | 0.197 |

*Table 13.4 Institution-level Pearson correlations between machine learning predictions with 50% used for training and actual scores for articles 2014-18, following Strategy 1 (averaged across 10 iterations) and aggregated by institution for UoAs 1-11 and 16 (Thelwall et al., 2023).*

| UoA | Actual vs machine learning predicted average score | Actual vs machine learning predicted total score |
|---|---|---|
| 1:Clinical Medicine | 0.895 | 0.998 |
| 2:Public Health, Health Services and Primary Care | 0.906 | 0.995 |
| 3:Allied Health Professions, Dentistry, Nursing & Pharmacy | 0.747 | 0.982 |
| 4:Psychology, Psychiatry and Neuroscience | 0.844 | 0.995 |
| 5:Biological Sciences | 0.885 | 0.995 |
| 6:Agriculture, Food and Veterinary Sciences | 0.759 | 0.975 |
| 7:Earth Systems and Environmental Sciences | 0.840 | 0.986 |
| 8:Chemistry | 0.897 | 0.978 |
| 9:Physics | 0.855 | 0.989 |
| 10:Mathematical Sciences | 0.664 | 0.984 |
| 11:Computer Science and Informatics | 0.724 | 0.945 |
| 16:Economics and Econometrics | 0.862 | 0.974 |

Hyperparameter tuning systematically searches a range of input parameters for machine learning algorithms, looking for variations that improve their accuracy. Whilst this marginally increases accuracy on some UoAs it marginally reduces it on others, so has little difference overall (Figure 13.3). The tuning parameters for the different algorithms are in the Python code (https://doi.org/10.6084/m9.figshare.21723227). The same architectures were used for the tuned and untuned cases, with the tuning applying after fold generation to simulate the situation that would be available for future REFs (i.e., no spare data for separate tuning).

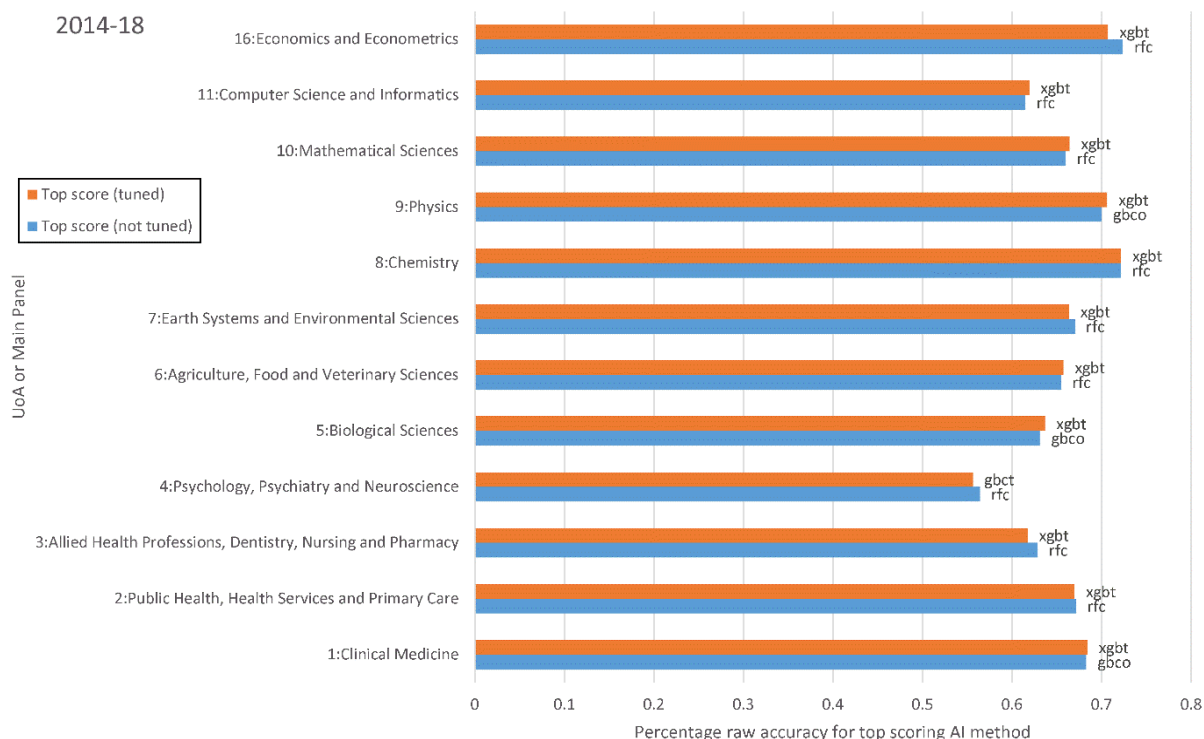

*Figure 13.3 The percentage accuracy for the most accurate machine learning method with and without hyperparameter tuning (out of the main six), trained on **50%** of the 2014-18 articles and **Input set 3: bibliometrics + journal impact + text; 1000 features in total**. The most accurate method is named (Thelwall et al., 2023).*

## 13.2.2 Accuracy for high prediction probability subsets

The methods used to predict article scores report an estimate of the probability that these predictions are correct. If these estimates are not too inaccurate, then arranging the articles in descending order prediction probability can be used to identify subsets of the articles that can have their REF score estimated more accurately than for the set overall.

The graphs below for the UoAs with the most accurate predictions (Figure 13.4) can be used to read the number of scores that can be predicted with any given degree of accuracy. For example, setting the prediction probability threshold at 90%, 500 articles could be predicted in UoA 1. The graphs report the accuracy by comparison with sub-panel provisional scores rather than the machine learning probability estimates. The graphs confirm that higher levels of machine learning score prediction accuracy can be obtained for subsets of the predicted articles. Nevertheless, they suggest that there is a limit to which this is possible. For example, no UoA can have substantial numbers of articles predicted with accuracy above 95% and UoA 11 has few articles that can be predicted with accuracy above 80%. If the algorithm is trained on a lower percentage of the articles, then fewer scores can be predicted at any high level of accuracy.

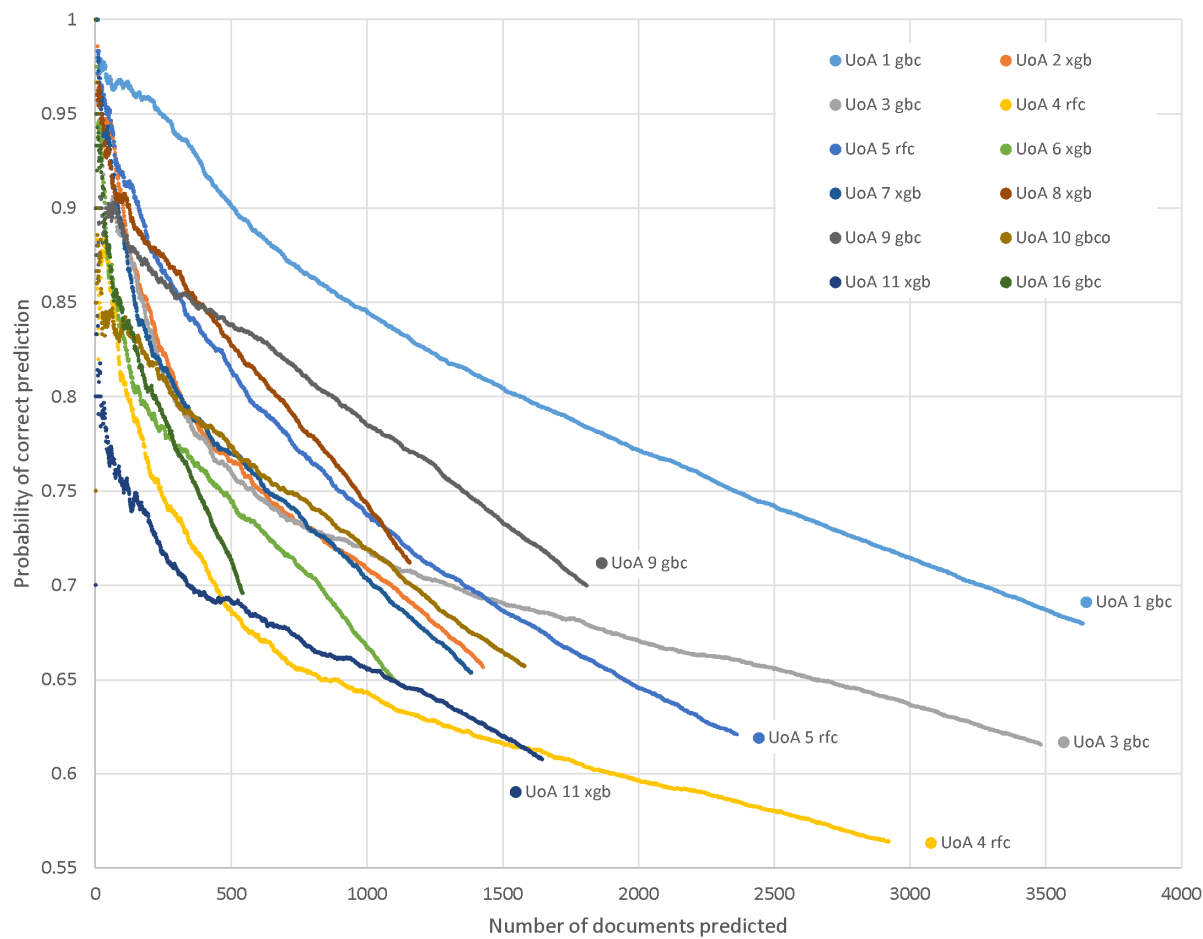

*Figure 13.4 Probability of a machine learning prediction (best machine learning method at the 85% level, trained on 50% of the data 2014-18 with 1000 features) being correct against the number of predictions for UoAs 1-11, 16. The articles are arranged in order of the probability of the prediction being correct, as estimated by the AI. Each point is the average across 10 separate experiments (Thelwall et al., 2023).*

### 13.2.3 HEI-level accuracy

For the UK REF, as for other national evaluation exercises, the most important unit of analysis is the institution because the results are used to allocate funding (or a pass/fail decision) to institutions for a subject rather than to individual articles or researchers (Traag & Waltman, 2019). At the institutional level, there can be non-trivial score shifts for individual institutions, even with high overall accuracy. UoA 1 has one of the lowest average score shifts (i.e., change due to human scores being partly replaced by machine learning predictions) because of relatively large institutional sizes, but these are still non-trivial (Figure 13.5). The score shifts are largest for small institutions, because each change makes a bigger difference to the average when there are fewer articles, but there is also a degree of bias, in the sense of institutions that then to benefit or lose out overall from the machine learning predictions.

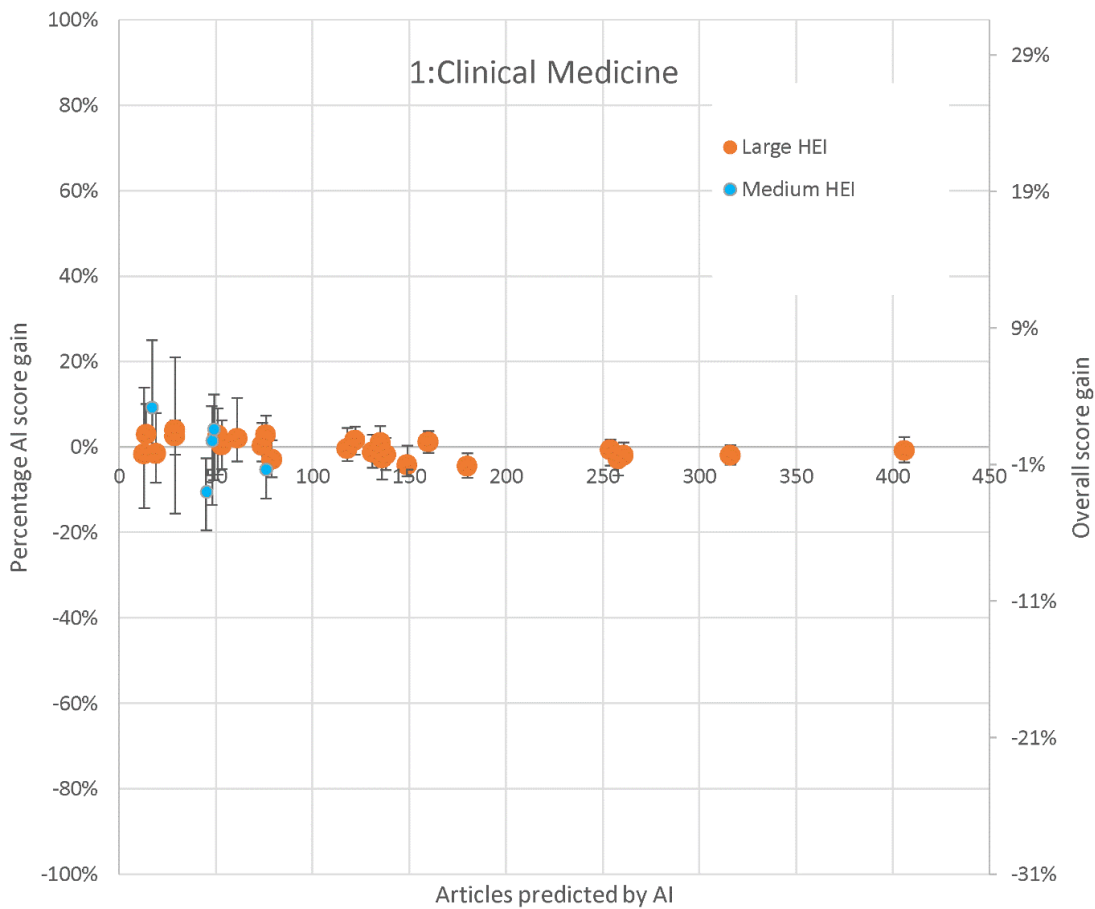

*Figure 13.5 The average REF AI institutional score gain on UoA 1-11, 16 for the most accurate machine learning method, trained on **50%** of the 2014-18 data and **bibliometric + journal + text inputs, after excluding articles with shorter than 500 character abstracts**. UoAs 1,2,6-10,16 have 65%-72% raw accuracy. AI score is a financial calculation, sometimes called research power (4\*=100% funding, 3\*=25% funding, 0-2\*=0% funding). Overall gain includes human classified articles (right hand axis for the same data) but not non-article outputs. Error bars indicate the highest and lowest values from 10 iterations. Source: Figure 4.1.2.1 of (Thelwall et al., 2022).*

Bias occurs in the predictions, even for UoAs with the highest level of accuracy. For example, in most UoAs, larger HEIs, HEIs with higher average scores, and HEIs submitting more articles to a UoA tend to be disadvantaged by machine learning score predictions (Figure 13.6). This is not surprising because, other factors being equal, high scoring HEIs would be more likely to lose from an incorrect score prediction. This is because they would have a higher proportion of top scoring articles (which would always be downgraded by errors). Similarly, larger HEIs tend to submit more articles and tend to have higher scores.

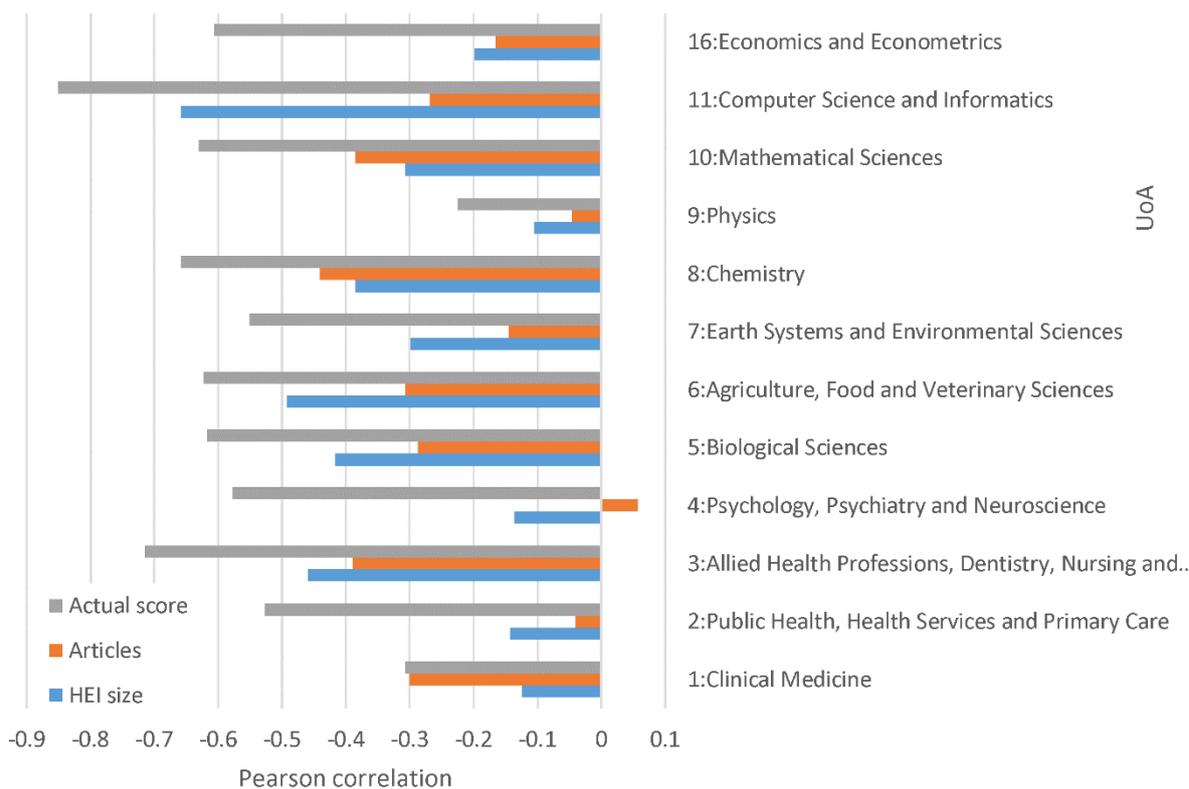

*Figure 13.6 Pearson correlations between institutional size (number of articles submitted to REF), submission size (number of articles submitted to UoA) or average institutional REF score for the UoA and average REF AI institutional score gain on UoA 1-11, 16 for the most accurate machine learning method, trained on* **50%** *of the 2014-18 data and* **bibliometric + journal + text inputs, after excluding articles with shorter than 500 character abstracts**. *UoAs 1,2,6-10,16 have 65%-72% raw accuracy. Prediction gain is a financial calculation (4\*=100% funding, 3\*=25% funding, 0-2\*=0% funding). Note: maximum score shift difference between institutions of 4% for institutions with at least 200 articles. Source: Figure 4.1.3.1 of (Thelwall et al., 2022).*

### 13.2.4 Accuracy on Scopus broad fields

If the REF articles are organised into Scopus broad fields before classification, then the most accurate machine learning method is always gbco, rfc, rfco or xgbo. The highest accuracy above the baseline is generally much lower in this case than for the REF fields, with only Multidisciplinary having accuracy above the baseline above 0.3, with the remainder being substantially lower (Figure 13.7). The lower accuracy is because the Scopus broad fields are effectively much broader than UoAs. They are journal-based rather than article-based and journals can be allocated multiple categories. Thus, a journal containing medical engineering articles might be found in both the Engineering and the Medicine categories. This interdisciplinary, broader nature of Scopus broad fields reduces the accuracy of the machine learning methods, despite the field normalised indicators used in them.

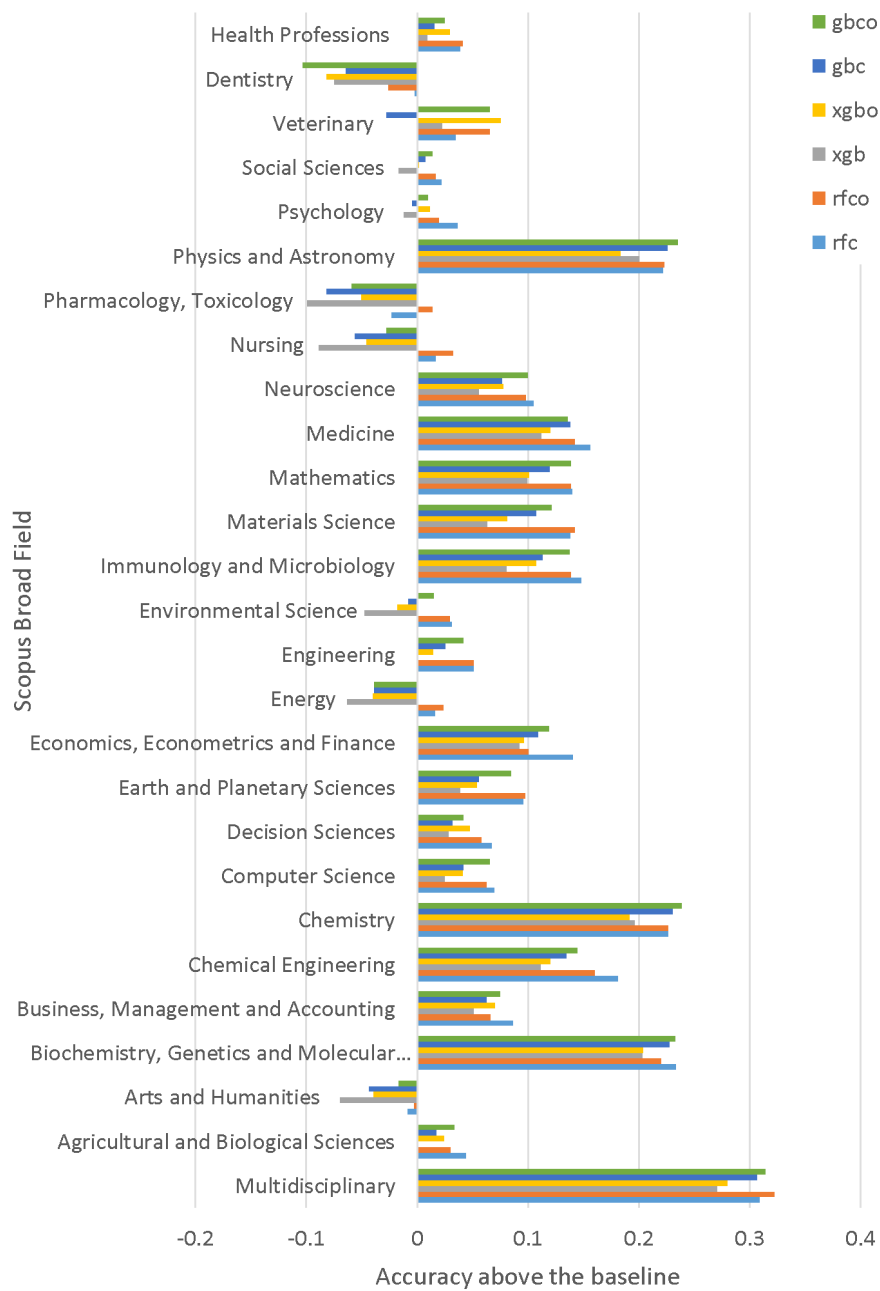

*Figure 13.7 The percentage accuracy above the baseline on Scopus broad fields for the three most accurate machine learning methods and their ordinal variants, trained on **50%** of the 2014-18 Input Set 3: Bibliometrics, journal impact and text, after excluding articles with shorter than 500-character abstracts, zero scores or duplicate within a Scopus broad field (Thelwall et al., 2023).*

## 13.3 Limitations

The results are limited to articles from a single country and period. These articles are self-selected as presumably the best works (1 to 5 per person) of the submitting UK academics over the period 2014-2020. The findings used three groups (1*-2*, 3*,4*) and finer grained outputs (e.g., the 27-point Italian system, 3-30) would be much harder to predict accurately because there are more wrong answers (26 instead of 2) and the differences between scores are smaller. The results are also limited by the scope of the UoAs examined. Machine learning predictions for countries with less Scopus-indexed work to analyse, or with more recent work, would probably be less accurate. The results may also change in the future as the scholarly landscape evolves, including journal formats, article formats, and citation practices. The accuracy statistics may be slightly optimistic due to overfitting: running multiple tests and reporting the best results. This has been mitigated by generally selecting strategies that work well for most UoAs, rather than customising strategies for UoAs. The main source of overfitting is probably in the choice of machine learning algorithm for the reported results, since six similar algorithms tended to perform well and only the most accurate one for each UoA is reported.

The practical usefulness of machine learning predictions is limited by a lack of knowledge about the reliability of the human scores. For example, if it was known that REF reviewing team scores agreed 85% of the time then machine learning predictions that are at least 85% accurate might be judged acceptable. Nevertheless, assessing this level of accuracy is impossible on REF data because each score is produced by two reviewers only after discussion between themselves and then the wider UoA group, supported by REF-wide norm referencing. Because of this, comparing the agreement between two human reviewers, even if REF panel members, would not reveal the likely agreement rate produced by the REF process. The REF score agreement estimate of 85% mentioned above was for articles in a UoA that seemed to have "accidentally" reviewed multiple copies of the same articles, giving an apparently natural experiment in the consistency of the overall REF process.

This chapter has not considered practical issues, such as whether those evaluated would attempt to game a machine learning prediction system or whether it would otherwise lead to undesirable behaviour, such as targeting high impact journals or forming citation cartels. Thus, great care must be taken over any decision to use machine learning predictions, even for more accurate solutions than those discussed here.

## 13.4 Summary

The results in this chapter show that machine learning predictions of article quality scores on a three-level quality scale are possible from article metadata, citation information, author career information and title/abstract text with up to 72% accuracy in some broad fields for articles older than two years, given enough articles for training.

Substantially higher levels of accuracy may not be possible because tacit knowledge is needed about the context of articles to properly evaluate their contributions. Whilst academic impact can be inferred to some extent through citations, robustness and originality are difficult to assess from citations and metadata, although journals may be partial indicators of these in some fields and original themes can in theory be detected (Chen et al., 2022b). The tacit knowledge needed to assess the three components of quality may be more important in the fields (or UoAs) in which lower machine learning prediction accuracy was attained. Higher accuracy may be possible with other inputs included, such as article full text (if cleaned and universally available) and peer reviews (if widely available).

It also seems likely that machine learning prediction of research quality scores is irrelevant for the arts and humanities (perhaps due to small article sets), most of the social sciences and engineering, and is weak for some of the remaining areas. It would be interesting to try large language models such as ChatGPT for journal article quality classification (see next chapter), although they would presumably still need extensive scored training data to understand the task well enough to perform it. It is not clear whether large language models could be reasonably effective at detecting originality, rigour, or impact in any or all fields. For example, it seems unlikely that they could detect non-obvious methods flaws or differentiate between realistic and unrealistic future impact claims. For originality, there are many ways in which research can be original (e.g., Sanchez et al., 2019) and large language models might not be able to differentiate between these and irrelevant types of originality, such as linguistic diversity.

Currently, the only practical use of machine learning quality scoring in important research assessments seems to be to provide quality estimates to peer reviewers alongside estimated probabilities that the predictions are correct. This information could help the peer reviewers to make their judgements in situations of uncertainty, such as when their topic knowledge did not cover the paper assessed. This would presumably improve the accuracy of the human classifications without saving much time. This is primarily applicable to the health and physical sciences as well as for some social sciences. Since machine learning systems need data to work, these predictions would only become available after about half of the human scores had been assigned, which is a complicating factor for organising such a system.

Substantially improved predictions seem to need more understanding of how peer review works, and close to universal publication of machine readable clean full text versions of articles online so that full text analysis is practical. Steps towards these would therefore be beneficial and this might eventually allow more sophisticated full text machine learning algorithms for published article quality to be developed, including with deep learning if even larger datasets can be indirectly leveraged. From ethical and unintended consequences perspectives, however, the most promising future application of all forms of artificial intelligence is in support of reviewers' judgements rather than replacing them.

# 14 Large language models for research quality prediction: Theoretical issues

The capabilities of Large Language Models (LLMs) like ChatGPT (Thelwall, 2024), Claude, DeepSeek, and Gemini (Thelwall, 2025b) for tasks like computer programming, proof reading, supporting grant and paper writing (Parrilla, 2023)**,** and supporting literature reviews (Borger et al., 2023) have led to questions about whether they may be better than traditional machine learning for many research evaluation tasks. This is supported by evidence that ChatGPT can provide useful peer review advice on unpublished manuscripts in some fields (Liang et al., 2024). This chapter discusses how ChatGPT and Google Gemini can be used to predict the quality scores that human experts would give to academic outputs.

## 14.1 Types of LLM

A LLM is a computer system with a specific learning configuration (a transformer type of deep learning algorithm) that has been trained with an enormous collection of documents and can identify likely continuations of user-input text based on this. For example, if a user entered "I am going to the" then the system would know that "shops" or "party" are reasonably likely continuations but not "milky way" (unless it was science fiction) or "it". End users usually see LLMs generating continuations of text, and these are types of Generative AI (sometimes shortened to GenAI). A GenAI LLM fed with "I am going to the" would form a reasonable continuation of any specified length, such as "I am going to the store. I need to pick up some groceries for dinner." These systems usually contain random parameters, so the same prompt the next time might generate a completely different response, like "I am going to the park. I want to enjoy some fresh air and relax for a while."

The well-known LLMs in early 2025 included ChatGPT, Gemini, DeepSeek, and Claude. These are Generic LLMs with an extra training stage that customises them to follow user instructions, in a process known as Reinforcement Learning from Human Feedback (RLHF) (Christiano et al., 2017; Ouyang et al., 2022). Post-RLHF LLMs can answer questions and follow instructions to complete a wide variety of tasks. For example, a user could ask, "What is RLHF?" and might get the response, "**RLHF** stands for **Reinforcement Learning from Human Feedback**. It's a machine learning technique used to fine-tune AI models by incorporating human preferences into the training process. This approach is widely used to make large language models (like ChatGPT) more helpful, aligned with human values, and less likely to produce harmful or nonsensical outputs." Because of the RLHF stage, the system would probably not give a response that does not explicitly answer the question, such as "What is RHLF? What are LLMs? What are Transformers?".

Users can customise post-RLHF LLMs with additional fine tuning RLHF stages or by describing a specific task. As explained below, tasks can be explained in system instructions. After ingesting these instructions, the LLM will be prepared to complete the new task. Here the focus is on research quality.

## 14.2 LLMs for research quality score prediction

An LLM for research quality score prediction is a generic LLM with additional instructions for this task. These would normally be registered as system instructions (Figure 14.1), although they could also be entered as part of a complex user prompt or as part of the setup for a custom GPT for ChatGPT. This chapter almost exclusively deals with research quality prediction LLMs: either ChatGPT or Google Gemini configured to predict the research quality scores that experts would give academic outputs.

At the time of writing (early 2025), LLMs including ChatGPT and Gemini allowed tasks to be described separately in "system prompts" that are designed for this role. The idea is that the LLM is first set up to conduct a particular type of task and then is separately fed the information to process for the task. This procedure is currently supported by Applications Programming Interfaces rather than the web interface of LLMs, although ChatGPT has a web near-equivalent in the form of a capability to configure a custom GPT with task description information.

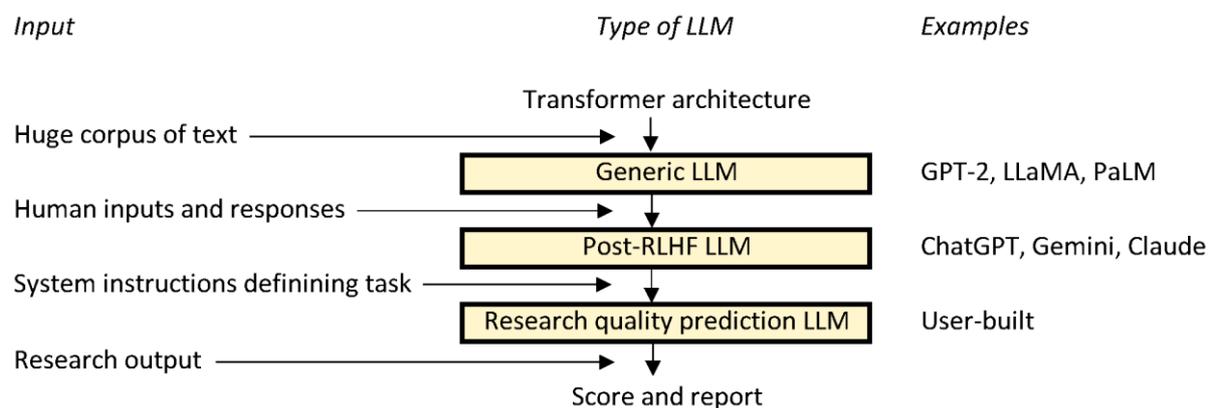

*Figure 14.1. Overview of the construction of a research quality prediction LLM (source: author).*

LLM-based systems have more goal flexibility than bibliometrics. This is because all citation-based indicators reflect an aspect of scholarly impact despite this being only part of one of the three common dimensions of research quality (Langfeldt et al., 2020), and altmetrics tend to reflect a combination of scholarly and societal impacts. In contrast, LLM-based systems can have goals specified by the research evaluators. These goals are by the system instructions (or natural language prompts) that describe the task.

This goal specified for an LLM might be to score an article for rigour alone, for originality alone, or for scholarly or societal impacts. More usefully, however, the goal could be specified as all core dimensions of research quality. This would align the LLM task directly with the evaluation needs and avoid one of the major criticisms of citation analysis: that it primarily reflects scholarly impact. Of course, specifying a goal does not guarantee that the LLM will be able to effectively evaluate a text for that goal. For example, an LLM might be asked to assess the rigour of pure maths paper but be completely unable to make sense of the logic and formulae in it.

## 14.3 Prompt engineering and system prompts

When designing an LLM-based system, a key task is the construction of system prompts and/or user prompts to generate the most useful answers. For example, the user prompt, "Evaluate this paper" might work reasonably well with ChatGPT but there are more effective versions with a higher chance of producing useful and accurate information. For example, a system prompt might include information about what to look for in the evaluation and what type of score to give, if any, and some norm referencing information. When designing prompts, the task goal should be considered to ensure that the LLM is asked to produce information that aligns with the needs of the evaluation.

An example of a system prompt for the task of assessing journal article research quality is given in Table 14.1. After this system prompt, the LLM could be fed with a simple user prompt like, "Score this", followed by the text (or document upload) of the article to be assessed.

Table 14.1. A system prompt for ChatGPT defining and describing a research evaluation task. It is slightly adapted from the descriptions of quality levels that are included as part of the guidance for the UK's Research Excellence Framework 2021 (Thelwall, 2025a). Indentation and bullet points have been added to aid human readability but are not included in the computer instructions.

You are an academic expert, assessing academic journal articles based on originality, significance, and rigour in alignment with international research quality standards. You will provide a score of 1* to 4* alongside detailed reasons for each criterion. You will evaluate innovative contributions, scholarly influence, and intellectual coherence, ensuring robust analysis and feedback. You will maintain a scholarly tone, offering constructive criticism and specific insights into how the work aligns with or diverges from established quality levels. You will emphasize scientific rigour, contribution to knowledge, and applicability in various sectors, providing comprehensive evaluations and detailed explanations for its scoring.

    Originality will be understood as the extent to which the output makes an important and innovative contribution to understanding and knowledge in the field. Research outputs that demonstrate originality may do one or more of the following: produce and interpret new empirical findings or new material; engage with new and/or complex problems; develop innovative research methods, methodologies and analytical techniques; show imaginative and creative scope; provide new arguments and/or new forms of expression, formal innovations, interpretations and/or insights; collect and engage with novel types of data; and/or advance theory or the analysis of doctrine, policy or practice, and new forms of expression.

    Significance will be understood as the extent to which the work has influenced, or has the capacity to influence, knowledge and scholarly thought, or the development and understanding of policy and/or practice.

    Rigour will be understood as the extent to which the work demonstrates intellectual coherence and integrity, and adopts robust and appropriate concepts, analyses, sources, theories and/or methodologies.

The scoring system used is 1*, 2*, 3* or 4*, which are defined as follows.

- 4*: Quality that is world-leading in terms of originality, significance and rigour.
- 3*: Quality that is internationally excellent in terms of originality, significance and rigour but which falls short of the highest standards of excellence.
- 2*: Quality that is recognised internationally in terms of originality, significance and rigour.
- 1* Quality that is recognised nationally in terms of originality, significance and rigour.

The terms 'world-leading', 'international' and 'national' will be taken as quality benchmarks within the generic definitions of the quality levels. They will relate to the actual, likely or deserved influence of the work, whether in the UK, a particular country or region outside the UK, or on international audiences more broadly. There will be no assumption of any necessary international exposure in terms of publication or reception, or any necessary research content in terms of topic or approach. Nor will there be an assumption that work published in a language other than English or Welsh is necessarily of a quality that is or is not internationally benchmarked.

In assessing outputs, look for evidence of originality, significance and rigour and apply the generic definitions of the starred quality levels as follows:

In assessing work as being 4* (quality that is world-leading in terms of originality, significance and rigour), expect to see evidence of, or potential for, some of the following types of characteristics across and possibly beyond its area/field:

- a primary or essential point of reference;
- of profound influence;
- instrumental in developing new thinking, practices, paradigms, policies or audiences;
- a major expansion of the range and the depth of research and its application;
- outstandingly novel, innovative and/or creative.

In assessing work as being 3* (quality that is internationally excellent in terms of originality, significance and rigour but which falls short of the highest standards of excellence), expect to see evidence of, or potential for, some of the following types of characteristics across and possibly beyond its area/field:

- an important point of reference;
- of considerable influence;
- a catalyst for, or important contribution to, new thinking, practices, paradigms, policies or audiences;
- a significant expansion of the range and the depth of research and its application;
- significantly novel or innovative or creative.

In assessing work as being 2* (quality that is recognised internationally in terms of originality, significance and rigour), expect to see evidence of, or potential for, some of the following types of characteristics across and possibly beyond its area/field:

- a recognised point of reference;
- of some influence;
- an incremental and cumulative advance on thinking, practices, paradigms, policies or audiences;
- a useful contribution to the range or depth of research and its application.

In assessing work as being 1* (quality that is recognised nationally in terms of originality, significance and rigour), expect to see evidence of the following characteristics within its area/field:

- an identifiable contribution to understanding without advancing existing paradigms of enquiry or practice;
- of minor influence.

The output from a prompt can be requested to be a simple score or a fuller report on the article, with the latter being more effective (Thelwall, 2025a). Table 14.2 gives an example of a research quality evaluation report from ChatGPT. These reports often start with a summary of the article assessed and then discuss the three quality components separately, before justifying the final score. Repeating a prompt to ChatGPT gives a different report each time and sometimes also different scores.

Table 14.2. An example of a report generated by ChatGPT after being configured with the system prompt in Table 14.1 and being asked to score a research article based on its title and abstract.

---

**Score: 2\*\*\***

**Originality: 2\* (Quality that is recognised internationally)\*\***

The article presents a useful analysis of social media (Twitter) data concerning global discussions around COVID-19, specifically focusing on cures, treatments, and vaccines. The originality lies in its exploration of international differences in public discourse as influenced by the pandemic, which is quite timely and relevant. However, while the study engages with a relevant topic, it largely operates within an established framework of social media research and public health discourse without introducing significantly innovative methodologies or theoretical advancements. Therefore, it qualifies as an incremental advance rather than a groundbreaking exploration.

**Significance: 2\* (Quality that is recognised internationally)\*\***

The work holds significance as it highlights international variances in public interest and misinformation, offering insights that could inform public health strategies and social media monitoring. Yet, while the findings are of some relevance to current discourse around health communication during the pandemic, the implications appear more descriptive, lacking the depth necessary for profound influence or significant shifts in understanding or policies. The mention of countries with a lower human development index engaging more about cures could lead to a useful dialogue, but this alone does not elevate the work to a higher significance point.

**Rigour: 2\* (Quality that is recognised internationally)\*\***

The article demonstrates intellectual coherence with a structured approach to analyzing Twitter data over a specific timeframe. The findings are supported by the analysis of trends, yet the authors themselves acknowledge methodological limitations that could impact the validity of their conclusions. The lack of robust discussion around these limitations may affect readers' confidence in the rigor of the study. While the analysis is logically presented and grounded in relevant literature, the reliance on Twitter data without a more comprehensive methodology (such as mixed methods or longitudinal data) diminishes its rigor. This work does fulfill the criteria for international recognition but does not exemplify the robust analytical depth expected at higher quality levels.

**Conclusion:\*\***

In summary, the article makes a recognisable contribution to the discourse on social media's role during the pandemic, particularly in capturing aspects of public sentiment toward COVID-19 treatments and vaccines across countries. However, the originality, significance, and rigor of the claims can be enhanced. These enhancements may include broader methodological frameworks, deeper analysis of the impact of misinformation, and a more thorough investigation into the implications of the findings for public health messaging and policy.

---

The best system prompt for a research quality score prediction task seems likely to be essentially the same instructions that are given to the human evaluators for the task, if they are reasonably detailed about the evaluation aspect. In other situations, a new system prompt would need to be written, and it should be informed by the human reviewer instructions for similar tasks, as well as the goals of the new evaluation.

## 14.4 LLM inputs and copyright

LLMs need to be fed information about research outputs to make a quality prediction about them, but they do not necessarily need to be fed with the entire output. Whilst the obvious way to use an LLM would be to configure it for a research evaluation task and then feed it with outputs to evaluate, this may infringe copyright, might not be technically possible for the system used, and may not generate the best results. A core problem for overall judgements of academic research is that articles are relatively large chunks of text, which is challenging for current LLMs to analyse holistically (Carabantes et al., 2023). LLMs may also not be able to interpret the graphs in papers and may therefore miss important issues with them and not be able to "understand" the paper (Biswas et al., 2023). The following are possible extracts from research outputs that may be fed into an LLM system for a quality score prediction.

- *The title alone*. This would give little useful information about the article (unless the LLM could cross-reference the title with information about the article in its training data or look it up online).
- *The title and abstract*. This would give a succinct summary of the key points of the output.
- *The full text without the tables, figures, and references*. This would give the main text of the article without the text aspects that an LLM might find difficult to interpret because it is complex.
- *The full text without the figures and tables*.
- *The full text without the figures*.
- *The full text document*. This would give the information necessary for a human expert evaluation. Some systems do not allow this and only accept plain text input.

Intuitively, only article full texts are reasonable inputs for LLM *evaluations* of academic outputs. Surprisingly, however, the best score *predictions* from ChatGPT have been obtained with title and abstracts inputs (i.e., the second one in the list above) (Thelwall, 2024, 2025a). The reason that ChatGPT is better able to exploit concise author summaries in abstracts to make a prediction may be that the extra information in full texts confuses it with less relevant information that it cannot interpret well. This may change in the future with more powerful LLMs. There is some evidence that the multimodal Google Gemini works best with full text document inputs (Thelwall, 2025b), so the overall pattern for LLMs is not clear. All experiments into this issue have been small scale so far.

Copyright law is an important factor in the choice of inputs. Anyone using a LLM needs to check that they are legally allowed to input their chosen documents. In the UK, for example, the law allows titles and abstracts to be fed into AI systems for research purposes, but not necessarily any other parts of an article. Thus, an evaluation may need to get permission from the authors or other copyright holders, if practical, or restrict itself to documents, or document parts, that have copyright conditions allowing AI processing.

If using a public LLM that can learn from its inputs, such as the web interface for ChatGPT at the time of writing, an additional copyright consideration is that the system may infringe copyright conditions that require acknowledgement, such as required by the CC BY licence. This could be infringed by the system partially regurgitating the information entered in the future, but without acknowledging the source. This issue does not apply if the LLM promises not to learn from the user inputs, as was the case with the ChatGPT API and Gemini interfaces in the UK at the time of writing. It also does not apply to private LLMs that run offlie and are not shared with others.

## 14.5 Fine tuning LLMs

One unknown at the time of writing (early 2025) is whether LLMs can benefit from specific research quality score training data. As illustrated in Chapter 13, traditional machine learning algorithms typically need to be fed with lots of correct answers (usually 1000+) to learn the patterns that they need to make predictions with. LLMs have already been pre-trained with billions of input documents for language modelling and an unknown amount of human feedback to help them learn to give good answers to questions. It is not clear whether they would also benefit from additional training data in the form of sets of academic documents and associated expert quality scores or reviews. It seems *likely* that large training sets are not necessary for LLMs because (a) they are pre-trained and general, (b) articles are highly varied, complicating the training process, and (c) the training data would probably need to include full reports rather than just scores and it is unclear whether a LLM would be possible to learn much from a small sample of long reports.

To illustrate how training data might be used, suppose that 100 articles with expert quality scores are available for this. An experiment might be set up on a "leave one out" basis: feeding the LLM with 99 of the inputs and scores and requesting a judgement on the last one and then repeating this for each of the 100 applications, resetting the LLM each time so that it did not know the correct answer. After 100 experiments, this would produce 100 LLM judgements. This experiment might need to be carried out on a private version of the LLM, so that the articles would not be leaked into the public domain. At the end of this, the expert judgements for the 100 predictions could be compared with the LLM predictions. If the results were better (e.g., a higher correlation with the expert scores) than directly predicting the scores without the training stage, then this would suggest that the training was useful.

The above experimental design with a training set pre-supposes that no development set is used. In reality, a development set is needed to try out different prompts or sequences of prompts for the LLM. Thus, ideally a second 100 articles would be needed to test for the most effective prompting strategy. If a prompting strategy is agreed in advance, however, such as from prior experience, then this development set would not be needed.

## 14.6 Assessing the value and validity of LLM outputs

LLMs are difficult to assess qualitatively because the plausibility of their incorrect answers (e.g., for ChatGPT and Gemini) means that it is difficult to identify why they give either correct or incorrect research quality score predictions. They are also tricky to assess quantitatively due to the lack of appropriate benchmark judgements, such as human quality scores or decisions, to evaluate them against, and the potential for hidden bias.

*Plausibility of correct answers*: LLMs seem to excel at generating reasonable looking responses to prompts, including by "hallucinating" information (Carabantes et al., 2023). An example of a hallucination is an invented reference added to the end of an otherwise believable section of text (Orduña-Malea & Cabezas-Clavijo, 2023; Wilby & Esson, 2023). To test this for research evaluation, I asked ChatGPT to write reports (a) recommending publication and (b) recommending rejection of some of my published articles and its responses included quite convincing opposite conclusions for the same article. Thus, LLM responses currently (early 2025) cannot be trusted and can be seductively misleading.

*Appropriate benchmark judgments*: Since qualitative analyses of plausible reports are difficult, empirical evidence is needed to assess the accuracy and biases of any LLM before it could be used in evaluations. For this, the judgements should ideally be private so that the LLMs would not have read them already on the web and could not "cheat" by knowing the answer before the question was posed. Thus, to test an LLM, an experiment would be needed where an LLM was queried with appropriate prompts and then its predictions compared against the "ground truth" of private expert judgements, as was carried out for traditional machine learning in a previous chapter. This would rule out the few contexts in which comprehensive research judgements are publicly available, such as F1000Research (all peer review reports and judgements are published, including for submissions that are not approved) and the SciPost journal system (reports and quality judgements published). REF2021 scores would be perfect for this, as reported in previous chapters, but unfortunately this data had to be deleted before any experiments were conducted by the scientometric team that analysed it. In practice, direct or indirect public scores have to be used for many tests because of the lack of sufficiently large-scale private quality score data. In particular, F1000Research and SciPost scores have been used (Thelwall & Yaghi, 2024), as have the public aggregate REF quality scores (Thelwall, 2025b) as indirect quality indicators. Such evaluations must explicitly state the possibility that the LLM has "cheated" through potential access to the correct scores. Unfortunately, it is not possible to know whether any document forms part of the training corpus of a LLM.

*Potential for hidden bias*: If asked for quality scores, machine learning algorithms including LLMs can leverage indirect indicators of quality. These are factors that tend to occur more often in higher (or lower) quality research even though they are irrelevant to it. This could include the nationality of the researcher, the quality of the grammar, or even stylistic devices that are more common in higher (or lower) quality research. The first two are problems if an LLM gives/guesses a poor score for a document partly because of the nationality of the author. As an example of strange associations between stylistic devices and research quality, the pronoun "we" is more likely to be used in higher quality medical research, "here we" in higher quality health, psychology, and biology research, and "the purpose of" in higher quality social work and social policy research (Thelwall et al., 2023). Thus, even if an LLM makes reasonably accurate quality predictions, this could be partly because of hidden biases like learned stereotypes and linguistic markers of quality. Using such a system might exacerbate existing biases in society and perhaps also introduce conservatism by recommending higher scores to people from countries or with writing styles that had been successful in the past. Thus, users of LLMs for evaluation should always seek to minimise the possibility of bias, check for bias when evaluating a system for future use and remain watchful for biases when it is employed.

## 14.7 Theoretical and ethical considerations for LLMs in research evaluation

Perhaps the most important current theoretical issue when using LLM quality score predictions for research evaluations is that they use text associations to predict scores rather than evaluating an article. Whilst a human expert would be expected to read, understand, and assess an article before giving it a score, LLMs instead read an article (or part of it) and then guess a quality score from it from associations with the system instructions and their original large collection of input texts. This is obviously true when a LLM is asked to score an article based only on its title or based only on its title and abstract. It is also true if the LLM is fed with the full text for two reasons: this is fundamentally how LLMs work; and experiments so far have shown that the predictions do not improve much (Gemini) or are worse (ChatGPT) with full text than with titles and abstracts. Thus, current LLMs effectively almost completely ignore everything except the title and abstract when fed with a research article.

Because LLM scores are predictions or guesses, this creates a theoretical difference with citation-based indicators because the latter are based on accumulated tangible evidence of scholarly impact from each citation. The difference is not clear-cut because of the many citations that reflect little or no scholarly influence. Similarly, LLM predictions based on author's titles and abstracts are not pure guesses because the claims in journal article abstracts are peer reviewed along with the rest of the article, so are likely to be reasonable descriptions of the contents of an article.

Despite the above caveats, when using an LLM to support research evaluation, the theoretical situation is the same as for citations: LLM scores do not evaluate the quality of research articles any more than citation counts do. The situation for both is therefore the same: their value as indicators rests primarily on their accuracy and the extent to which they are influenced by systematic biases. Thus, empirical evidence of their correlation with expert scores is essential for LLMs, as for citation-based indicators, and can support their use in the same evaluative contexts. This empirical evidence for LLMs is discussed in the next chapter.

A second theoretical issue is that the use of LLMs for research evaluations may cause systemic effects, such as authors or journal editors trying to craft abstracts to get higher scores. Whilst citation gaming is already recognised, crafting LLM-friendly abstracts could emerge as an unwanted side-effect. In the worst case, this could detract from the value of abstracts as communication devices, consume the time of authors trying to customise for LLMs and perhaps waste the time of reviewers who might then need to check abstracts carefully for validity. It could also undermine the effectiveness of LLM-based scoring in the future. The seriousness so of this issue remains to be seen, even if LLMs become more widely used for research evaluation.

In terms of ethics, authors may not want to be evaluated by AI and so an evaluator should consider whether they ought to be given the right to opt out. Authors may also not want their works to be uploaded to public AI systems even if they have assigned copyright to a publisher that allows it.

## 14.8 Summary

This chapter introduced the idea of using LLM quality score predictions to support research evaluation. It also argued that customising system instructions allows LLM to be used to tailor score predictions to specific goals. This is a theoretical advantage over citation-based indicators, which seem to primarily reflect scholarly impact. The chapter also discussed the types of LLMs that could be used, and, in broad terms, how to configure LLMs for quality score predictions, in terms of system instructions and the choice of inputs. Copyright and ethical issues are also important for all AI-based evaluations since evaluators may not have the right to upload publications to LLMs even when they have legal access to them.

As argued above, LLM-based scores are predictions or guesses rather than genuine evaluations, but this does not rule out their use as indicators any more than it does for citation counts. If they are known to provide results that correlate positively with expert scores and do not have too large unwanted systematic biases, then they may play a similar role to citation-based indicators. The next chapter covers possible applications and some empirical evidence of correlations with expert scores.

# 15 Large language models for research quality prediction: Applications and empirical evidence

This chapter discusses two potential journal-related applications of LLMs (desk rejections of article submissions and support for peer review) and then introduces evidence that ChatGPT and Google Gemini can be used to give research quality score predictions for academic outputs, publication venues, and research impact claims. Within a year of the first experiments with ChatGPT, it was already clear that both ChatGPT and Google Gemini could give score predictions that were more powerful in most fields than those obtainable from citation analysis or traditional AI. In addition, LLMs can be applied when an article is published, whereas citation analysis needs a two or three year delay whilst citations accrue. Thus, LLMs are in a technically strong position to take over the role of citation analysis in support of some research evaluation tasks.

## 15.1 Desk rejecting journal submissions

Although this book primarily covers indicators to support the post-publication evaluation of journal articles, it is useful to consider two peer review applications of LLM-based scores since they logically feed into journal article evaluation. Peer review is at the heart of the academic research system, whether as a gatekeeper against the publication of poor or irrelevant work or evaluating the importance and rigour of publications. This section, like most of the rest of this book, focuses on the most common type of output, academic journal articles. Unlike all other chapters, it also discusses research quality evaluations of papers submitted to journals as part of the editorial quality assurance process.

From an AI for publishing quality control perspective, journal articles have a small number of gatekeepers, journal editors, that see a large volume of submissions and can therefore become more knowledgeable about LLMs for their tasks and could potentially exploit them in some contexts. In contrast, journal article reviewers would be less well placed to understand the limitations of LLMs for research evaluation because they individually assess less articles. This may change if LLMs become better understood and something like LLM-literacy becomes widespread. LLMs can also be used to rank journals (Thelwall & Kousha, 2025), but this is not discussed here.

Journal editors spend considerable time reading submissions and desk rejecting those that are deemed unworthy of peer review due to being flawed, low quality, out of scope, or otherwise unsuitable (e.g., Ansell & Samuels, 2021). For this judgement, the bibliometric-driven approach of the previous chapter would not work because the submissions would be unpublished and therefore uncited and without any journal information, losing the most powerful indicators. LLMs seem likely to be the best available AI solution, perhaps through making desk rejection recommendations to editors, with rationales. Some editors may already be using ChatGPT for this task.

As discussed above, experimental evidence would be needed to assess the accuracy of LLMs for desk rejection. It seems plausible that they could be reasonably effective at identifying clearly poor submissions but would not be good at making fine-grained decisions about more borderline cases. They may also be biased against author nationality or grammatical structures, which editors should look out for and ideally submit only anonymised manuscripts to any LLM. Nevertheless, editors may wish to harness ChatGPT for submission triage without prior systematic checking if they are careful to check its recommendations, because even plausible explanations for decisions might be wrong. They should also look out for biases in the results, and perhaps also cross-reference the results with their own opinions to check for their own biases.

Of course, any use of LLMs by editors should be restricted to environments where the reviewed texts are siloed from other LLM users so that confidentiality and copyright are not broken. They should also consider whether consent should be obtained from authors for AI processing of their work. Overall, whilst it must be tempting for editors of large journals to use LLMs to support desk rejections, any decision to do so must be carefully thought through.

## 15.2 Peer review of journal submissions that are not desk rejected

Peer review of journal submissions that have not been desk rejected is a more difficult problem for LLMs than that of desk rejection for two reasons. First, the quality of the non-rejected articles should be more homogeneous, so quality assessment is more difficult. Second, peer review is multi-faceted, with detailed expert knowledge of different kinds needed to be applied. Thus, for example, whilst an LLM can easily produce plausible peer reviews and might even be able to correct grammar and guess at the presence of common problems (e.g., no mention of a development set, cross-validation, or overfitting precautions in a machine learning paper), they cannot be expected to detect non-trivial methodological flaws.

Despite the above concerns, there might be some useful roles for LLMs in peer review (e.g., Carabantes et al., 2023; Biswas et al., 2023). For example, there is early evidence that ChatGPT can make non-trivial recommendations and be more useful than some peer reviewers (Liang et al., 2024). In addition, editors might use LLMs as an additional opinion if they are not sure of the quality of reviews produced by their selected referees. It seems safer for editors to do this than reviewers because editors see many submissions and so should be better placed than reviewers to critically analyse LLM suggestions.

Some editors and many reviewers are probably already asking LLMs to write reviews for them, but the time saved by reviewers may be at the expense of lower quality reviews and a higher chance of missing key flaws. Again, the problem is that LLM outputs can be very plausible, even when completely wrong, and an overworked reviewer may not notice subtle invalid points. Of course, peer reviewers are humans with their own biases and knowledge limitations and make mistakes, so this argument does not mean that LLMs are always or usually worse than experts (as shown by: Liang et al., 2024), but without evidence, turning over the peer review quality control function to AI would be a substantial risk.

Again, any use of LLMs by reviewers should be restricted to environments where the reviewed texts are siloed from other users so that confidentiality and copyright are not broken. Moreover, the ethical and systemic effects of a decision should be carefully considered in advance.

## 15.3 Post-publication quality predictions for journal articles

As the previous chapters have argued, it could be efficient to employ LLMs or other AI for post-publication quality score prediction of articles, for example as part of national research evaluation exercises. This might also be useful for other tasks where the quality of sets of publications that have passed peer review is of interest, such as internal university-level departmental assessments and journal quality assessment. Of course, published articles also need to be assessed by scholars as part of their literature review, for promotion, and job applications and for internal university quality management purposes. For these issues, traditional machine learning with bibliometrics might be expected to be more accurate than LLMs because they can systematically take bibliometric information into account in a structured and informed way through the careful curation of inputs. As will be seen below, however, the opposite is true.

A small case study has suggested that individual scores from ChatGPT-4 are weak indicators of research quality (correlation r=0.28 with expert scores), although it has some ability to differentiate between research that is weak and research that is not (Figure 15.1). Interestingly, ChatGPT often gives different scores for the same paper if given the identical prompt repeatedly. In one test, averaging this correlation over 15 iterations improved the correlation between ChatGPT and expert scores to r=0.51 (Thelwall, 2024), which is higher than expected for bibliometric data or tradition AI.

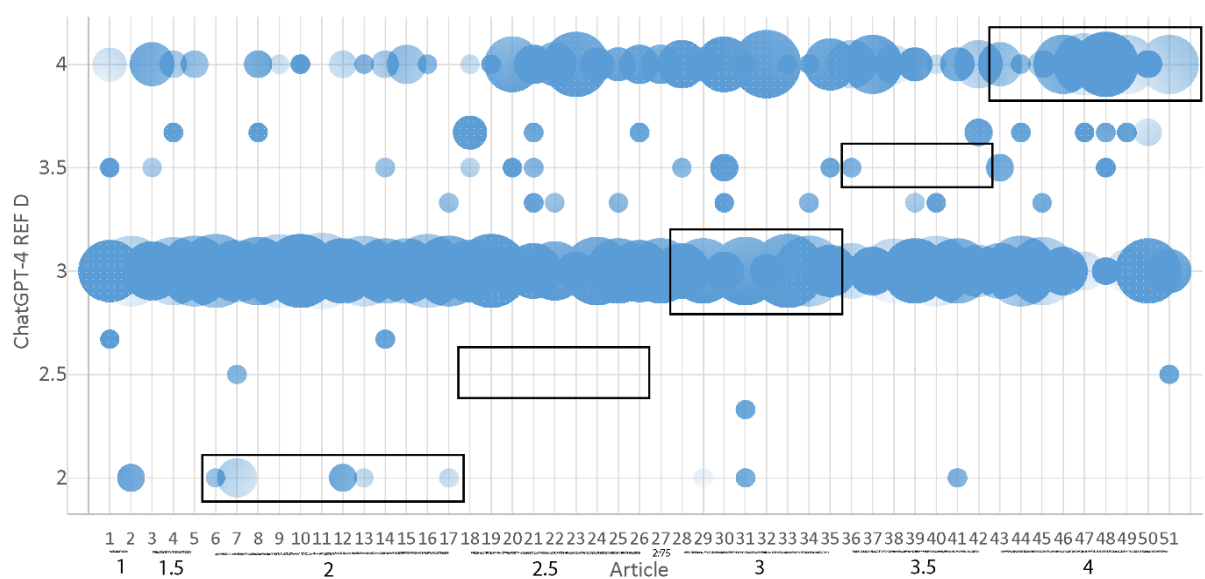

*Figure 15.1. The range of REF star ratings given by the REF D GPT against the author's prior evaluation of the REF score of 51 of his open access articles. The area of each bubble is proportional to the number of times the y axis score was given by ChatGPT to the x axis article. My REF scores are marked on the x axis and circles inside the boxes represent scores that are the same for me and ChatGPT (Thelwall, 2024).*

A larger scale study focusing on the UK has suggested that ChatGPT 4o-mini's quality score estimates may correlate more strongly with expert judgements for journal articles in most fields than a bibliometric indicator (MLNCS) and traditional machine learning (Figure 15.2) (Thelwall & Yaghi, 2024). Similar results for the same dataset have been obtained with Google Gemini 1.5 Flash (Thelwall, 2025b). This confirms that LLMs have surpassed citation-based indicators and traditional AI as the most powerful research quality indicators for journal articles. A limitation of these studies is that they have used public evidence of research quality in the form of departmental quality scores. Whilst this information can be linked to the outputs assessed, it is not clear that LLMs can do this because the score information is encoded in a tabular format and is separate from the outputs assessed.

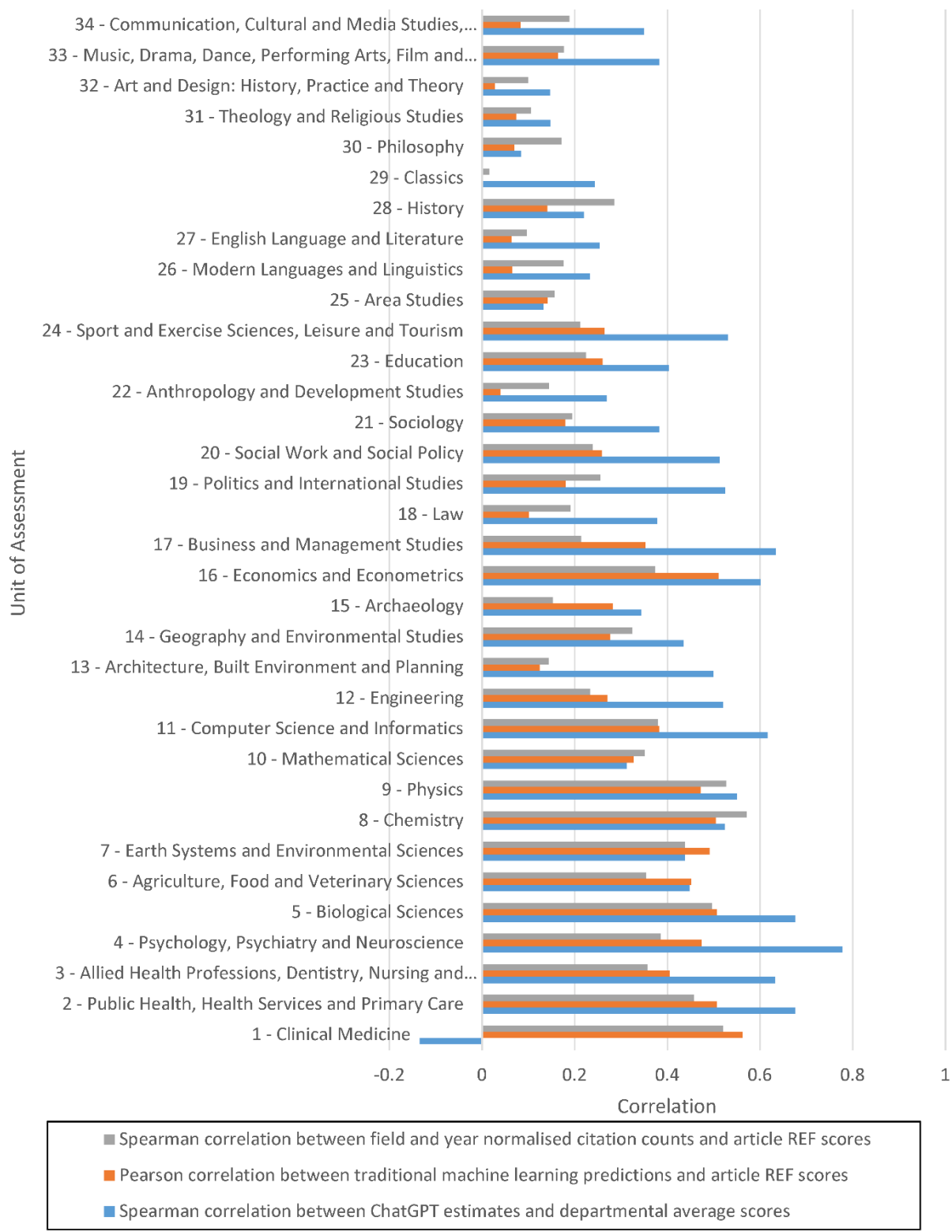

*Figure 15.2. Correlations between various indicators and either expert quality scores or departmental quality scores (Thelwall & Yaghi, 2024). Comparisons between sources are indicative rather than definitive because of the different sample types and quality evidence types.*

An important and perhaps unexpected consideration is that ChatGPT scores vary by field and year, so some normalisation for these is needed if evaluating documents from multiple fields or years (Thelwall & Kurt, 2024). There are only small differences between years so year normalisation can be achieved by subtracting an amount from each year that would make the mean ChatGPT score for each year the same. Field normalisation tends to be substantial, however, but can be achieved in the same way as for field normalised citation indicators: by dividing each score by the field average score. As with field normalised citation indicators, this would give an indicator that is 1 for a score that is average for the field and year of publication.

## 15.4 Journals and conferences

ChatGPT average scores can also be used to rank academic journals and conferences as an alternative to citation-based ranking and expert rankings. Whilst the use of any publication value ranking in research evaluation is discouraged on the basis that the quality of outputs should be individually assessed, some countries find them useful as a crude instrument to encourage researchers to aim for higher reputation conferences and journals. Ranking journals based on average ChatGPT scores has the advantage over citation-based rankings that the results can be based on the most recent year of the journal or conference, not needing to wait for citations to accrue. From a theoretical basis, comparing citation-based rankings and expert rankings of journals or conferences with average ChatGPT scores gives additional insights and evidence into the relationship between the three.

A large scale test of average ChatGPT scores for articles in journals compared to their expert ranks from three countries found a positive correlation of mostly moderate strength in 24 out of 25 broad fields, but they were not clearly superior to citation-based rankings for older journals. Positive results have also been obtained for full peer reviewed conference papers in conference-oriented academic fields, in the sense that the average score for all papers in a conference correlates positively with expert conference rankings (Thelwall, 2025d). These results tend to corroborate the value of LLM-based quality scores for academic outputs.

## 15.5 Research impact claims

Academics in some countries are increasingly asked to demonstrate that their work has led to non-academic benefits, such as healthcare improvements, new products, technological innovations, or beneficial changes in professional practice (Adam, 2018). For example, in Australia and the UK, academics have to write Impact Case Studies (ICSs) periodically. In the UK, they are 7-page evidence-based structured narrative claims for the societal impact of their research. These claims are then graded, with the different aspects receiving scores between 1* and 4*, with scores of 3* and 4* being rewarded by funding (Watermeyer & Hedgecoe, 2016).

ICSs through citation analysis because they are assessed before publication and are not naturally citable academic documents. They are also difficult to assess quantitatively through comparisons of the numbers contained in them (e.g., how many people were affected by the changes) because the quantities vary between ICSs and the depth of the impact reported is usually subjective. Thus, they are scored by human assessors without any attempt to provide systematic quantitative support.

At the time of writing there had been one investigation of whether ChatGPT could predict ICS scores (Kousha & Thelwall, 2024). It found that when complete ICSs were entered into ChatGPT with system instructions describing the evaluation criteria, the predicted score was almost always 4*, so the predictions were not useful. In contrast, when only the title and summary were entered with the same system instructions, the score predictions were more varied and tended to correlate positively (between 0.18 and 0.56 for the 34 Units of Assessment) with a proxy for the expert scores. These correlations are not high enough for the scores to replace expert evaluations or to make a serious contribution to them, but they might be useful in more minor roles, such as to arbitrate between expert assessors that cannot agree on a score.

It is strange that the score predictions are only useful when complete ICSs are withheld and only titles and summaries are entered. Perhaps LLMs score predictions are partly based on accumulating evidence so that a large amount of evidence can outweigh the fact that not all of it is convincing. If true, this would be an unwanted property.

## 15.6 Converting LLM results into actionable information

The tests of ChatGPT in research evaluation contexts so far have primarily used correlation, showing that ChatGPT scores associate positively with expert quality judgements or a proxy for these. Nevertheless, the ChatGPT results operate on a different scale, tending to avoid low scores. This is true for both journal articles and ICSs. If the rank order is all that is needed (e.g., to identify the 10% of lowest scoring or highest scoring articles) then this does not cause any problems.

In practical applications of LLMs, their scores may need to be transformed into a human expert scale before use. This could be achieved with a lookup table to convert LLM outputs into a more standard human range or with a transformation function (Thelwall, 2025a). A lookup table could work like the hypothetical example in Table 15.1. Here, if the average of 30 ChatGPT predictions for an article was 2.72 then the table would convert this into a prediction that an expert would allocate a score of 2* to the paper. This type of lookup table can be created by cross-referencing average ChatGPT scores with expert scores for a sample of research outputs of the type to be assessed.

*Table 15.1. An example of a lookup table to convert ChatGPT scores to predicted expert scores (source: author).*

| ChatGPT score range | Predicted expert quality score |
|---|---|
| 1-2.5 | 1* |
| 2.51-2.80 | 2* |
| 2.81-3.03 | 3* |
| 3.04-4 | 4* |

## 15.7 Summary

The empirical evidence so far shows that LLMs (both ChatGPT and Google Gemini) give more powerful indicators of research quality than citation-based indicators and give similarly powerful score predictions for Impact Case Studies. LLMs may also be useful to support peer review and desk rejection decisions, if care is taken. As the above discussion has argued, however, the use of LLMs for research evaluation is not straightforward because of the potential for bias and the difficulty in systematically evaluating the accuracy of LLM judgements in most contexts due to a lack of benchmark data. Importantly, LLM scores for individual outputs have little meaning because they work on a different scale to experts. To have some use, multiple scores (e.g., at least five) would be needed. These should then be averaged and converted to a human expert scale before use.

As for citation-based indicators, LLM scores may still be able to play a role in supporting decisions about LLMs as part of research evaluations when the decision makers evaluate LLM suggestions carefully or use LLMs enough to be able to judge the value of their responses. Of course, ethical considerations (e.g., informed consent), copyright, and systemic effects should be considered before using them for important research evaluations.

Like bibliometric indicators, LLMs are likely to be particularly useful when applied on a large scale. They have the distinct advantage that they can be applied to documents as soon as they have been published, without having to wait several years to allow time for citations to accrue. LLMs may also have the potential to extend technology supported research assessment beyond journal articles to other scholarly outputs, at least in a minor supporting role.

# 16 Assessing outputs other than academic journal articles

The book focuses on journal articles, but much citation analysis also applies to full text conference papers and potentially also to scholarly monographs. Large Language Models also have applications to other types of scholarly outputs (including impact case studies, as discussed in the previous chapter), and this section briefly discusses these possibilities.

## 16.1 Conference papers

Major computer science and computational linguistics conferences are like journals in that they publish regularly (usually annually), employ peer review (very strict for some conferences) and are subject based. Conference papers in these fields are typically the primary outputs of researchers (e.g., Qian et al., 2017). More generally, conference papers seem to be more important in engineering-related fields than in others, at least judging by the indexing of conferences in the quality-controlled Web of Science (Michels et al., 2014). On the other hand, in nearly all other fields, conference papers are not the primary outputs and are less strictly reviewed. Moreover, only abstracts are reviewed for many conferences in some fields. Thus, in computer science and computational linguistics, but not most other fields, the chapters above and the sections about LLMs apply.

A partial caveat to the above is that conference proceedings may be indexed less systematically than journal articles in the main citation indexes, leading some to use Google Scholar as a more comprehensive source of computing citations (Frachtenberg, 2023). This is undesirable since Google Scholar can easily be gamed by academics (Delgado López-Cózar et al., 2014), so a traditional citation index is recommended for evaluations where gaming is a risk. A second partial caveat is that average citation rates do not seem to be important for conferences and self-reported rejection rates seem to be viewed as better evidence of a conference's quality (e.g., Godoy et al., 2015; Lee, 2019). Thus, the chapter on journal citation rates has less relevance for computer science conference papers, even if average conference citation rates are calculated. Moreover, traditional AI may be less effective for predicting computing conference paper quality than it was for predicting computing journal article quality because it may not be able to effectively substitute conference citation rates for journal citation rates as one of the inputs.

Whilst citation counts and LLMs might be used to support the evaluation of full text conference papers, the importance of conferences might be assessed through expert rankings (as published for computer science by the Australian Computing Research & Education website and by some countries), in addition to average citation rates, or paper rejection rates.

An investigation of the extent to which citation rates, average ChatGPT scores, and expert based rankings of engineering conferences agree found that expert conference rankings agreed better with citations than with average ChatGPT quality score predictions (Thelwall, 2025). Despite the issues mentioned above that may limit the value of citation rates for conferences, this suggests that either citation rates are better than ChatGPT for assessing conferences or that expert conference rankings are influenced by citation-based indicators.

## 16.2 Book chapters

Book chapters are important outputs in some social sciences, arts and humanities. Whilst they are like journal articles in length and format, and are subject to peer review, this quality control seems to be less systematic than for journals, and edited books are rarely published periodically. Moreover, chapters in many edited volumes synthesise previous research (e.g., handbooks) rather than introducing primary work, or may target students rather than researchers.

Despite the above caveats, most of the theory and techniques described in this book for journal articles could be applied to book chapters, although the results are likely to be weaker. The inclusion of selected books within traditional citation indexes has made book chapter citation indexing possible (Torres-Salinas et al., 2013), including for LLM processing through chapter abstracts. Given the likely weakness of traditional bibliometrics, LLMs might be a relatively attractive option for book chapter evaluations, especially if they can be used to help decide on the intended audience of a chapter and whether it contains primary research.

## 16.3 Edited books

Book editors may consider their edited volumes to be primary research outputs. These are complex for a research evaluation perspective because they are based on the work of others. Moreover, editing a collection of chapters is typically a senior researcher role (Ossenblok et al., 2015) so a bibliometric analysis of edited volumes might be unfair against the editors since they are probably mainly compared against other successful researchers. This makes it more difficult to demonstrate an above average performance. For these reasons, any type of bibliometric analysis of edited books seems undesirable. Nevertheless, there may be some scope for using LLMs to support peer review evaluations of them. A technical limitation for LLMs is that edited books are typically very long, which may complicate their analysis process.

## 16.4 Monographs

Monographs are primary research outputs in the humanities and some social sciences (e.g., Kulczycki & Korytkowski, 2020; Verleysen & Ossenblok, 2017). They tend to be peer reviewed, as organised by the publishers, but the process may be less systematic and thorough than for journal articles because publishers deal with a wider range of topics and may be less expert than journal editors. Even more than edited books, monographs can target different audiences. For example, an academic's history book might target undergraduate students (possibly as a textbook), postgraduate students, the public, or other scholars.

A problem with book citation analysis is that citations to books tend to come from other books but they seem to be much less comprehensively included in traditional citation indexes than are journals. OpenAlex may break out of this limitation, however. Whilst it is possible to get wider coverage of book citations from Google Books, this is difficult to do on a large scale (Kousha & Thelwall, 2009). There have been some interesting citation analyses of books using online information about them, however (e.g., Zhou, 2022). Overall, it seems that the relative scarcity of books compared to journal articles, the variety of purposes for books, and the lack of comprehensive citation indexing for books all mitigate against effective citation analysis for them. It is not clear whether LLMs would be more useful, however, because book field classifications and abstracts/blurbs would be needed on a large scale to benchmark or scale LLM quality score predictions enough to give useful information, but there is some early promising evidence of a weak relationship between ChatGPT scores and citations for books (Thelwall & Cox, 2025). As for book chapters, however, a strategy that exploited LLMs to guess a book's audience and then evaluate its value for that audience might have some value.

## 16.5 Other research outputs

Academics produce a diverse range of research outputs, from artistic performances to scientific instruments. It seems unlikely that any type of citation analysis would be useful for systematic evaluations of rare output types, other than to provide evidence to support a narrative claim that an output had made an impact. For example, someone might report that their software had been cited 1200 times as evidence that it was very useful. LLMs may be useful to help evaluate some of these output types, but a bespoke strategy would presumably be needed for each one and extreme caution would be needed given the demonstrated potential for ChatGPT to produce plausible but unhelpful research evaluations for journal articles, and the need for benchmarking to convert ChatGPT quality score predictions into human scale scores.

## 16.6 Grant applications

Although grant applications are rarely published or cited, LLMs are already used by at least one funder to help evaluate grants. The la Caixa Foundation noticed that some of the proposals that they received were relatively easy to reject and developed an LLM system to try to identify them. This predicted whether human reviewers would reject each proposal. When the LLM flagged a proposal as being likely to be rejected, two human experts were then asked to check the decision and if both agreed with it then the proposal was rejected without the full human review. This saves reviewer time by moderately reducing the number of proposals needing full expert review. The funder tested their system extensively, obtained consent from applicants, and used a human-in-the-loop approach both for its value and to comply with national law (Carbonell Cortés et al., 2024). This seems like an excellent example of a practical application of LLM evaluation to save time.

More generally, a research funder wishing to use an LLM to evaluate submissions would first need to test it by comparing its scores (from suitable prompts) to the scores of its normal human evaluators and panels. There are several alternative strategies that could be used for this.

One obvious strategy would be to use archives of previously submitted proposals and their scores as a benchmark collection to test the LLM for the extent to which it (or different prompts for it) agrees with human experts. This could be biased, however, because funding bodies typically publish information about funded projects but hide information about those that are unfunded. Thus, an LLM could work out from a random sample of submitted projects that the unknown ones were likely to be low quality and that the known ones were higher quality. Alternatively, an LLM could be fed a list of successful projects and asked to rate them, comparing these ratings against the unpublished ratings of the projects, assuming that the applicants were not told their ratings or were not allowed to publish them. This could still be unfair if higher scoring projects produced more or better outputs in the public domain that the LLM knew about. A fairer test would be to task the LLM to estimate the quality of the rejected proposals, given that there would be no information about them in the public domain. Nevertheless, the better rejected proposals might have been reworked and submitted successfully elsewhere, so there is still no guarantee that the LLM could not "cheat" In any case, any evaluation that did not use the full set of proposals would be artificial and risk giving misleading results.

The safest way to evaluate an LLM for grant proposals seems to be to run "live" tests on proposals that were evaluated before the scores were shared with the applicants. This would allow the experiment to use a private expert score benchmark set and to be unbiased.

The critical disadvantage here is that such a sample would probably be too small to give statistically significant information.

## 16.7 Researcher CVs

CVs are important for appointments and promotion, and may need to be submitted with funding applications. Whilst citations and altmetrics might be reported for the individual elements of a CV (e.g., Piwowar & Priem, 2013), they would not be appropriate to evaluate it overall.

Employers may be particularly tempted to use LLMs for the initial vetting of CVs for academic jobs or roles with many applicants. The potential for automatic CV analysis is increasing through standardised online public formats, including ORCID to some extent (Wang et al., 2024) and Current Research Information Systems (van Leeuwen et al., 2016). Recruitment-associated AI already has a poor track record, however, such as for learning gender bias when targeting job adverts at likely candidates (Drage & Mackereth, 2022). Given that CVs are about people and cannot therefore be effectively anonymised, extreme care would be needed if AI is used in any aspect of recruitment. For example, a possible use might be to desk reject candidates with specific criteria that would have a very high degree of confidence for disbarring them. For this an LLM could be asked to reject candidates with no prior experience of teamworking. Focusing on necessary skills in this way rather than overall judgements might help to minimise bias.

For grant applications, there seems to be a move away from achievement-based traditional CVs to skills-based narrative CVs (e.g., Bordignon et al., 2023; Chawla, 2022). LLMs may be able to help with the latter type through queries requiring synthesis, such as through prompts asking for the skills demonstrated to be matched with the skills required for a project. This might help to overcome the disadvantage of narrative CVs that they are time consuming to evaluate.

Of course, any attempt to use any form of AI with CVs should use extreme caution to avoid introducing bias or alienating applicants, as well as to comply with local laws.

## 16.8 Impact case studies

As discussed in the previous chapter, Impact Case Studies (ICS) are evidence-based narrative claims about the non-academic impacts of research teams. In the UK, they are important academic outputs that are evaluated nationally as part of the Research Excellence Framework. From REF2021 scores, 25% of UK universities' block grants were decided by their ICS scores (£0.5 billion per year in 2024). Like grant applications and CVs, the UK ICS are designed purely for evaluation rather than for disseminating research. Like CVs, they may contain numerical evidence (e.g. the amount of money saved, or profit made, or the number of lives enhanced as the result of the research: Lim, 2020) but can also be supported by written testimonials by those affected (Bandola-Gill & Smith, 2022; Greenhalgh & Fahy, 2015).

ICS can't be evaluated through citations because they are rarely published before evaluations, and they are not primary research outputs. Whist the references in ICS could be analysed with citations, the focus is on non-academic impacts so this would be a counter-intuitive approach. Evaluators may well compare the numbers reported in the ICS, such as for many people were affected, but LLMs seem an obvious choice as evaluation tool since they are text based and relatively self-contained. ChatGPT has a weak to moderate ability to distinguish between higher scoring and lower scoring ICS (Kousha & Thelwall, 2024).

## 16.9 Summary

Outputs other than journal articles are probably assessed only qualitatively in most cases, although citation analysis can support the research evaluation of conference papers and expert rankings might support the evaluation of both conference papers (via conference reputation/quality) and books (via publisher prestige). In addition, quantitative indicators might be included within some outputs (e.g., CVs, impact case studies, grant applications) that might help evaluators. There now seems to be some possibility to harness LLMs to support these evaluations too, although the care taken by La Caixa (Carbonell Cortés et al., 2024) points to the need to proceed cautiously, with testing, and with due care and attention to ethical and legal (and systemic effect) considerations.

# 17 Perverse incentives and systemic effects

Evaluation strategies that are sensible theoretically may not work in practice. Imagine that a prize is awarded for excellent teaching based on student feedback, but the winning teacher is successful by letting their pupils go home early in return for perfect scores. Here the prize has created a perverse incentive to weaken teaching by shortening it. More generally, a *perverse incentive* is a reward that can be gained by activities that run counter to the goal of the system offering the incentive. Perverse incentives are serious problems in situations where the behaviour evaluated is complex, as in the case of academic research (Bouter, 2015). This is exacerbated when a complex phenomenon is partly assessed by simple indicators, since they can only reflect part of the desirable behaviour. Thus, bibliometrics always carry the risk of generating perverse incentives when they are influential within financial or reputational reward systems. In contrast, one of the advantages of peer review is that, however inaccurate and inefficient it is, it does not seem to create perverse incentives for those analysed, although the system that it is part of might do (e.g., by undervaluing research outputs or other contributions that are difficult to peer review).

Of course, perverse incentives are part of the wider issue of research integrity (Moher et al., 2020; Shaw & Satalkar, 2018). Bibliometrics more generally can undermine integrity by creating formal or informal targets for scholars (e.g., publications, citations, LLM-friendly abstracts) that may encourage them to conduct unethical behaviour, such as plagiarism, image copying and other shortcuts to bypass the time needed to generate research (Bouter, 2015).

## 17.1 Not reading the text

Perhaps the most insidious perverse incentive of bibliometrics is that they encourage evaluators to take shortcuts in evaluations by not reading the text(s) evaluated but estimating their value based on a bibliometric indicator, such as the journal impact factor, the article citation count, the ChatGPT score prediction, or the author's citation record. Whilst this is perhaps unavoidable in some contexts, such as evaluating the CVs of hundreds of job applicants, the existence of a quick time saver in the form of bibliometrics is a perverse incentive to do a poor job of evaluation. It also seems to be a perverse incentive to avoid thinking about the real purposes of academic research but to equate research quality with high citation-based indicator scores.

The San Francisco Declaration of Research Assessment (DORA, 2020) can be interpreted as an argument against the use of Journal Impact Factors as a perverse incentive to avoid reading journal articles that need to be evaluated.

## 17.2 Gaming

Bibliometric indicators have the potential to be manipulated or gamed, in the sense of altered by activities primarily designed to change the indicators rather than to improve research quality. If a journal editor attempted to increase the journal's impact factor by soliciting higher quality research, then this would be legitimate but if they encouraged authors to cite other papers in the same journal then this would be gaming because the mechanism is not one for improving research (e.g., Caon, 2017; Siler & Larivière, 2022). At the journal level, gaming can vary in legitimacy and transparency (Falagas & Alexiou, 2008). For example, a journal might publish large numbers of commentaries or editorials in the belief that they inflate JIFs by adding to the JIF numerator (i.e., increasing the journal's citations), without increasing the JIF denominator (the number of citable items, which may be restricted to standard journal articles). This might be seen as legitimate in the field if the editorials and commentaries are valued, or gaming if they are not. At the other extreme, a journal editor calling for review articles that exclusively cite articles in the journal would not be seen as legitimate by many.

Gaming can also occur for individual scholars. For example, there may be 'citation cartels' in the sense of groups of scholars or journals that agree to cite each other's work even when not very relevant (Perez et al., 2019). Excessive self-citing is another form of gaming and involves a scholar citing their own articles more than necessary for the argument in the citing paper (Szomszor et al., 2020). A future form of gaming might be crafting LLM-friendly abstracts if LLMs become used for evaluations.

There are many natural and artificial checks against gaming. A natural check is that if scholars in a field judge an activity to be gaming then the scholarly reputation of those involved will suffer. Citation indexes like Scopus and the Web of Science also have artificial checks in the form of calculations designed to detect artificial manipulation. They regularly take punitive action, such as delisting journals for one or more years both to warn about the behaviours and to remove tarnished journals from their systems.

## 17.3 Incentivising homogeneity

A generic disadvantage of using citation-based indicators in research evaluation is that, to some extent, they probably encourage some types of homogeneity whenever they have at least moderate influence. At one extreme, an evaluation exclusively based on citations to journal articles would discourage the production of the wide variety of other output types that are essential to science, such as monographs, chapters, artworks, software, and databases. At the other extreme is the UK REF, which encourages all types of research and uses article-level citation indicators as a minor supplementary source of information for peer reviewers judging articles. This level of influence is very small and seems unlikely to disincentivise submitting non-article outputs.

Incentivising homogeneity in research is undesirable because some types of research outputs are core to some fields or useful in others and so their absence would undermine science. An example of this is biodiversity data: many researchers systematically monitor animal and plant species, recording the results in shared datasets that can be used to track geographic and temporal trends (Proença et al., 2017). If this time consuming and often expensive activity is discouraged because it is not an efficient way to generate impactful journal articles, then this might lead to less comprehensive or lower quality biodiversity data. Another example is performing arts: if scholars that write articles about performances are rewarded better than those that stage them or appear in them then academia's experience of performance arts will become narrower through a loss of relevant expertise.

Incentivising publication in specific journals, whether through impact factors or ranked lists, can also incentivise homogeneity by encouraging academic to conform to the requirements or norms of the selected journals. This can be particularly difficult for junior academics and push them away from their core interests (Malsch & Tessier, 2015).

## 17.4 Focusing on publishing

A corollary to the above is that the use of bibliometrics can push an agenda in which only publishing counts, especially if researchers are given bibliometric targets for promotion and tenure. This can be a substantial perverse incentive in the Global South (Lebel & McLean, 2018; Kraemer-Mbula et al., 2020; Mills et al., 2023), but is a problem everywhere.

## 17.5 Management studies: A cautionary tale

There are some cases where bibliometric perverse incentives have contributed to substantial damage to a field. In management studies, it has been claimed that many researchers have focused on publishing in prestigious journals, using journal impact factors as a key indicator, and have prioritised publishing citable content at the expense of developing useful knowledge for the profession. This may have resulted in the perceived top journals focusing on increasingly esoteric issues with little relevance to management practice. This may even have fed through into higher education, with graduates mastering knowledge that they rarely use in their future work (Tourish, 2020). Although this claim is partly contested (Bamberger, 2020), it seems plausible and is a warning to all fields about what can happen when journal impact or prestige become proxies for field goals.

Whilst management studies seems like an extreme example, many others have claimed that their fields have been undermined by high impact factor journals being respected at the expense of useful scholarship. The replication crisis in psychology (Scheel, 2022) where a substantial proportion of research results cannot be reproduced is an example. There have also been suggestions that other fields do not engage practical problems closely enough, perhaps for the same reason (e.g., accounting: Rajgopal, 2021).

## 17.6 Summary: Considering perverse incentives

Perverse incentives are a serious problem for all evaluative uses of bibliometrics and should be considered whenever citations are proposed for use in evaluations. This is a systemic effect of evaluations in the sense that the evaluation can influence the system it is evaluating (Rushforth & Hammarfelt, 2023). If bibliometrics or LLMs are thought to only be of minor use in an evaluation, then a judgement should be made about whether their value outweighs the perverse incentives that their uptake may generate. The REF solution is slightly different but also seems valid: bibliometrics are accepted for some fields and in a limited role in support of peer review but otherwise they are banned. The explicit messaging about this to evaluators sends the strong signal that bibliometrics are thought to have no value in most fields and limited value in others. This might still generate small perverse incentives, but these are at least partially offset by messaging that counteracts any individual academic's strong beliefs in the value of citation-based indicators.

# 18 Summary

This book has discussed the extent to which citation-based indicators, altmetrics and artificial intelligence can be used to help assess the overall quality of academic research, with a primary focus on journal articles. It has introduced a range of indicators together with the practical issues for constructing and interpreting them carefully. It has also reported primary evidence of their value as research quality indicators, including as part of machine learning AI, mostly taken from a UK context. The book has also introduced Large Language Models for research quality prediction, reporting early evidence that they are generally more powerful than citation-based indicators and traditional AI. This makes LLMs a contender to take over the role of citations in future research evaluations.

To reiterate an important point made in Chapter 1 (illustrated in Figure 1.1), the value of citations and LLMs for research evaluation depends on whether the information that they contain are relevant to the overall goals of the research evaluation. The usefulness of these approaches in practice also depends on the availability and affordability of the human experts that are the primary judges of research quality.

## 18.1 Book themes: Reflexivity, disciplinary differences and appropriate formulae

Although the book primarily focuses on applying and interpreting citation-based indicators, altmetrics, and AI, it should be clear from all the chapters that it does not uncritically endorse their use. When considering whether and how to use indicators it should always be remembered that they are imperfect, can be misleading, and do not capture all the important work of scholars. For example, producing articles for international journals may not be a priority for arts and humanities, some social sciences, and Global South scholars, for different reasons. In these and other contexts, the use of any citation-based indicators may be counterproductive to research evaluation goals. Moreover, in most contexts the adoption of quantitative indicators may be demoralising to those doing valuable work that is difficult to quantify.

An important thread running through the book is that the value of citation-based indicators varies substantially between fields from almost complete irrelevance in the arts, humanities, and some social sciences to moderately strong evidence of research quality in some contexts for health, life, and physical sciences. Thus, it is likely that individual scientists have a misleading impression of the value of citations in academia based on experience within their own field. The graphs in this book showing the substantial field differences in the relationship between article-level citation indicators and research quality scores provide evidence of the overall pattern that can be used to align personal experience with that of science overall. This seems particularly important for policy makers and high-level managers who may bring their intuitions about the value of citation-based indicators from their disciplinary backgrounds and then apply it to contexts in which these intuitions are misleading. This evidence may also be useful for interdisciplinary panels of research assessors when there are disagreements about the extent to which bibliometric information should be taken into account within assessments of candidates, applicants, or bodies of research.

Another key issue is the use of appropriate formulae to convert citation counts into indicators. In this book I have argued for the MNLCS as a field-independent, time-independent, skew corrected indicator for the average citation rate of a set of articles. This is still a minority position, with simpler formulae being more common and other experienced scientometricians preferring the MNCS or percentiles for their greater simplicity. I think that all three approaches are reasonable but prefer the greater precision of the MNLCS formula.

A minor point in the book but personal hobby horse is that I avoid calling citation-based indicators "metrics", preferring "indicators". This is because calling something a metric suggests a reasonable degree of accuracy, but the same is not true for an indicator. This is partly a perspective issue: a citation-based formula might be an accurate measure of the citation rate of a group of articles but in all fields would be a weaker and more biased approximation to their average quality or impact and is therefore only an indicator of quality or impact. Nevertheless, avoiding the term "metric" reduces the possibility that anyone is misled into believing that citation-based indicators ever accurately measure research quality.

## 18.2 Responsible use of bibliometric indicators and LLM scores

Despite the above misgivings, the overall message of this book is that citation-based indicators can reasonably be used in most areas of scholarship outside the arts and humanities and perhaps the Global South as research quality indicators if they are (a) employed in a way that the perverse incentives generated do not outweigh the advantages of using them, and (b) they are used "humbly" in the sense of being mindful of their limitations. In most cases, this means using them to support human judgements rather than to replace them, in line with the Leiden Manifesto (Hicks et al., 2015), Metric Tide (Wilsdon et al., 2015), and EC (CoARA, 2022). The exceptions, where human judgement cannot be the primary evidence, are likely to be large scale policy-informing or theoretical bibliometric analyses where the value of the results is not high enough to justify the extensive manual labour needed for sufficient peer review. In these cases, analysts need to be explicit about the limitations of the bibliometric indicators in their reports.

The altmetrics chapter of the book shows that some altmetrics can have value as early quantitative indicators of article quality in some fields, although it is not clear that this is ever due to them being indicators of non-scholarly impacts. They are strongest in health fields, and the physical sciences and weakest in the arts and humanities. Of all the altmetrics, Mendeley readers have the most value as (early) quality indicators, but Tweet/X counts may also be useful as early attention indicators. Because of their potential for manipulation, altmetrics are not useful for research evaluations when those evaluated are aware of this in advance. Thus, their main value seems to be for early formative assessments of the research quality or future impact of journal articles.

In contrast to altmetrics, traditional machine learning AI (excluding LLMs) seems to be reasonably accurate at estimating the average research quality of articles authored by departments in the physical, life and health sciences, although it is inaccurate for individual articles. The use of machine learning for research quality prediction goes beyond citation-based indicators, which primarily reflect scholarly impact. The main practical limitation of traditional machine learning for quality prediction is that it cannot easily completely replace human assessment because a substantial number of articles need to be assessed to serve as the training data for any system. Moreover, in the absence of concrete information about inter-judge agreement rates, it seems likely that academics will be resistant to the use of machine learning for important research assessments based on score changes due to some inaccurate predictions. Nevertheless, there seems to be scope for using AI to support peer review by providing experts with predicted scores and confidence levels, if the inputs into the system, including journal impact, are not considered to provide significant perverse incentives.

Whilst LLMs seem to have a greater ability to estimate article quality than citation-based indicators and traditional AI approaches, there is still (early 2025) little information about the nature of their biases and limitations. Thus, extreme caution is advised when they are used until the situation is clearer. Their strongest current value might be for research that is too recent to have attracted enough citations to assess. They seem to be preferable to altmetrics for this task, since altmetrics have many known substantial biases and weaker correlations with expert judgement.

Finally, it is vital that bibliometricians and those using citation-based indicators, altmetrics or AI for research evaluation only apply them when appropriate and are always clear about their imperfections. This means understanding these limitations in the context of a research assessment's goals. In almost all circumstances there will be important research goals that do not influence citations and altmetrics as well as important research goals that citations and altmetrics do not reflect well. In almost all circumstances there will be field differences in the value of indicators that also affect the fairness of the results. Human judgements must therefore be used instead of bibliometrics as the primary or sole evidence in most cases. At the same time, expert decisions and interpretations can also be biased, inaccurate, and disputed so there is no perfect solution for the task of research evaluation. Whilst it is an important goal for future investigations to gain more insights into the limitations of expert review for research evaluation, this should not prevent users of bibliometrics from being honest and open about their limitations.

## 18.3 The future for citation analysis and LLMs

Despite the many valid criticisms of citation analysis and its applications in various contexts, it seems likely to continue for the foreseeable future because in some cases it is better than peer review. When there is a lack of expertise or time to evaluate research outputs, then citation-based formulae may be appropriate. Policy makers may even consider widely disparaged formulae, such as versions of the Journal Impact Factor, to be useful as part of national strategies to encourage academics to publish in competitive journals to improve overall quality or to benchmark quality against other similar countries. In some contexts, the advantages of citation-based incentives may be thought to outweigh their disadvantages.

The Mertonian concept of citations as acknowledging relevant prior work seems to be intuitive for working scientists and may underpin the value that many give to citation-based formulae. Even though there are many flaws with this theory, and individual scientists may well notice highly cited poor quality work and little cited excellent work, the incremental accumulation of citations, at least some of which acknowledge the contributions of prior work, mirrors the accumulation of knowledge in the more hierarchical areas of science (e.g., medicine, life and physical sciences) and may therefore also seem like a scientific approach to assessing science, especially in comparison to qualitative peer review. These factors seem likely to see the informal use of citation-based indicators continue in science for the foreseeable future.

Despite the evidence presented for the superiority of LLMs over citations as research quality indicators in some contexts, it seems unlikely that they will rapidly supplant them. Research evaluation needs to be conservative because it supports important decisions, because more evidence about biases and practical uses is needed, and because they lack the intuitive connection to research value that citations have. It seems likely that LLM use will start on a small scale and increase slowly as evidence accumulates about their value and the factors that influence scores. Further in the future, if new generations of LLMs perform substantially better on full texts than on titles and abstracts then this would also give a boost to their potential by suggesting a capability to analyse the content of articles. This would be especially valuable if open access easily machine-readable full texts become the norm for science and copyright conditions allowed their processing for research evaluation purposes. The combination of these three, if they all occur, may then see LLMs become the default choice for supporting research evaluation.